\documentclass[12pt]{report}
\usepackage{glddod,graphicx,time}
\usepackage{amssymb}

\begin{document}
\title{ { \Huge GLD Detector Outline Document }
% \footnote{The version released at LCWS2006 will be found at \newline
% http://ilcphys.kek.jp/gld/documents/dod/glddod.pdf}
\vglue 12pt
Version 1.2% 
\vglue 12pt
\date{July 13, 2006}
% {\it Draft }
}
\author{ {\Large GLD Concept Study Group } }

\maketitle

%%%%%%%%%%%%%%%%%%%%%%%%%%%%%%%%%
\newpage
%
%  Author List
%

%\begin{Large}
%List of Authors\\
%\end{Large}
%\begin{center}
%\fbox{
%\parbox{16cm}{
%{\it Notes to potential authors}\\
%Those in the lcdds mailing list, lcdds@ilcphys.kek.jp, 
%are the author of this document, 
%unless notified explicitly to editors.
%We'd like to announce this statement to the authors 
%by the end of February.
%\par
%Attached below is a tentative author list, based on the 
%lcdds@ilcphys.kek.jp maililng list.  
%For addition to or removal from the author list, please send a request by e-mail to 
%akiya.miyamoto@kek.jp or other editors.
%}
%}
%\end{center}
%\vglue 12pt

\begin{center}
\noindent
Koh~Abe$^{78}$,
Koya~Abe$^{60}$,
Toshinori~Abe$^{78}$,
Shinichiro~Ando$^{59}$,
Laci~~Andricek$^{30}$,
Kazuaki~Anraku$^{78}$,
Dennis~C.~Arogancia$^{33}$,
Eri~Asakawa$^{43}$,
Yuzo~Asano$^{80}$,
Yoichi~Asaoka$^{78}$
% ...

\vglue 12pt
(GLD Concept Study Group)
\end{center}
\vglue 12pt
\footnotesize
\noindent
Postal address to contact:\\
% ...
$^{30}$ Max-Planck-Institut fuer Physik, Fohringer Ring 6, D-80805, Munchen, Germany \\
% ...
$^{33}$ Mindanao State University, Iligan Institute of Technology, 9200 Iligan city, Philippines \\
% ...
$^{43}$ Ochanomizu University, 1 Ohtsuka 2-1, Bunkyo-ku, Tokyo 112-8610, Japan  \\
% ...
$^{59}$ Tohoku University, Aoba, Aramaki, Aoba-ku, Sendai 980-8578, Japan \\
$^{60}$ Tohokugakuin University, 1-13-1, Chuo, Tagajo, Migagi 985-8537, Japan \\
% ...
$^{78}$ University of Tokyo, 7-3-1 Hongo, Bunkyo-ku, Tokyo 113-0033, Japan \\
% ...
$^{80}$ University of Tsukuba, Institute of Applied Physics, Tsukuba, Ibaraki 305-8571, Japan \\
% ...
\normalsize

%%%%%%%%%%%%%%%%%%%%%%%%%%%%%%%%%
\newpage
%
%  Preface
%

\begin{center}
\begin{Large}
Preface
\end{Large}
\end{center}

\vglue 12pt

This report describes the GLD detector outline for International Linear Collider.
The study was initiated by the call for Detector Outline Documents
by World Wide Study of Physics and Detectors for Future Linear Colliders in 2004,
following the international consensus for joint efforts to realize International
Linear Collider (ILC).

\vglue 12pt

ILC provides collision between electron and positron at the energy
scale where the origin of masses and the true nature of vacuum
are expected to be uncovered and
new particles relevant to cosmology may also be discovered.
With their  initial states both in energy and helicity well
defined and background processes
low in general, it provides unique opportunities to discover
tiny signals of new particles and unveil the underlining physics.

\vglue 12pt

Progresses of high energy physics have established a modern view of
ultra-microscopic world; three generations of elementary
fermions belonging to the group of $SU(3)\times SU(2) \times U(1)$ with
their forces being governed by the gauge principle.
Particle masses are considered to be generated by
the spontaneous symmetry breaking of $SU(2)\times U(1)$ symmetry which is
cause by the yet-to-be-found Higgs boson.
According to current understanding, it is expected that the Higgs
boson will be discovered at LHC. And if found,
precise knowledge of its properties such as mass and couplings to particles
are crucial for our understanding of vacuum and mechanism of mass generation.
The presence of the Higgs boson, however,  poses a  new problem known as the hierarchy problem.
It  indicates that there may exist new physics at TeV scale where SUSY is one candidate thereof..
Also, the standard model does not provide candidates for the dark matter
which is thought to account for about one quarter of the mass of universe.
Recent studies suggest  that the dark matter particle based on the SUSY scenario may be found
in the ILC energy region.
The goal of the GLD experiment
is to carry out these measurements at precisions only possible at ILC.

\vglue 12pt

In order to perform this physics program, the detector should
have unprecedented precision in measurements of jets and charged particles,
efficient quark identification capability, and should have a good hermetic
coverage of the interaction point.
Through detector studies for linear collider in Asia, Europe,
and North America,  we have developed a detector
concept that consists of a large calorimeter and a gaseous central tracker placed in a
moderate magnetic field, both electro-magnetic and hadron
calorimeters being placed inside the magnetic coil
to have enough hermeticity and good jet energy resolution.
Details of the detector concept as well as expected performances are
described in the following chapters.

\vglue 12pt

The study of GLD concept was kicked off at the time of 7th ACFA workshop, November 2004.
The group is formed as an inter-regional team
lead by contact persons, two from each region; Hitoshi Yamamoto,
Hwanbae Park from Asia, Ron Settles, Mark Thomson from Europe, Mike Ronan from Europe and Graham Wilson from North America.
Since the kickoff, the concept has been developed and benchmarked 
through e-mail communications, at TV meetings and a series of meetings
held at workshops such as 8th ACFA workshop, Snowmass2005, ECFA workshop at Vienna.
Those who joined the GLD mailing list are listed as authors of this document.
Optimizations of detector parameters and studies of detector technologies are still 
in progress and this document is to summarize the current status of our study.
New participation to our activity is highly welcomed.
The home page of the group is available at http://ilcphys.kek.jp/gld.

%%%%%%%%%%%%%%%%%%%%%%%%%%%%%%%%%
\newpage
\tableofcontents

%%%%%%%%%%%%%%%%%%%%%%%%%%%%%%%%%
\clearpage
\pagestyle{headings}
\pagenumbering{arabic}
\setcounter{page}{1}
%%%%%%%%%%%%%%%%%%%%%%%%%%%%%%%%%
\chapter{Description of the Concept}
\label{PART_concept}
%
% GLD Concept
%
\section{GLD Concept}
\label{SectionConcept}

\subsection{Introduction}
\label{SubsecConIntro}
The physics gaol of the International Linear Collider (ILC)
project ranges over a wide variety of processes in
a wide energy region of $\sqrt{s}$ 
from $M_Z$ to 1~TeV~\cite{Con_ACFA, Con_TESLA, Con_SM01}.
In experiments at the ILC, it is essential to reconstruct
events at fundamental particle (leptons, quarks, and gauge
bosons) level. Most of interesting physics processes include
gauge bosons ($W$ or $Z$), heavy flavor quarks ($b$ and $c$),
and/or leptons ($e, \mu, \tau$) as direct products
of $e^+ e^-$ collisions or
as decay daughters of heavy particles (SUSY particles,
Higgs boson, top quark, etc.). The detectors at the ILC
have to have capability of efficient identification
and precise measurement of four-momenta of these fundamental
particles. 
In order to satisfy these requirements, the detector
must have the following performances:
\begin{itemize}
\item good jet-energy resolution to separate W and Z
in their hadronic decay mode,
\item efficient jet-flavor identification capability,
\item excellent charged-particle momentum resolution, and
\item hermetic  coverage which gives high veto
efficiency against \mbox{2-photon} background.
\end{itemize}

General purpose detectors for ILC experiments
will be composed of a vertex detector, a central 
tracker, an intermidiate tracker (if necessary), 
a calorimeter system, a solenoid coil, an iron
flux return yoke interleaved  with muon detector,
and forward (small angle) calorimeters.
The world-wide concensus of the performance goal 
for the detector system corresponding to the items
listed above  are~\cite{Con_WWS};
\begin{itemize}
\item jet energy resolution of 
$\delta E_j/E_j = 30\%/\sqrt{E_j \ \rm{(GeV)}}$,
\item impact parameter resolution of 
$\delta_b \leq 5\oplus \frac{10}{p\beta \sin^{3/2}{\theta}} \ (\mu$m)
for jet flavor tagging,
\item transvertse momentum resolution of 
$\delta p_t/p_t^2 \leq 5 \times 10^{-5} (\rm{GeV/c})^{-1}$
for charged tracks at high momentum limit, and
\item hermeticity down to 5~mrad from the beam line.
\end{itemize}
In order to achieve these performances, we propose a
large detector model based on a large gaseous tracker,
named ``GLD''.

\subsection{Basic Design Concept of GLD}
\label{SubsecConBasic}
The basic design of  GLD has a calorimeter
with fine segmentation and large inner radius to
optimize it for ``Particle Flow Algorithm (PFA)''.
Charged tracks are measured by a large gaseous
tracker, presumably a Time Projection Chamber (TPC),
with excellent momentum resolution.
The TPC also has good pattern 
recognition capability which is
advantageous for efficient reconstruction of
$V^0$ particles such as $K^0$, $\Lambda$, and
new unknown long-lived particles, and for
efficient matching between tracks measured by the  TPC
and hit clusters in the calorimeter.
The solenoid magnet is located outside of the
calorimeter. Because the detector volume is huge,
a moderate magnetic field of 3 Tesla has been chosen.

Jet energy resolution is one of the most important
issues for ILC detectors. Precise mass reconstruction and
separation of $W$ and $Z$ in their hadronic decay mode
are essential in many physics channels.
The PFA  (Particle Flow Algorithm)
is a method to get the best jet-energy resolution.
In this method,
each particle in a jet is measured separately;
charged particles by the tracker, photons by
the EM (electromagnetic) calorimeter ECAL, 
and neutral hadrons by the hadron calorimeter HCAL.
The ultimate PFA performance can be
achieved by complete separation of charged-particle
hit clusters from neutral hit clusters in the
calorimeter. Actual jet energy resolution
is dominated by a contribution from
confusion between charged and neutral clusters.
Optimization of algorithm and calorimeter
design for PFA is necessary to get
better jet energy resolution.
In all three detector concepts (SiD, LDC, and GLD),
optimization for PFA is the major concern.

\begin{figure}[h]
\begin{center}
\includegraphics[width=9cm]{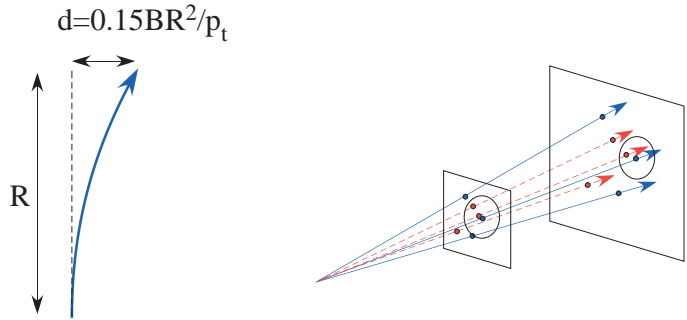}
\end{center}
\caption{Cluster separation proportional to $BR^2$ (left), or
proportional to $R^2$ (right).}
\label{Con_clssep}
\end{figure}

In order to avoid the confusion and to get
good jet energy resolution, separation of
particles in the calorimeter is important.
Therefore, the calorimeter should have a
small effective Moliere length,
fine segmentation, and a large distance
from the interaction point. Stronger
solenoid field is preferable  to
spread out the charged particles more.
The figure of merit which is often quoted
for the cluster separation
in ECAL is expressed as
$BR_{\rm{in}}^2/R_M^{\rm{eff}}$,
where
$B$ is the solenoid field, $R_{\rm{in}}$ is the
inner radius of the barrel ECAL and
$R_M^{\rm{eff}}$ is effective Moliere length
of the ECAL. However the things are not so
simple. Even with $B=0$, photon energy
inside a certain distance from a charged track
in the ECAL scales as $\sim R_{\rm{in}}^{-2}$
(see Figure~\ref{Con_clssep}).
In any case, larger inner radius of the calorimeter
is favorable for achieving good PFA performance.
     
The outer radius of the main tracker (TPC) is also
large in GLD. Consequently the lever arm of the
tracking is long and the number of sampling can be large.
Therefore, we can expect  an excellent momentum
resolution for the charged particles
($\delta p_t / p_t^2 \propto 1/BL^2\sqrt{n_{\rm{sample}}}$),
and good particle identification ($\pi /K/p$)
capability by $dE/dx$.
The relatively low magnetic field of GLD is advantageous
for the track reconstruction of low $p_t$ charged particles,
and subsequently for vertex-charge determination, PFA, 
and so on.

\subsection{Baseline Design}
\begin{figure}
\begin{center}
\includegraphics[width=16cm]{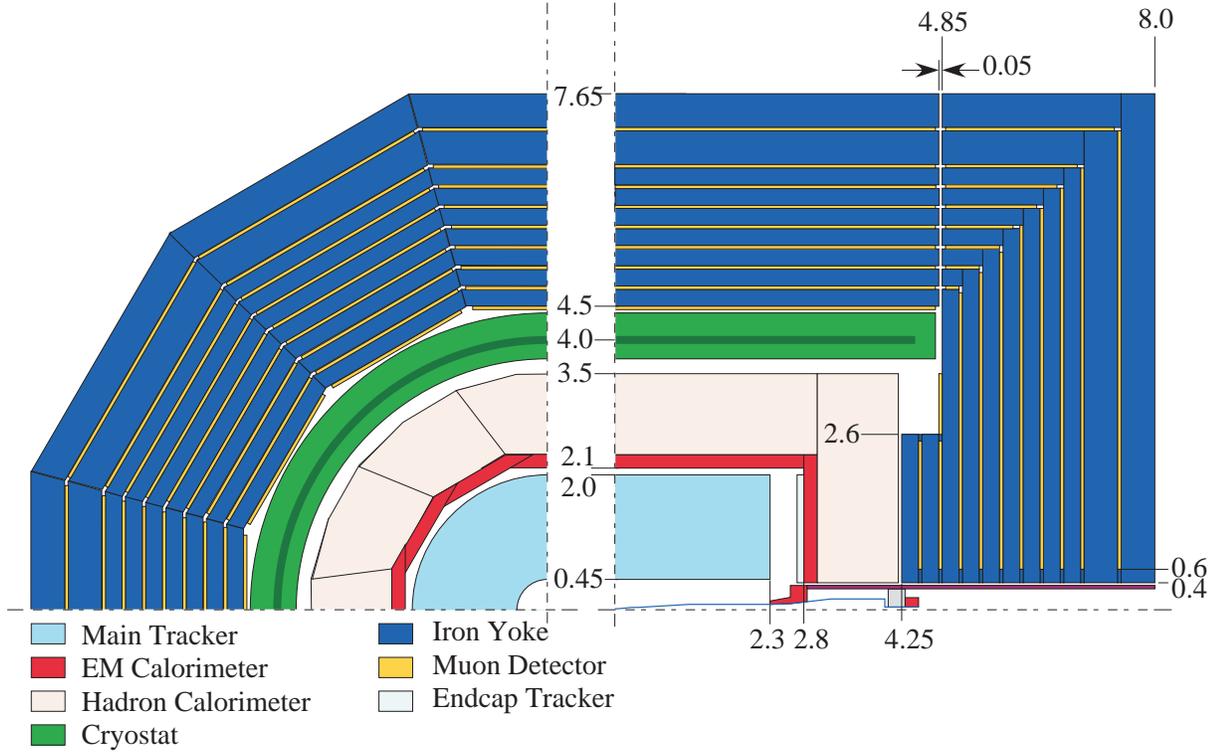}
\end{center}
\caption{\label{Con_GLD}Schematic view of two different quadrants 
of GLD Detector. The left figure shows the $r\phi$ view 
and the right shows the $rz$ view. Dimensions are in meter. The vertex detector
and the silicon inner tracker are not shown here.}
\end{figure}

\begin{figure}
\begin{center}
\includegraphics[width=10cm]{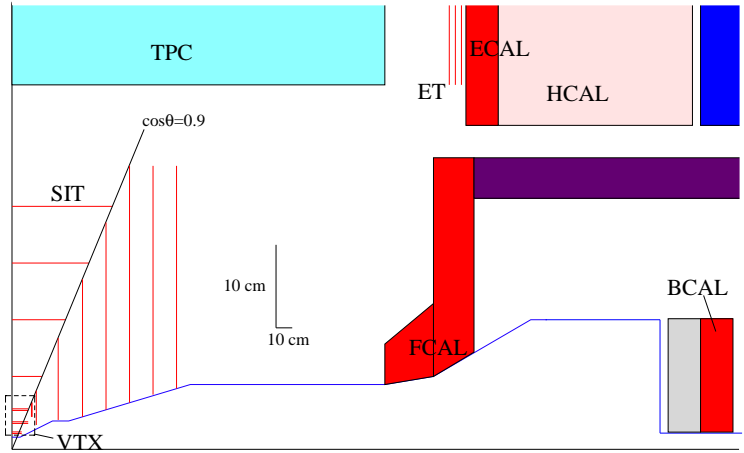}
\end{center}
\caption{\label{Con_GLDIR}Schematic view of the inner and forward
detectors of GLD. The horizontal scale and the vertical scale is 
not same as indicated in the midle of the figure.}
\end{figure}

Figure~\ref{Con_GLD} shows a schematic
view of two different quadrants of the baseline design of GLD as of
March 2006. 
The inner and forward detectors are schematically
shown in Figure~\ref{Con_GLDIR}.
The baseline design has the following sub-detectors:
\begin{itemize}
\item a large gaseous central tracker, presumably TPC,
\item a large-radius medium/high-granularity ECAL with
tungsten-scintillator sandwich structure,
\item a large-radius thick ($\sim 6\lambda$)
medium/high-granularity HCAL with
lead-scintillator sandwich structure,
\item forward EM calorimeters (FCAL and BCAL) down to 5~mrad,
\item a precision silicon micro-vertex detector,
\item silicon inner (SIT) and endcap(ET) trackers,
\item a beam profile monitor in front of BCAL,
\item a muon detector interleaved with iron plates of the return yoke, and
\item a moderate magnetic field of 3~T.
\end{itemize}
The iron return yoke and barrel calorimeters have 
dodecagonal shape (24-sided shape for the outside
of HCAL) rather than octagonal shape
in order to reduce unnecessary gaps between
the muon system and the solenoid, between 
HCAL and the solenoid, and between TPC and ECAL.

In addition to the baseline configuration, the following 
options are being considered.
Silicon tracker between TPC and
EM calorimeter in the barrel  
region is proposed to improve the momentum
resolution still more.
It is also suggested that a TOF counter in front of
the EM calorimeter can improve the particle identification
capability, but this function could be included
in the EM calorimeter.

MDI (Machine Detector Interface) issues, 
as well as the physics requirements,
give impact on the detector design.
Beam background has to be taken into account 
for the design of ILC detectors. The beam pipe
radius and inner radius of the vertex detector 
of GLD have been determined based on the 
consideration of pair background 
(see Section~\ref{SectionVTX}).
The configuration of FCAL and BCAL of GLD has been
chosen so that the back-scattered photons 
produced by the dense core of pair background at BCAL
do not hit the TPC drift volume directly.

There are three options for the beam crossing angle;
2~mrad, 14~mrad, and 20~mrad.
In case of 20~mrad crossing angle, a dipole
magnetic field could be implemented inside the
detector in order to cancel the transverse
field component of the solenoid magnet
for the incoming beam
and make the electron and positron beams
collide vertically head-on.
This dipole field could be produced by
a so-called
detector-integrated dipole (DID).
Because DID doubles the transeverse field component
for the outgoing beam, it could cause a background
problem in SIT  due to backscattering from BCAL.
In case of 14~mrad crossing angle, use of anti-DID
which has reverse polarity of the DID is considered
to make the beams collide vertically head-on.
In this case, SIT is free from the backscattering 
problem.

Another parameter which affects the detector design
is $L^*$, the distance between the interaction point and 
the front surface of the final quadrupole magnet.
We assume a large $L^*$ for GLD 
($L^*=4.6$~m).

\begin{figure}
\begin{center}
\includegraphics[width=10cm]{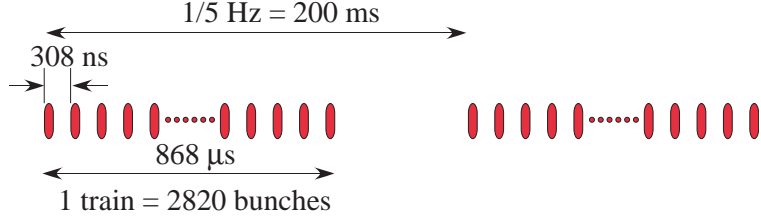}
\end{center}
\caption{\label{Con_beam}Time structure of the ILC beam.}
\end{figure}

Time structure of the ILC beam gives impact
on the requirment for the performances of 
some sub-detectors.
Figure~\ref{Con_beam} shows the time structure of 
the ILC beam. In the nominal option of the ILC
accelerator design, 2820 bunches of electrons
and positrons make a train with 307.7~ns time
intervals between bunches, and the trains are
repeated at a rate of 5~Hz.
In order to untangle the event overlap
in one train, bunch-identification (time stamping)
capability is  necessary for the silicon trackers
and calorimeters.

The parameters of the detector
and the sub-detector technologies of GLD are modified
from time to time based on  considerations on the
detector performance,  development of more realistic
detector design, considerations on 
cost issue, requirements from accelerator side, 
and so on. It is unrealistic to do the simulation 
study again and again for every modifications.
Therefore the detector model implemented in the
detector simulator does not necessarily 
reflect all the modification of the baseline
design. In this report, the detector model
assumed in Chapter~\ref{PART_performance} (Sections 
on the performance) is based on an older version than
what is described in Chapter~\ref{PART_concept}
and Chapter~\ref{PART_subsystem}. 
The cost estimation will be done
based on the baseline design described in this section.

\subsection{Overview of Sub-detectors}
In this subsection, we will describe each sub-detector
very briefly. The detail of the detector sub-system
will be described in Chapter~\ref{PART_subsystem}.
The parameters of the sub-detectors are listed in
Table~\ref{Con_trkparam} for trackers and in 
Table~\ref{Con_calparam} for calorimeters.
The parameters listed in this table are based on 
the latest baseline design and not necessarily
consistent with the parameters given in 
Part~\ref{PART_performance} (Table~\ref{PP_JUPITERGeom}).

\begin{table}
\begin{center}
\caption{Parameters of trackers of GLD. The spatial resolutions
are values to be implemented in the simulator and not necessarily
segmentation size devided by $\sqrt{12}$. The thickness $t$ includes
the support structure converted into silicon-equivalent thickness.
The configuration of first two layers of VTX depends on the 
machine parameter option (see Section~\ref{SectionVTX}).}
\begin{tabular}{|l|c|c|c|c|l|}
\hline
    &  $R$ (cm) & $Z$ (cm) & $\cos{\theta}$ & $t$ ($\mu$m) & Resolution  \\
\hline
\hline
     & 2.0 & 6.5 & 0.9558 &  & \\
     & 2.2 & 6.5 & 0.9472 &  & \\
     & 3.2 & 10.0 & 0.9524 &  &\\
VTX  & 3.4 & 10.0 & 0.9468 &  80 &5~$\mu\rm{m}^2$  pixel \\
     & 4.8 & 10.0 & 0.9015  &    & $\sigma = 2\ \mu$m   \\
     & 5.0 & 10.0 & 0.8944  &  & \\
     & 4.0--5.8 & 12.0 & 0.9003--0.9487 &  &\\
     & 4.0--5.8 & 12.2 & 0.9031--0.9502 &  &\\
\hline
     & 9.0 & 18.5  & 0.8992  &      & R-$\phi$: 50~$\mu$m strip pitch, \\
BIT  & 16.0 & 33.0 & 0.8998  & 560  & $\sigma = 10\ \mu$m \\
     & 23.0 & 47.5 & 0.9000  &      & $Z$: 100~$\mu$m strip pitch, \\
     & 30.0 & 62.0 & 0.9002  &      & $\sigma = 50\ \mu$m \\
\hline
     & 2.4--7.6 & 15.5 & 0.8979--0.9882 &      & \\
     & 3.2--14.0& 29.0 & 0.9006--0.9940 &      & \\
     & 3.7--21.0& 43.5 & 0.9006--0.9964 &      & \\
FIT  & 4.7--28.0& 58.0 & 0.9006--0.9967 & 560  & $\sigma = 25\ \mu$m  \\
     & 5.7--38.0& 72.5 & 0.8857--0.9969 &      & \\
     & 6.6--38.0& 87.0 & 0.9164--0.9971 &      & \\
     & 7.6--38.0& 101.5 & 0.9365--0.9972 &     & \\
\hline
     & & 270.0 & 0.7964--0.9864 & & \\
ET   & 45.0--205.0 & 274.0 & 0.8007--0.9868 & 560 & $\sigma = 25\ \mu$m  \\
     & & 278.0 & 0.8048--0.9872 & & \\ 
\hline
TPC  & 45.0--200.0 & 230.0 & 0.7546 (full) 
& --  & $\sigma_{r-\phi}=50$--150~$\mu$m  \\
     &               &       & 0.9814 (min)  
&   & $\sigma_z=0.5$~mm   \\
\hline  
\end{tabular}
\label{Con_trkparam}
\end{center}
\end{table}
\begin{table}
\begin{center}
\caption{Parameters of calorimeters. }
\begin{tabular}{|l|c|c|c|c|l|}
\hline
    &  $R$ (m) & $Z$ (m) & Structure & $X_0$ & $\lambda$  \\
\hline
\hline
ECAL & 2.1 -- 2.3 & 2.8 & W/Scinti./gap & 26  & 1.0  \\
     & 0.4 -- 2.3 & 2.8 -- 3.0  & 3/2/1(mm) $\times 33$ layers  &  &   \\
\hline
HCAL  & 2.3 -- 3.5 & 3.0 & Pb/Scinti./gap & 165  & 5.7 \\
     & 0.4 -- 3.5 & 3.0 -- 4.2 & 20/5/1(mm) $\times 46$ layers  &   &  \\
\hline
FCAL  & (0.08 -- 0.36) & (2.3 -- 2.85) & W/Si & & \\
\hline
BCAL   & 0.02 -- 0.36 & 4.3 -- 4.5 & W/Si or W/Diamond & & \\
\hline
\end{tabular}
\label{Con_calparam}
\end{center}
\end{table}
\subsubsection{Vertex Detector}
Very good impact parameter resolution
for charged tracks  is required
at ILC for efficient
jet-flavor identification. The target value of the
impact parameter resolution is
$$\sigma_b=5\oplus \frac{10}{p\beta \sin^{3/2}{\theta}} \ (\mu m).$$
In order to achieve this resolution, the Si pixel
vertex detector has to have excellent point resolution and
thin wafer thickness.
          
For the baseline design  of the vertex detector,
we envisage fine pixel CCDs (FPCCDs) as the sensors.
The inner radius is 20~mm and the outer radius is 50~mm.
It consists of three layers of doublets where a doublet
is made by two sensor layers with 2~mm distance.
             
In FPCCD option, pixel occupancy is expected less than 0.5\%
for the inner most layer (R=20~mm) at B=3~T for the
ILC nominal machine parameters~\cite{Con_raubenheimer}.
The hit density is, however, as high as 40/$\rm{mm}^2$.
Therefore, very thin wafer (much less than 100~$\mu$m)
is required in order to keep wrong-tracking
probability due to multiple scattering reasonably low~\cite{Con_sugimoto05b}.
The R\&D effort on the wafer thinning is
very important, as well as the fabrication
of the small pixel sensors.

\subsubsection{Silicon Trackers}
%\subsubsection*{Silicon Inner Tacker}
\noindent
\underline{Silicon Inner Tracker}

The silicon inner tracker is located between the vertex detector
and main tracker. It consists of the barrel inner tracker (BIT)
and the forward inner tracker (FIT). 

The roles of the barrel inner tracker are to
improve the linking efficiency between the main tracker
and the vertex detector, and  to reconstruct and measure
momenta of low $p_t$ charged particles. Time stamping capability
to separate bunches (307.7~ns  or 153.8~ns interval) is
necessary as well as good spatial resolution.
                                                                               
Silicon strip detectors will be used for the BIT.
Four layers of silicon strips are being considered for
stand-alone tracking capability.
The innermost and outermost layers of the BIT are located at
the radii of 9~cm and 30~cm, respectively.

%
%\subsubsection*{Silicon Forward Tracker} 
%\noindent
%\underline{Silicon Forward Tracker} 

Forward silicon tracker (FIT) should cover the angular range
down to $\sim 150$~mrad which corresponds to the
coverage of the endcap calorimeter.
The technologies used for the FIT 
depends on the track density of jets and
the background level (beam background
and 2-photon background). Detailed 
simulation study is necessary to determine the technology.
We assume silicon pixel sensors for the first three layers 
and silicon strip sensors for the other four layers.

%
%\subsubsection*{Silicon Endcap Tracker}
\noindent
\underline{Silicon Endcap Tracker}

Several layers of silicon strip detectors are placed
in the relatively large gap between the TPC
and the endcap EM calorimeter. This endcap silicon tracker (ET)
improves momentum resolution for charged particles
which have small number of  TPC hits.
Another role of the ET is to  improve
matching efficiency between TPC tracks and shower clusters
in the EM calorimeter.
This function is  important particularly
for low momentum tracks.

\subsubsection{Main Tracker}
A large gaseous tracker will be used for GLD
as the main tracker. In the baseline design,
a TPC (Time Projection Chamber) 
with 40~cm inner radius and 200~cm outer
radius is assumed. The maximum drift length
%in z-direction is 235~cm.
in z-direction is 230~cm.
                                                                              
The requirement for the  performance of the TPC in GLD
is to achieve the momentum resolution of
$\delta p_t / p_t^2 < 5\times 10^{-5}$
combined with the silicon inner tracker  and 
the vertex detector  at the high $p_t$ limit.
                                                                               
TPCs have been used in a number of large collider experiments in the past
and have performed excellently. These TPCs were read out by multi-wire
proportional chambers (MWPCs).
The thrust of  R\&D is to develop a TPC based on novel
micro-pattern gas detectors (MPGDs), which promise to have better point
and
two-track resolution than wire chambers and to be more robust
in high backgrounds than wires.
Systems under study at the moment are
Micromegas\cite{Con_mm} meshes and
GEM (Gas Electron
Multiplier)\cite{Con_gem} foils.
Both operate in a gaseous atmosphere and are
based on
the avalanche amplification of the primary produced electrons. The
gas amplification occurs in the large electric fields in
MPGD microscopic structures with sizes of the order of 50 $\mu$m.  MPGD lend
themselves naturally
to the intra-train un-gated operation at the ILC, since, when operated
properly, they
display a significant suppression of the number of back-drifting ions.
                                                                               
\subsubsection{Calorimeter}
As mentioned in Section~\ref{SubsecConBasic},
the  calorimeter  of GLD should  have
large radius and
fine 3D segmentation   in order to get excelent
jet energy resolution by PFA.
The target value of the jet energy resolution is
$$ \sigma (E_j)/E_j = 30\% / \sqrt{E_j(GeV)}. $$

%\subsubsection*{EM Calorimeter} 
\noindent
\underline{EM Calorimeter} 

The EM calorimeter (ECAL) should have small effective
Moliere length in order to suppress the shower spread
and minimize the deterioration of the jet-energy resolution
due to  confusion of $\gamma$s and charged tracks.
For this reason, tungsten will be used for the absorber
material.
                                                                               
Because the size of EMCAL is quite large
($\sim 100\ \rm{m}^2$/layer), it may not be
practical to use silicon pad as the sensor
due to cost. Therefore, the baseline
design adopts scintillator strips or tiles with
wavelength-shifter fiber readout. 
As the photon sensor, the use of MPPC (Multi-Pixel
Photon Counter)  is considered.
It consists of 30 sampling layers of
tungsten/scintillator with the
thicknesses of 3~mm/2~mm and 1~mm gap for readout.
The effective segmentation cell size is 1~cm$\times$1~cm 
with orthogonal strips.
                                                                               
%
%\subsubsection*{Hadron Calorimeter} 
\noindent
\underline{Hadron Calorimeter}
 
The hadron calorimeter (HCAL) of GLD,
as a baseline design,
consists of  46 layers
of lead/scintillator
sandwiches with 20~mm/5~mm thickness
and 1~mm gap for readout.
This configuration is thought as a
``hardware compensation'' configuration which
gives the best energy resolution for a single particle.
The effective cell size is 1~cm square to be
achieved by   1~cm ~x 20 ~cm strips
and 4 ~cm ~x 4~cm tiles.
As the photon sensor, the use of MPPC 
is considered to read scintillating
lights through a wave length
shifting fiber.
Another option of ``digital hadron calorimeter''
is also considered for HCAL so as to reduce the cost of 
read out electronics.
For the digital HCAL, the base line design
consists of scintillator strip may
have shower overlap problem. With a realistic PFA model,
we need to clarify this,
so as to determine the optimal width and length of the strips.

%
%\subsubsection{Forward Calorimeters}
\noindent
\underline{Forward Calorimeters}

The forward calorimeter of GLD consists of two parts:
FCAL and BCAL. The z-position of FCAL is close to
that of endcap ECAL, and it locates outside of
the dense core of the pair background in R direction.
BCAL is located just in front of the final
quadrupole magnet ($\sim 4.5$~m). The inner radius
of FCAL and BCAL depends on the machine parameters.
In case of small crossing angle of 2~mrad, the inner
radius of the BCAL can be as small as 20~mm
and the minimum veto angle
for the electrons of 2-photon processes is $\sim 5$~mrad.
                                                    
Since BCAL is hit by the dense core of the pair background,
it creates a lot of backscattered $e^\pm$ and photons.
A mask made by low-Z material with the same inner radius
as the BCAL should be put in front of BCAL to absorb
low energy  backscattered $e^\pm$. The z-position of FCAL
should be chosen so that FCAL works as a mask for
the backscattered photons from
BCAL and they cannot hit TPC directly.
                                                                               
Technology of FCAL and BCAL is still open question.
For FCAL, W/Si sampling calorimeter will work well.
For BCAL, more radiation hard sensors, such as
diamond, would be the option.
\subsubsection{Muon System}
The muon detector of GLD
is not required to work as a tail catcher, because
calorimeters of GLD has the thickness close to 7 interatcion length
which is thick enough to contain hadron showers. Therefore, the baseline
design of GLD has just 9 or 10 layers of muon detectors interleaved
with the iron return yoke, each layer being consists of 
two-dimensional array of
scintillator strips with wavelength-shifter fiber
readout by MPPC.
\subsubsection{Detector Magnet and Structure}
The detector magnetic field is generated by a
super-conducting solenoid with correction winding
at both ends. The radius of the coil center is
4.0~m and the length is 8.9~m. 
Additional serpentine
winding for the detector integrated dipole (DID)
might be necessary to compensate the
radial component due to finite crossing angle.                      
The integrated field uniformity at the tracker region
with this configuration satisfies
$$ \biggl| \int_{0}^{z_{max}} \frac{B_r}{B_z} dz \biggr| < 2~\rm{mm}$$
without DID. This value is
good enough for TPC.
The total size
of the iron structure has a height of 15.3~m and
a length of 16~m. The thick iron  return yoke is 
required to keep leakage field low enough. The 
requirement for the leakage field 
from the accelerator side is less than 50~Gauss
on the beamline at $z=10$~m. 

%
% DAQ subsubsection included in 2006/5/8
%
\subsubsection{Data Aquisition}
The main goal of the data acquisition (DAQ) system is
to take data of interesting events efficiently in the
presence of several orders of magnitude higher backgrounds.
Although the minimum bias event rates are expected
to be lower than the hadron colliders, the data size
will be large due to the huge readout channels to measure
physics processes with the required accuracy. The bunch
structure of the ILC operation conditions leads to the
proposal of an event building system without any hardware
trigger and of some pipelinings to achieve a dead-time
free DAQ system.

Since the DAQ system depends on the final design and also
on the rapid development of the technologies,
the system presented here is still conceptual,
showing possible options and technologies forecast.

\subsection{Optional Sub-detectors}
\subsubsection{Silicon Outer Tracker}
The performance goal of the tracking system
has long been thought as
$\delta p_t / p_t^2 = 5 \times 10^{-5}$
at high $p_t$ limit~\cite{Con_WWS}.
This value comes from a consideration of
Higgs mass measurement error in
$e^+ e^- \rightarrow Z H, \ Z \rightarrow
\mu^+ \mu^- $ that the error should be
dominated by beam energy spread and
beam strahlung. Recently, however,
reconsideration based on new beam parameters
suggests better resolution than
$5 \times 10^{-5}$ could give better
physics outputs. 
In order to get better momentum resolution,
putting a high resolution silicon tracker
outside the TPC in the barrel region is an option 
of GLD.
The performance and feasibility of this
option should be studied in case the better
momentum resolution is required.

\subsubsection{Particle Identification}
Determination of heavy quark sign (quark-antiquark tag)
plays an important role in physics study at ILC.
Measurement of angular distribution and 
left-right asymmetry in $e^+ e^- \rightarrow b \ \bar{b}$
could reviel the existence of extra dimension~\cite{Con_Hewett}.  
Differential cross setion of 
$e^+ e^- \rightarrow \tilde{\chi}^+_1 \tilde{\chi}^-_1 $
has to be measured to study the property of chargino,
and charge of charm and bottom quark in 
$ \tilde{\chi}^{\pm}_1 \rightarrow W^{\pm} \tilde{\chi}^0_1$,
$W^{\pm} \rightarrow c \  \bar{b}$ or $\bar{c} \ b$ has to be 
measured to determine the sign of the mother chargino.
Vertex charge measurement is an approach widely used
at SLD and LEP experiments. However, kaon charge 
identification would increase the efficiency significantly.

In GLD, $K/\pi$ separation can be achieved to some extent
by $dE/dx$ measurement in the TPC. Recently $dE/dx$ resolution
better than 3\% was suggested using ``digital TPC''. 
If this resolution is realized, we will have a fairly good
efficiency in $K/\pi$ separation above 2~GeV/c. 
However, there is an efficiency gap between 0.9~GeV/c and
2~GeV/c in $K/\pi$ separation by $dE/dx$. 
This gap can be filled by TOF measurement in front of
ECAL with a resolution of $\sim 100$~ps.
If efficiency loss due to this gap is non-negligible,
the first layer of  ECAL should have the TOF measurement
capability.

 % to be prepared by Y.Sugimoto
\clearpage

%%%%%%%%%%%%%%%%%%%%%%%%%%%%%%%%%
\chapter{Detector Sub-systems}
\label{PART_subsystem}

%input mdi/mdi.tex    % to be prepared by T.Tauchi

%
% Vertex detector
%
\section{Vertex Detector}
\label{SectionVTX}
\subsection{Introduction}
In order to achieve the performance goal of 
$\sigma_b = 5 \oplus 10/p\beta \sin{\theta}^{3/2}$~$\mu$m, 
the vertex detector should have very thin 
layer thickness ($<100 \ \mu$m/layer) 
and small inner radius.
Compared with other two detector
concepts, GLD has relatively low magnetic field
of 3~T (LDC and SiD  have 4~T and 5~T, respectively).
The impact of the magnetic field on the vertex detector
design is the radius of the innermost layer.
Higher magnetic field confines the pair-background
in a smaller radius, and the beam pipe and the
vertex detector can be put closer to the beam line.
So the GLD vertex detector has to have slightly
larger inner radius $R_{VTX}$ to keep background hit density
same as other detector concepts.

At ILC, 2820 bunches of electron and positron beam 
make collisions successively with 307.7~ns bunch intervals. 
This succession of 2820 bunches is called ``train'', 
and trains are repeated at a rate of 5~Hz. 
Due to low energy electron/positron beam background, 
the hit rate of the innermost layer of the vertex detector 
is estimated to be $\sim 1.5 \rm{\ hit/cm^2}$ at R=2.0~cm 
and B=3~T for one bunch crossing (BX). 
If the hits are accumulated for one train, 
the hit density becomes very high, 
and the pixel occupancy for a pixel detector 
with 25~$\mu$m pixel size exceeds 10\%, 
which is not acceptable.

One method to keep the pixel occupancy acceptable level ($\sim 0.5$\%) 
is to read out the sensors more than 20 times in one train. 
Another method is to increase the number of pixels 
by factor of $\sim 20$ useing very fine pixels. 
In this fine pixel option, the data can be read out
in 200~ms interval between trains and very fast readout
is not necessary.

\subsection{Baseline Design}
\subsubsection{Fine Pixel CCD}
As the vertex detector for the ILC experiment,
a lot of sensor technologies are proposed but
non of them seems to be demonstrated to work
satisfactorily at ILC. For the moment, 
we assume fine pixel CCD (FPCCD) option for the 
baseline design of the GLD vertex detector.
It does not mean that the standard pixel options
are rejected for GLD, of course. 

For the FPCCD vertex detector~\cite{Vtx_sugiLCWS05, Vtx_sugiSM05}, 
we use very fine 
($\sim 5$~$\mu$m) pixel CCDs. By increasing the
number of pixels by a factor of about 20 compared
with standard pixel sensors, the pixel occupancy
will be less than 1\% even if the hit signal is
accumulated during a whole train of 2820 bunches.
In order to suppress the number of hit pixels
due to diffusion in the epitaxial layer, 
the sensitive layer of the FPCCD has to be fully 
depleted. 
 
A big challenge of the FPCCD vertex detector is
the high hit density due to pair-background hits.
Although the pixel occupancy is satisfactorily low, 
the hit density is as high as 40~hits/$\rm{mm}^2$
at B=3~T and R=20~mm with the machine parameter of
the ``nominal option''~\cite{Con_raubenheimer}. 
This high hit density could cause
tracking inefficiency if the multiple scattering
effect is large. 
When a signal hit candidate on a layer is searched for
by extrapolating signal hits of outer layers,
the background hits cause misidentification
probability $p_{mis}$. For a normal incident track,
$p_{mis}$ is given by
\begin{eqnarray}
p_{mis} &=& 2\pi \sigma R_0^2, \\
R_0     &=& d\theta_0 \nonumber
\end{eqnarray}
where $\sigma$ is background hit density
of the inner layer,
$d$ is the distance between inner and outer layers,
and $\theta_0$ is the multiple scattering angle
by the outer layer. The angular and momentum
dependence of $p_{mis}$ is
$p_{mis} \propto p^{-2}\sin^{-4}{\theta}$,
where $p$ is the momentum and $\theta$ is
the polar angle.

\begin{figure*}
\centering
\includegraphics[height=8cm]{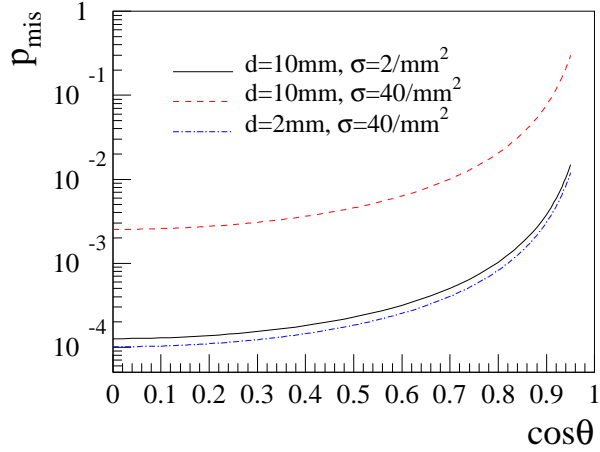}
\caption{Misidentification probability  of signal hit  with
background hit as a function of $\cos{\theta}$ for several
layer configurations. The layer thickness
of 50~$\mu$m and the particle momentum of 1~GeV/c are assumed.}
\label{Vtx_pmiss}
\end{figure*}

The misidentification probability for 1~GeV/c
particles is plotted
as a function of $\cos{\theta}$ in Figure~\ref{Vtx_pmiss}
assuming the layer thickness of 50~$\mu$m Si.
As can be seen from this figure, misidentification
probability quickly goes up in the forward region.
If the distance between inner two layers is
10~mm, $p_{mis}$ is nearly 30\% at $\cos{\theta}=0.95$
with the background hit density of $40/{\rm mm}^2$.
To reduce the misidentification probability,
the distance between inner two layers should be
small. If the distance is 2~mm,  $p_{mis}$ is
as small as the case of $d=10$~mm and $\sigma=2/{\rm mm}^2$,
which is expected when the sensor is read out 20 times per train.

\begin{figure*}
\centering
\includegraphics[height=8cm]{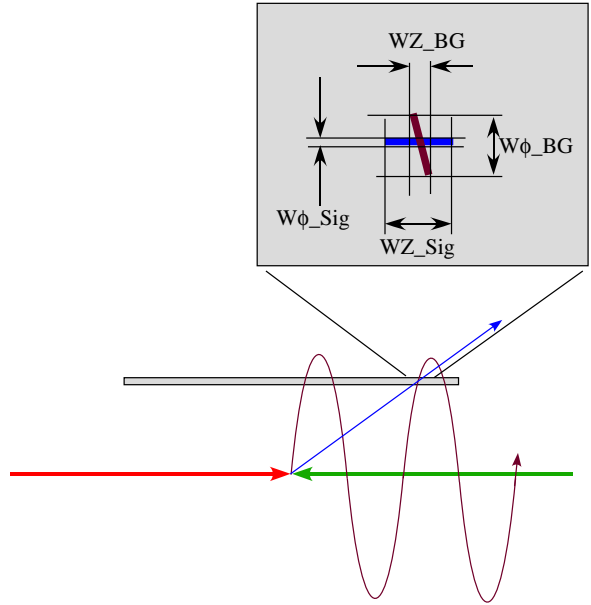}
\caption{Hit cluster shape of a high $p_t$ track (blue)
and a pair-background track(brown).}
\label{Vtx_cls}
\end{figure*}

Another way to reduce  $p_{mis}$ more is background rejection
using hit cluster shape of the FPCCD.
The momentum spectrum
of the pair-background particles hitting the
innermost layer of the vertex detector has a peak
around 20~MeV/c at 3~T magnetic field.
Therefore, the incident $\phi$ angle of
background particles to  the sensor plane is
quite different from
that of large $p_t$ signal particles.
As a consequence, the hit clusters of
background particles have larger spread
in $\phi$ direction and smaller spread
in $z$ direction than what is expected for the large $p_t$
particles as shown in Figure~\ref{Vtx_cls}.
Background rejection of about factor 20 is
expected for large $\cos{\theta}$ region 
where the misidentification probability 
becomes large~\cite{Vtx_sugiSM05}.

\subsubsection{Layer Configuration}
\begin{figure}
\centering
\includegraphics[height=5cm]{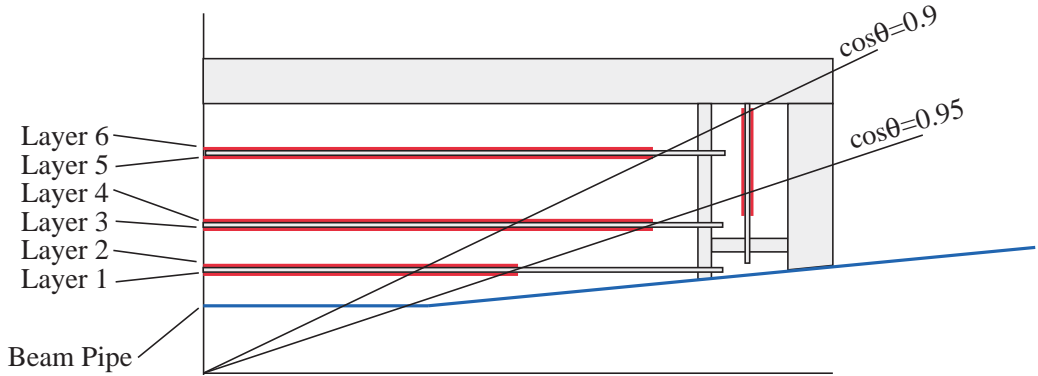}
\caption{Schematic design of the FPCCD vertex detector.}
\label{Vtx_schema}
\end{figure}

The baseline design of the vertex detector is 
schematically shown in Figure~\ref{Vtx_schema}.
Two sensor layers  put in proximity make a
doublet to reduce the misidentification
probability.
CCD wafers will be thinned down to 50~$\mu$m and 
glued on both sides of 2-mm-thick plates 
made of rigid foam such as 
reticulated vitreous carbon (RVC) foam  or 
silicon carbide (SiC) foam.
The density of  RVC can be  as small as 3\% 
of graphite, which corresponds to 0.016\%$X_0$
for 1~mm thickness.  

The angular coverage is $\cos{\theta}<0.9$ with
6 barrel layers and $\cos{\theta}<0.95$
with 4 barrel layers plus 2 forward-disk layers.
Because we need end-plates with sizable material budget
to support barrel laddes, the role of the 
forward-disk layers may be less important for low momentum
tracks.
The whole ladders are supported from outer shell 
made by berylium or CFRP, and confined in a cryostat made
by low-mass material (polystyren foam for example).
The CCDs are operated at
low temperature in order to keep dark current 
and charge transfer inefficiency due to radiation
damage reasonably low. 

\begin{figure*}
\centering
\hspace*{2.5cm}
\includegraphics[height=50mm]{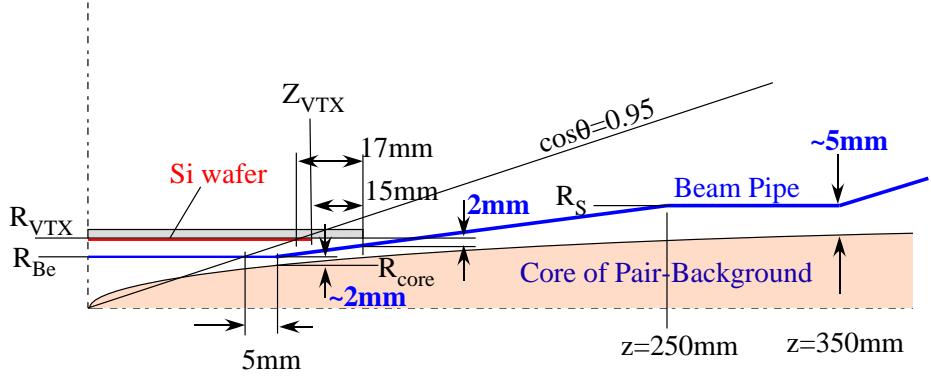}
\caption{Model for estimation of radii of the beam pipe
and innermost layer of the GLD vertex detector.}
\label{Vtx_Rmin}
\end{figure*}

The inner radius of the vertex detector is determined
by a consideration on beam background. We have estimated
the possible smallest radii of the beam pipe and
the innermost layer of the vertex detector
based on a simple model.
The model we have used is shown in Figure~\ref{Vtx_Rmin}.
The minimum radii of the beam pipe and the first layer
of the vertex detector were determined using following 
design criteria:
\begin{itemize}
 \item The dense core of the pair background should not
hit the beam pipe. It should have $\sim 5$~mm clearance
at $z=350$~mm and $\sim 2$~mm clearance at the junction
of the central beryllium part and the conical part.
\item The silicon wafer is 2~mm longer than what is
required to cover $|\cos{\theta}| <  0.95$.
\item The ladder length is longer than the silicon wafer by 15~mm.
The clearance between the ladder and the conical part of the
beam pipe is 2~mm.
\end{itemize}

\begin{figure*}
\centering
\includegraphics[height=58mm]{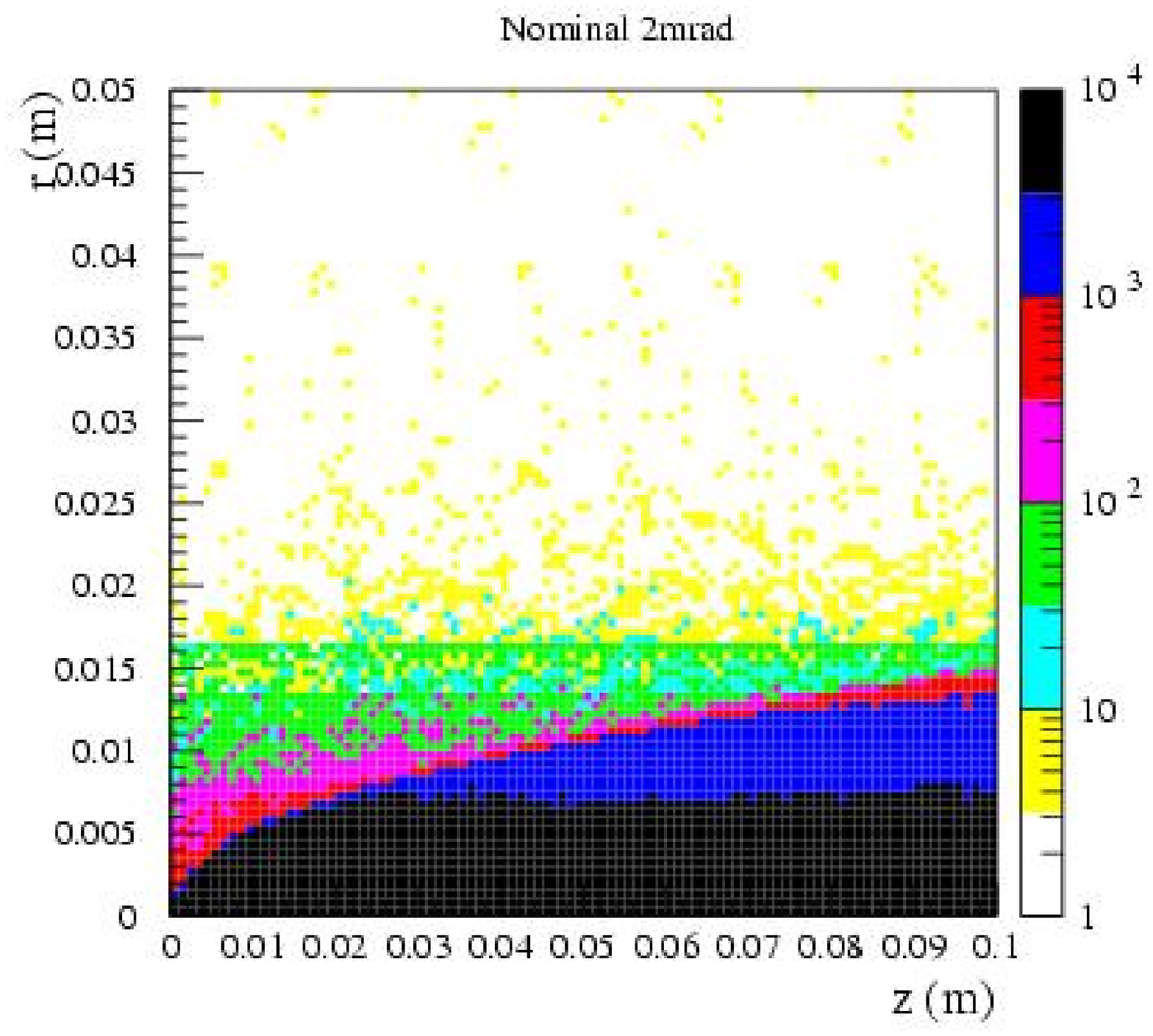}
\hspace*{-7mm}
\includegraphics[height=58mm]{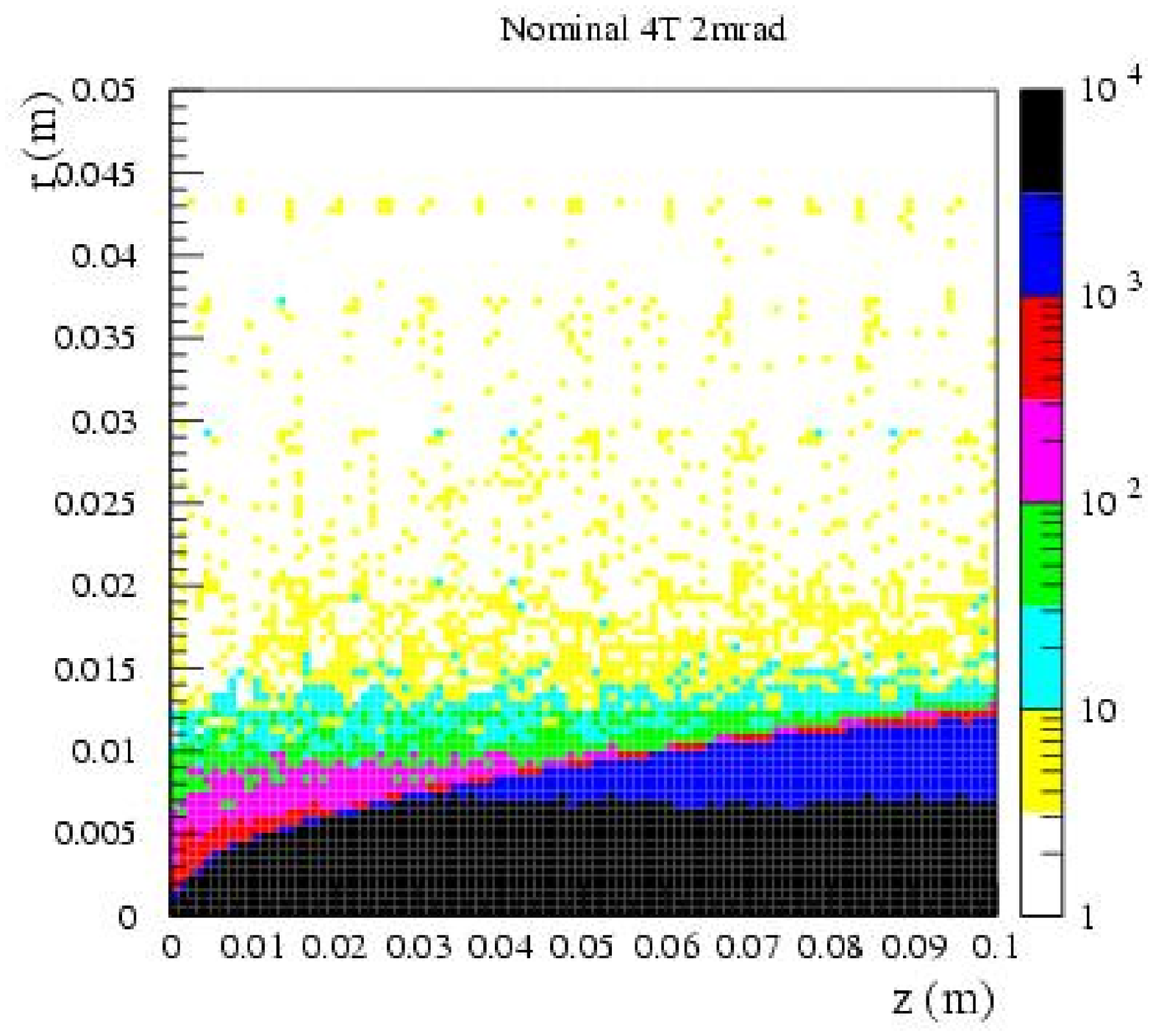}
\hspace*{-7mm}
\includegraphics[height=58mm]{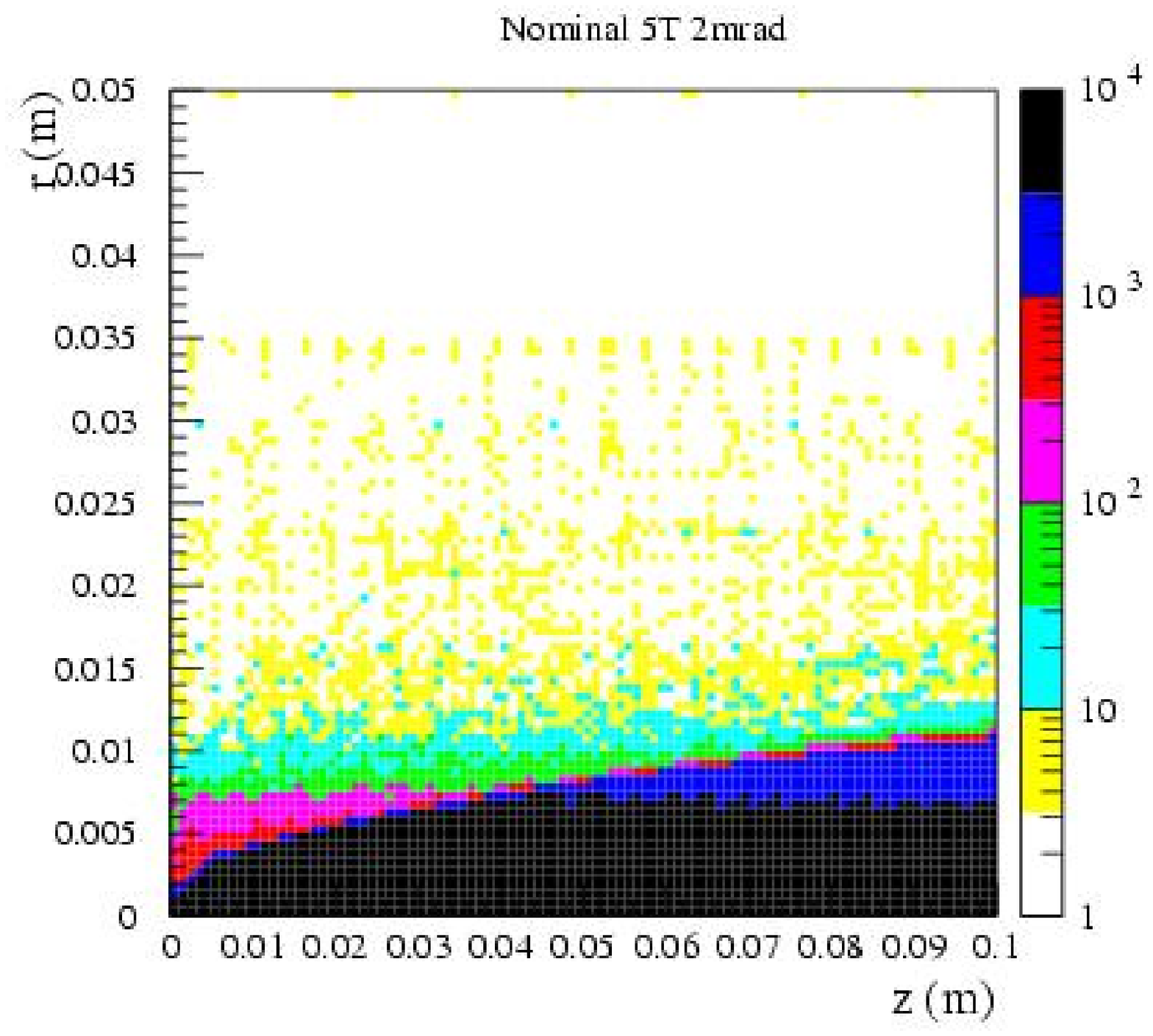}
\caption{Pair-background track density ($/\rm{cm}^2/\rm{BX}$)
with the nominal ILC machine parameter at 500~GeV
and 2~mrad crossing
angle in 3~T (left), 4~T (center), and 5~T (right)
magnetic field.}
\label{Vtx_nominal}
\end{figure*}

\begin{figure*}
\centering
\includegraphics[height=58mm]{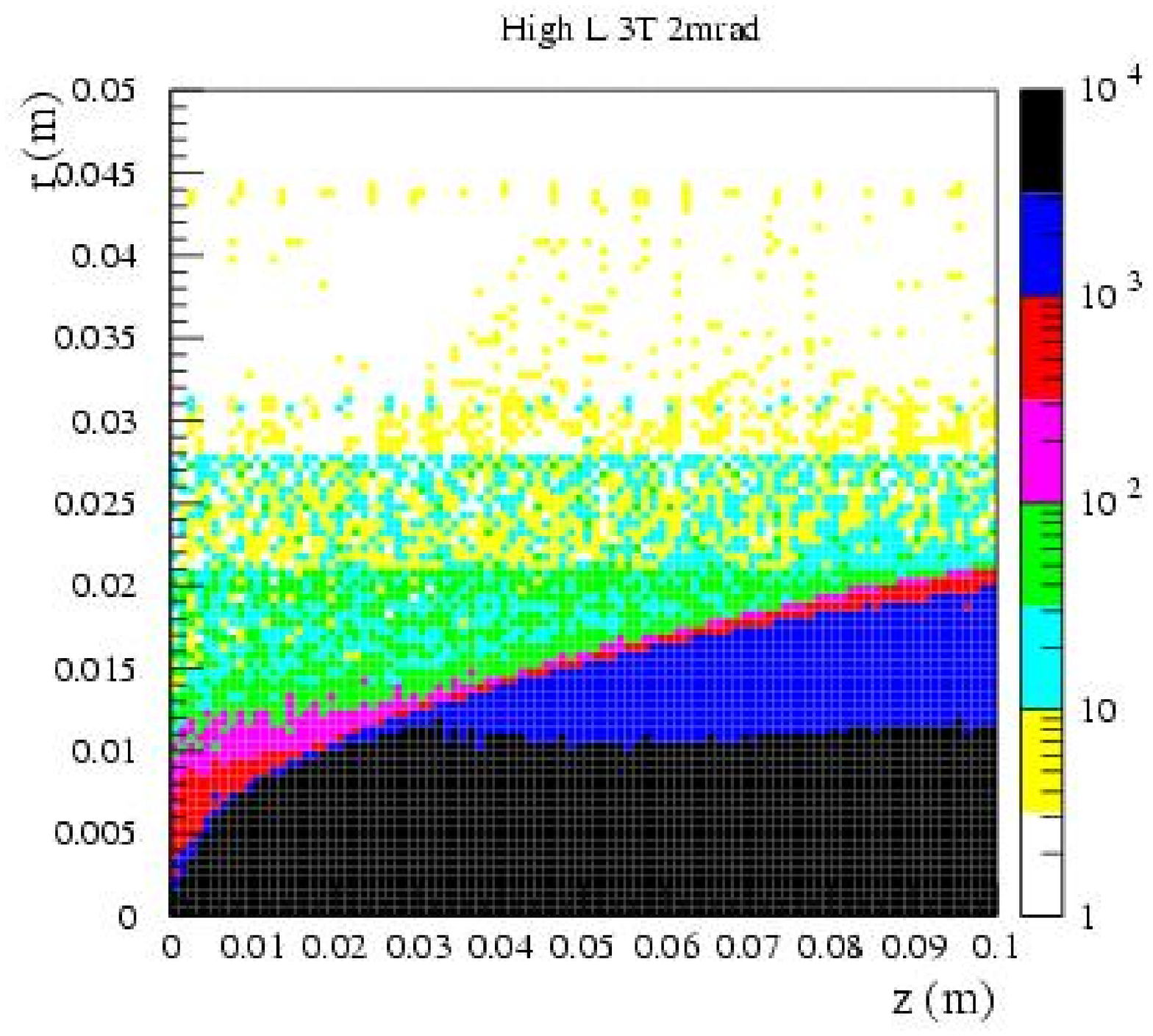}
\hspace*{-7mm}
\includegraphics[height=58mm]{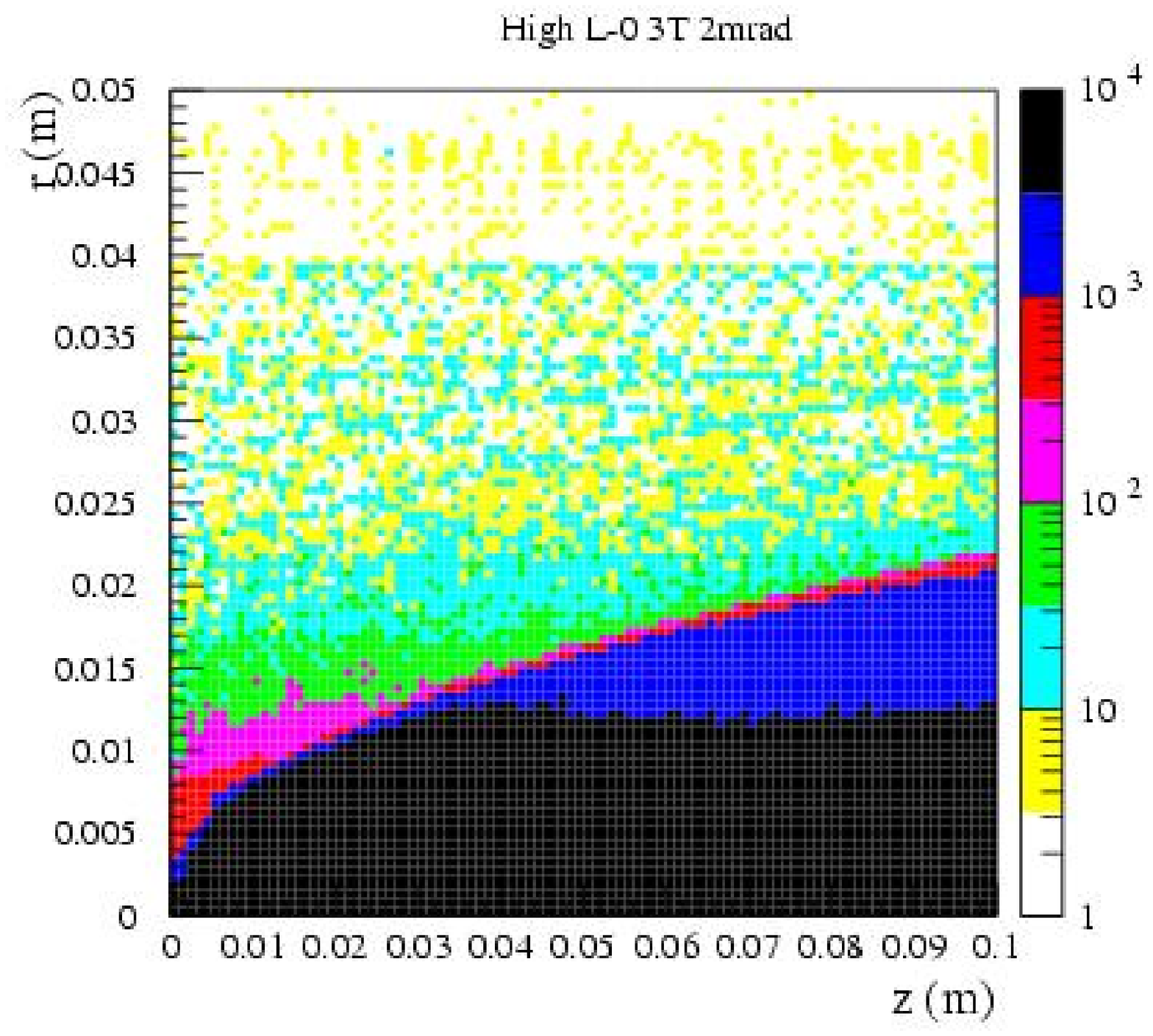}
\hspace*{-7mm}
\includegraphics[height=58mm]{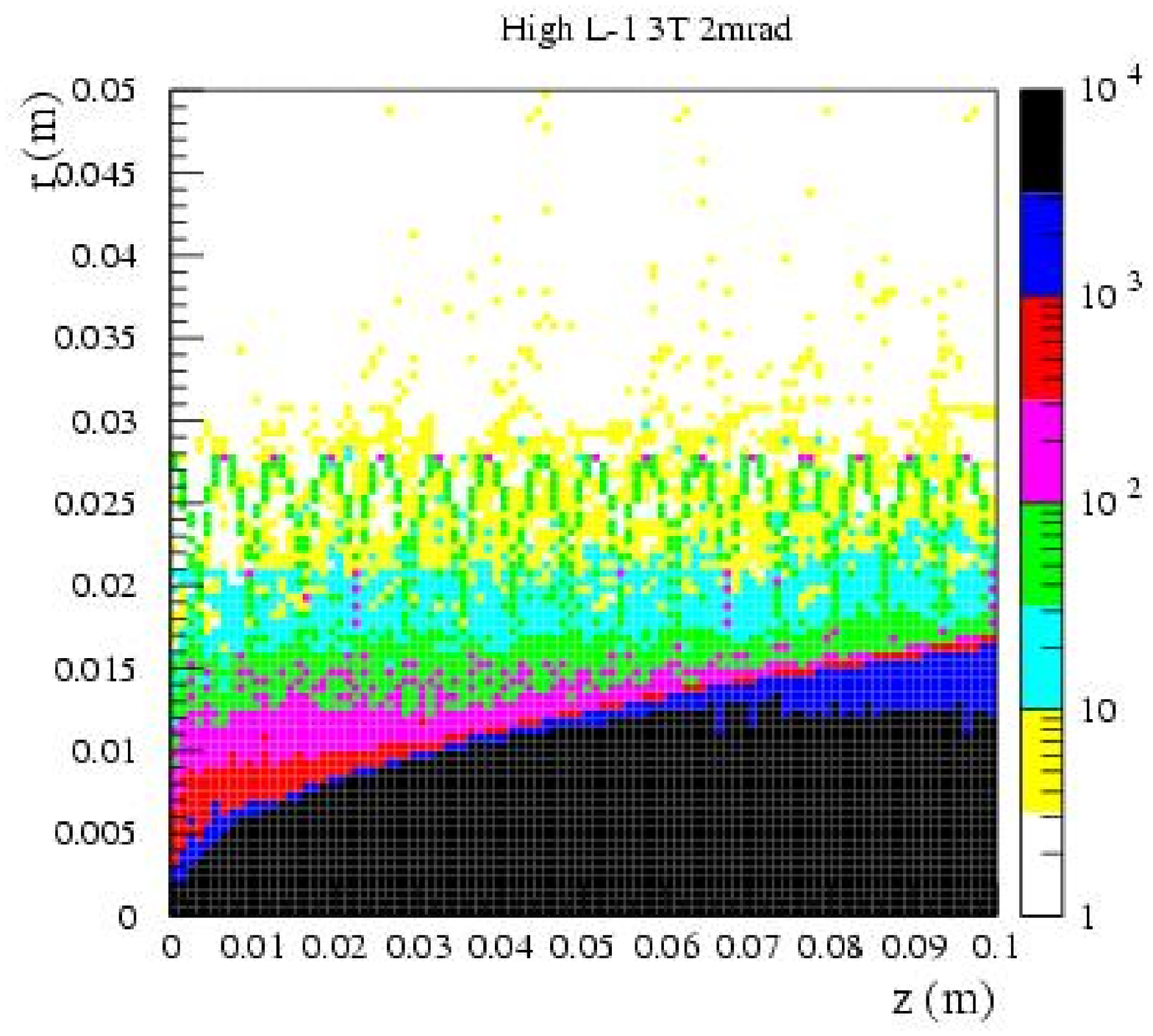}
\caption{Pair-background track density ($/\rm{cm}^2/\rm{BX}$)
with the high luminosity option of ILC machine parameter
at 500~GeV (left),
1~TeV (center)~\cite{Con_raubenheimer}, and the new high luminosity
option at 1~TeV~\cite{Vtx_seryi}
(right). The crossing angle is 2~mrad
and the magnetic field is 3~T for all three cases.}
\label{Vtx_options}
\end{figure*}

The simulation for pair background was done using CAIN
for various ILC parameter sets.
The track density of the pair background
in $z$-$r$ plane is shown in
Figure~\ref{Vtx_nominal} with the nominal ILC parameter 
set~\cite{Con_raubenheimer} and crossing angle of 2~mrad 
for 3, 4, and 5~T magnetic field. 
Figure~\ref{Vtx_options} shows the track density distribution
for high luminosity option of ILC parameters~\cite{Con_raubenheimer}.
The distribution of the dense core of the pair-background
tracks with the original
high luminosity option
is significantly broader than that with the nominal
option. Recently, A.~Seryi proposed new high-luminosity
parameter sets~\cite{Vtx_seryi}. These new high luminosity parameter
sets give less and narrower pair background as can be seen
from Figure~\ref{Vtx_options}.

The beam-pipe parameters and minimum
radius of the vertex detector
$R_{VTX}$ determined  by the design criteria and the
background simulation described above are summarized
in Table~\ref{Vtx_results}. We can see that minimum $R_{VTX}$
strongly depends on the machine parameter option.
\begin{table}
\begin{center}
\caption{The beam-pipe parameters and minimum $R_{VTX}$
for several machine options and different magnetic
field $B$ based on a certain design criteria. High Lum-A1 and
High Lum-A2 stand for the new high luminosity parameter sets
proposed by A.~Seryi.
See Figure~\ref{Vtx_Rmin} for the definition of each parameter.}
\begin{tabular}{|l|c|c|c|c|c|}
\hline 
$E_{CM}$ & Option & B (T) &
$R_{core}\ (mm)$ & $R_{Be}\ (mm)$ &
$R_{VTX}\ (mm)$  \\
\hline
500 GeV & Nominal  & 3  & 10.5  & 12.5  & 16.6    \\
\hline
 &  & 4  & 9 & 11 & 14.9  \\
\hline
 &  & 5  & 7.5  & 9.5 & 13.2  \\
\hline
 & High Luminosity & 3  & 16.5   & 18.5 & 24.1 \\
\hline
 &  & 4 & 13.5  & 15.5 & 20.2  \\
\hline
 &  & 5 & 12  & 14 & 18.4  \\
\hline
1~TeV & Nominal  & 3  & 11  & 13 & 17.3  \\
\hline
 & High Luminosity & 3 & 18.5  & 20.5 & 25.8  \\
\hline
 & High Lum-A1 & 3 & 13 & 15 & 19.4  \\
\hline
 & High Lum-A2 & 3 & 11.5  & 13.5  & 17.8 \\
\hline
\end{tabular}
\label{Vtx_results}
\end{center}
\end{table}

From this study, we choose the radii of Be beam pipe
($R_{Be}$) and innermost layer of the vertex detector
($R_{VTX}$) as 15~mm and 20~mm, respectively, as the baseline.
We also considered two options for different
background conditions as
listed in Table~\ref{Vtx_params}.
The baseline configuration can be used even for  High-Lum-A1 option
at 1~TeV. The ``small-R'' configuration can be used only 
for the nominal machine option at 500~GeV, and somewhat risky.
The ``large-R'' configuration is compatible with
the high luminosity option at 500~GeV.
\begin{table}
\begin{center}
\caption{Design parameters of the beam pipe radius
($R_{Be}$), the radius ($R_{VTX}$) 
and the half length ($Z_{VTX}$) of
the first layer of the vertex detector.}
\begin{tabular}{|l|c|c|c|}
\hline 
Configuration  & $R_{Be}\ \rm{(mm)}$ & $R_{VTX}\ \rm{(mm)}$ 
  & $Z_{VTX}\ \rm{(mm)}$  \\
\hline
Baseline & 15 & 20 & 65 \\
\hline
Small R  & 13 & 17  & 55  \\
\hline
Large R & 19 & 24  & 75   \\
\hline
\end{tabular}
\label{Vtx_params}
\end{center}
\end{table}

\subsubsection{Signal Readout}
Each wafer of the FPCCD will have multi-port
readout in order to reduce the readout time
and to reduce the effect of charge transfer
inefficiency (CTI) caused by radiation damage.
A readout ASIC consisting of amplifiers, 
correlated double samplers, and analog-to-digital
converters (ADCs) will be put on both ends of a ladder.
Because the FPCCD option can achieve an excellent
spatial resolution even with digital readout,
few bits will be enough for the ADCs.

The data size of the vertex detector is 
dominated by the contribution from 
the pair-background hits.
The total number of pixels is as large as  
$\sim 1\times 10^{10}$ (10~G pixels). 
If the pixel data is consisted of 
34~bit address plus 5~bit analog data, the data size
for 1\% pixel occupancy  
becomes about 5~Gbits per train.
Actually the pixel occupancy of the outer layers
is much less than 1\%, and  the dada size will be
much smaller than this vaue. 
The data size derived from the 
expected number of hits for the nominal option of the machine
parameters is less than 0.5~Gbits per train. 
Therefore, a small number of optical fiber 
cables with few Gbps throughput will be enough for 
the data transfer.

\subsection{Possible Options}
In this report, we assume FPCCD as a technology for
the vertex detector of the baseline GLD design,
but this is just an assumption. There are a lot of
candidate technologies studied all over the world.
The options are; column parallel CCD (CPCCD), 
CMOS monolithic active pixel sensor (MAPS),
depleted FET (DEPFET), pixel sensor based on 
SOI technology (SOI), CMOS pixel sensor with
registers in each pixel (CAP/FAPS), in-situ
storage image sensor (ISIS), and fine pixel CCD (FPCCD).
Among them, CAP, FAPS, ISIS, and FPCCD accumurate
the signal during a train and are read out in between
trains. 
  
The R\&D efforts for these technologies will 
be continued for  several years. The technology
choice for the vertex detector in future 
will be done by real ``collaboration groups''
of the ILC experiment based on the results
of the R\&D.

\subsection{R\&D needed}
Development of sensors and demonstration of
their performance satisfying the requirements
as the vertex detector for ILC experiments
are the highest priority R\&D issues for
all sensor technologies. Study of wafer thinning
technique and development of their support structure
are also indispensable.
Other R\&D items are; development of the front-end
readout ASIC, minimization of power consumption, 
data compression and the back-end electronics,
cooling system with minimum material,
and development of thin beam pipe.

    % to be prepared by Y.Sugimoto
\clearpage

%
% Silicon Trackers
%
\section{Silicon Inner Tracker}
\label{SectionSIT}

\subsection{Detector Features}

The silicon inner tracker (SIT) is considered 
to improve a momentum resolution 
and reconstruction efficiency of long-lived particles with the vertex detector and help pattern recognition in linking the tracks found in the Time 
Projection Chamber (TPC) 
with the tracks found in the vertex detector.
The SIT consists of the barrel inner tracker (BIT) in the barrel region
and the forward inner tracker (FIT) in the endcap region.
Four cylindrical BIT layers are located between the vertex detector and the TPC (Figure~\ref{Con_GLDIR}).

\begin{figure}[h]
\begin{center}
\includegraphics[width=1.0\textwidth]{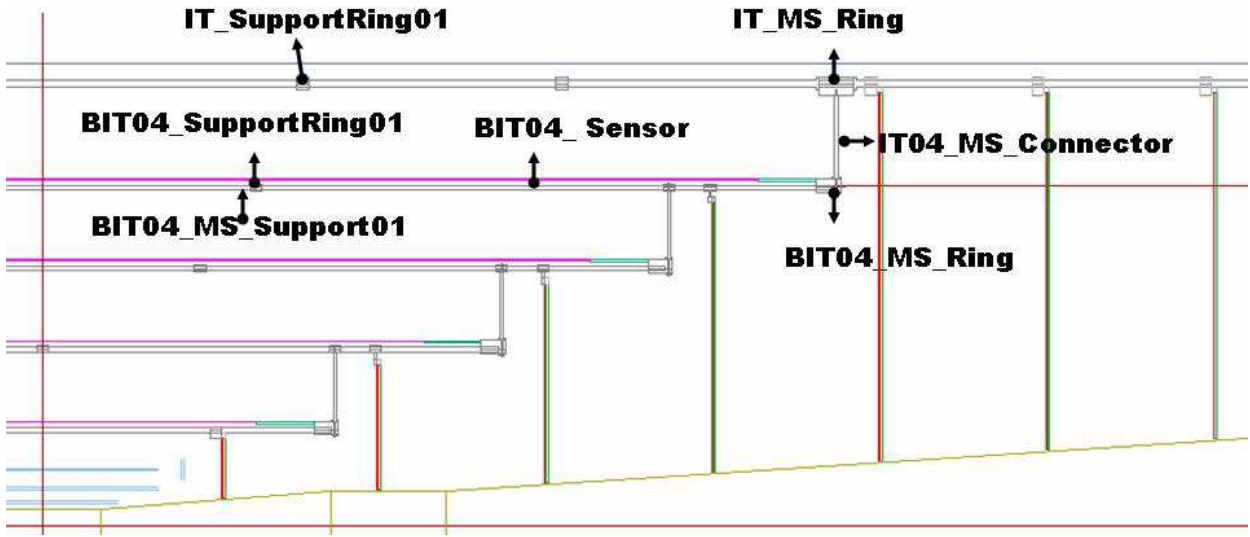}
\end{center}
\caption{\small A layout of the silicon inner tracker (SIT) consisting of 
four cylindrical in the barrel (BIT) and seven planes in the endcap (FIT).}
\label{layout}
\end{figure}

The four layers will consist of the double-sided silicon strip detectors 
with 10 $\mu$m spatial resolution in r$\phi$ direction. 
Seven plane disks perpendicular to the beam direction are positioned 
as the FIT. 
The four inner planes with any pixel-based sensors and the remaining three 
planes with silicon strip detectors (modest resolution) are being considered.
The SIT design of the BIT and the FIT is shown in Figure~\ref{layout}.

\begin{table}
\caption{\small Parameters for the BIT.}
\label{bit}
\begin{center}
\begin{tabular}{lccc} \hline \hline
BIT &half Z(cm) & R(cm) &sensor size(cm$^2$) \\\hline
layer1 &18.5 &9 &5~$\times$~5 \\
layer2 &33.0 &16 &5~$\times$~5 \\
layer3 &47.5 &23 &5~$\times$~5 \\
layer4 &62.0 &30 &9~$\times$~9 \\

\hline
\end{tabular}
\end{center}
\end{table}

Some of key mechanical parameters associated with the BIT and FIT design 
are listed in Table ~\ref{bit} and Table ~\ref{fit}, respectively. 

\begin{table}
\caption{\small Parameters for the FIT.}
\label{fit}
\begin{center}
\begin{tabular}{lccc} \hline \hline
BIT &half Z(cm) & R$_{min}$(cm) &R$_{max}$(cm) \\\hline
layer1 &15.5 &2.4 &7.6 \\
layer2 &29.0 &3.2 &14.0 \\
layer3 &43.5 &3.7 &21.0 \\
layer4 &58.0 &4.7 &28.0 \\
layer5 &72.5 &5.7 &38.0 \\
layer6 &87.0 &6.6 &38.0 \\
layer7 &101.5 &7.6 &38.0 \\

\hline
\end{tabular}
\end{center}
\end{table}

%input TABLE%
%input TABLE%
%input TABLE%

\subsection{Performance of the SIT} %new paragraph

\begin{figure}[htb]
\begin{center}
\includegraphics[height=6cm]{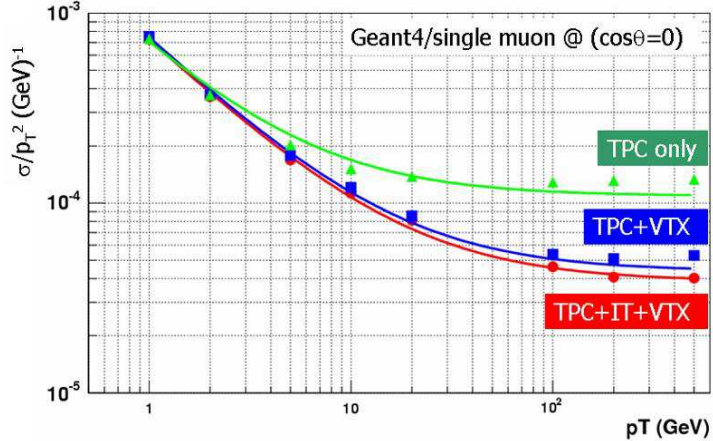}
\end{center}
\caption{\small Momentum resolution as a function of the momentum for a polar angle $\theta$ = $90^\circ$ of single muon for different combination of subdetector.}
\label{resolution}
\end{figure}

The SIT has been designed to have nearly the full solid angle coverage and stand alone track finding and reconstruction.
The momentum resolution of the tracking system 
is required to be $5~\times~10^{-5}~$(GeV/c)$^{-1}$~\cite{Con_TESLA}.
The resolutions of $7~\times~10^{-5}~$(GeV/c)$^{-1}$ for the vertex detector alone 
and $1.4~\times~10^{-4}~$(GeV/c)$^{-1}$ for the TPC alone are achievable 
and the addition of the silicon layers in the space between 
the vertex and the TPC detectors improves the resolution 
to the required precision of $5~\times~10^{-5}~$(GeV/c)$^{-1}$ 
as shown in Figure~\ref{resolution}.

The multiple Coulomb scattering is dominant in the low momentum region 
and addition of the SIT does not help to improve the momentum resolution.

\begin{figure}[h]
\begin{center}
\includegraphics[height=6cm]{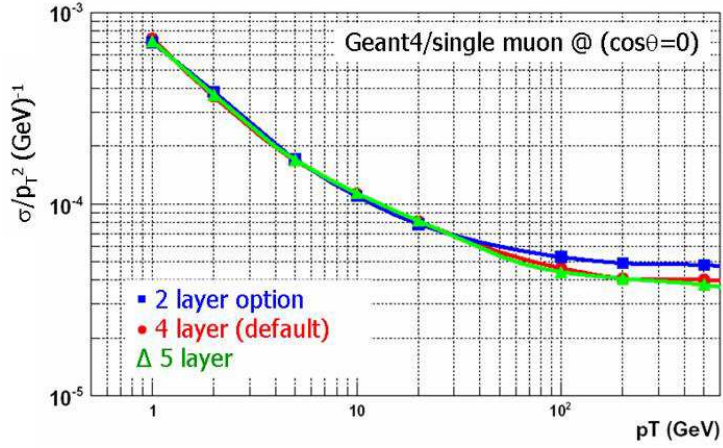}
\end{center}
\caption{\small Momentum resolution as a function of transverse momentum for the different number of layers option in SIT.}
\label{lowmomentum}
\end{figure} 
Figure~\ref{lowmomentum} shows the momentum resolution as a function of the momentum for the different number of layers in the SIT and the simulation
result shows that four layers 
is reasonable choice to achieve the required momentum resolution.

The linking and reconstruction efficiency of charged tracks from physics events should be studied.
Since the angular resolution in the detector performance is important 
in the endcap regions, the polar angular resolution as a function 
of the polar angle should be studied.
This is important in the bremsstrahlung energy spectrum 
because error on the effective center of mass energy is given by error 
on collinearity distribution of Bhabha events.

\subsection{Technologies} %new paragraph

The Double-sided Silicon Strip Detector (DSSD) will be used for the BIT and 10~$\mu$m resolution in r$\phi$ is required. 
The SIT consist of 4 layers of DSSD.
Since z measurement is needed to improve the track finding efficiency, 
50~$\mu$m resolution in z is largely sufficient. The physical dimension of 
the sensor will be 50mm $\times$ 25mm with 300 $\mu$m thickness. There will be 
511 $n^+$ strips with 50 $\mu$m pitch, and 511 $p^+$ strips with 100 $\mu$m 
pitch. This type of the DSSD sensors~\cite{SIT_dssd} are already used for 
the silicon vertex detectors in high energy experiments~\cite{SIT_high}.

The inner four planes and the outer three planes of the FIT will consist 
of pixel and strip detectors, respectively. Among the silicon pixel sensor
technology, ATLAS pixels with a pixel of 50 $\mu$m $\times$ 300 $\mu$m 
can be used~\cite{SIT_atlas}.
The requirements of the strip detectors for the FIT are somewhat 
loose compared with those for the BIT.
25 $\mu$m resolution is required and this resolution can be achieved with a strip pitch of 100 $\mu$m and a readout pitch of 300 $\mu$m.

\subsection{Detector Conceptual Designs}

\begin{figure}[h]
\begin{center}
\includegraphics[height=5cm]{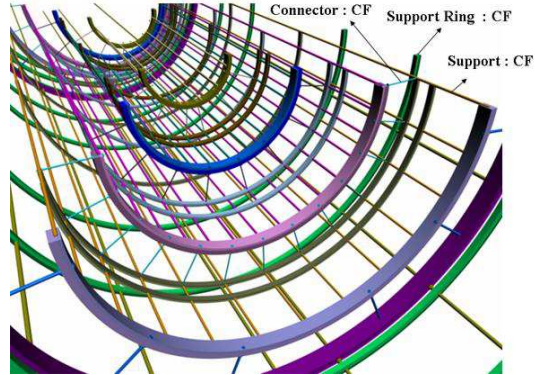}
\end{center}
\caption{\small A conceptual layout of the BIT and FIT support structure. The space frame and the support ring consist of carbon fiber struts. The support shell and the support legs are made from composite materials.}
\label{concept}
\end{figure} 

Figure~\ref{concept} shows a conceptual layout of a possible mechanical support structure.
In this design we have to consider that the SIT is not only mechanically 
very rigid but also as thin as possible.
The readout electronics will be located at the very end of the BIT layers 
in order to minimize the material in front of the TPC.
The SIT is mechanically independent of the TPC 
and the whole TPC can be withdrawn from the detector. 

In a recent measurement it has been shown that the Lorentz angle 
in silicon causes a broadening of the clusters of about 180 $\mu$m 
for electrons and about 40 $\mu$m for holes for 300 $\mu$m thick detectors 
and a magnetic field of 3 T if the strip are parallel to the B-filed.
Not to be limited by this effect one has to use the p-side of the detectors for the r$\phi$-measurement and the n-side for z.
Table~\ref{parameter} shows parameters in the conceptual design of the BIT.

\begin{table}
\caption{\small The number of channels between sensor types.}
\label{parameter}
\begin{center}
\begin{tabular}{lccccc} \hline \hline
BIT &sensor area &\# of sensor in a ladder &\# ladder &\# sensor &total area(mm$^2$)\\\hline
layer1 &50~$\times$~50 &4 &24 &96 &240000\\ 
layer2 &50~$\times$~50 &7 &48 &336 &840000\\ 
layer3 &50~$\times$~50 &10 &64 &640 &1600000\\ 
layer4 &90~$\times$~90 &7 &24 &168 &1360800\\ 

\hline
\end{tabular}
\end{center}
\end{table}

\begin{figure}[h]
\begin{center}
\includegraphics[height=5cm]{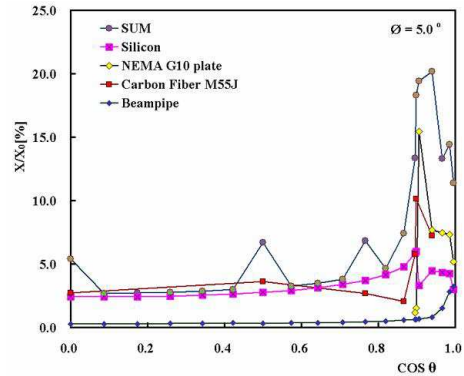}
\end{center}
\caption{\small Material in units of radiation lengths as a function of 
the polar angle up to the end of the vertex detector and SIT.}
\label{radiationLength}
\end{figure} 

With the assumption of overlap of 1.6 mm between neighboring detectors and silicon sensor size of 
5 $\times$ 5 cm or 9 $\times$ 9 cm, the total number of sensors in a ladder on each layer and total 
number of sensors in a layer are given. In the pixel planes of the FIT electronic detectors 
of 0.5 cm$^{2}$ are bonded to 
a detector chip allowing great flexibility in the layout of the modules.
In the strip planes of the FIT the strip disks are preferably built with 
double-sided detectors to minimize the material.
The material budget as a function of the polar angle up to 
the end of the SIT is shown in Figure~\ref{radiationLength}

\subsection{R\&D Program and Needed}

\begin{figure}[h]
\begin{center}
\includegraphics[height=5cm]{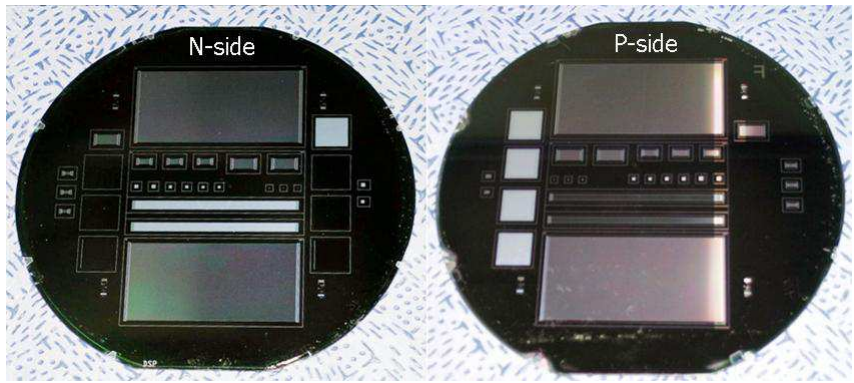}
\end{center}
\caption{\small P-side and N-side of DSSD and test patterns.}
\label{sensor}
\end{figure} 

The R\&D program of the DSSD with DC-type is ongoing and AC-type is just started. 
Also single sided silicon strip sensors are under study. 
The DC-type of DSSD design is as follows. 
The sense strips on one side of the layer are orthogonal to the ones on 
the other side. This way one plane measures x and y coordinates of point, 
where ionizing particle goes through, on its two 
different sides~\cite{SIT_lutz}. 
The p-side has two metal layers; one layer for implantation strip 
and one for readout strip. The each of sensor side consists of 512 
sensor strips with 50 $\mu$m and 100 $\mu$m pitch on n-side and p-side, 
respectively. Each strip is bias strip and the readout pads are made large 
for bonding. The sensor size has an area of 13.3 cm$^2$ and 2.6 cm long 
readout strips. 

\begin{figure}[h]
\begin{center}
\includegraphics[height=5cm]{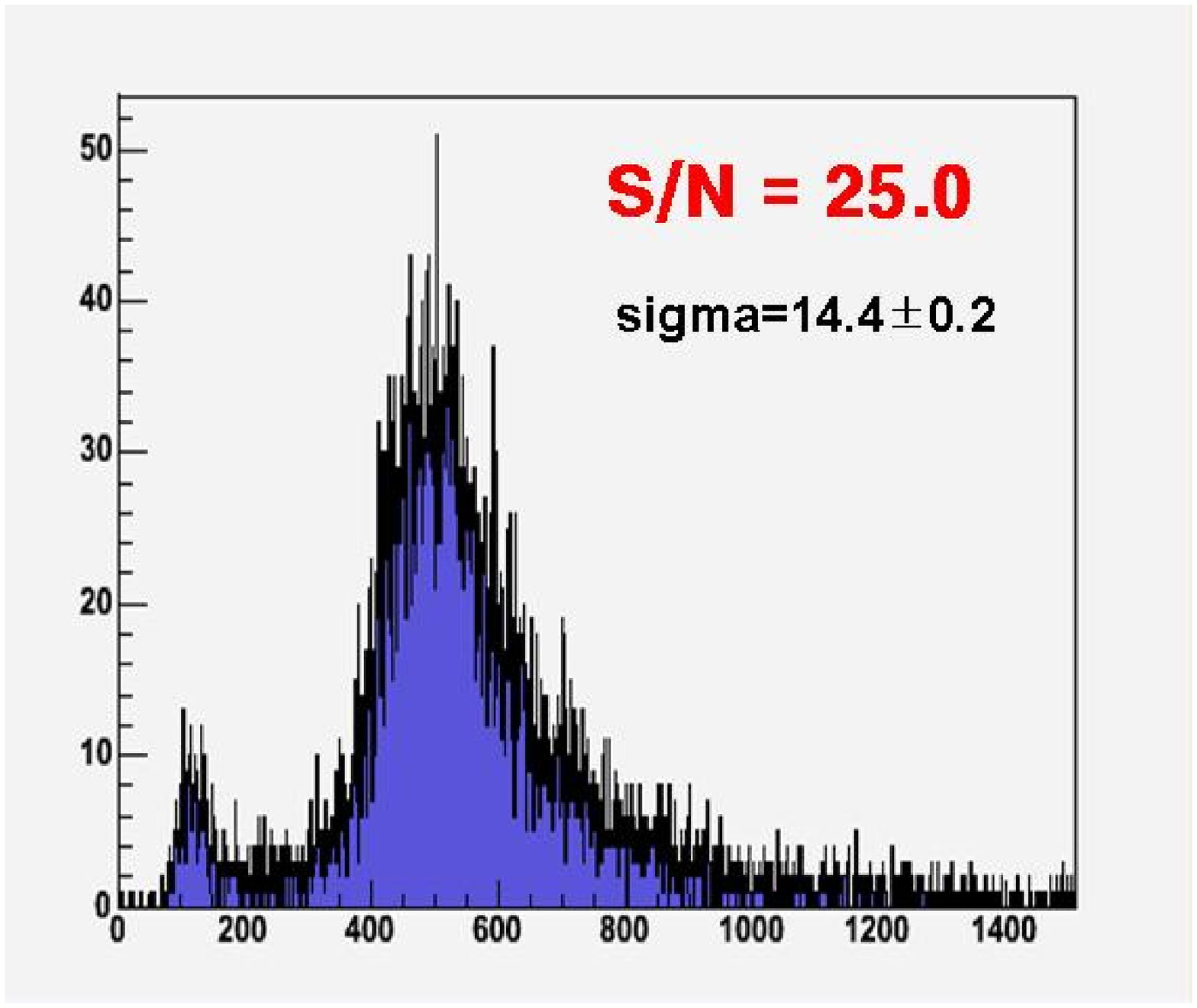}
\end{center}
\caption{\small Signal to noise ratio of DSSD with $^{90}$Sr beta source.}
\label{sovern}
\end{figure} 

Two batches of 25 wafers were processed in ETRI~\cite{SIT_etri} on 127 
mm diameter, FZ, 380 $\mu$m thick, $<$100$>$-oriented, 
n-type silicon wafers, with resistivity 
of $>$ 5 k$\Omega$cm. A total of 11 masks steps were needed for the 
sensor fabrication process: 5 and 6 masks for n-side and p-side, 
respectively. To make the junction depth of the n
-side deeper we started with n$^+$ implantation and then p$^+$ implantation 
is followed and Si$_3$N$_4$ instead of SiO$_2$ is employed as 
the isolation material at the second batch run.
Figure~\ref{sensor} shows pictures of the n-side and p-side of the fabricated 
double-sided silicon strip sensors which are fabricated in 5-inch 
fabrication line in Korea. 
The leakage current and capacitance of the prototype strip sensor is being 
measured. It shows that the leakage current level of the single 
strip sense is 
from 8nA$-$20nA up to the full depletion voltage and values of the bulk 
leakage current level and the capacitance are about 1 $\mu$A and 50 pF, 
respectively. This measurement provides us information of 
the bulk characteristics and quality of the fabricated sensor.

We used $^{90}$Sr beta source for the radioactive source test purpose. 
After the full depletion voltage was applied for the prototype, 
we measured the beta source
signal with test readout electronics. 
We measured noise level of the silicon sensor 
without the beta source and the beta source is then put on the top 
of the silicon sensor in a dark box. 
The beta source signal was clearly seen and 
signal to noise is obtained to be 25 as shown in Figure~\ref{sovern}.

The full R\&D of the AC-type of DSSD and single sided silicon strip 
is necessary. Also comparison between the DC-type and AC-type should be done.
The VA-chip was used for current R\&D for the front-end readout electronics. 
However, it can't be used for the ILC environment. 
We need to study which front-end readout ASIC is fit for the 
SIT. Also zero suppression and back-end electronics 
need to be studied.

    % to be prepared by H.J.Kim
\clearpage

% 
% TPC 
% 
% Change log: 
%      20-March-2006 : R.Settles put changes to the Version 1.0 document.
%      22-March-2006 : A.Miyamoto put minor English corrections and uploaded to plone server

\section{Main Tracker} 
\label{SectionTPC} 
%-- to be prepared by R.Settles and A.Sugiyama 

% overall introduction here 

%%[AM]A Time Projection Chamber is has been chosen as central
A Time Projection Chamber is chosen as central 
tracker for the GLD detector concept at the ILC. An R\&D 
program is underway to develop the technology and prove the feasibility 
of a high-performance TPC required for this application. The use of 
new micro-pattern gas devices (MPGD) is an attractive possibility for the 
gas 
amplification. The decision for using an MPGD device, either the Gas 
Electron Multiplier (GEM) or the Micromegas technique, 
has included a 
comparison with the multi-wire proportional chamber (MWPC) technology 
used in past TPCs in a number of large collider experiments. 
For future TPCs, the MPGD technology promises to have better point 
and two-track resolution than wire chambers and to be more robust 
in high backgrounds. 

\subsection{The Basic Concept of the LC-TPC} 
\label{whytpc} 
General arguments for a TPC as main tracker are as follows. 
\begin{itemize}
\item 
The tracks can be measured with a large number of ($r\phi$,$z$) 
space points, 
so that the tracking is continuous and the efficiency remains close to 
100\% for 
high multiplicity jets and in presence of large backgrounds.
\item 
It presents a minimum of material to particles 
crossing it. This is important for getting the best possible performance 
from the electromagnetic calorimeter, and to minimize 
the effects due to 
the $\sim$10$^3$ beamstrahlung photons per bunch crossing which 
traverse the detector.
\item
The comparatively moderate $\sigma_{singlepoint}$ and double-hit 
resolution are compensated by the continuous tracking and 
the large volume which can be filled with 
fine-granularity coverage.
\item
The timing is precise to 2ns (corresponding to 
50 $\mu$m/ns 
drift speed of tracks hooked up to 
the $z$-strips of a silicon inner detector with 100$\mu$m pitch), 
so that 
tracks from different bunch crossings or from cosmics can 
readily be distinguished via time stamping. 
\item
To obtain good momentum resolution 
and to suppress backgrounds near the vertex, the TPC has to operate in a 
strong magnetic field. It is well suited for this environment 
since the electrons drift 
parallel to $\vec{B}$, which in turn improves the two-hit 
resolution by compressing 
the transverse diffusion of the drifting electrons 
(FWHM$_{T}\leq$ 2 mm for Ar-10\%CH$_4$ gas and a 3T magnetic field).
\item
Non-pointing tracks, e.g. for V$^0$ detection, are an important 
addition to the particle flow measurement and help in 
the reconstruction of physics signatures in many 
standard-model-and-beyond scenarios.
\item
The TPC gives good 
particle identification via the specific energy loss, dE/dx, which is 
valuable for many physics analyses, electron-identification and 
particle-flow applications.
\item 
The TPC will be designed to be robust and at the same time
easy to maintain so that
an endplate readout chamber can
readily be accessed or exchanged in case of 
accidents like beam loss in the detector. 
\end{itemize}

Two additional properties of a TPC 
will be compensated by proper 
design.
\begin{itemize}
\item
The readout endplanes and electronics present a 
small but non-negligible amount 
of material in the forward direction. 
The goal is to keep this below 30\%X$_0$.
\item
The $\sim$50$\mu$s memory time 
integrates over background and signal events from 160 ILC bunch 
crossings 
at 500 GeV for the nominal accelerator configuration. 
This is being compensated by 
designing for the finest possible granularity: 
the sensitive volume will consist of 
several$\times 10^9$ 3D-electronic readout voxels 
(two orders of magnitude better than at LEP). 
It has been estimated to result in an 
occupancy of the TPC of less than 
$0.5\%$ from beam backgrounds and gamma-gamma 
interactions\cite{Con_TESLA}. 
(See below for further discussion.) 
\end{itemize}

\subsection{Design issues.} 
There are many aspects for the layout of the LC detector and 
its subdetectors. The detector has to be designed globally 
to cover all possible physics channels, and the roles of the 
subdetectors in reconstructing many of these channels are 
highly interconnected. 
For the TPC, the issues are performance, size, endplate, electronics, 
gas, alignment and robustness in backgrounds. 

% $\bullet $ 1.{\underline{Resolution expected/needed}} 
\subsubsection{Resolution expected/needed}

The requirements for a TPC at the ILC are summarized 
in Table~\ref{TPC_parameters}. 
\begin{table}[h] 
\caption{ 
Typical list of performance requirements for a TPC at the ILC detector.} 
\label{TPC_parameters} 
\begin{center} 
%\hline 
\begin{tabular}{| ll |} 
\hline
Size&$\phi = 4.1$m, L $= 4.6$m\\ 
Momentum resolution &${\delta (1 / {p_t})} \sim 10^{-4}$/GeV/c 
(TPC only; $\times$ 2/3 when IP included)\\ 
Solid angle coverage & Up to at least $\cos\theta \sim 0.98$\\ 
TPC material budget& $ <0.03 {\rm X}_0$ to outer field cage in $r$\\ 
& $< 0.30 {\rm X}_0$ for readout endcaps in $z$\\ 
Number of pads & $\sim$ 1.3$\times$10$^6$ per endcap\\ 
Pad size/Number of pad rows& $\sim$ 1mm$\times$6mm/$\sim$200\\ 
$\sigma_{\rm single point}$ in $r\phi$ & $\sim 100 \mu$m (average over driftlength)\\ 
$\sigma_{\rm single point}$ in $rz$ & $\sim 0.5$~mm \\ 
2-track resolution in $r\phi$ & $< 2$~mm \\ 
2-track resolution in $rz$ & $< 5$~mm \\ 
dE/dx resolution & $< 4.5$~\%\\ 
Performance robustness & $>$ 95\% tracking efficiency (TPC only), 
$>$ 98\% overall tracking 
\\ 
Background robustness & Full precision/efficiency in backgrounds of 10-20\% 
occupancy, 
\\ 
&whereby simulations estimate $< 0.5$\% for nominal backgrounds. \\ \hline
\end{tabular} 
\end{center} 
% \caption{\label{TPC_parameters} 
\end{table} 

The main question to answer is: what should 
the resolution be for the overall tracking? This will define how many silicon layers are needed.  
According to various studies, that overall momentum resolution of ${\delta (1 / {p_t})}\sim 5 \times 10^{-5}$/GeV/c 
will be sufficient, as defined mainly by the e$^+$e$^-$$ \rightarrow HZ \rightarrow H\ell\ell$ channel used for measuring the Higgs production rate (see ref.\cite{Con_WWS}, for example).
 
This resolution is achievable with inner-silicon tracking and a 
TPC performance given in Table~\ref{TPC_parameters}. If for physics reasons, 
the overall tracking accuracy should be better, 
a larger TPC and/or more silicon layers 
should be envisaged\cite{bssnowmass}. 

%$\bullet $ 2.{\underline{Endplate}} 
\subsubsection{Endplate} 

As stated in the introduction, MPGDs are the default technologies 
for the gas amplification since 
they promise better performance than the MWPC. 
Systems mainly under study are 
Micromegas \cite{Con_mm} meshes and 
GEM \cite{Con_gem} foils. 
Both\cite{CERNmpgd} operate in a gaseous atmosphere and are 
based on 
the avalanche amplification of the primary produced electrons. The 
gas amplification occurs in the large electric fields within the 
MPGD microscopic structures with sizes of the order of 50$\mu$m. MPGDs lend 
themselves naturally 
to the intra-train un-gated operation foreseen for 
the ILC, since, when configured 
properly, they 
display a significant suppression of the number of back-drifting ions. 
In addition a gating plane will be foreseen for inter-train gating in 
order to have a safety factor in case of unexpected backgrounds (see below). 

\begin{figure}[h] 
\begin{center} 
% \begin{tabular}{c} 
%\includegraphics*[clip,width=0.3\textwidth,height=0.48\textwidth,rotate=90] 
\includegraphics*[clip,width=0.8\textwidth] 
{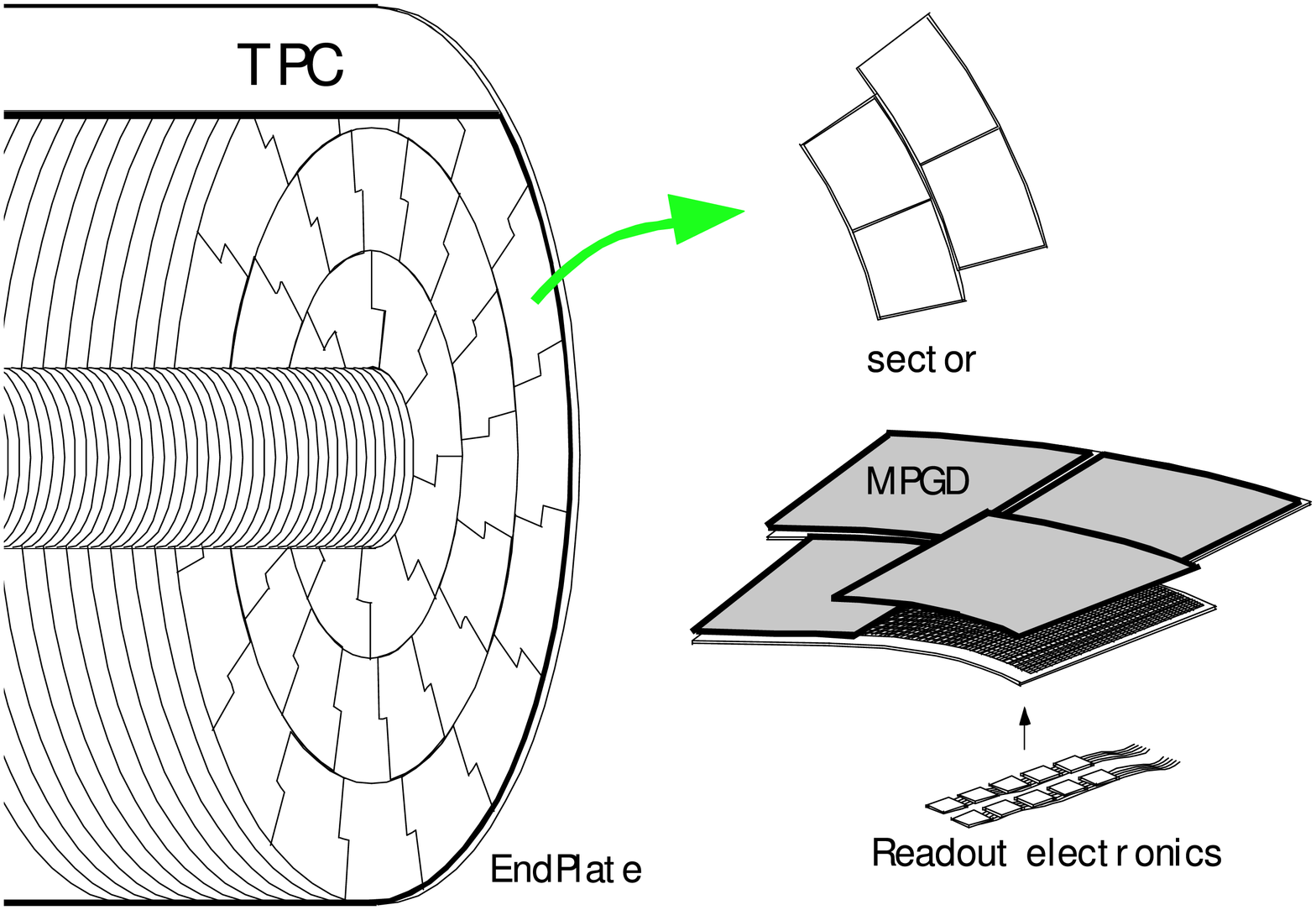} 
%{dummy1.eps} 
% \end{tabular}{c} 
\caption{Ideas for the layout of the TPC endplates.} 
\label{fig:endplate} 
\end{center} 
\end{figure}

The two TPC endplates have a surface of about 10~m$^2$ of sensitive 
area each. 
The layout of the endplates, i.e. 
conceptual design, stiffness, division into sectors and 
dead space, has been started, for instance as shown in 
Figure~\ref{fig:endplate}. 
In this example the question arises as to 
how to make odd-shaped MPGDs if needed. 
In general, the readout pads, their 
size, geometry and 
connection to the electronics and the cooling 
of the electronics, are all highly correlated design 
tasks related to the endplates. 
As stated in Section~\ref{whytpc}, the material budget for the 
endcap and its effect on ECAL for the particle-flow measurement 
in the forward direction must be minimized. 
More details are covered in the next item. 

% $\bullet $ 3.{\underline{Electronics}} 
\subsubsection{Electronics}

For the readout electronics, one of the important issues 
is the density of pads that can be accommodated while 
guaranteeing a thin, 
coolable endplate. The options being studied are (a) 
a standard readout (meaning, as in previous 
TPCs) of several million pads 
or (b) a pixel readout of a few hundred times more by using CMOS techniques. 

\noindent {\bf (a) Standard readout:}\\
Pad sizes under discussion are, 
for example, 2 mm times 6 mm (the TDR size\cite{Con_TESLA}) or 
1 mm times 6 mm which has found to be better as a result of 
our R\&D experience (see below). 
A preliminary look at the FADC-type approach using 130 nm technology 
indicates that even smaller sizes like 1 mm times 1 mm might 
be feasible (in which case charge-spreading would not be needed). 
In all of these cases there 
are between 1.5 and 20 million pads to be read out. 
An alternative to the FADC-type is the TDC 
approach (see \cite{prc}\cite{eudet}) 
in which time of arrival and charge per pulse 
(via time over threshold) is measured. 
In case the material budget requires larger pads, then the 
resistive-foil technique\cite{madhu} is an option
to maintain the point resolution. 

\noindent {\bf (b) CMOS readout: }\\
A new concept for the combined gas amplification and readout 
is under development. In this concept\cite{prc} the MPGD is 
produced in wafer using post-processing technology on top of a 
CMOS pixel readout chip, thus forming a thin integrated device of 
an amplifying grid and a very high granularity endplate, with 
all necessary readout electronics incorporated. This concept 
offers the possibility of pad sizes small enough to observe 
individual single electrons formed in the gas and count the number 
of ionization clusters per unit track length, instead of measuring 
the integrated charge collected. 
Initial tests using Micromegas\cite{pixelref1} 
and GEM\ foils\cite{pixelref2} 
mounted on the Medipix2 
chip provided 2-dimensional images of 
minimum ionizing track clusters. 
A modification of the Medipix2 chip (called Timepix) to measure also 
the drift time is under development\cite{eudet}. Also a 
first working integrated grid has been produced\cite{pixelref3}. 

% $\bullet $ 4.{\underline{Chamber gas}} 
\subsubsection{Chamber gas}

This issue involves (a) gas choice, 
(b) ion buildup and (c) ion feedback. \\

\noindent {\bf (a) Gas Choice}\\
The choice of the gas for a TPC is an important and central 
parameter. 
Gases being investigated are variations of standard TPC gases, e.g.,\\
\hspace*{36pt}Ar(93\%)CH${}_4$(5\%)CO${}_2$(2\%)--``TDR'' gas,\\ 
\hspace*{36pt}Ar(95\%)CH${}_4$(5\%)--``P5'' gas, \\ 
\hspace*{36pt}Ar(90\%),CH${}_4$(10\%)--``P10'', \\ 
\hspace*{36pt}Ar (90\%)CO$_2$(10\%),\\ 
\hspace*{36pt}Ar (95\%)Isobutane(5\%) and\\ 
\hspace*{36pt}Ar(97\%)CF$_4$(3\%). 
%Ar(93\%)CH${}_4$(5\%)CO${}_2$(2\%)--``TDR'' gas,\\ 
%Ar(95\%)CH${}_4$(5\%)--``P5'' gas, \\ 
%Ar(90\%),CH${}_4$(10\%)--``P10'', \\ 
%Ar (90\%)CO$_2$(10\%),\\ 
%Ar (95\%)Isobutane(5\%) and\\ 
%Ar(97\%)CF$_4$(3\%). 

When choosing a gas a number of requirements have to be taken into account. 
The $\sigma_{\rm single point}$ resolution achievable in $r\phi$ 
is dominated by the transverse diffusion, which 
should be as small as possible. Simultaneously a sufficient number of 
primary electrons should be created for the point 
and dE/dx measurements, and the drift velocity 
at a drift field of a few times 100~V/cm should be about 
5~cm/$\mu$s or more. The hydrogen component of hydrocarbons, 
which traditionally are used as quenchers in TPCs, have a 
high cross section for 
interaction with low energy background neutrons which will 
be crossing the TPC at the ILC\cite{Con_TESLA}. Thus the concentration 
of hydrogen in the quencher should be as low as possible, to minimize the 
number of background hits due to neutrons. An interesting 
alternative to the traditional gases is a Ar-CF$_4$ mixture. These 
mixtures give drift velocities around $8-9$~cm/$\mu$s at drift 
field of $200$~V/m, have no hydrocarbon content and have a 
reasonably low attachment coefficient at low electric fields. 
However at intermediate fields ($\sim$5-10~kV/cm), as are present in the 
amplification region of 
a GEM\ or a Micromegas\, the attachment increases drastically, thus 
limiting the use of this gas to systems where the intermediate field 
regions are of the order of a few microns. This is the 
case for Micromegas, but its use has not been tested thoroughly 
for a GEM-based chamber. 
Whether CF4 is an appropriate 
quencher for the LC~TPC is not yet known and is being tested as a 
part of 
our R\&D. \\

\par

\noindent {\bf (b) Ion Build-up}\\
Ion build-up at the surface of the gas-amplification 
plane and in the drift volume. 

\begin{itemize}
\item
At the surface of the gas-amplification plane 
vis-a-vis the drift volume, 
during the bunch train of about 1~ms and 3000 bunch crossings, 
there will be 
few-mm thick layer of positive ions built up due to the incoming charge, 
subsequent gas amplification and ion back drift. 
An important property of MPGDs is that they suppress naturally 
the back drift 
of ions produced in the amplification stage. 
This layer of ions will reach a density of some fC/cm$^3$ 
depending on the background conditions during operation. Intuitively 
its effect on the coordinate measurement 
should be small since the drifting electrons incoming to the 
anode only experience this environment during the last few mm of drift. 
In any case, the TPC is planning to run with the lowest possible 
gas gain, meaning a few times 10$^3$, in order to minimize this 
effect. 

\item
In the drift volume, 
a positive ion density due to the primary ionization 
will be built up 
during about 1s (the time it takes for an ion to drift 
the full length of the TPC), will be higher near the cathode 
and will be of order fC/cm$^3$ 
at nominal occupancy ($\sim 0.5$\%). The tolerance on the 
charge density will be established by our R\&D program, 
but a few \t\ fC/cm$^3$ is orders of magnitude 
below this limit. 
\end{itemize}

\par

\noindent {\bf (c) Ion back drift and gating}\\

In order to minimize the impact of 
ion feeding back into the drift volume, 
a required back drift suppression of about 
$1/{\rm gasgain}$ has been used as a rule-of-thumb, 
since then the total charge 
introduced into the drift volume is about the same as 
the charge produced in the primary ionization. 
Not only have these levels of back drift suppression 
not been achieved during our R\&D program, 
but also this rule-of-thumb is misleading. 
Lower back drift levels will be needed since these ions would drift 
as few-mm thick sheets 
through the sensitive region 
during subsequent 
bunch trains. Even if a suppression of $1/{\rm gasgain}$ 
is achieved, the overall charge within the sheets will be 
the same as in the drift volume so that 
the density of charge within a sheet will be one to 
two orders of magnitude greater than 
the primary ionization in the total drift volume. 
How these sheets would affect the track reconstruction 
has to be simulated, but to be on the safe side 
a back drift level of $<< 1/{\rm gasgain}$ will be desirable. 
Therefore, since the back drift can be completely eliminated by a 
gating plane, a gate should be foreseen, to guarantee a 
stable and robust chamber operation. 
The added amount of material for a gating plane is 
small, $< 0.5$\%X$_0$ average thickness. 
The gate will be 
closed between bunch trains 
and remain open throughout one full train. 
This will obviate the 
need to make corrections to the data for such an ``ion-sheets effect'' 
which could be necessary without inter-train gating. 

% $\bullet $ 5.{\underline{The field cage}} 
\subsubsection{The field cage}

The design of the field cage 
involves the geometry of the potential rings, the resistor chains, 
the central HV-membrane, the gas container and 
a laser system. 
These have to be laid out for 
sustaining at least 100kV at the HV-membrane and a minimum of material. 
Important aspects for the gas system are purity, 
circulation, flow rate and overpressure. 
The final configuration depends on the gas mixture, which is discussed above, 
and the operating voltage which must also take into account the 
stability under operating conditions due to fluctuations in temperature 
and atmospheric pressure. For alignment purposes (see next two items) 
a laser system will be foreseen, either integrated in the 
field cage\cite{star} or not\cite{aleph}.

% $\bullet $ 6.{\underline{Effect of non-uniform field}} 
\subsubsection{Effect of non-uniform field}

\begin{itemize}
\item
Non-uniformity of the magnetic field of the solenoid will be 
by design within the tolerance of 
$\int_{\ell_{\rm{drift}}} \frac {B_r}{B_z} dz < 2$mm 
used for previous TPCs. This homogeneity is achieved by 
corrector windings at the ends of the solenoid. 
At the ILC, larger gradients 
could arise from the fields of the 
DID (Detector Integrated Dipole) or anti-DID, which 
are options for handling the beams inside the detector 
in case a larger crossing-angle optics is chosen. 
This issue was studied intensively at the 2005 Snowmass 
workshop\cite{rswwlcnote}, where it was shown 
that the TPC performance will not be degraded 
if the B-field is mapped 
to 10$^{-4}$ relative accuracy and the 
calibration procedures outlined in 
Section~\ref{align} are followed.
Based on past experience, the field-mapping 
gear and methods should be able to accomplish 
this goal. 
The B-field 
should also be monitored since the DID or 
corrector windings 
may differ from the configurations mapped; 
for this purpose the option 
a matrix of hall plates 
and NMR probes mounted on the outer surface 
of the field cage is being studied. 

\item
Non-unformity of the electric field can arise from the 
field cage, back drift ions and primary ions. 
For the first, the fieldcage design, the non-uniformities can be minimized 
using the experience gained in past TPCs. 
For the second, as explained 
above, the back drift ions can be minimized at the MPGD plane using 
low gas gain 
and eliminated entirely in the drift volume using gating. 
The effect due to the third, 
the primary ions, is due to backgrounds 
and is irreducible. As discussed above, 
the maximum allowable electrostatic charge density 
has to be established, but studies by the STAR 
experiment\cite{startdr} indicate that up to 
1 pC/cm$^3$ can be tolerated, whereas 
at nominal occupancy 
%($\sim 0.5$\%) 
it will be of order fC/cm$^3$. 
This will be 
revisited by the LC TPC collaboration by simulation and by 
the R\&D program below. 
\end{itemize}

%$\bullet $ 7.{\underline{Calibration and alignment}} 
\subsubsection{Calibration and alignment}
\label{align}
The tools for solving this issue are Z peak running, 
the laser system, the B-field map, 
a matrix of hall plates and NMR probes 
and the silicon layers outside the TPC. 
In general about 10/pb of data at the Z peak 
will be sufficient during commissioning to 
master this task, and typically 1/pb during 
the year may be needed depending on the 
background and energy of the ILC machine. 
A laser calibration system will 
be foreseen which can be used to understand both 
magnetic and electrostatic effects, while a 
matrix of hall plates and NMR probes may supplement the B-field 
map. The $z$ coordinates determined by the 
silicon layers inside the inner field cage of the TPC 
were used in Aleph\cite{wwviennawc} for 
drift velocity and alignment measurements, 
were found to be extremely effective and will thus 
be included in the LC~TPC planning. The overall 
tolerance is that systematics have to be 
corrected to 30$\mu$m throughout 
the chamber volume in order to guarantee 
the TPC performance, and this level has already 
been demonstrated by the Aleph TPC\cite{rswwlcnote}. 

%$\bullet $ 8.{\underline{Backgrounds and robustness}} 
\subsubsection{Backgrounds and robustness}

The issues here are the primary-ion charge buildup (discussed 
above) and the track-finding efficiency in the presence 
of backgrounds, which will be discussed here. 
There are 
backgrounds from the accelerator, 
from cosmics or other sources and 
from physics events. The main source is the accelerator, which 
gives rise to gammas, neutrons and charged particles being 
deposited in the TPC at each bunch crossing\cite{rsstmalo}. Preliminary 
simulations of these under nominal conditions\cite{Con_TESLA} 
indicate an occupancy of 
the TPC of less than about 0.5\%. This level 
would be of no consequence for the LC TPC performance, 
but caution is in order here. 
The experience at LEP was that the backgrounds 
were much higher than expected at the beginning 
of the running (year 1990), but after the 
simulation programs were improved 
and the accelerator better understood, they 
were much reduced, even negligible at the end (year 2000). 
Since such simulations have to be tuned to the 
accelerator once it is commissioned, the backgrounds 
at the beginning could be much larger, so the 
 LC TPC should be prepared for much more occupancy, 
up to 10 or 20\%. The TPC performance at these 
occupancy levels will hardly deteriorate due to 
its continuous, high 3D-granularity tracking 
which is still inherently simple, robust and very 
efficient with the remaining 80 to 90\% of 
the chamber. 

\subsection {R\&D Program} 
To meet the above goals and to better understand 
the technologies, a number of institutes\cite{lctpcgroups} 
have joined together as LC-TPC groups, 
with the goal of sharing information and experience in the process of 
developing a TPC for the linear collider and of providing common 
infrastructure and tools to facilitate these studies. 

The R\&D goals are as follows: 

\begin{itemize}

\item
Operate MPGDs in small test TPCs and compare with MWPC gas 
amplification to prove that they can be 
used reliably in such devices. 

\item
Investigate the charge transfer properties in MPGD structures 
and understand the resulting ion backflow. 

\item
Study the behavior of GEM and Micromegas with and without 
magnetic fields. 

\item
Study the achievable resolution of a MPGD-TPC for 
different gas mixtures and carry out ageing tests. 

\item
Study ways to reduce the area occupied per channel of the readout 
electronics by a factor of at least 10 with a minimum of material budget. 

\item
Investigate the possibility of using silicon readout techniques or 
other new ideas for 
handling the large number of channels. 

\item
Investigate ways of building a thin field cage to meet the 
requirements at the ILC. 

\item
Study alternatives for minimizing the endplate mechanical thickness. 

\item
Devise strategies for robust alignment. 

\item
Pursue software and simulation developments needed for understanding 
prototype performance.
\end{itemize}

This R\&D work is proceeding in three phases: 
\begin{itemize} 
\item[(1)] Demonstration Phase: Finish the work on-going related to many 
items 
outlined in the preceding paragraph using ``small'' ($\phi \sim 30$cm) 
prototypes, built and tested by many of the LC-TPC groups\cite{lctpcgroups}. 
This work is providing 
a basic evaluation of the properties of a MPGD TPC and 
demonstrating that the requirements outlined at the 
beginning of this section can be met. 
\item[(2)] Consolidation Phase: Design, build and operate a ``Large 
Prototype'' 
(LP). By ``Large'' is meant $\sim$1m diameter so that the detector is significantly 
larger than the current prototypes, so that: first iterations of 
TPC design-details for the LC can be tested, larger 
area readout systems can be operated and tracks with 
a large number of points are available for analyzing the various effects. 
\item[(3)] Design Phase: Start to work on an engineering design for 
aspects of the final detector. This work in part will overlap 
with the work for the LP, 
but the final design can only start after the LP R\&D results 
are known. 
\end{itemize} 

\subsection{What R\&D tests have been done?} 
\subsubsection{Overview of what has been learned by the LC TPC 
groups\cite{lctpcgroups}.} 
Several of the findings have been mentioned in the sections above. 
Up to now during Phase(1)

\begin{itemize}
\item
3 to 4 years of MPGD experience has 
been gathered,
\item
gas properties have been rather well understood,
\item
diffusion-limited resolution is being understood, 
\item
the resistive foil charge-spreading technique has demonstrated,
\item
CMOS pixel RO technology has been successfully demonstrated and
\item
design work is starting for the LP. 
\end{itemize}

\subsubsection {An example, the small-prototype TPC tests at KEK.} 
To provide a comparison and explore the potential improvements using MPGDs 
a small prototype chamber, initially with an MWPC endplate, was built at the 
Max-Planck-Institut f\"ur Physik at Munich. This chamber was 
commissioned at MPI, tested using cosmics at DESY in 
their 5T magnet and subsequently was exposed to 
many beam tests at KEK using MWPC, GEM, Micromegas\ and resistive-foil technologies. 
The chamber will be called MPT, for MultiPrototype-TPC, in the following. 

\begin{figure} 
\begin{center} 
% \begin{tabular}{c} 
%\includegraphics*[clip,width=0.3\textwidth,height=0.48\textwidth,rotate=90] 
%\includegraphics*[clip,width=1.\textwidth] 
%{tpc/tpc-pi2.ps} 
\begin{tabular}{cc}
\includegraphics*[width=0.5\textwidth]{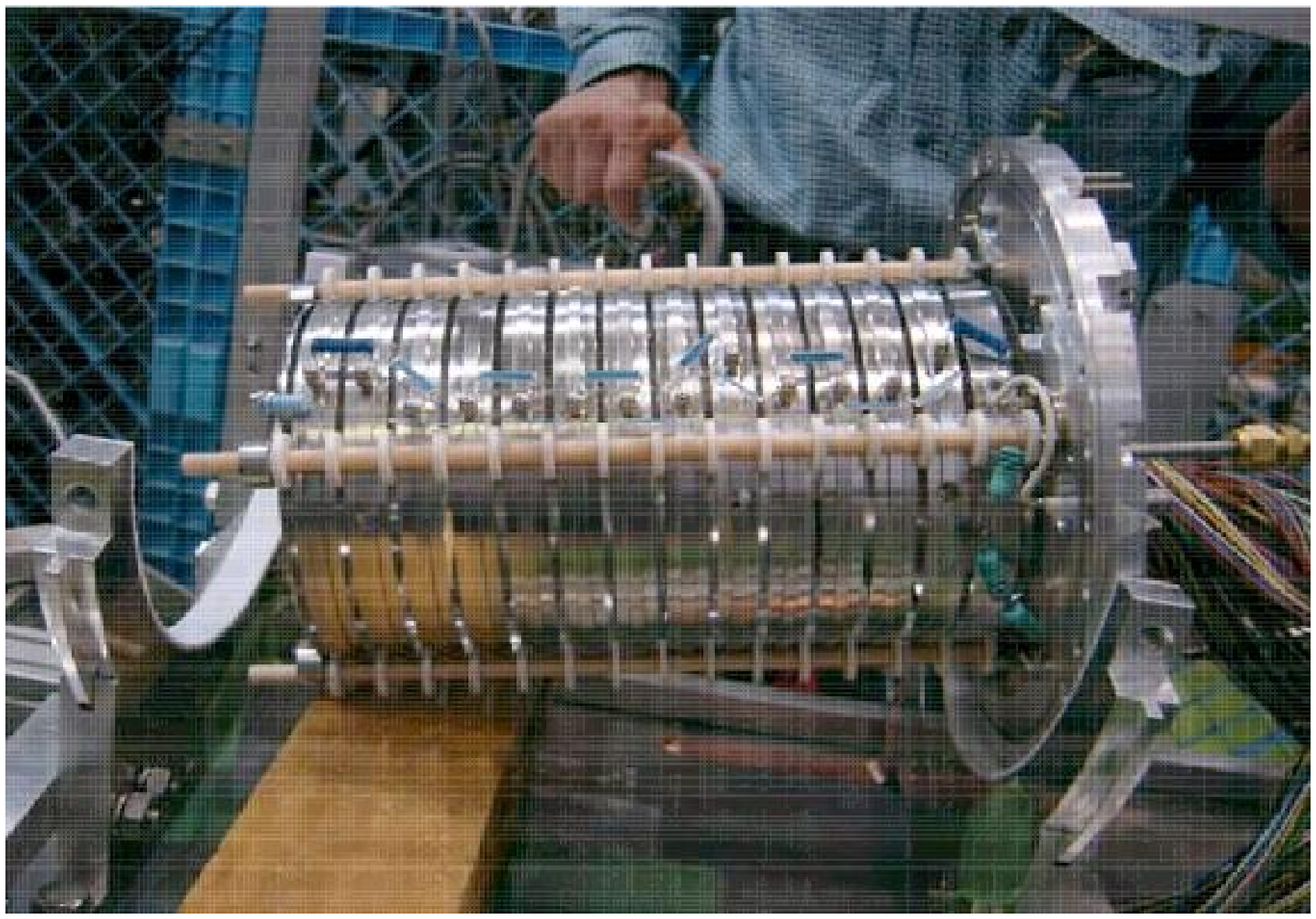}&
\includegraphics*[width=0.5\textwidth]{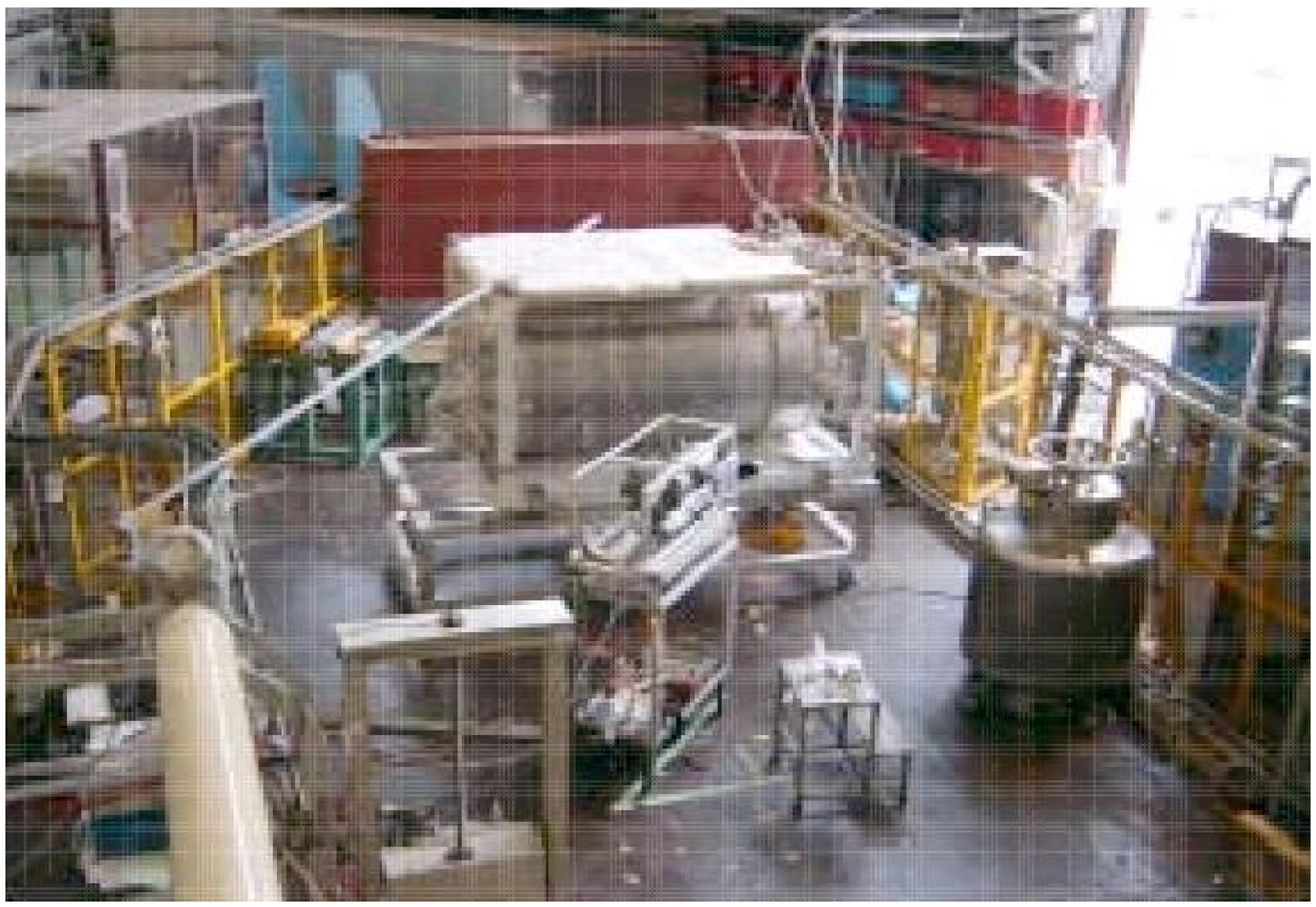}\\
\end{tabular}
%{dummy2.eps} 
% \end{tabular}{c} 
\caption{The MPT (left), used for measurements with 
cosmics in the DESY 5T magnet and in a test beam in a 1.2T magnet at KEK (right).} 
\label{fig:tpctechnical} 
\end{center} 
\end{figure}

The MPT is shown in Figure~\ref{fig:tpctechnical}. 
For the MWPC version, the MPT had 
significantly reduced wire-to-wire and wires-to-pads 
spacing to increase the achievable resolution and 
two-track separation. 
The wire readout consisted of a plane of sense wires 
with a $20$$\mu$m diameter and spaced with a $2$mm pitch. 
The sense-wire plane was placed 
$1$~mm above the pad plane onto which the 
signal was induced. 
Potential wires as used in previous TPC to define the 
structure of the electrostatic cell around the sense wire 
did not exist in this chamber. 
The pads were readout by 
the Aleph TPC electronics. 

\begin{table}[h] 
\begin{center} 
\begin{tabular}{|l|l|l|l|}\hline 
Gas Amplif. method & Gas & Pad Pitch( mm$^2$) & Magnetic Field( T )\\ \hline 
MWPC & TDR & 2.3$\times$6.3 & 0, 1, 4(Cosmic) \\ \hline 
GEM & TDR, P5 & 1.27$\times$6.3 & 0, 1 \\ \hline 
MicroMegas & Ar/C$_4$H$_{10}$ & 2.3$\times$6.3 & 0, 0.5, 1\\ 
~~w/ and w/o resistive foil& & & \\ \hline 
\end{tabular} 
%\hline 
\end{center} 
% \caption{\label{TPC_parameters} 
\caption{ 
Some of the parameters used for the MPT beam and cosmic ray tests.} 
\label{MPT_KEK} 
\end{table}

%Other things\\ 
%--Common setup for beam test (magnet,electronics,DAQ,analysis)\\ 
%--MWPC-TPC\\ 
%--GEM-TPC\\ 
%--Micromegas-TPC\\ 
%--Micromegas(GEM)-Resistive-Foil TPC\\ 
%--Comparison analytic calculation$\leftrightarrow$simulation\\ 
%--Comparison of simulations using 
%Maxwell$\leftrightarrow$3DGarfield\\ 
%--Comparison to parameterized simulation\\ 
Three different devices of gas amplification were tested 
with a combination of gas mixtures and  
read-out pad planes mounted in the MPT 
using a 4$GeV/c$ $\pi^-$ test beam; examples are listed in
Table~\ref{MPT_KEK}. 
The diffusion constant($C_D$) is 
a key parameter to determine a single-point resolution 
and was measured from using the behavior 
of signal-charge spread as a function of 
drift distance. 
The point resolution is naively parameterized by 
$\sigma_{r\phi}=\sqrt{\sigma_0^2+C_D^2/N_{eff}\times z}$, 
where, $\sigma_0$ is mainly determined by 
diffusion during gas amplification and 
by the ratio between signal-charge spread 
and geometrical pad size,
$N_{eff}$ is 
the effective number of electrons contributing to the resolution 
as determined by  
effects of statistics, gain fluctuations and efficiency in amplification,
and $z$ is the drift distance. 
The data obtained have been analyzed and compared with Monte Carlo simulation, 
preliminary results have been presented at recent conferences related 
to the ILC and the final results 
will be published in the near future. 

\subsubsection{Next R\&D steps} 
As stated above, following tests using small prototype TPCs 
(Phase(1)), 
a Large Prototype (LP) TPC will be built and tested 
(Phase(2)). This will 
be carried out in conjunction with the EUDET program\cite{eudet} 
for LC detector R\&D. 
The LP work is expected to take the next four years and will be 
followed by Phase(3), the design of the LC TPC 

Finally, to overview briefly the planning 
for the LC~TPC and the LP, the tasks have been broken down 
into ``Workpackages" as follows, and the tasks have been 
distributed among the LC~TPC groups\cite{lctpcgroups}. 
%\begin{verbatim} 

\begin{enumerate} 
\item Workpackage Mechanics 
\begin{enumerate} 
\item LP design (including endplate structure) 
\item Field cage, laser, gas 
\item GEM panels for endplate 
\item Micromegas panels for endplate 
\item Pixel panels for endplate 
\item Resistive foil for endplate 
\end{enumerate} 
\item Workpackage Electronics 
\begin{enumerate} 
\item "Standard" RO/DAQ sytem for LP 
\item CMOS RO electronics 
\item Electronics, power switching, cooling for LC TPC 
\end{enumerate} 
\item Workpackage Software 
\begin{enumerate} 
\item LP software + simulation/reconstruction framework 
\item TPC simulation, backgrounds 
\item Full detector simulation/performance 
\end{enumerate} 
\item Workpackage on Calibration 
\begin{enumerate} 
\item Field map for the LP
\item Alignment 
\item Distortion correction 
\item Radiation hardness of materials 
\item Gas/HV/Infrastructure for the LP 
\end{enumerate} 
\end{enumerate} 
%\end{verbatim} 

    % to be prepared by R.Settles and A.Sugiyama
\clearpage

%
% Calorimeter
%
\section{Calorimeter}
%\label{SectionCAL}
%-- to be prepared by T.Takeshita

%\documentclass[a4paper,12pt]{article}
%\usepackage[dvips]{graphicx}
%\usepackage{amssymb}

%\begin{document}

%\setcounter{section}{6}
%\section{Calorimeter}

\subsection{Introduction}
\begin{figure}
\begin{center}
\includegraphics[width=8cm]{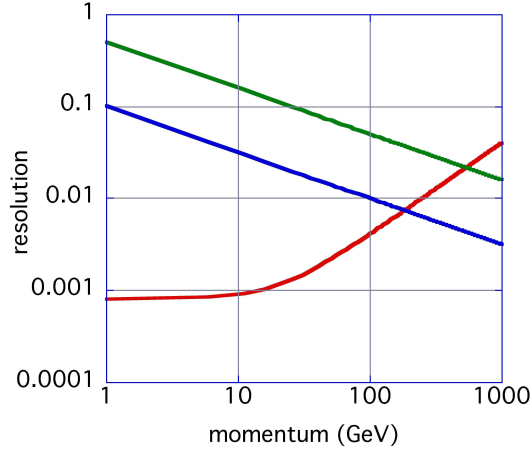}
\end{center}
\caption{Resolutions as a function of pt or Energy. The red line
shows for Tracker of $\sigma_{P_t}/P_t = 1 \times 10^{-4} p_t$,
the blue line is for ECAL of $15\% /\sqrt{E}$
and the green line is for HCAL
of $50\%\sqrt{E}$. }
\label{Resolutions}
\end{figure}

The  final state 
in the next generation electron positron collider, ILC,  
will be dominated by the jets. 
At the high energy collisions, 
collimated multi-jets will emerge 
as the  final states of quark, gluon and weak bosons. 
As we search for fundamental particles and interactions, 
the intrinsic interactions must be clarified 
from the final jets environment. 
The aim of the detector is to identify and separate 
the source of each jet. 
The collimated jet makes the detector configuration 
difficult to separate the particles in a jet, 
so that fine granularity is required. 
The information of jet-particle interactions 
in the calorimeter will help to resolve a jet. 
Not only the transverse granularity 
but also the longitudinal fine segmentation is indispensable. 
The larger the detector, the easier to resolve jet-particles. 
Thus the large detector makes the things simpler 
with less probability of overlapping of particle clusters. 
In order to resolve jet, 
the most important detector 
is the calorimeter by using PFA. 
With the help of PFA, 
one can analyze the jet components 
by ECAL for gamma contents and by HCAL 
for neutral hadron contents. 
The precision of jet energy determination is 
one of the measure of the calorimeter. 
The jet energy resolution is expressed 
as a sum of charged , gamma and neutral-hadron components.
$$
\sigma_j^2 = \sum\sigma_c^2 + \sum\sigma_{\gamma}^2 + \sum\sigma_{nh}^2,
$$
where $\sigma_j$, $\sigma_{\gamma}$, $\sigma_{nh}$
denote the energy resolution for a jet, 
a charged particle, a photon and a neutral hadron, respectively. 
The energy resolution for photon and neutral 
hadron are supposed to be scaled with $1/\sqrt{E}$. 
Thus these two equations hold. 
$$
\frac{\sigma_{\gamma}}{E_{\gamma}} =  \frac{a_{\gamma}}{\sqrt{E_{\gamma}}}
\ \ {\rm and} \ \ 
\frac{\sigma_{nh}}{E_{nh}} = \frac{a_{nh}}{\sqrt{E_{nh}}},
$$
where two parameters are independent of energy, 
$a_{\gamma}=0.15$ and $a_{nh}=0.5$.
The momentum resolution of tracker is expressed as 
$$
\frac{\sigma_c}{p_T} = 5 \times 10^{-5} p_T.
$$ 
In Figure~\ref{Resolutions}, 
these three resolutions are compared 
as a function of momentum or energy. 
At the lower momentum region, 
resolution for charged particle by tracker is 
much smaller than that for photons  and neutral hadrons by calorimeter. 
Thus we can neglect the contribution 
from the charged particles in the resolution.
Here the momentum resolution assumed is 
$0.4 \times 10^{-5} p_t \oplus 0.8 \times 10^{-3}$ 
(adding in quadrature) in red line, 
the energy resolution of ECAL is assumed to be 
$15\%/\sqrt{E}$ in blue line,
and for HCAL, $50\%/\sqrt(E)$ in green line. 

The square of jet
energy resolution is expressed as 
$$
\frac{\sigma_j^2}{E_J^2}
=
\frac{a_{\gamma}^2 \sum E_{\gamma}
+
a_{nh}^2 \sum E_{nh}}
{E_j^2}
$$
The average photonic energy in jet is
25~\% and that for neutral hadron energy is 10~\%. 
$$
\frac{\sum E_{\gamma}}{E_j} \sim 0.25\ \ {\rm and}
\ \ \frac{\sum E_{nh}}{E_J} \sim 0.10.
$$
Finally,
\begin{eqnarray}
\frac{\sigma_j}{E_J}& = & 
\frac{\sqrt{0.25 a_{\gamma}^2+0.1a_{nh}^2}}{\sqrt{E_j}} \nonumber \\
%& = & \frac{\sqrt{0.25 (0.15)^2+0.1(0.50)^2}}{\sqrt{E_j}} \nonumber \\
& = & \frac{\sqrt{0.056_{(\gamma)}+0.25_{(nh)}}}{\sqrt{E_j}} \nonumber \\
%& = & \frac{\sqrt{0.0306}}{\sqrt{E_j}} \nonumber \\
& = & \frac{0.175}{\sqrt{E_j}}. \nonumber 
\end{eqnarray}
The ideal jet energy resolution is about
17.5~\%. 
It is shown that the jet energy
resolution is dominated by the energy
resolution of neutral hadrons, although
the average energy of neutral hadron is
as small as 10~\% in a jet. 
%In the formula, however, the energy
%resolution of a jet is almost
%independent from the neutral hadron
%energy resolution. 
There is another
more important factor to deteriorate
the jet energy resolution. 
That is the
shower overlapping of multi-particle jet. 
As the energy of jet is bigger,
the overlapping effect becomes
more serious. 
The overlapping effect might
be reduced by optimizing the detector
segmentation and layer. It should be
clarified by hadron simulation,
however, there are no reliable
simulator. We have to make a prototype
to test these simulators. 
There is an
effect of neutron which will give
delayed signal after the jet
production, due to the slow neutron
generated from hadronic interaction in
the HCAL. 
The neutron may carry some
part of jet energy. In order to measure
the neutron contribution in jets , we
need proton rich active material to be
sensitive against the
neutrons. 

%%[AM]In order to measure photons precisely, we set
%%[AM]the ECAL. To measure the neutral
%%[AM]hadrons, we set HCAL behind the ECAL. 
%%[AM]The PFA needs for both ECAL and HCAL, 
%%[AM]finer granularity which had been
%%[AM]employed for recent experiment such as
%%[AM]LEP and LHC. 
ECAL is used to measure photons precisely.  
HCAL is placed behind ECAL to measure neutral hadrons.
In order to obtain the high jet energy resolution by PFA, 
both ECAL and HCAL need to have finer granularity than 
those employed for recent experiments such as LEP and LHC. 
The photon shower in ECAL and neutral hadron shower in HCAL 
should be separated and distinguished from neighboring charged particles. 
The shower clustering algorithms and fine granularity 
are crucial to realize this performance.
However, developments of algorithms are in progress,
and the granularities of calorimeters are yet to be optimized.
\par
\begin{figure}
\begin{center}
\includegraphics[width=8cm,angle=-90]{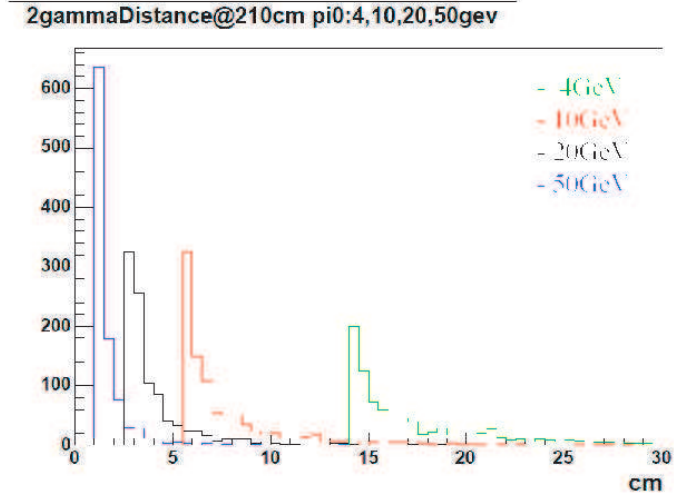}
\end{center}
\caption{$\pi^0$s with certain momentum 
are generated and decayed into two photons that 
are measured at the front surface of ECAL ($R=210$~cm) 
with their distances. 
This distance is in the horizontal axis. 
The incident $\pi^0$ momentum are 4, 10, 20 and 50~GeV. 
It is found that the $\pi^0$ of 50 GeV 
or less energy can be detected separately with the granularity 
of 1~cm by 1~cm.}
\label{Pi0distance}
\end{figure}

The average distance of two photons from $\pi^{0}$ decay
is plotted as a function of incident momentum for GLD ECAL 
in Fig~\ref{Pi0distance}. 
It is shown that at least the size of the minimum detector cell 
is smaller than 1 cm up to 50 GeV $\pi^0$. 
On the other hand, neutral hadron clustering is difficult, 
since it would contain some hadrons 
whose interaction length is much longer 
than radiation length for electromagnetic shower case. 
The cluster size of HCAL becomes bigger 
than the ECAL. 
This indicates rather bigger segmentation allowed for HCAL, 
as far as the algorithms works fine with respect to ECAL case.

\subsection{Electromagnetic calorimeter}
\subsubsection{Baseline design}
In order to optimize the GLD detector concept for the Particle Flow
Algorithm (PFA), the electromagnetic (EM) calorimeter
is required to have:
\begin{itemize}
\item
Large inner radius to separate particles with each other,
\item
Small effective Moliere radius to resolve near-by EM
     showers,
and
\item
Fine granularity both in transverse and longitudinal directions.
\end{itemize}

We propose 
a tungsten-scintillator sandwich calorimeter to fulfill these requirements.
Tungsten is chosen as the absorption material 
because of its small Moliere radius (9~mm).
Scintillator
is chosen as the active detector material.
The scintillator 
is read out by the tile/fiber technique first used for
the CDF plug upgrade EM calorimeter.
The tile/fiber system has the following features:
\begin{itemize}
\item Essentially crack-less hermeticity,
\item Capability for very fine longitudinal segmentation,
\item Capability for fine transverse segmentation, and
\item Less expensive than silicon pads.
\end{itemize}
The last item is important to realize a very large EM calorimeter.
A layer of the EM calorimeter has a 3~mm thick tungsten layer,
a 2~mm thick scintillator layer, and a 1~mm thick space for services,
resulting in one-layer thickness of 6~mm and
an effective Moliere radius of 18~mm.
Both barrel and endcaps of the EM calorimeter have 30 such layers,
corresponding to a total thickness of $26 X_0$.

\begin{figure}
\begin{center}
\includegraphics[height=18cm]{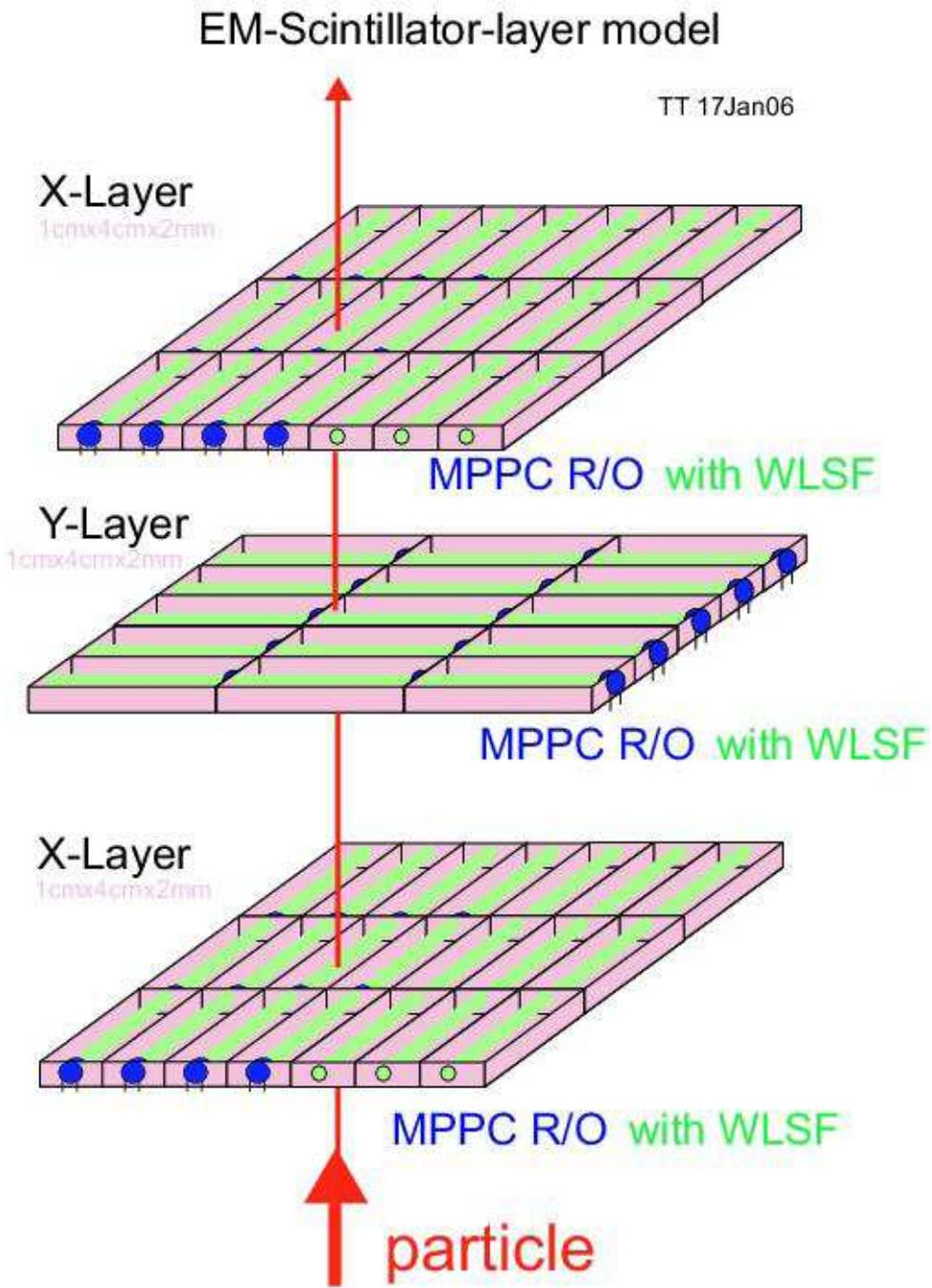}
\end{center}
\caption{Schematic view of a super-layer of the strip array EM calorimeter.}
\label{StripArray}
\end{figure}

We have two options to realize fine transverse segmentation.
\begin{enumerate}
\item 
{\bf Strip array}:
This is the baseline configuration
where a scintillator layer is composed of short strips.
A super-layer 
is schematically shown in Figure~\ref{StripArray},
where two tungsten layers and
two scintillator layers ($X$- and $Y$-layers) make a super-layer.
The scintillator strips in $X$- and $Y$-layers are arranged to
orthogonal with each other.
The width of the strips is 1~cm, about a half of the effective Moliere
radius of the EM calorimeter.
This strip array is expected to have  
an effective granularity of 1~cm $\times$ 1~cm.
To suppress ghost images which may appear when multi-particles
hit a small area of the strip array,
strips should not be long. 
The strip length is tentatively 4~cm
and will be optimized by simulation studies.
\item
{\bf Small tiles}:
This is a backup solution. A scintillator layer is composed of
small square tiles. This configuration is free from
the ghost image problem but has less granularity than the strip-array
configuration. The tile size is tentatively 2~cm $\times$ 2~cm,
and will be optimized by simulation studies.
\end{enumerate}

The number of the scintillator strips/tiles amounts to about 10 millions.
To keep the fine granularity, 
signals of all the  strips (tiles)
must be individually read out.
Photons from each scintillator strip (tile) are
read out via a WLS fiber embedded in its straight groove.
A very compact photon detector
such as SiPM\footnote{Here we collectively use the name `SiPM' 
(Silicon Photomultiplier) as the photon detectors
developed by Russian groups.}/MPPC\footnote{MPPC (Multi-Pixel Photon Counter)
is the name of the photon detector being developed by Hamamatasu
Photonics Co. }
is directly attached to the WLS fiber at the end of the strip.
The photon detector has a high gain ($\sim 10^6$) with no pre-amplifier.
The electric signals of the photon detectors are then transported
with thin flex cables to the calorimeter end where the signals are digitized. 
To keep the cost of the EM calorimeter reasonable,
the photon detectors and readout electronics must be cost-effective. 

\begin{figure}
\begin{center}
\includegraphics[width=6cm]{cal/Figs/EMC-Tilecal-mppc.epsf}
\includegraphics[width=6cm]{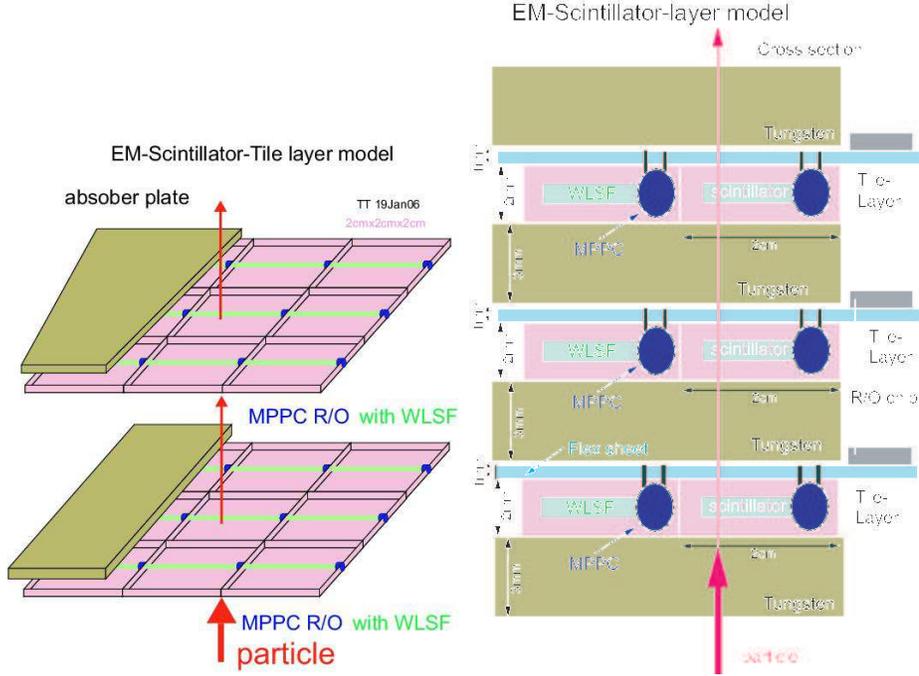}
\end{center}
\caption{Schematic view of the layout of the small-tile EM calorimeter.}
\label{SmallTiles}
\end{figure}

The parameters of the EM calorimeter are listed in Table~\ref{EMCtable}.

\begin{table}[h]
\caption{Parameters of the EM calorimeter.}
\label{EMCtable}
\begin{center}
\begin{tabular}{|l|c|}\hline 
Barrel Inner Radius & 210~cm \\
Barrel Outer Radius & 228~cm \\
Endcap Inner Radius & 40~cm \\
Endcap Face Z-location & 280~cm \\ 
\hline
Angular Coverage & \\
\ \ Full thickness & $|\cos\theta| \leq 0.990$ \\
\ \ Partial thickness & $|\cos\theta| \leq 0.991$ \\ 
\hline
Thickness & \\
\ \ Single layer & $0.86 X_0$ \\ 
\ \ 30 layers & $26 X_0$ \\
\hline
Transverse segmentation & \\
\ \ Strip-array option & 1~cm $\times$ 4~cm (tentative) \\
%\ \ Effective granularity & 1~cm $\times$ 1~cm  \\
\ \ Small-tile option &  2~cm $\times$ 2~cm (tentative) \\
\hline
\end{tabular}
\end{center}
\end{table}
%\subsubsection{Expected performance}
%\input{cal_performance}

\subsection{R\&D studies}
\subsubsection{Test beam results of strip-array ECAL}
An EM calorimeter test module, made of scintillator strips and
lead plates, was constructed and tested with beam at KEK-PS ~\cite{NIM557}.
The purpose of the test beam was to study EM shower development
extensively as well as to evaluate the performance of the
energy, position and angle measurements for incident electrons.

\begin{figure}[htbp]
  \begin{center}
    \includegraphics[width=0.8\textwidth]{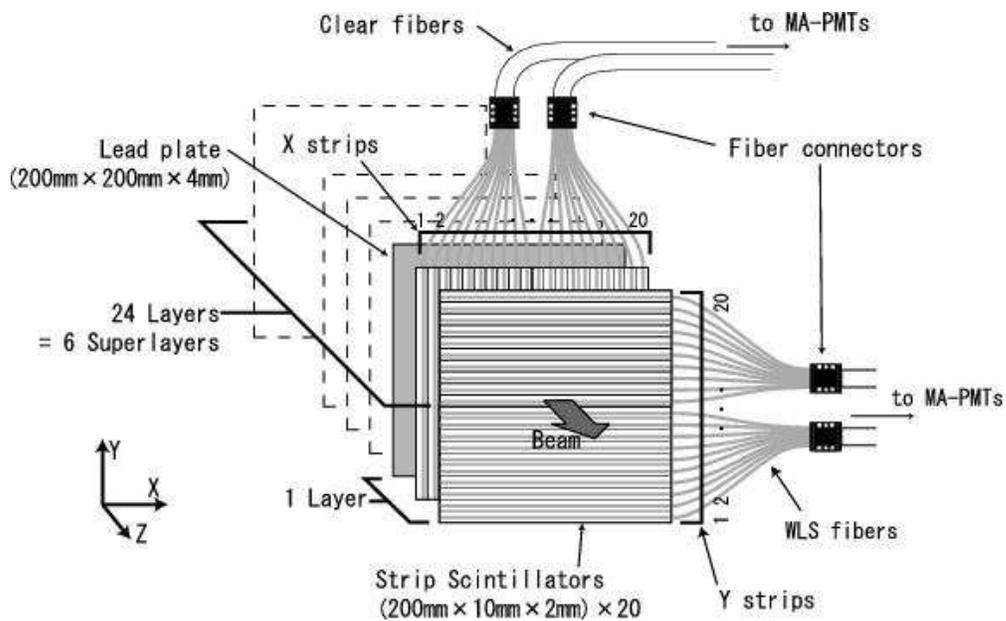}
  \caption{Schematic view of the strip-array EM calorimeter module.}
  \label{tem-module}
  \end{center}
\end{figure}

%%[AM] Instead  of present tense, past tense is used when reporting the beam test results
The test module (Figure~\ref{tem-module}) consisted of 24 sampling
layers, each of which had a lead plate (4 mm thick) and two
scintillator layers (2 mm thick each) with a cross section of
20 cm $\times$ 20 cm. In each scintillator layer, twenty strips
(1~cm $\times$ 20~cm) were laid out so as to make the two
neighboring layers orthogonal ($X$ and $Y$). Signal was read out
with multi-anode PMTs per super-layer which was defined as four
successive $X$- or $Y$-layers. In the beam test, a tracking system
was placed upstream of the calorimeter in order to reconstruct
a track of the incident particle.

By using electron beam of energies from 1 to 4 GeV/c,  performance of the calorimeter was
evaluated. The energy resolution ($sigma_{E}/E$ ) was found to be 
$((13.10 \pm 0.12)/\sqrt{E} \oplus (0.00^{+0.73}_{-0.00}))$ \%, which
was consistent with the Monte Carlo simulation.
The position resolution for the $x$ coordinate ($\sigma_X$) at the second layer (around the shower maximum) was
$((4.5 \pm 0.0)/\sqrt{E} \oplus (0.0 \pm 0.2))$ mm,
 For 4 GeV/c electrons,  the $\sigma_{X}$
becomes 2.2 mm.

\begin{figure}[htbp]
  \begin{center}
    \includegraphics[width=0.7\textwidth]{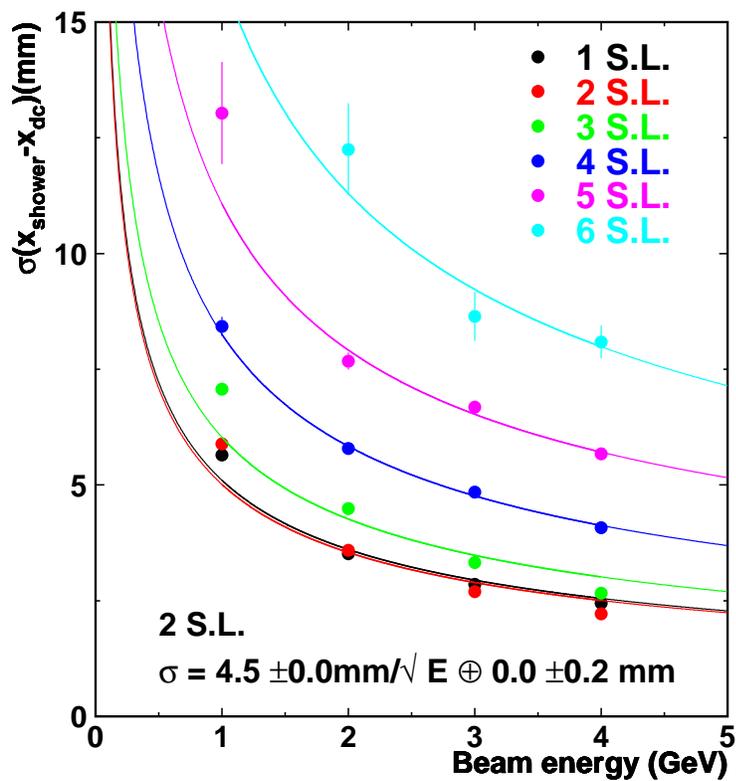}
  \caption{Position resolution as a function of the electron beam energy.}
  \label{tem-posres}
  \end{center}
\end{figure}

With the positions measured in the first four superlayers,
incident angle of the electron beam was also measured. This measurement
is important especially for detecting non-pointing photons. Obtained
angular resolution (~$\sigma$~) for 1 to 4 GeV/c electrons was
$(4.8 \pm 0.1)^{\circ} /\sqrt{E} \oplus (0.0 \pm 0.5)^{\circ}$,
and its linearity had no significant bias against the incident angle.

The response uniformity was examined by MIP signals and electrons.
At the first superlayer, the response uniformity for the response
summed over strips was  measured, and the uniformity results were  2.4\%
for the shorter (1 cm-width) direction and 1.5\% for the longer
(20 cm-length) direction. With electron beams, the uniformity
of the total energy measurement was found to be 1.1 \%.

The lateral shower profiles were also studied in detail.
We introduced $I(x)$, which was defined as the energy
deposit integrated between minus infinity and $x$,
divided by the total energy deposit:
\[
  I(x) = \int_{-\infty}^{x} \rho(x') dx' \bigg/
         \int_{-\infty}^{\infty} \rho(x') dx'
\]
where $\rho$ represents the shower density.
The origin of $x$ was set, event by event, to the particle incident
position determined by the extrapolation of the reconstructed track.

\begin{figure}[htbp]
  \begin{center}
    \includegraphics[width=0.7\textwidth]{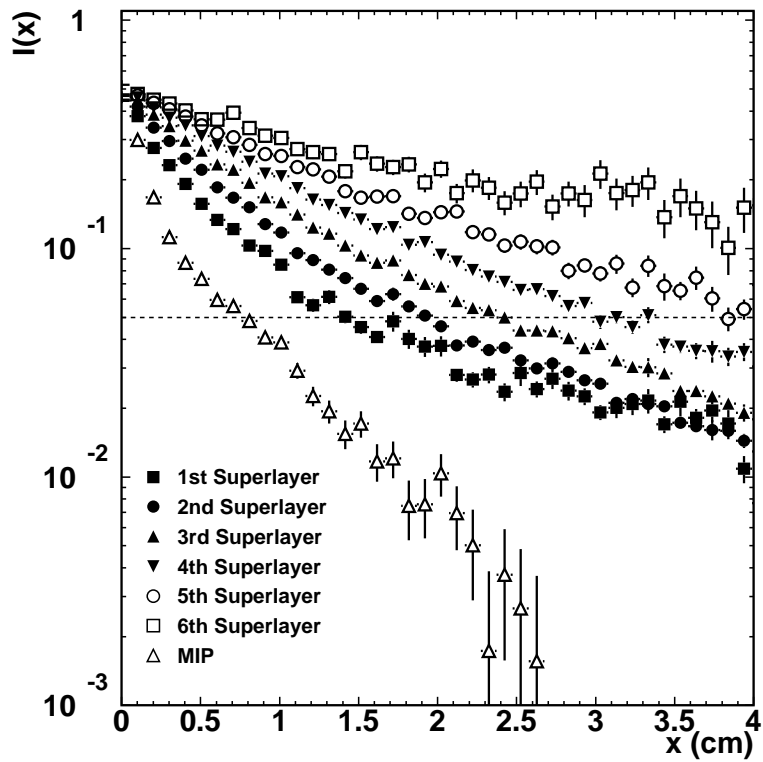}
  \caption{Integral lateral shower profiles at each superlayer
    for 4 GeV/c electrons and at the second superlayer for MIPs.}
  \label{tem-spread}
  \end{center}
\end{figure}

Figure \ref{tem-spread} shows the $I(x)$ distributions of each of the
superlayers for 4 GeV/c electrons and that of the second superlayer
for MIPs. The spread of the MIP signal mainly originated from light leakage between
strips and cross-talks in the multi-anode PMT. On the other hand,
the spreads for electrons mostly came from the EM shower spread
in the calorimeter. Taking into account these detector effects,
we can reproduce the spreads for the electron and also the expected
effective Moli\`ere radius of 27 mm with the GEANT simulation.

\subsubsection{Test beam results of small-tile ECAL}

We performed the beam-test at the KEK in March 2004 and tested
the prototype of the tile/fiber structure EM calorimeter.

Prototype module was a tile/fiber sampling-type EMCAL
using 4~mm-thick lead absorber and 1~mm-thick plastic scintillator,
total thickness is 17~$X_{0}$.
Scintillator layer was separated by 4~cm $\times$ 4~cm tile size
without the tile corner part to compose the mega-tile structure.
WLS fiber was inserted in each tile groove and read out by MAPMT.

%==== Figures of the module setup ============================
\begin{figure}[hcbt]
 \begin{center}
  \includegraphics[width=\textwidth]{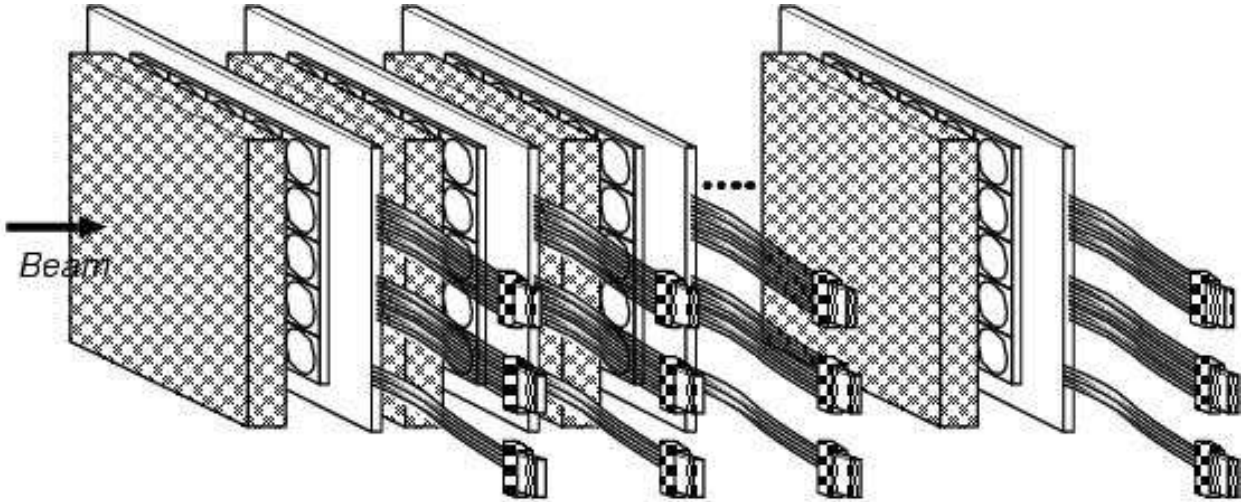}
  \caption{Front side view of the calorimeter module.
           Each layer was composed of alternately aligned lead sheet and scintillator}
  \label{Module}
  \end{center}
\end{figure}

%\begin{figure}[hcbt]
% \begin{center}
%  \includegraphics[width=0.8\textwidth]{megatile_5x5_square.eps}
%  \caption{One sheet of square mega-tile structure scintillator,
%           each tile was separated by gap and inseted WLS fiber to the groove} 
%  \label{Megatile}
% \end{center}
%\end{figure}

%===== Energy resolution section =============================
\subsubsection{Energy resolution}

We used 1 to 4 ~GeV/c electron beam for the energy resolution study. We achieved $16.0~\% / \sqrt{E} \oplus 3.5~\%$ energy resolution at center tile and the deviation from the linearity was less than 2.5\%.
Position resolution by reconstructing electron shower event was also checked and achieved 
% $\sigma_{x} = 5.4~mm $ for X-axis and 
$\sigma_{x}$ = 5.0~mm$/\sqrt{E} \oplus$ 4.9~mm.

%---- Figure of the Energy resoulution ----------------------
\begin{figure}[htbp]
\begin{center}
 \includegraphics[width=0.8\textwidth]{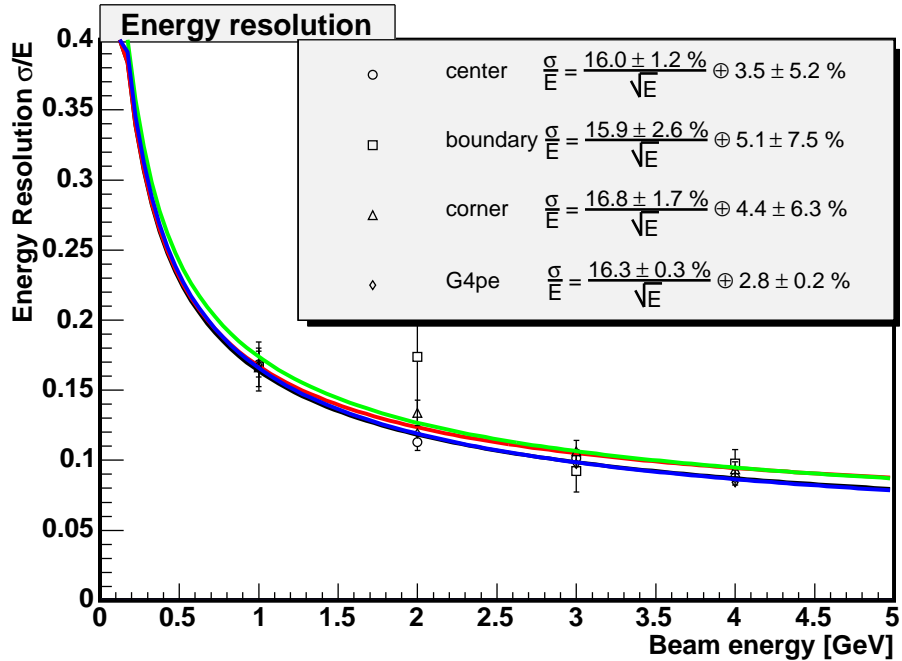} 
\end{center}
 
 \caption{Energy resolution of each tile center($\bigcirc$),
 tiles boundary($\Box$), tile corner($\triangle$) electron
 incident events and the simulation results($\Diamond$). }
 
 \label{energy_resolution}
\end{figure}

%\subsubsection{Position resolution}
\subsubsection{Unifromity mapping}

We tested the response uniformity by MIP like signals at the inside of the tile and boundary.
Typical response uniformity at the center of tile was checked by
2~mm $\times$ 2~mm mesh size
and good light containment was observed at the inside of the fiber groove.
Also we checked responces to the MIP like signal 
in the  central flat part ( -1.5~cm $\sim$ 1.5~cm )
along the x-axis and  
achieved $1.60~\%$ response uniformity in RMS.

Also we checked the response uniformity along the x-axis (-3~cm $\sim$ 3~cm) for electron signals integrated along the longitudinal direction with 4~mm $times$ 20~mm mesh size. We achieved 2.3\% uniformity.

\begin{figure}[htbp]
 \begin{center}
  \includegraphics[width=0.7\textwidth]{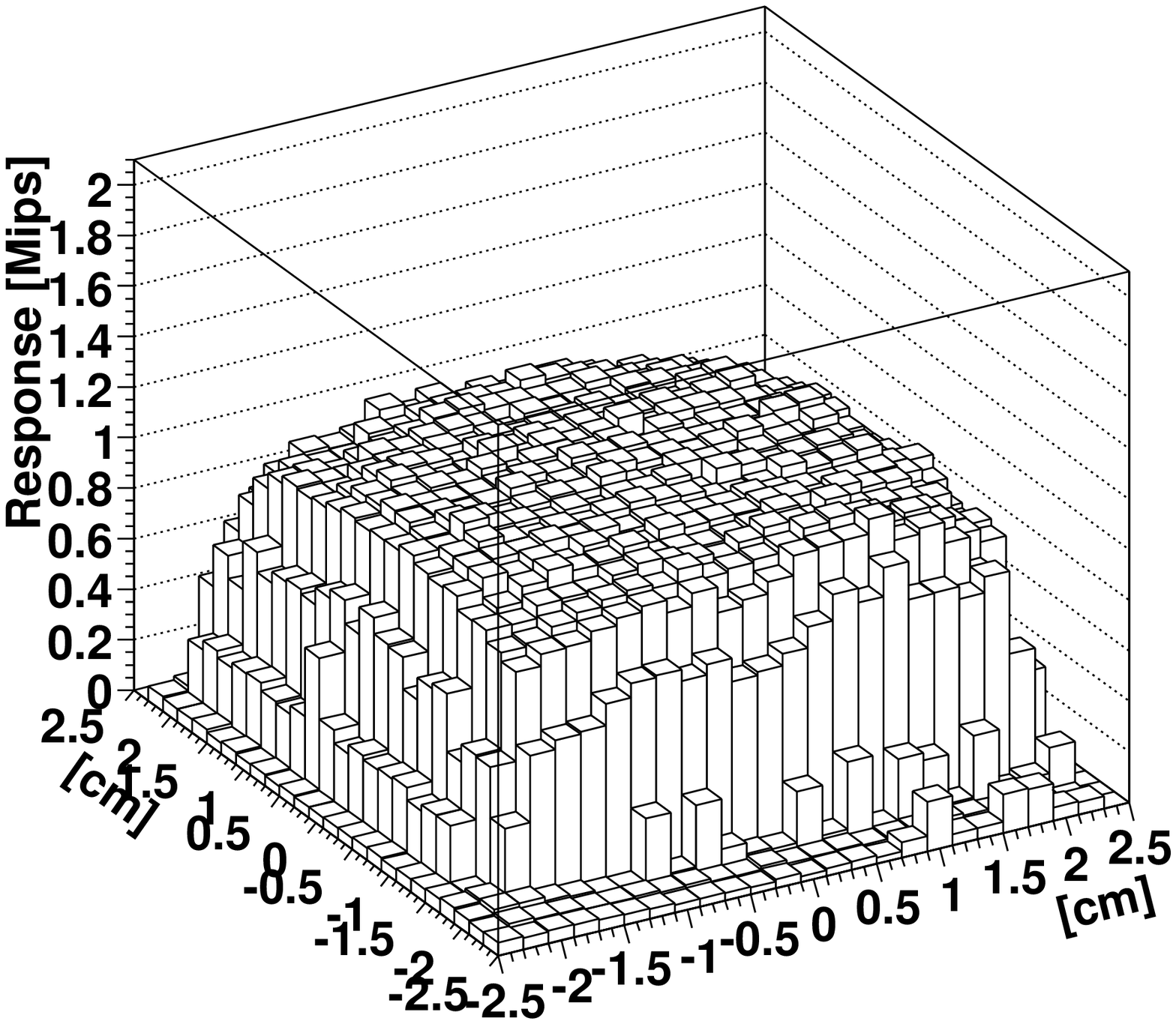}
  \caption{Response mapping by MIP like particles from one tile}
\end{center}
\end{figure}
 \begin{figure}[htbp]
  \begin{center}
\includegraphics[width=0.7\textwidth]{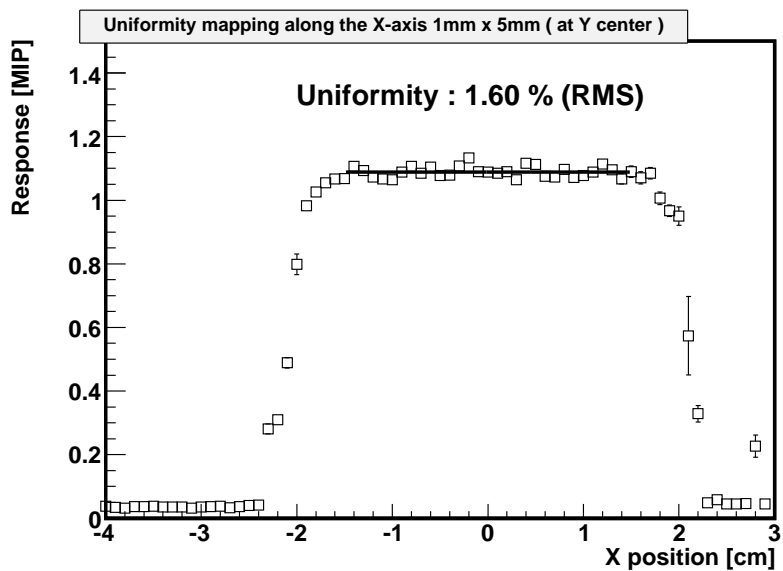}
  \caption{Uniformity mapping along the X-axis. The uniformity at the central region from  -1.5~cm  to 1.5~cm was 1.6\%.}
 \end{center}
 \end{figure}

\begin{figure}[htbp]
 \begin{center}
 \includegraphics[width=0.7\textwidth]{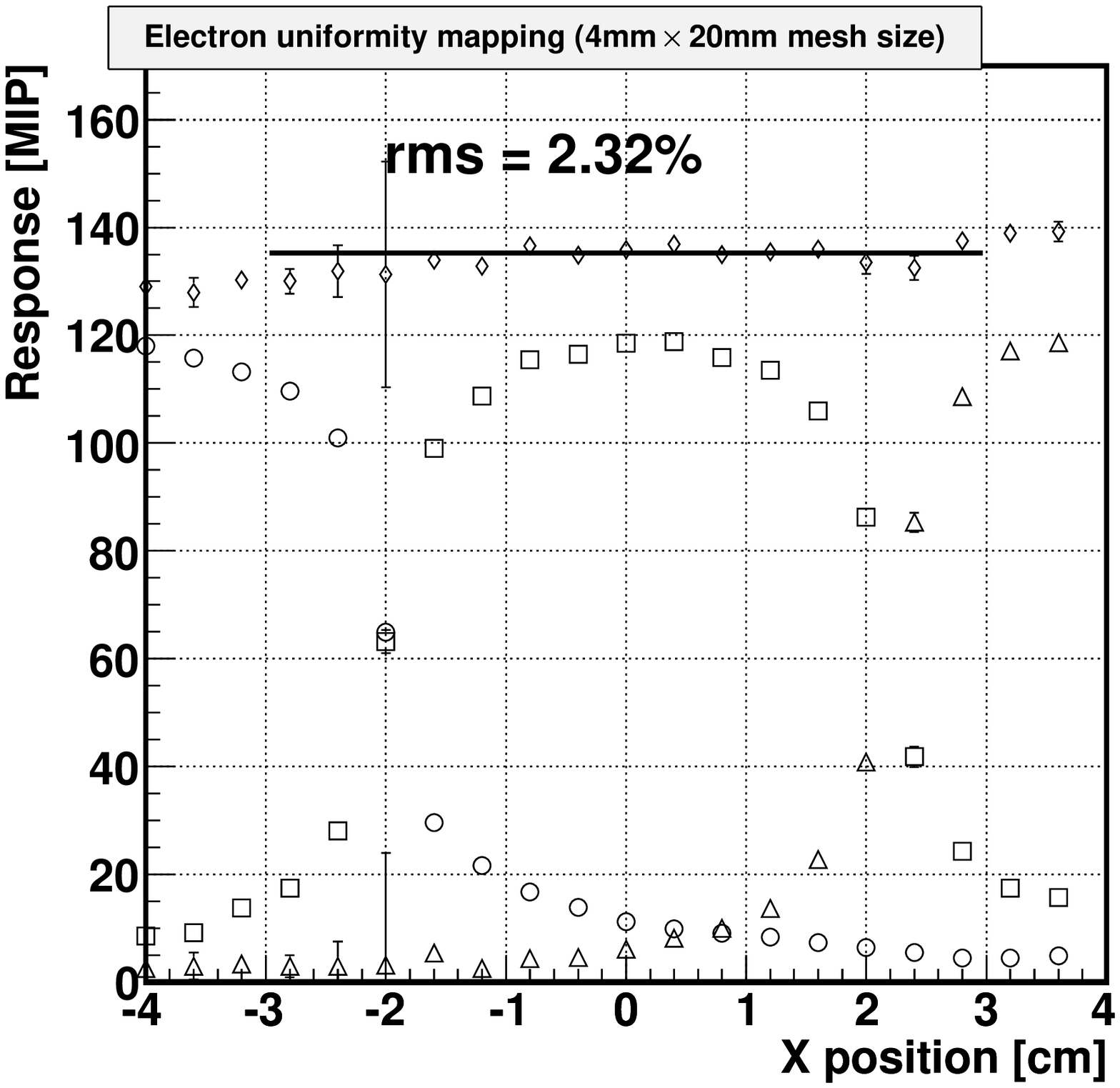}
 \caption{Electron mapping analysis along the x-axis. MIP signal was
  merged along the longitudinal direction and meshed 4~mm $\times$
  20mm(y). Uniformity achieved  was 2.3\% at the region of (-3cm $\to$ 3cm).}
  \end{center}
\end{figure}

%\subsubsection{Summary}

%=============================================================
%  Additional documents include from extenal file
%=============================================================
%\include{main}

%\end{document}
%=============================================================
%  End of document
% ============================================================
\subsubsection{Absorber plates}
\begin{figure}[h]
\begin{center}
\includegraphics[width=8cm]{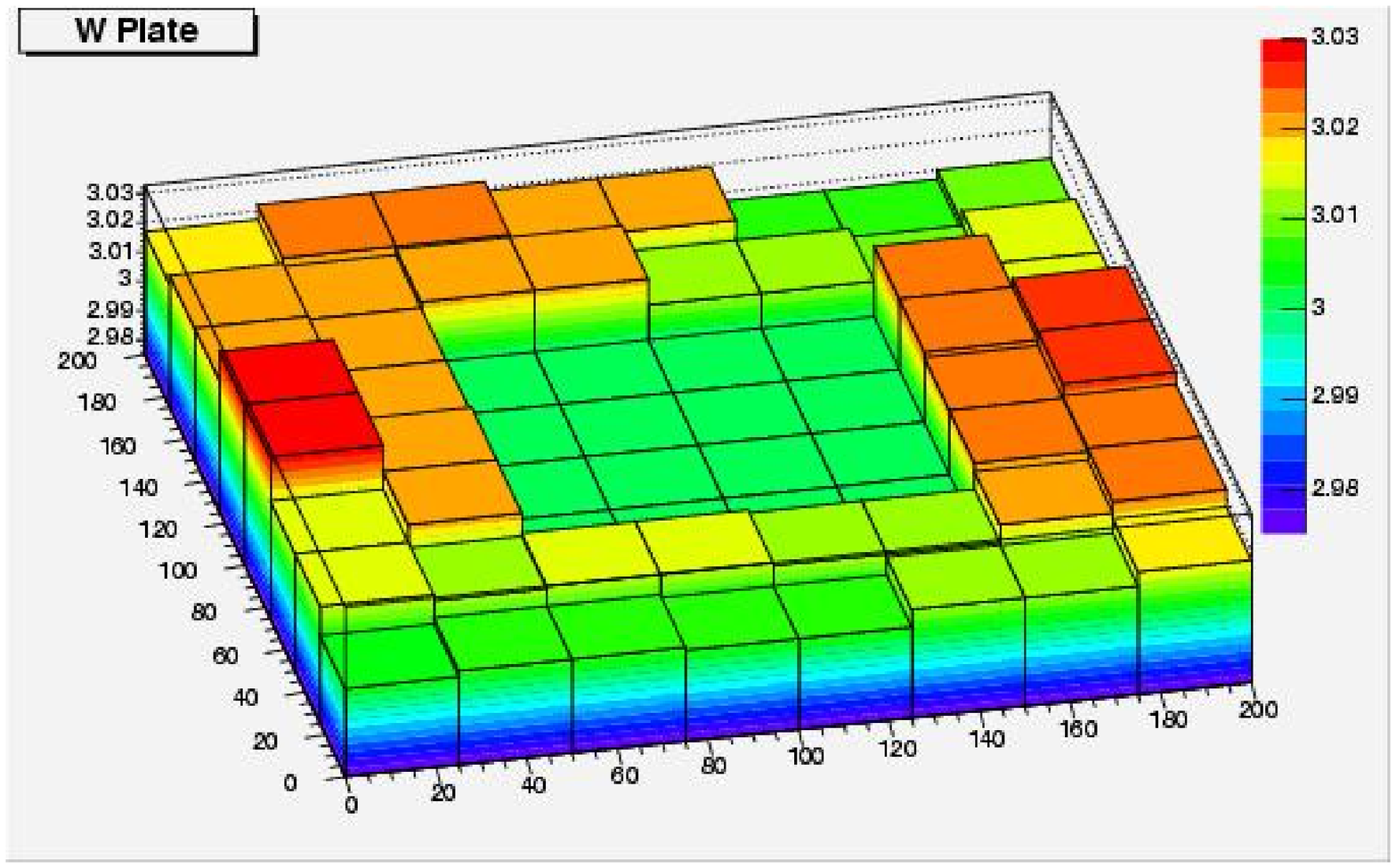}
\end{center}
\caption{Thickness of the tungsten-alloy plate measured by a digital 
micrometer. 
The measurement could not be made at the middle part of the plate}
\label{TungstenThickness}
\end{figure}

As a test sample, a tungsten-alloy plate with a dimension of 
200~mm $\times $ 200~mm $\times$ 3~mmt 
was produced by a Japanese company. 
Instead of pure tungsten, tungsten-alloy (W:Ni:Cu=95.0:3.4:1.6)
was used for much easier
manufacturing and handling.
The density of the tungsten-alloy 
was 18.20~g/cm$^3$ and the degree of hardness was
310~Hv.
As the uniformity of the absorber material is important for good energy
resolution,
the thickness of the plate was measured at many points
using a digital micrometer.
The result is  
shown in Figure~\ref{TungstenThickness}.
The average, minimum, and maximum thickness 
was 3.016~mm, 3.006~mm, and 3.032~mm, respectively,
and the RMS was 0.007~mm (0.23~\%).
If the uniformity of the thickness can be kept at this level,
the effect to the energy resolution should be negligible.

%For the GLD EM calorimeter we need a total of about 200~tons of
%tungsten-alloy plates.
%According to the Japanese company, cost reduction 
%down to about 10~JPY/g would be possible, if the plates
%are mass-produced. Therefore a rough cost estimate of the tungsten-alloy
%plates is 2 B JPY.
%For further cost reduction, 
%mass production of the plates 
%at Korean or Russian companies is also investigated.

\subsubsection{Scintillator strips/tiles}
\begin{figure}
\begin{center}
\includegraphics[width=16cm]{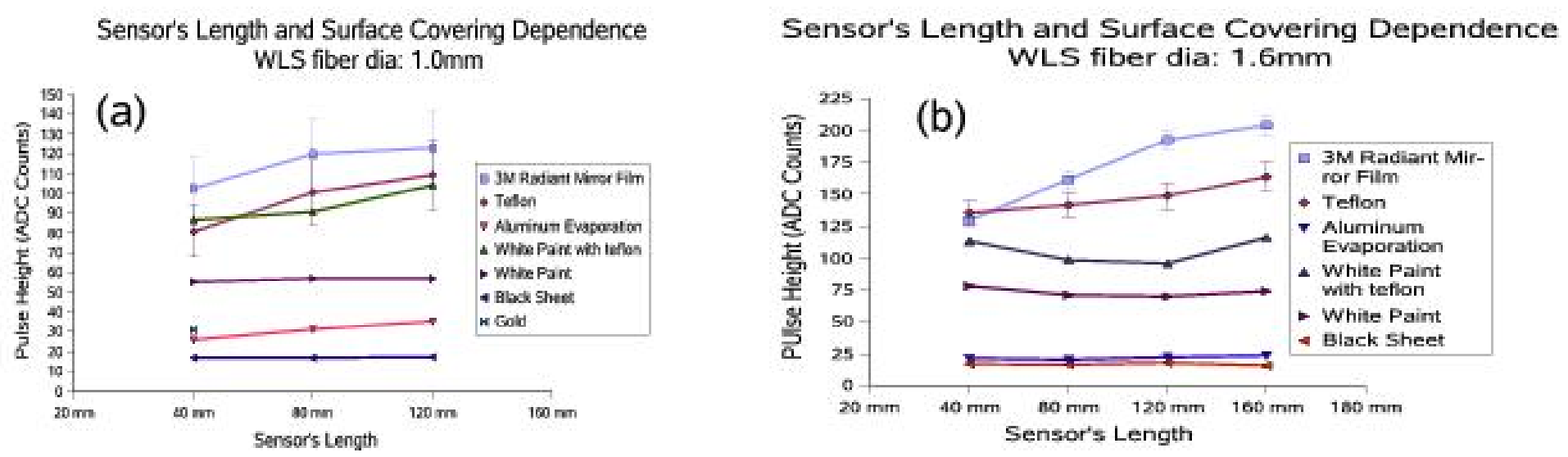}
\end{center}
\caption{Strip length and surface covering dependence
on pulse height through (a) 1.0~mm diameter WLS fiber and
(b) 1.6~mm diameter fiber.}
\label{Edith09}
\end{figure}

Small plastic scintillators of different sizes, 
thickness and wrapping reflectors are systematically studied. 
A 16-channel photomultiplier tube (PMT) is used as read out
through the 1.0~mm and 1.6~mm diameter wavelength shifting (WLS) fiber. 
The scintillating light generated by particles from the collimated $^{90}$Sr source 
is collected from the scintillator 
by the WLS fiber and converted into electrical signals 
at the PMT. 
%These electrical signals are then 
%being passed through NIM and CAMAC modules and 
%are analyzed using the ROOT macro analysis package. 
The best light yield scintillator 
is determined by comparing 
the measured pulse height of each 10$\times$40$\times$2~mm$^2$ 
strip scintillator covered with 3M reflective mirror film, 
teflon, white paint, black sheet, gold sputtering, 
aluminum evaporation and white paint+teflon. 
The position, length and thickness dependence 
of the 3M reflective mirror film 
and teflon wrapped strip scintillators are measured. 
The number of photons being emitted by the scintillator 
is also determined. 
As shown in Figure~\ref{Edith09}
the 3M radiant mirror film-wrapped scintillator 
has the greatest light yield with an average of 9.2 photoelectrons.
It is observed that light yield slightly 
increases with scintillator's length, 
but increases to about 100~\% 
when WLS fiber diameter is increased from 1.0~mm to 1.6~mm. 

\begin{figure}
\begin{center}
\includegraphics[width=12cm]{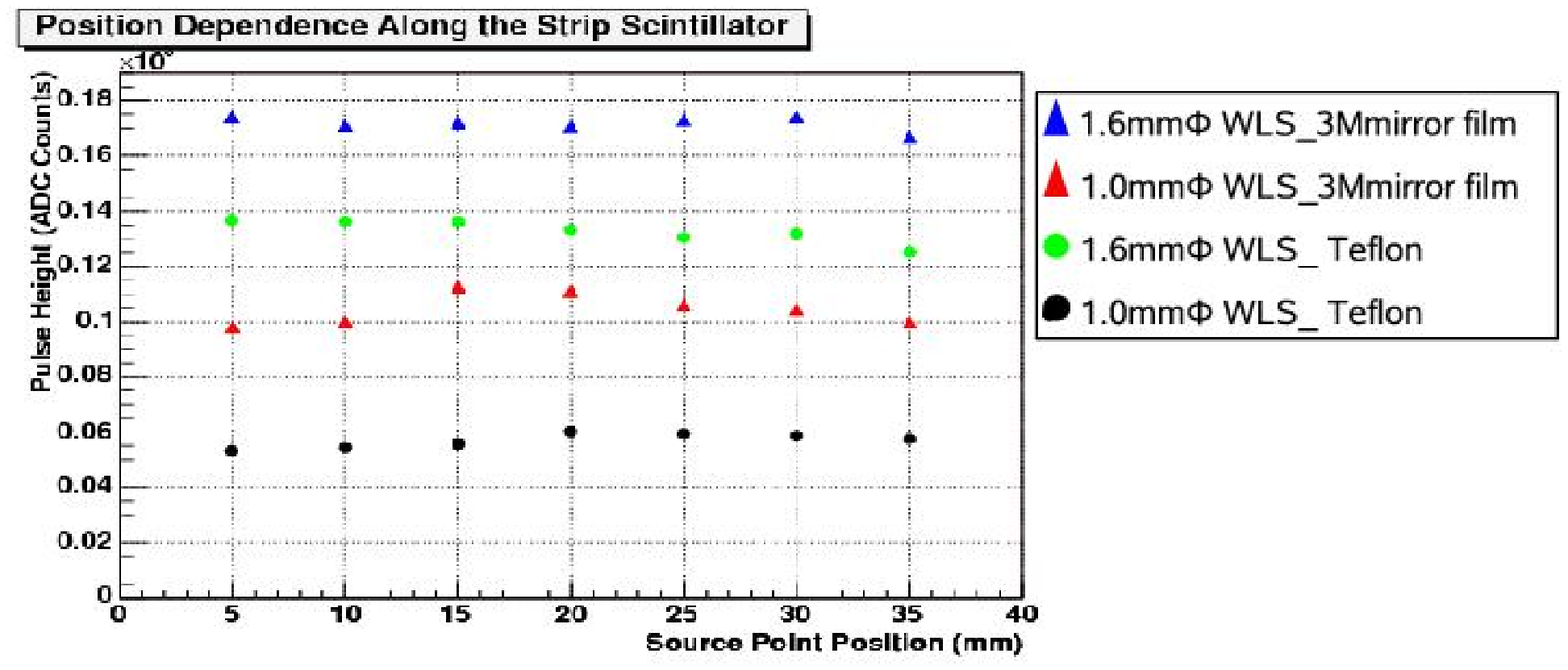}
\end{center}
\caption{Position dependence of pulse height along the
10$\times$40$\times$2~mm$^3$ strip 3M radiant mirror film wrapped
scintillator.}
\label{Edith10}
\end{figure}

\begin{figure}
\begin{center}
\includegraphics[width=12cm]{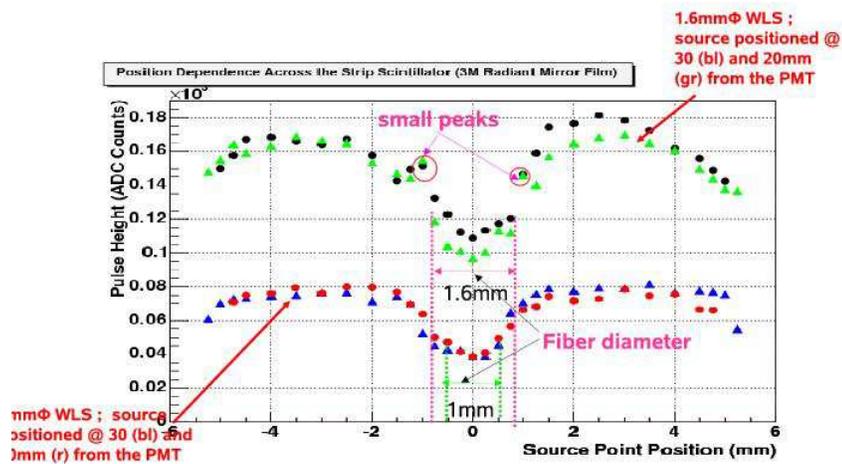}
\end{center}
\caption{Position dependence of pulse height across the
10$\times$40$\times$2~mm$^3$ strip 3M radiant mirror film wrapped
scintillator.}
\label{Edith11}
\end{figure}

Position dependence of the pulse height
of the
10$\times$40$\times$2~mm$^3$ scintillator strip wrapped with
3M radiant mirror film was measured.
Figure~\ref{Edith10} shows no significant change
in pulse height along the scintillator strip.
This means that there is uniformity of the light transmission
from the sensor to the PMT along the scintillator strip.
Figure~\ref{Edith11}
reveals the position dependence
on pulse height across the strip scintillator.
A  dip  is observed across the strip which is 40~\%
of the maximum pulse height that corresponds
to the scintillator s thickness (300~$\mu$m)
just below the 1.6~mm fiber.
Small peaks are noticeable near the fiber with 1.6~mm WLS fiber.

\subsubsection{Photon sensors}
As the whole calorimeter is designed to be located inside the
superconducting solenoid,
photon detectors should be operational in strong magnetic field.
In the case of crystal calorimeter, the light yield is large and popular
PIN silicon photo diode or APD can be used. 
However, 
in the case of sampling calorimeter, the light yield is relatively poor
and high-gain high-sensitivity 
photon detectors are needed. 
The photon detectors must also be very compact to be used for the
fine-segmented calorimeter. 

\begin{figure}
\begin{center}
\includegraphics[width=8cm]{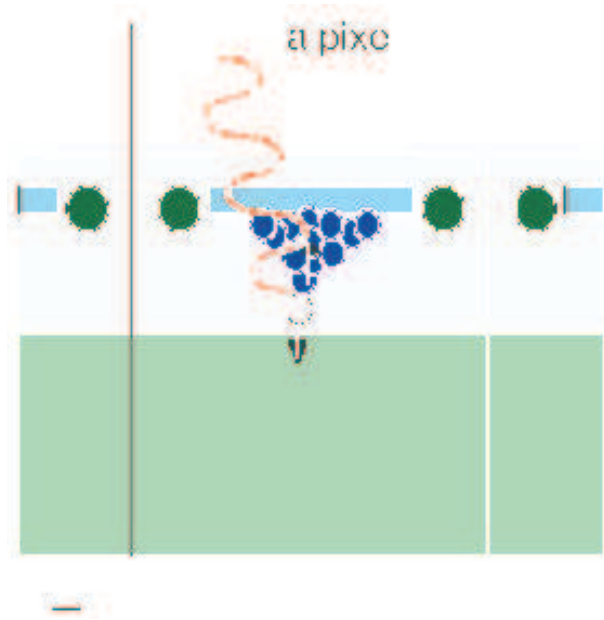}
\end{center}
\caption{Schematic view of SiPM/MPPC structure.}
\label{MPPCStructure}
\end{figure}

A new silicon photon detector, Silicon Photomultiplier (SiPM),
has been developed by several Russian Groups. 
SiPM consists of many micro APD pixels being operated in limited Geiger
mode,
and the output signal is a sum of the signals from pixels fired by
photons.
A schematic view of the structure is shown in
Figure~\ref{MPPCStructure}.
A similar photon detector is also being developed with a name of
Multi-pixel Photon Counter (MPPC) by Hamamatsu Photonics.
Evaluations of some samples are in progress:
SiPMs with 1156 pixels and
MPPCs with 100, 400 and 1600 pixels.
%MPPCs with more pixels are not available yet.
%Pulse shapes of the SiPM/MPPC are shown in Figure~\ref{NoisePulseShape}. 
%The data were taken with no light source, and the observed signals
%are caused by thermal electrons.
%The first thick peak corresponds to a signal of single pixel fired.
%\begin{figure}
%\begin{center}
%\includegraphics[width=10cm]{cal/Figs/NoisePulseShape.eps}
%\end{center}
%\caption{Pulse shapes of a SiPM with 1156 pixels (upper left),
%an MPPC with 100 pixels (upper right), 
%and an MPPC with 400 pixels (lower left).  
%Note that the vertical and horizontal scales are different
%with one another.}
%\label{NoisePulseShape}
%\end{figure}

The new silicon photon detectors 
are potentially ideal for tile/fiber readout
because of the following characteristics.
\begin{itemize}
\item
The silicon photon detectors are insensitive to magnetic field at all.
\item 
They are very compact ($100 \sim 1000$~pixels/1~mm$^2$)
and can be directly attached to the WLS fiber at the end of the
scintillator strip (tile).
No clear fibers are needed to transport the scintillation light 
to the photon detector.
\item
They have high gain ($\sim 10^{6}$) at bias voltage
of $25 \sim 80$~V.
High gain amplification is not necessary in readout electronics.
\item
They have photon counting capability, good at measurement
of low intensity light. 
The photon counting capability of an MPPC with
1600 pixels is demonstrated in Figure~\ref{PhotonCounting}
% ,where up to $\sim$30-th peak is clearly visible. 
\begin{figure}
\begin{center}
\includegraphics[width=10cm,angle=-90]{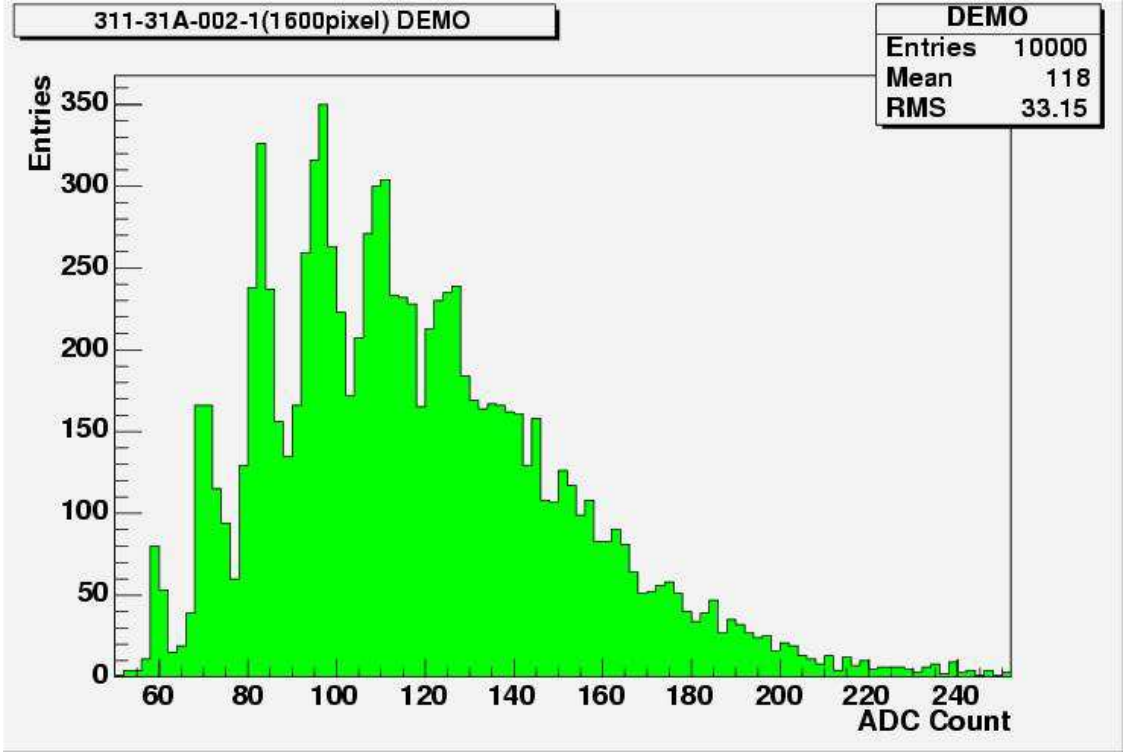}
\end{center}
\caption{Pulse height distribution of an MPPC with 1600 pixels
with LED light.}
\label{PhotonCounting}
\end{figure}
\item
The signal output is fast enough for the timing measurement.
%According to Ref.~\cite{SiPM},
A time resolution of 30~ps is obtained for 10 photoelectrons 
with a SiPM.
\item 
The thermal noise rate
of one photo-electron (one pixel) 
equivalent signal is rather high, typically about 1~MHz.
This is not a serious problem because the bunch crossing timing
is exactly known at the collider and a narrow gate can be used 
for charge integration. 
\end{itemize}
However, SiPM/MPPC is a new and developing photon detector
and there are many R\&D items to be performed.
\begin{itemize}
\item
According to a result of a test bench measurement where
a 2~mm-thick scintillator strip, a 1~mm$\phi$ WLS fiber, and
a conventional PMT ($QE \sim 20$~\%) were used, 
the average number of photoelectrons was 2.5.
To have more photon statistics, we will use 1.5~mm$\phi$ diameter WLS
fibers. Therefore SiPM/MPPC with larger size to fit to the fiber
diameter should be developed.
\item
Photon detection efficiency ($PDE$) of SiPM/MPPC
is defined by the following formula:
$$
PDE = \epsilon_{geom} \times {QE} \times \epsilon_{G} ,
$$
where $\epsilon_{geom}$ is the geometrical efficiency 
(a fraction of active pixel area to the total device area),
${QE}$ is the wavelength-dependent quantum efficiency,
and $\epsilon_{G}$ is the probability for a carrier created in active
pixel area to initiate a Geiger-mode discharge.
The photon detection efficiency of the existing SiPM/MPPC is typically
20\%.
Higher $PDE \sim50$~\% is desired to have more photon statistics.
\item
Dynamic range of SiPM/MPPC is limited by the total number of pixels
in the device, $N_{tot}$.  Neglecting cross talk effects, the relation of the output signal
(the number of fired pixels, $N_{pix}$) and 
the number of photoelectrons, $N_{pe}$, is expressed as follows:
$$
N_{pix} = N_{tot}\exp(-N_{pe}/N_{tot}).
$$ 
For hadronic calorimeter and muon detector,
$N_{tot} \sim 1000$ may be sufficient.
For the EM calorimeter
where electromagnetic showers make dense energy deposit in localized
     region,
much more pixels is needed. According to a Monte Carlo simulation,
at least $N_{tot} \sim 5000$ is necessary for the EM calorimeter
not to degrade the energy resolution.
\end{itemize}

We measure the basic properties such as
\begin{itemize}
\item
Gain and linearity,
\item
Noise rate
\item
Photon detection efficiency and
\item
Cross talks.
\end{itemize}
They are dependent on bias voltage and temperature.

To understand the performance of SiPM/MPPC, it is also important
to measure 
\begin{itemize}
\item
pixel-to-pixel difference and
\item
position-dependent behavior in a pixel.
\end{itemize}
We perform such measurements with a YAG laser system in Niigata
university, 
where the minimum spot size and the position precision of the laser beam
are 2~$\mu$m and 2~$\mu$m, respectively.
An improved test bench with a new laser system is being built at KEK
dedicated to the measurements.
%In addition, 
%we perform uniformity measurement with a YAG laser system 
%The wavelengths are 532~nm and 1064~nm.
%The minimum spot size and the position precision of the laser beam
%are 2~$\mu$m and 2~$\mu$m, respectively, well below 
%the pixel size of SiPM/MPPC.
%An example of the measurement is shown in Figure~\ref{MPPCPixelGain}
%and \ref{MPPCPixelEff},
%where gains and $PDE$s of 50 pixels 
%out of 100 MPPC pixels were measured for the first MPPC sample.
%This MPPC sample was found to have good uniformity in gain,
%but poor uniformity in $PDE$.
%The uniformity is further improved to level of $\sim 2$~\%,
%and the PDE has also been much improved in recent MPPC samples. 
%\begin{figure}
%\begin{center}
%\includegraphics[width=6.5cm]{cal/Figs/position100_PH.eps}
%\includegraphics[width=6.5cm]{cal/Figs/fluc_PH.eps}
%\end{center}
%\caption{Left:The gains of the 50 pixels of the first MPPC sample
%with 100 pixels and 
%right: the distribution of the 50 pixel gains.}
%\label{MPPCPixelGain}
%\end{figure}
%\begin{figure}
%\begin{center}
%\includegraphics[width=6.5cm]{cal/Figs/pe_100HPK_eff2pe.eps}
%\end{center}
%\caption{The photon detection efficiency 
%of the 50 pixels of the first MPPC sample
%with 100 pixels.}
%\label{MPPCPixelEff}
%\end{figure}

Some other properties are still to be tested. 
\begin{itemize}
\item
Long term stability,
\item
Radiation damage.
\end{itemize}
We need more samples to do
the tests.

We feed back our test results to the companies for further improvements.
Good communication between experimental groups and companies is essential.

\subsubsection{Electronics}
The calorimeter signals are sent from the photon detectors
to the the front-end electronics through thin cables. As the gain
of the new silicon photon sensor is high enough, we do not need
any preamplifier. The output signal is so fast that the timing
measurement can be performed with a good accuracy. An important thing
is that there are huge (several ten million) channels,
thus low power consumption and small dimension are very desirable.

The readout electronics should digitize an amplitude/current and
timing information of the signal. The timing measurement is
necessary to identify the bunch crossing, and the required
resolution of about 10 ns can be obtained without difficulty.
The about  1 ns time resolution will be required if we want to
distinguish the energy deposits of slow neutrons.
Moreover, with such accurate time resolution one may identify
cosmic-ray muons.
The time resolution of the optical system is dominated by that
of the wavelength shifting fibers, but it is still possible to
achieve it with the appropriate electronics.

The digitization will be done by either analog-to-digital converter (ADC)
with a  time-stamping capability or FADC. The dynamic range
for the energy measurement is limited by number of cells in the
photon sensor and the electronics. Note that the former has a
saturation effect even below the overflow. In order to discriminate
between a MIP and sensor noises clearly, A MIP signal should produce at least 10 hit cells at the photon sensor. On the other hand,
the pulse height at the shower maximum is expected to be around 200 times that of the MIP signal. Therefore, the number of cells should be more than 2000, and
the ADC or FADC is required to have 10-bit or more accuracy.

Another important role of the electronics system is to supply
power to the silicon photon detectors. The device has very narrow
range of the bias voltage for the operations ($<$ 1 V), thus
the power supply should be very stable. In addition, the operational
voltage range differs between pieces; it is more than a few V or
10 V in some cases. A possible solution would be to group the photon
sensors such that the voltage range in a group can be small
enough to be adjusted by, e.g. a digital-to-analog (DAC) converter
within 5 V.

The power supply as well as the ADC or FADC will be implemented
in an ASIC, which can have a buffer for storing data during
a bunch train and a gigabit Ethernet to send data to the
trigger/DAQ system. Some components for the ASIC exists, and
the integration for all features are under development.

%%\subsubsection{Calibration system}
%%\subsubsection{Mechanical structure}

\subsection{Hadron calorimeter}
\subsubsection{Baseline design}
\begin{figure}
\begin{center}
\includegraphics[width=12cm]{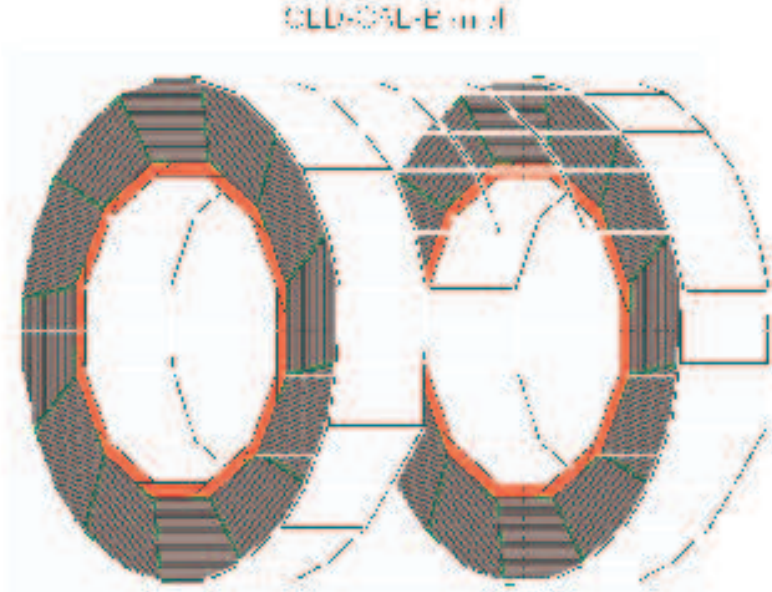}
\end{center}
\caption{Schematic view the GLD barrel calorimeter.}
\label{HCALfig0}
\end{figure}

\begin{figure}
\begin{center}
\includegraphics[height=20cm]{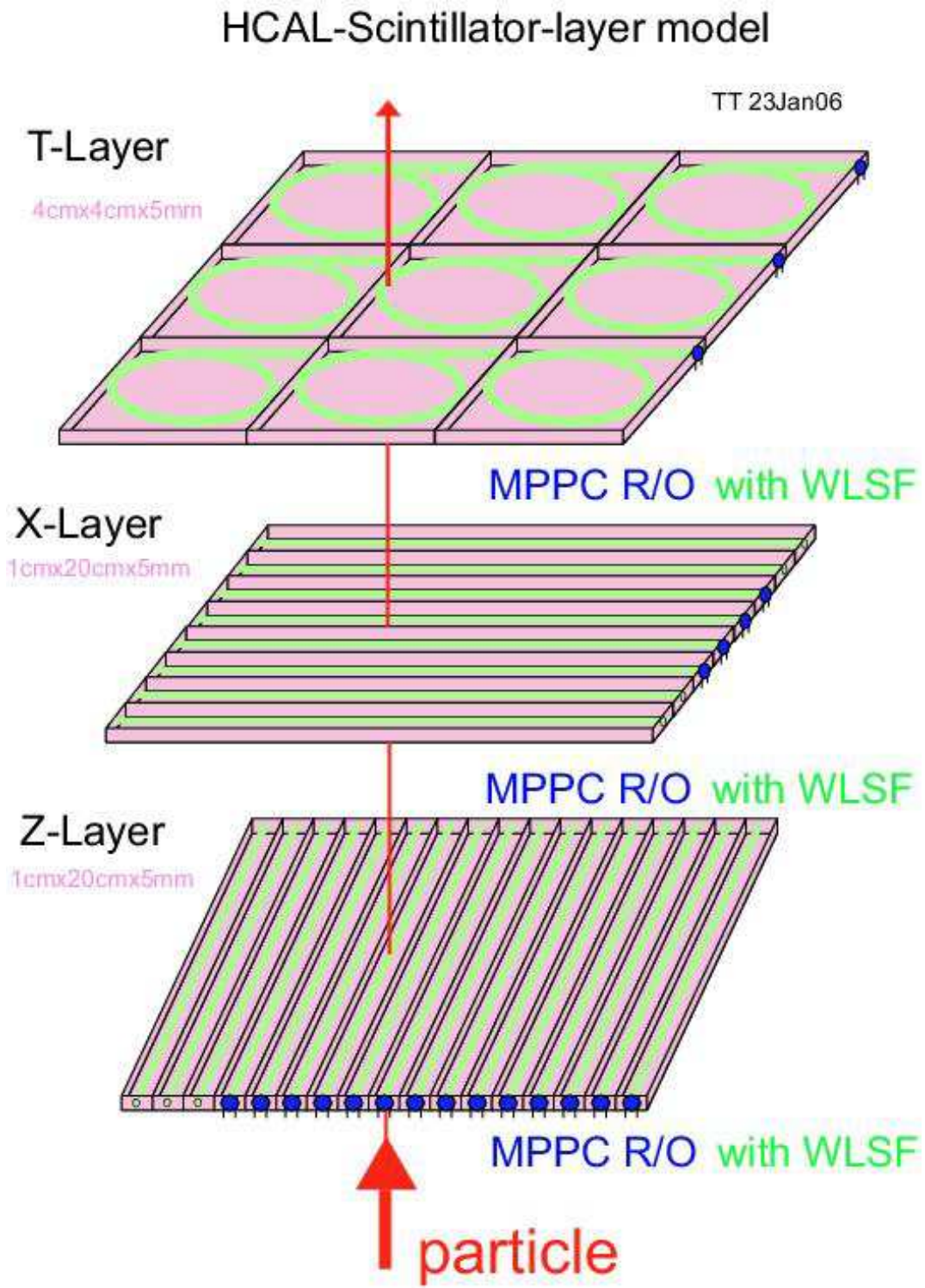}
\end{center}
\caption{Schematic view of GLD-HCAL super-layer is shown. 
It consists of three layers of scintillator and lead.}
\label{HCALfig1}
\end{figure}

%\input{hcaldesign}
%HCAL
Global structure of the GLD calorimeter detector is shown 
in Figure~\ref{HCALfig0}.
%Base line design
The GLD hadron calorimeter consists of absorber (lead plate)  
and plastic scintillator sandwich. 
The plastic scintillator has a shape  of strip, 
which has a straight gloove where a wave length shifting fiber (WLSF)
is embedded. 
This readout scheme gives freedom for the strip length and width,  
since the pulse height from the WLSF does not 
depend on  the position where a particle is passing by.  
As we are still trying to optimizing the strip length via simulation, 
we set the strip length to be 20~cm as the  starting point. 
The width of the strip is set to be 1~cm, similar to the ECAL case, 
in order to achieve effective granularity of 1~cm $\times$ 1~cm with orthogonal
strip layers (Figure~\ref{HCALfig1}).
.
The thickness of the scintillator is 
determined to have enough light output 
from minimum ionizing particles by newly developed photon sensors
which has a pixel structure. 
As the base line design,  the thickness is set to be 5~mm.

Lead is selected as the absorber material 
so as to achieve hardware  compensation with scintillator. 
Since the energy resolution is governed by fluctuation 
of electromagnetic shower to the purely  hadronic interaction, 
in order to minimize its fluctuation,  
electromagnetic shower output should be suppressed. 
According to our test beam results,
the hardware compensation is achieved by adopting the ratio of
lead : plastic scintillator = 9.1 : 2.

\subsubsection{Choice of active material}
Plastic scintillator is reliable and stable material 
for the particle detection. 
For a long time the plastic scintillator has been  used
to detect and measure not only the charged particles  but also
neutrons. 
One of the reasons to choose the plastic scintillator 
as the active media of the calorimeter sites on its reliability. 
Since we need fine granularity  as ever constructed before,  
we have to be careful for the uniform response of each sub component  
in numerous number of pieces. 
The plastic scintillator is one of the  candidates 
to satisfy this  requirement, since the strip structure  
has flexibility for its size 
by using a wave length shifting fiber as scintillation light read out.   
A problem might be an increase of number of channels 
of photon detector for the read out system. 
Recent  development of new type silicon photon sensor 
is quite promising  for our detector. 
The progress is already discussed in the ECAL section. 
The light signal of scintillator is fast enough
to measure  the timings, 
with which we are able to reject background hits and 
to recognize neutron hits in the calorimeter.

The active material which contains hydrogen is sensitive to  neutrons.
It is important to detect and measure neutrons
not only for the compensation 
but also for the total energy measurement of  hadrons.
Those neutrons appear at the tail of the hadron shower process from decays of excited nuclei.

Generations of their signals usually delays from a particle injection to the calorimeter. 
In simulation program used for evaluations of calorimeter performance, the nuclear excitation and its decays has to be taken into account properly.

\begin{figure}
\begin{center}
\includegraphics[width=12cm]{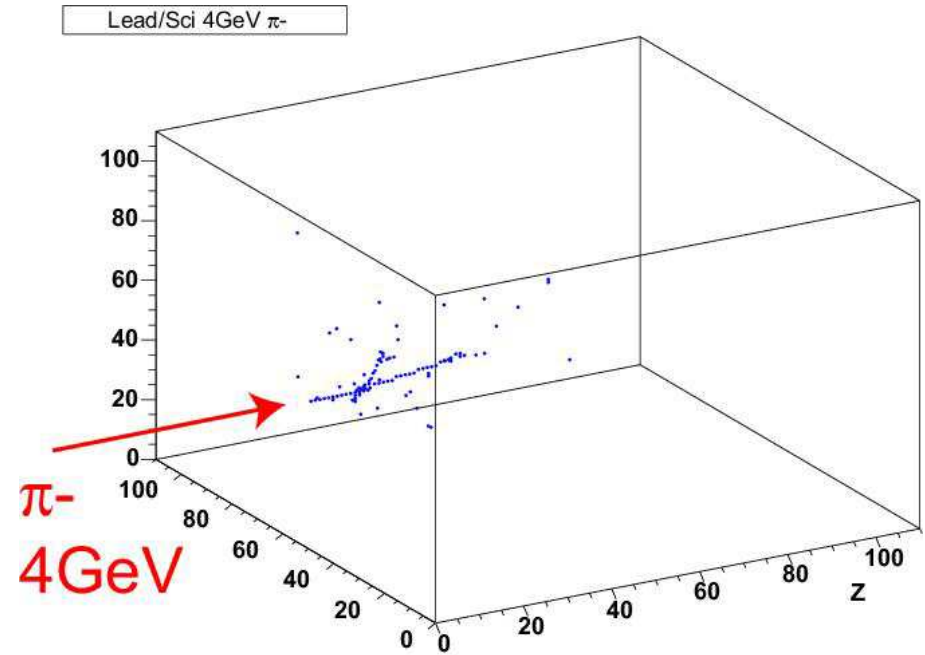}
\end{center}
\caption{An event induced by a pion of 4~GeV into model 
HCAL of 100~layers. The thickness of lead is 8mm and that of scintillator is 
2~mm  with 1~cm by 1~cm cells of 1~M pieces. Axis labels are in unit of cm.}
\label{HCALfig2}
\end{figure}

Being equipped with the the read out electronics of  time measurement capability, we  are able to recognize the neutron signal 
from other charged hadrons.   
It is studied by a simulation base on Geant4.6.2. 
In the simulation, single-pions is  injected into a test HCAL detector 
which consists of lead and plastic sandwich calorimeter of 100 layers. 
The thickness of the lead plates is 8~mm and 
that of scintillators is 2~mm, 
where the size of the  detector is 1~m$^3$.  
A typical shower eventinduced by a pion  is shown in Figure~\ref{HCALfig2}.
The hits are recorded in 1~cm $\times$ 1~cm cell of plastic scintillator 
of 2~mm  thickness, 
when the energy deposit exceeds 10\% of that of 
minimum ionization particles (MIPs). 
In the figure, satellite hits can be seen apart from the incident pion trajectory.

\begin{figure}
\begin{center}
\includegraphics[width=9cm]{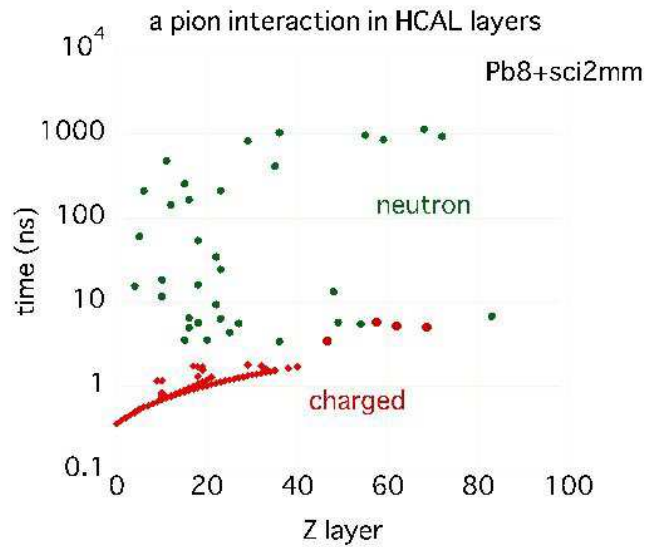}
\end{center}
\caption{Timing information (vertical) is ploted 
as a function of layer numbers. 
The injected pion and its charged daughter hits are shown 
by red points, 
where the neutron contribution is shown by green points. }
\label{HCALfig3}
\end{figure}
For the event shown in Figure~\ref{HCALfig2}, hit timing of each cell are 
shown as a function of layer numbers in Figure~\ref{HCALfig3}.
In the figure, some charged  components come 
from electrons and positrons produced by gammas in delayed decay of excited nucleus. These hits are also shown in red round marks in the figure. The figure shows a clear separation between 
neutron and charged components signals.  
According to this Geant 4 simulation, we can conclude that an inclusion or exclusion of neutron contribution in hadronic shower reconstruction are determined based on a timing information of hit cells.

\begin{figure}
\begin{center}
\includegraphics[width=9cm]{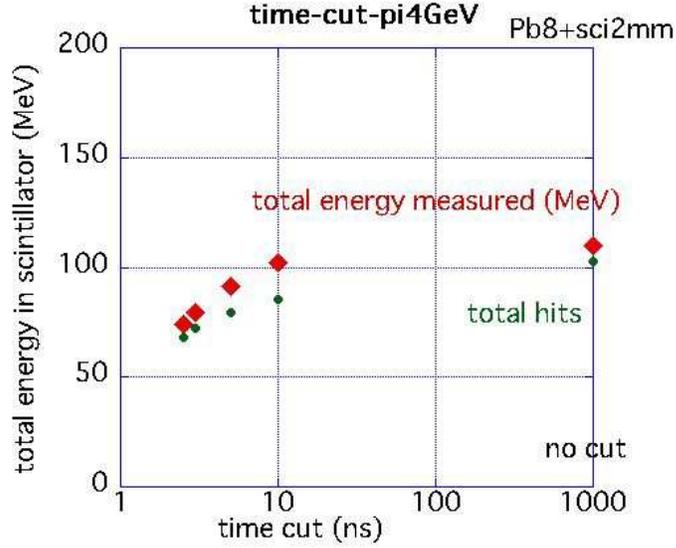}
\end{center}
\caption{The otal energy deposited in the scintillators (red diamond) and the total number of hits (green bullet) are plotted as a function of timing cut. At 1000 ns, the energy shows without any time cut. This shows about 20\% of energy is taken by slow neutrons. }
\label{HCALfig4}
\end{figure}

\begin{figure}
\begin{center}
\includegraphics[width=12cm]{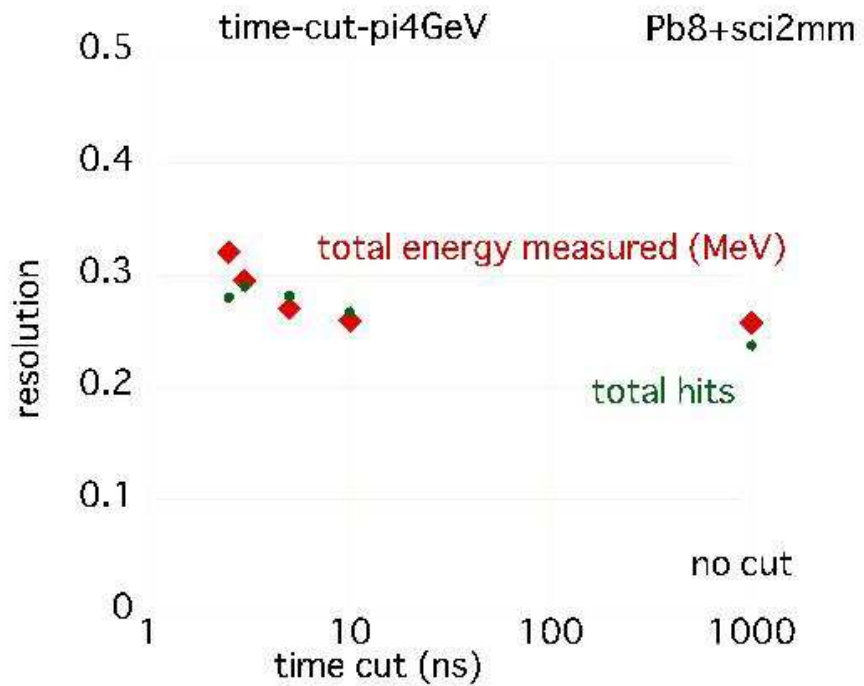}
\end{center}
\caption{The energy resolution of HCAL(red diamond) and the total number of hits (green bullet) are plotted as a function of timing cut. The point at 1000ns means without any time cut applied. This figure shows the energy resoluion becomes worse about 20\% if 5~nsec timing cut is applied to take out slow neutron components. }
\label{HCALfig5}
\end{figure}

\begin{figure}
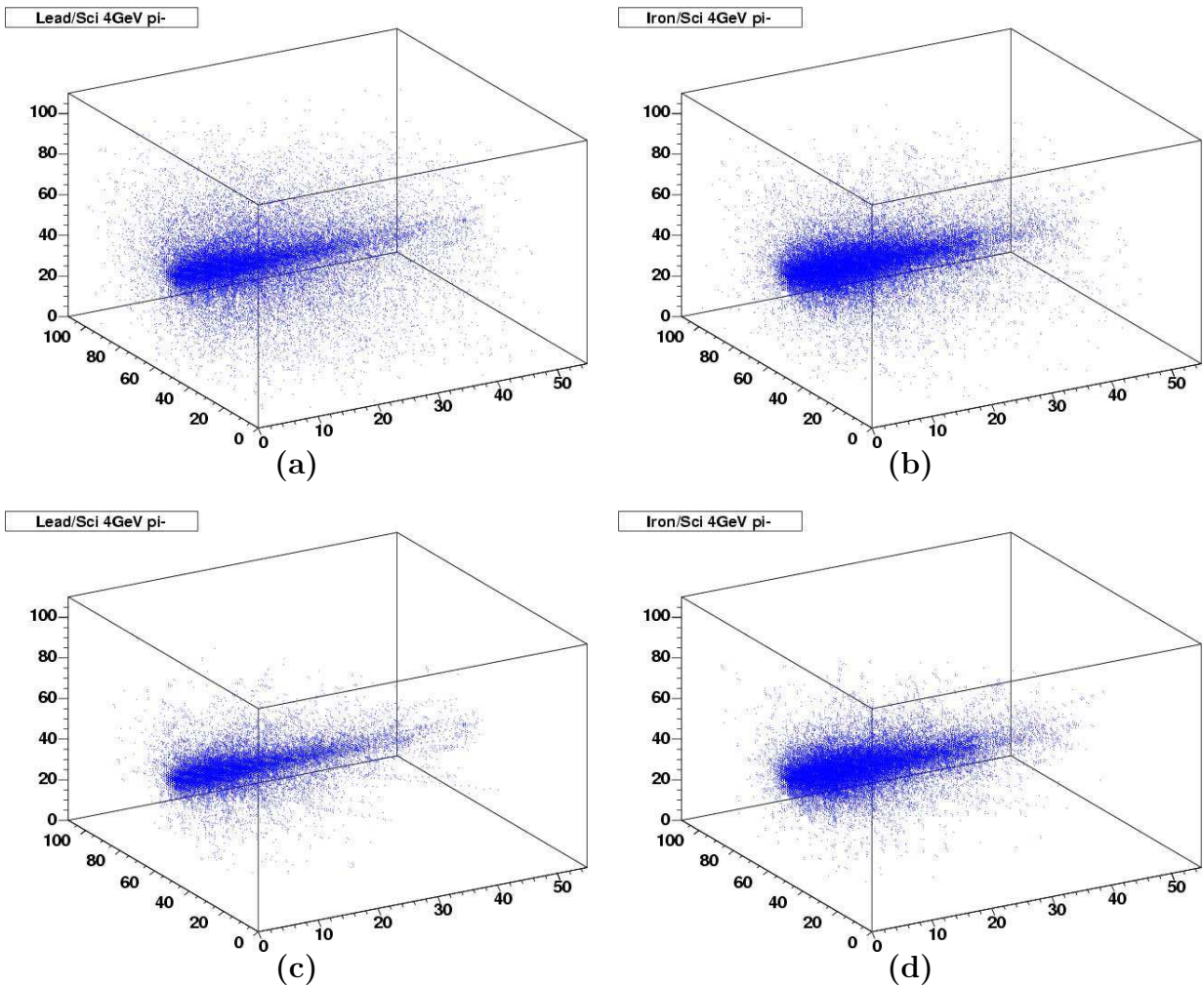

\begin{center}
%\includegraphics[width=9cm]{cal/Figs/HCALfig6.eps}
%\end{center}
%\caption{ The event displays with 1000 events overlapped are shown. The left ones are for the 
%case of lead absorber and the right ones are for the case of iron absorber. The neutron hits
% which scattered out from the interacting core region are more significant for lead absorber. }
\begin{tabular}{c c}
\includegraphics[width=6cm,angle=-90]{cal/Figs/lead_4_pi.epsi}
&
\includegraphics[width=6cm,angle=-90]{cal/Figs/iron_4_pi.epsi}\\
{\bf (a)} & {\bf (b)} \\
\includegraphics[width=6cm,angle=-90]{cal/Figs/lead_4_pi_5ns.epsi}
&
\includegraphics[width=6cm,angle=-90]{cal/Figs/iron_4_pi_5ns.epsi}\\
{\bf (c)} & {\bf (d)} 
\end{tabular}
\end{center}
\caption{ The event displays with 1000 events of 4~GeV/c pions overlapped are shown. 
Labels on axes are in unit of cm.
The upper left figure (a) and the upper right figure (b) is for the case of lead absorber 
and iron absorber, respectively, without the timing cut. 
The lower left figure (c) and the lower right figure (d) is for lead 
and iron absorber, respectively, with the timing cut of 5nsec. 
The neutron hits which scattered out from the interacting core region are more significant for lead absorber. }
\label{HCALfig6}
\end{figure}

The scintillator calorimeter can detect the neutrons effectively,  
while gas calorimeter can not. 
The energy measured by the scintillators 
are plotted as a function of timing cut in Figure~\ref{HCALfig4}.
As seen in the figure, the energy deposit after 5~nsec amounts to about 20\% of the total energy deposit.  It can be attributed to the contribution of slow neutrons.
On the other hand, the energy resolution becomes worse if we cut the slow components of signal at 5~nsec as seen in Figure~\ref{HCALfig5}.
The spread of hadron shower with and without the timing cut of 5~nsec is shown in Figure~\ref{HCALfig6} (a) and (c). For comparison, similar plots in the case of iron-absorber calorimeter are shown in Figure~\ref{HCALfig6} (b) and (d), where 
the thickness of the iron absorber and scintillator is 20~mm and 5~mm, respectively.
Thicker material thickness is adopted to simulate with a configuration similar to the calorimeter baseline desing.
Since the nuclear interaction lengths of iron and lead are similar, the ratio of absorber and sensor thickness are same for the both cases.  
A thousand of pions are injected and events 
are overdrawn  in a same figure. The outer shower cloud is due to the slow neutron component, whose spreads are determined by the nuclear interaction lengths of the absorbers. Since the nuclear interaction lengths of lead and iron are similar, the sizes of the outer shower clouds are similar. On the other hand, the size of the central part of shower in the lead absorber calorimeter is smaller than that of iron absorber calorimeter, because the radiation length of lead, which characterizes the size of the central part of shower, is much shorter than that of iron. This effect is more clearly seen when slow neutron components are cut off by applying a timing cut of 5~nsec in Figure~\ref{HCALfig6}.

\begin{figure}
\begin{center}
\includegraphics[width=15cm]{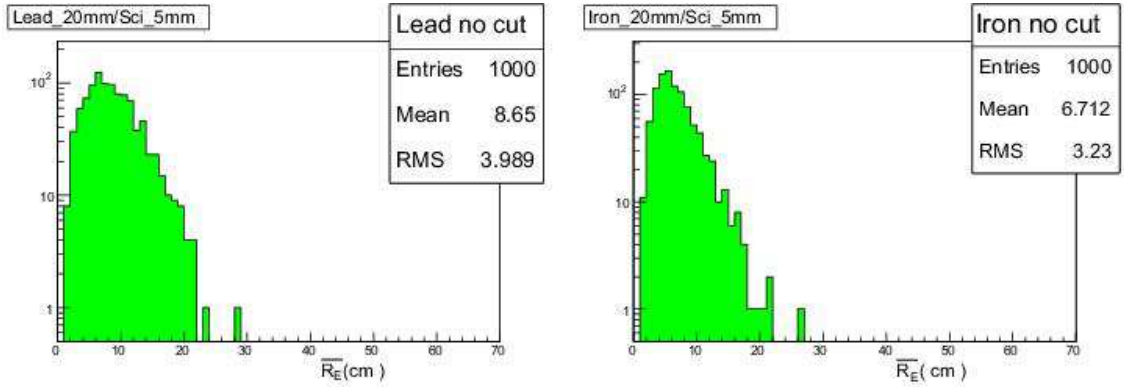}
\end{center}
\caption{Averaged energy weighted radius of showers in lead(left) and iron(right) calorimeter, respectively. No cut on cell timing is applied.}
\label{HCALfig7}
\end{figure}

\begin{figure}
\begin{center}
\includegraphics[width=15cm]{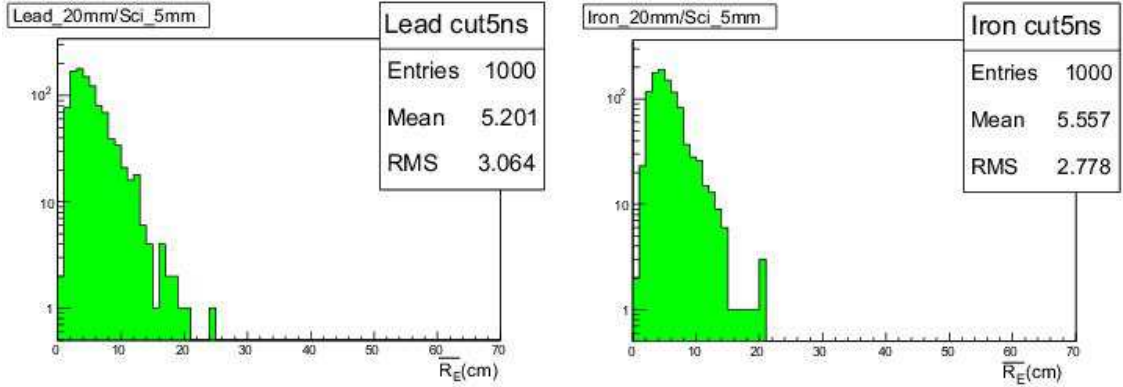}
\end{center}
\caption{Averaged energy weighted radius of showers in lead(left) and iron(right) calorimeter, repectively. Hit timing of each cell is required to be 
within 5~nsec from particle incident to the calorimeter. }
\label{HCALfig8}
\end{figure}

This is more clearly shown in Figures~\ref{HCALfig7} and \ref{HCALfig8}.
These figures show the averaged energy weighted radius of shower ($\bar{R_E}$) in lead and iron absorber calorimeter with (Figure~\ref{HCALfig8}) and without (Figure~\ref{HCALfig7}) cuts of 5~nsec timing.  The $\bar{R_E}$ is defined as follows:
$$
\bar{R_E} \equiv \sum_i { r_i e_i \over e_i } 
$$
where $r_i$ is a distance between the shower axis and a cell. $e_i$ is the 
energy deposit in the cell, and $i$ runs for all hits in the event.
Figure~\ref{HCALfig8} shows that the central part of shower in lead-based calorimeter is smaller than that of iron-based calorimeter.

We used the strip shaped scintillator for the calorimeter readout to achieve 
the effective readout granularity of 1~cm $\times$ 1~cm. 
This is correct as far as a passage of charged particle in x and y strips is concerned.
We are developing an algorithm 
to resolve the problem 
when multi-particles pass through  in small area. 
When  the analog information from a strip is available, 
we will be  able to recognize two particles with their pulse heights.
\subsubsection{Choice of absorber material}
We have two candidates  for the absorber materials of the  HCAL. 
One is lead which has good nature to have compensation  
in  reasonable dimensions, 
while the other is iron or steel whose  compensation condition is 
not far from the reality.
The compensation  means 
the total measured energies  with an HCAL 
by electrons and  hadrons have same amount.    
With lead, the compensation condition is  achieved 
at 9.1~mm lead vs 2~mm plastic scintillator thickness  
according to our test beam results \cite{NIM432}. 
The compensation  condition for iron or steel 
is not found by experiments, however,  R.~Wigmans expects it to 
be Fe/scintillator to be 15/1 in the  thickness,
while his prediction for lead is Pb/scintillaotr  to be  4/1.  
The compensation can not be achieved by Liquid Ar, 
since it does not contain hydrogen atoms. 
The crucial role of hydrogen in the  active material 
has been  demonstrated in L3 experiment at LEP.  

Lead  has shorter radiation length than that of iron, 
while the interaction  lengths are similar. 
As discussed already and shown in Figures~\ref{HCALfig6}, \ref{HCALfig7}, 
and \ref{HCALfig8}, the lateral spread of shower in a calorimeter using lead absorber is smaller 
than that using iron absorber.  The smaller lateral spread of shower is preferable for efficient particle flow analysis.
%The resultant hit distributions are shown in \ref{HCALfig6}, 
%the electromagnetic shower size in lead is smaller than  
%that of iron in pion injection case.   
%Neutron sensitivity of iron is  much smaller 
%than that of lead, which is shown in Figure~\ref{HCALfig9}
%\begin{figure}
%\begin{center}
%\includegraphics[width=9cm]{cal/Figs/HCALfig9.epsf}
%\end{center}
%\caption{ The energy measured in the scintillators are plotted as a function of timing cut for the iron based calorimeter. The point at 1000ns corresponds to the data without time cut.}
%\label{HCALfig9}
%\end{figure}

Here we discuss the calorimeter absorber from the detector point of view.
In the engineering point of view, the iron absober is preferred because 
mechanical strength of iron is much higher than that of lead. 
%It is easier for iron to make structure.
 
\subsection{R\&D studies}
\begin{figure}
\begin{center}
\includegraphics[width=13cm]{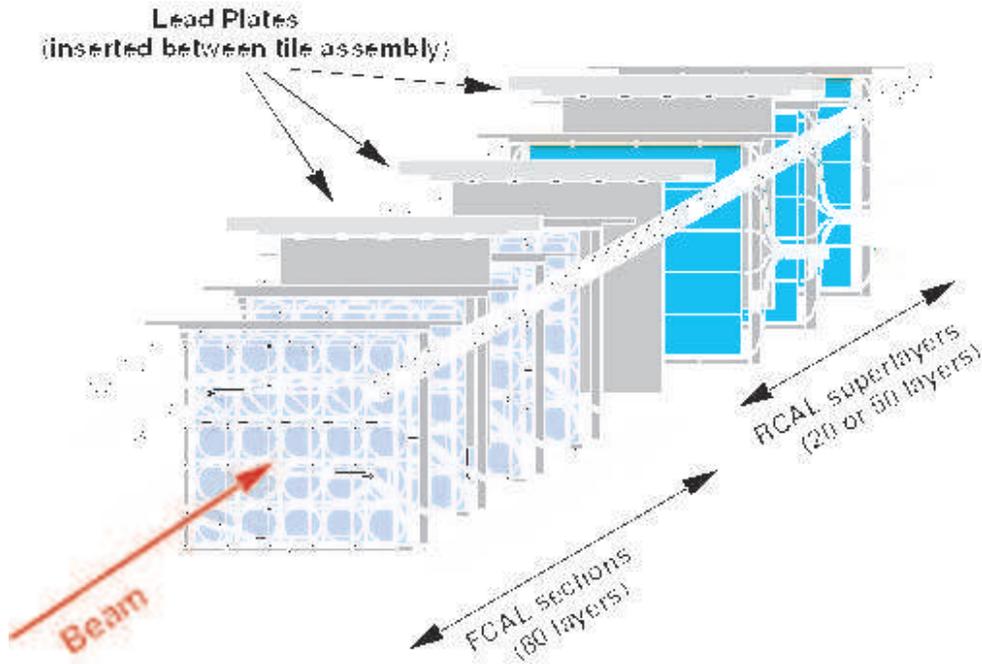}
\end{center}
\caption{ Schematic view of hadron calorimeter test module in 1999 at Fermilab  }
\label{HCALfig10}
\end{figure}

\begin{figure}
\begin{center}
\includegraphics[width=9cm]{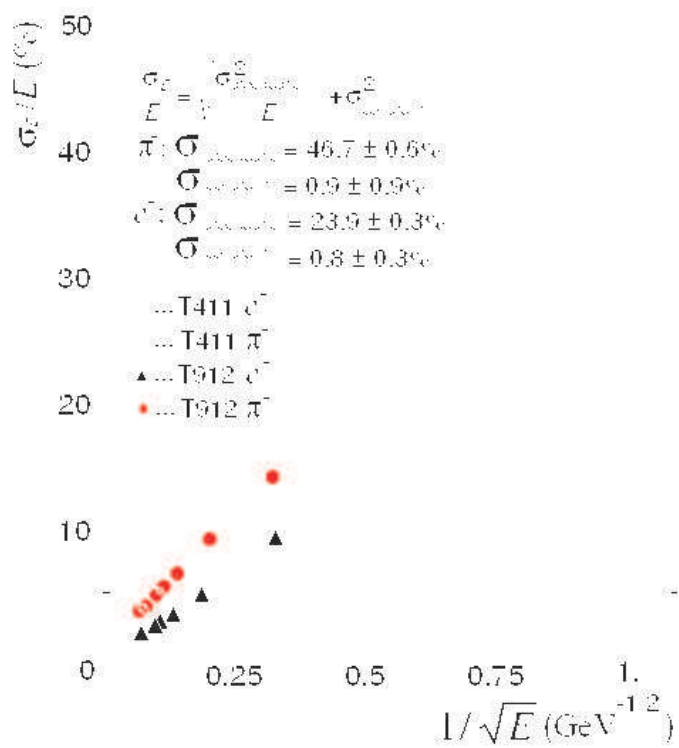}
\end{center}
\caption{ The energy resolutions for electrons (green triangle) and pions (red circle) as a function of incident energy. }
\label{HCALfig11}
\end{figure}

\begin{figure}
\begin{center}
\includegraphics[width=9cm]{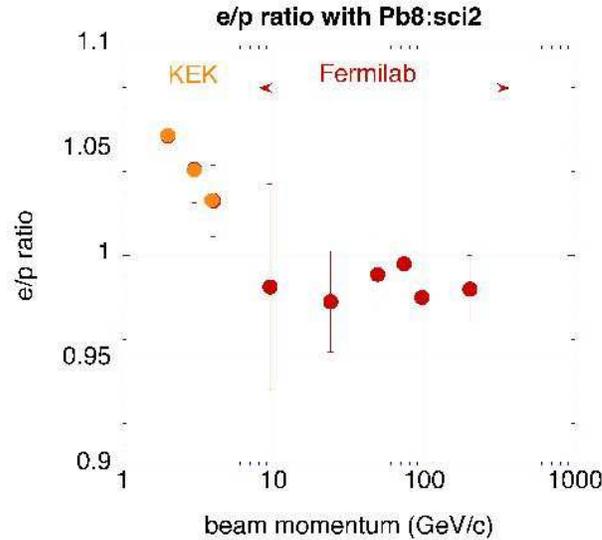}
\end{center}
\caption{ The e/pi ratios as a function of incident beam momentum. At high energy, it closes to 1.0. }
\label{HCALfig12}
\end{figure}

We have already studied the lead scintillator sandwich hadron
calorimeter in the test beam. 
The results have been published elsewhere \cite{NIM487}. 
With 20~cm by 20~cm tile, we have  constructed 1~m$^3$ HCAL, 
which were segmented in 100 (= 5 x 5x 4)  parts. 
The schematic view of the detector is shown in Figure~\ref{HCALfig10}.
A tile of 2~mm thick plastic scintillator were used 
for the  active material and 8~mm thick lead plates were 
the absorber in this HCAL. 
Wave length shifting fiber was embedded in a groove of a tile  
in order to read the scintillation light.  
The energy resolution  measured up to 200~GeV 
was plotted for pions and electrons in Figure~\ref{HCALfig11}.
The e/p ratio which is the measured energy ratio of electrons to 
pions is plotted in Figure~\ref{HCALfig12}.

\begin{figure}
\begin{center}
\includegraphics[width=9cm]{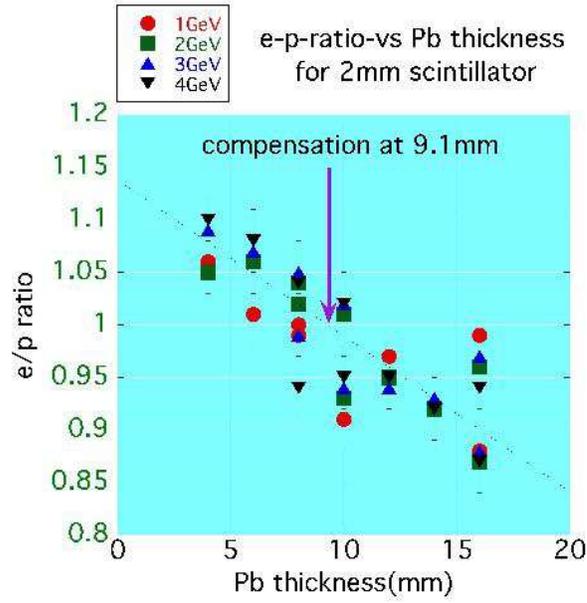}
\end{center}
\caption{ The e/pi ratios as a function of absorber (lead ) thickness in mm. The compensation condition is achieved at 9.1~mm of lead plate. }
\label{HCALfig13}
\end{figure}

Figure~\ref{HCALfig12} shows the hardware compensation 
at higher energy region,  
since the e/p ratio is very close to be 1.0. 
At lower energy region, 
we have measured the e/p ratio \cite{NIM432}
by changing  the volume ratio of lead and plastic scintillator. 
The result is  shown in Figure~\ref{HCALfig13}.

\begin{figure}
\begin{center}
\includegraphics[width=9cm]{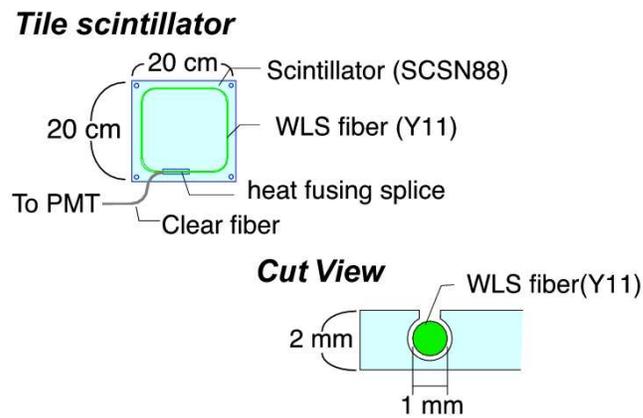}
\end{center}
\caption{ A tile scintillator and its cross section around the fiber groove.}
\label{HCALfig14}
\end{figure}

By fitting those data except 1 GeV data,  
we found that the  compensation is achieved at 9.1~mm 
of lead thickness when the plastic  scintillator is 2~mm thick. 
The expected  value of lead thickness by   R.~Wigmans is 8~mm 
for 2~mm scintillator.
The tile size was 20 by 20 cm$^2$ for those data. 
A schematic view of  groove for WLFS is shown in green 
in Figure~\ref{HCALfig14}. 
The thickness of  the plastic-scintillator is 2~mm, 
where we make groove of 1.2~mm to  embed the WLSF of 1~mm.    

With the size of 20~cm by 20~cm , we have experienced and tested the  HCAL.

\subsection{calibration system}
Unlike the gas chamber in the calorimeter, 
the stable response of  the plastic scintillator 
is well known.
However, we must have a system to monitor and calibrate 
the detector response.   
There are several ideas to check the response 
beside the test pulse into  the electronics circuit to test the
performance.
\begin{enumerate}
\item
{\bf Calibration with cosmic muons}
Suppose the size of a strip is 1~cm x 20~cm , 
expected number of cosmic muons passing by is   
0.1/cm$^2$/min $\times$ 20~cm$^2$ = 2/min.
This  number is obtained assuming that
the detector is located close to the ground. 
For the  calibration, we need about 100 muons for each strip. 
This will be  achieved only in 50 minutes.
\item
{\bf Calibration on $Z^0$ pole}
Since the cross section of $Z^0$ pole is very high, 
there is an idea to operate the collider 
at the $Z^0$ pole to collect calibration  particles, 
not only for HCAL but also for other detector components. 
Assuming an integrated luminosity of 1 fb$^{-1}$, 
we have 25 muons / strip. 
This number  is not sufficient for calibration, 
however, the high energy 45~GeV muons are much more reliable than cosmic ray muons. 
We keep this method as an option for calibration.
\item
{\bf Halo muons for the endcap HCAL calibration}
A large number of background muons are expected to be generated 
along the ILC beam line.
The number of the halo muons are so large 
that the muons  will be used for the calibration of HCAL 
in the endcap region, 
where the strips are orthogonal to the beam line.
\end{enumerate}
Relative monitoring of scintillation strips is discussed in the next  section.
\subsection{Missing R\&Ds}
\begin{enumerate}
\item			        
{\bf Relative  calibration scheme}
We need to have a stable light source 
to calibrate each scintillator strip, 
otherwise we can't follow the response of  all strips. 
While the beam or cosmic ray calibration discussed above
have limited statistics, 
the light source system 
to illuminate the  strip has no such statistical problems.
Systematic  
fluctuation of light source is rather problematic in this case.
Here we have an idea to feed laser or LED light  
to many strips simultaneously with clear fibers
(Figure~\ref{HCALfig15}). 
The problem is expected  at the light connection 
between the fiber and each strip point.  
Here we have found a good solution 
to reduce the number of light source  
which will be the main source of the systematic errors.
\begin{figure}
\begin{center}
\includegraphics[width=12cm]{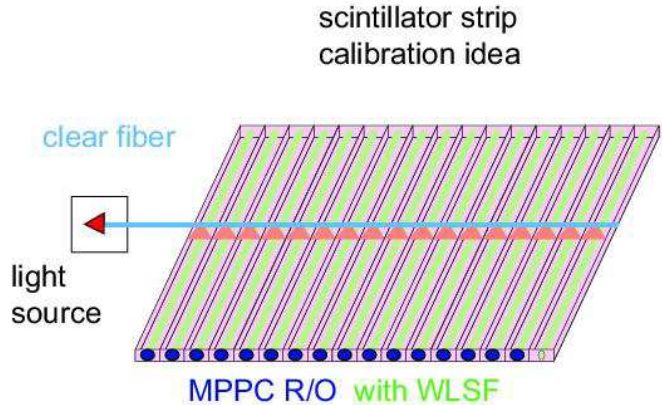}
\end{center}
\caption{ Schematic view of relative calibration of strips by a clear fiber.}
\label{HCALfig15}
\end{figure}

\item
{\bf Support structure by engineering study}
Since we suppose to construct the HCAL 
with lead and scintillator sandwich, 
we have to take into account 
the mechanical  structure of the lead plates. 
The pure lead plate is not stiff enough as the structure material, 
however, by adding Molybdenum , 
hardness increases and becomes stiff enough to  support itself.  
For real structure of HCAL, we have to study this  point furthermore.
\item
{\bf Other options}
\begin{itemize}
\item Square tile not strip array scintillator
\item Iron instead of lead as absorber
\end{itemize}
\end{enumerate}

    % to be prepared by T.Takeshita
\clearpage

%
% Muon System
%
\section{Muon detector}
\subsection{Introduction}
The main task of the muon detector is to identify muons.
The muon momentum is precisely determined by
the inner tracking devices, therefore the muon detector should
measure hit positions with modest accuracy in order to associate
them to the fitted track. However, an independent momentum measurement
of the penetrating particle by the muon system would be
preferable, especially for energetic or well-collimated jets,
for better matching with the extrapolated track or even better
identification of the secondary muons. Modest timing resolution
is also required to tag cosmic-ray muons and to identify the bunch
crossing as well as to reject backgrounds.

\begin{figure}[htb]
  \begin{center}
\includegraphics[width=0.8\textwidth]{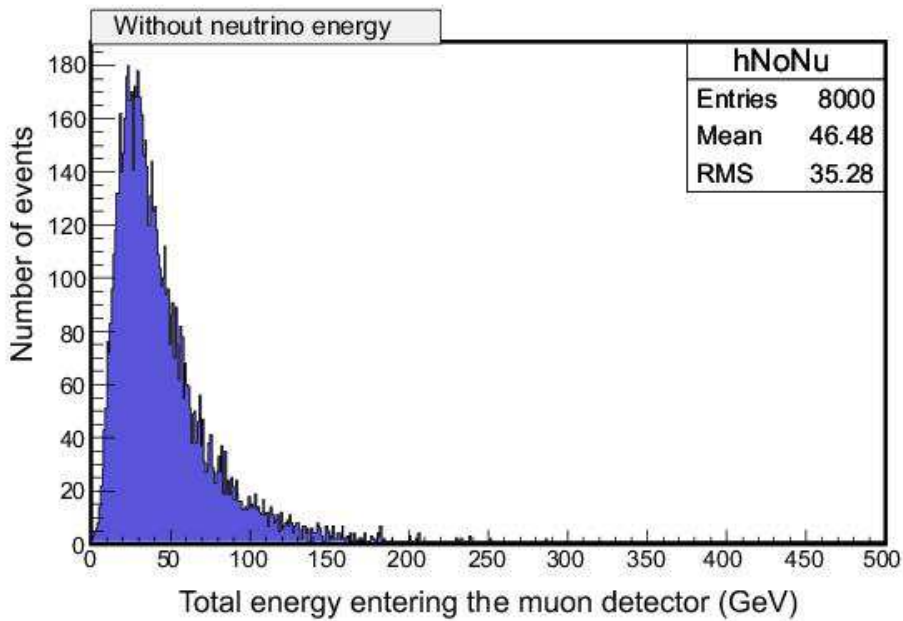}
\caption{Total energy of particles entering the muon detector, except neutrino, 
in each event
of $e^{+}e^{-}\rightarrow b\bar{b}$ at 500 GeV center of mass energy.}
    \label{muon-energy.eps}
  \end{center}
\end{figure}

A secondary task of the muon detector is to measure the energy
leakage of the hadron showers, {\it i. e.} work as the tail catcher.
The total energy of particles entering the muon detector was estimated by 
a Geant4 simulation for the case of the process 
$e^{+}e^{-}\rightarrow b\bar{b}$ at 500 GeV center of mass energy and shown 
in Figure~\ref{muon-energy.eps}. 
As seen in the figure, about 10\% of the total evern energy leaks into the muon detector.
Since the jet energy resolution is expected to have 30\%/$\sqrt{E}$
with the particle flow technique, the tail catching may become
more important than the previous experiments.

The muon detector will be located outside the hadron calorimeter
and the solenoid coil. The muon identifiers are interleaved between
the return yoke layers which function as absorbers in the muon
system. The magnetic field strength is about 3 Tesla and
the total iron thickness is about 2 m. The structure of the
return yoke will be optimized for better field strength
in both the bore and the fringe. On the other hand, the shape of the
muon system, number of active layers, segmentations and detector
technology will be chosen by evaluating the muon identification
efficiency and the energy measurement efficiency by
the Monte Carlo simulation and the hardware R \& D's. 

Although detector occupancy is supposed to be relatively low
in case of $e^{+}e^{-}$ collisions, we need to verify that
the muon detector, especially in the forward region, can cope with
dense jets or particles. Detector coverage and hermeticity of
the muon detector should be excellent for the good detection efficiency
and the measurements of the missing momentum with good accuracy.

\begin{figure}[htb]
  \begin{center}
    \includegraphics[width=0.7\textwidth]{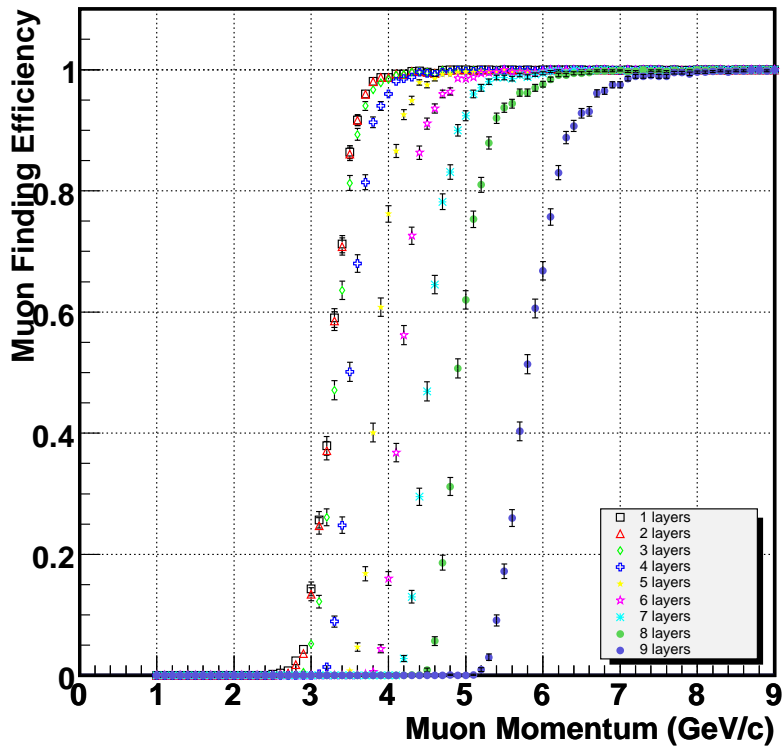}
    \caption{Muon detection efficiencies as a function of the muon
      energy in the baseline GLD configuration(see Figure~\ref{Con_GLD}). The results are
      obtained by using the Jupiter. The efficiency
      threshold is found to be 3.5 GeV requiring a hit in the
      first layer, or 6 GeV requiring hits in all layers. }
    \label{muon_eff}
  \end{center}
\end{figure}

The association of inner muon tracks with the segments of the
muon detectors suffers from energy loss and multiple scattering
in the calorimeter and the coil, and the return yoke itself.
The muon momentum is required to be more than 3.5 GeV to go
through the coil, and to be more than 6 GeV to fully penetrate
the return yoke in the barrel region , as shown in Figure~\ref{muon_eff}.
Hence, in the low momentum region muons have to be identified
by only the calorimeter, and just below 5 GeV the range measurement
in the muon system would help the energy measurement of the particle.

\begin{figure}[htb]
  \begin{center}
    \includegraphics[width=0.8\textwidth]{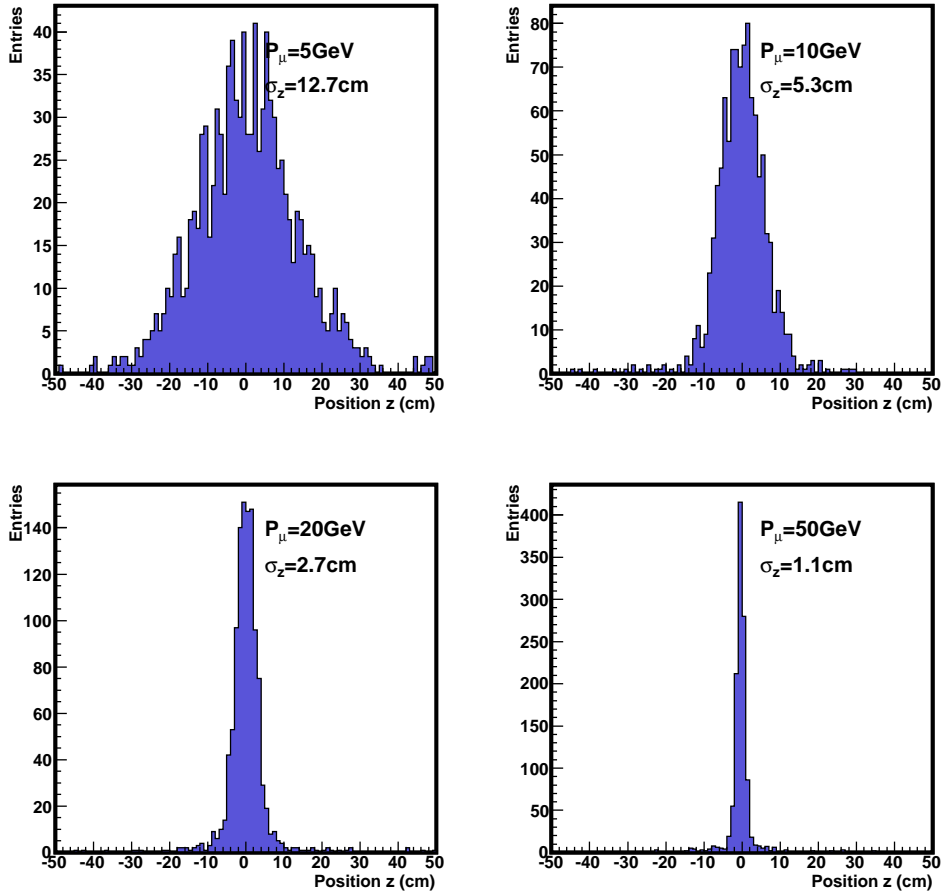}
    \caption{Distributions of the $z$-coordinate at the first layer
      for muons of $p_{\mu}$=5 GeV, 10 GeV, 20 GeV and 50 GeV.}
    \label{muon_spread}
  \end{center}
\end{figure}

The spatial resolution of the muon system needed for the track
matching should be less than the spread from the multiple
scatterings for muons, given that no other particles pass through
nearby. The spatial spread for single muons is obtained by the
full simulator, as shown in Figure~\ref{muon_spread}.
In the simulation, muons are generated at the interaction point
and emit into the direction of cos~$\theta$=0. The hit position is
taken from the $z$-coordinate of the muon at the first active layer.

\begin{figure}[htb]
  \begin{center}
    \includegraphics[width=0.7\textwidth]{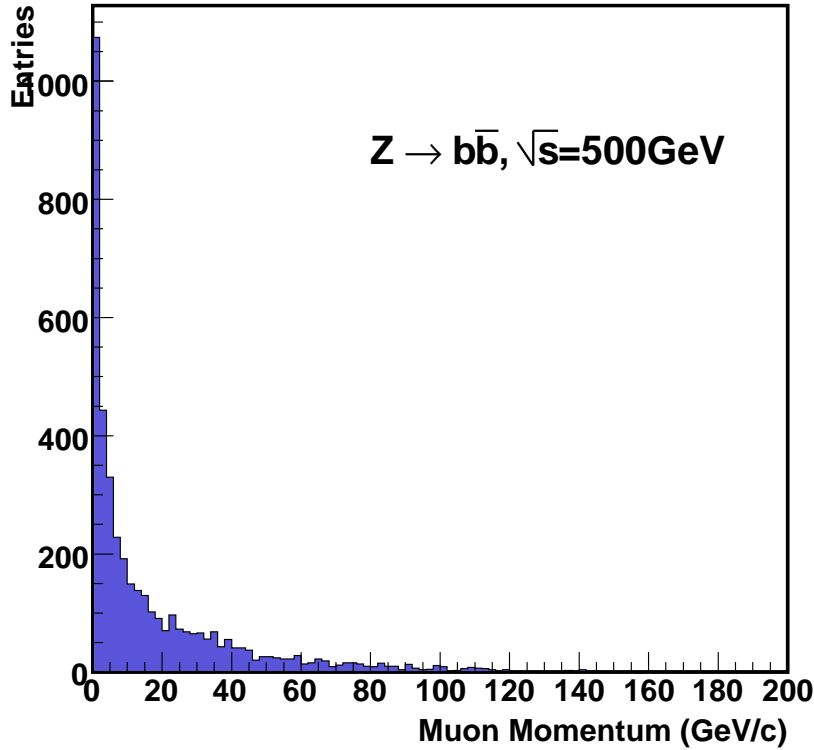}
    \caption{Muon momentum distribution at the front face of the
      muon detector for  $e^{+}e^{-} \to b\bar{b}$ at $\sqrt{s}$ = 500 GeV.}
    \label{muon_bbbar_momentum}
  \end{center}
\end{figure}

The relevant muon momentum region is examined by events including
jets, for example $e^{+}e^{-} \to b\bar{b}$ at $\sqrt{s}$ = 500 GeV
(Figure~\ref{muon_bbbar_momentum}). As seen in the histogram, the average
momentum is 20 GeV/$c$ and the majority of the muons are below
50 GeV/$c$. As the result, we conclude that the position resolution
of about 1 cm is sufficient for the good association of charged
tracks from the inner trackers.

%%%  This part removed until new figure become available.

%%The energy leakage should not be so large in the GLD detector
%%because of its thick calorimeter, but still sizable. 
%Figure~\ref{muon_bbbar_energy} shows the total leakage energy at the surface
%%of the muon detector in $b$-quark pair production events at
%%$\sqrt{s}$ = 500 GeV.

%\begin{figure}[htb]
%  \begin{center}
%    \includegraphics[width=0.7\textwidth]{muon/Figs/gammaZ2b_500.0GeV_totale_pre.eps}

%    \caption{Muon momentum distribution at the front face of the
%      muon detector for  $e^{+}e^{-} \to b\bar{b}$ at $\sqrt{s}$ = 500 GeV.}
%    \label{muon_bbbar_energy}
%  \end{center}
%\end{figure}

%%% Local Variables:
%%% mode: latex
%%% TeX-master: "calmuo_dod"
%%% End:

\subsection{Baseline design}
In the GLD detector, the muon detector covers a very large region
and is sandwiched between the return yoke. This implies that 
the technology should be inexpensive and reliable because it will
be difficult to access and replace, and we have to operate it

for more than a decade. Also, the detector must be tolerant of

the radiation and the aging. Therefore a well-established technology
must be chosen, and we adopt an scintillator-fiber technique as the

baseline design. Since this technology is also chosen for the

calorimeter, we will be able to save the cost for production

as well as R\&D's. Photon sensor will be a Geiger-mode photon

counter, which is being developed actively, can also be shared

with the calorimeter, and the read-out electronics is similar to

that of the calorimeter. Sensitivity to neutrons is also an issue
for the scintillator, as discussed in the calorimeter section,
and the timing information may be helpful again.

%\begin{figure}
%  \begin{center}
%    \includegraphics[width=0.7\textwidth]{muon/Figs/muon-view.eps}
%    \caption{Old layout of the muon detector.}
%    \label{muon_detector}
%  \end{center}
%\end{figure}

The cross-sectional view of the

muon detector is shown in Figure~\ref{Con_GLD}. It has an dodecagonal shape in the $r$-$\phi$ view.
There are nine active layers in both barrel and endcap regions,
and each active layer has two orthogonal layers of the
scintillator-bar arrays. The scintillator bar has a cross section
of 5 $\times$ 5 cm and the length varies according to the location.
A wavelength shifting (WLS) fiber is embedded into the straight
groove along the scintillator, and at the both ends of the groove
the silicon photon sensors are directly attached to the fiber.
The scintillator and the WLS fiber might be too long to propagate
photons to the sensors, and we may need to break the scintillator
into pieces to decrease the light attenuation. In this case
we still keep the detection efficiency high to minimize the dead space.
With the readout at both ends, we can obtain the rough hit position
along the scintillator by the time difference of the signals.

Detector calibration will be performed by using cosmic muons
for the barrel region and halo muons for endcap regions. Because
the scintillator size is large, we can calibrate the system
within a few \% level. Additional calibration system (LED etc.)
may be used for the backup.

%%% Local Variables:
%%% mode: latex
%%% TeX-master: "calmuo_dod"
%%% End:

\subsection{R\&D studies}
Since the scintillator-fiber scheme is well-established,
we concentrate on photon sensor for hardware R\&D's, and
it is already described in the calorimeter section.
The electronics for the muon detector is also almost
the same as that for the calorimeter.

Simulation studies have been performed to investigate
the following items: (1) hit density at the front of the
muon detector, (2) hit point smearing due to the material
in front of the muon detector and the return yoke.

%%% Local Variables:
%%% mode: latex
%%% TeX-master: "calmuo_dod"
%%% End:

\subsection{Missing R\&Ds}
In order to finalize the detector design, some parameters are
still to be optimized. They are number of layers,
scintillator thickness and width. More simulation studies for
the muon detection efficiency and the tail catching capability
are needed for tuning those parameters. %Figure~\ref{muon_detector2}
%shows the baseline design in which number of layers are ten.

%\begin{figure}
%  \begin{center}
%    \includegraphics[width=0.8\textwidth]{muon/Figs/muon-view2.eps}
%    \caption{Baseline design of the muon detector.}
%    \label{muon_detector2}
%  \end{center}
%\end{figure}

For the hardware study, as the scintillator bar and the WLS fiber
become very long, care should be taken for the optical quality
of the detector so that the photon sensors can receive enough photons.
Thus, the optical characteristics of the scintillator and the WLS fiber,
scintillator length and photon-detection efficiency of the photon
sensors should be tested and justified for real use.

Although the scintillator-fiber system is expected to perform well,
another technology is considered as an option, which is resistive

plate chambers (RPCs). Large area RPC systems have been used by
experiments at $B$-Factories. RPC has simple structure and
can be built with inexpensive cost. However, RPC needs a mixture
of gases which should be avoided if possible for stable operations.

%%% Local Variables:
%%% mode: latex
%%% TeX-master: "calmuo_dod"
%%% End:

  % to be prepared by T.Takeshita
\clearpage

%
% FCAL, BCAL and Pair-Monitor
%
\section{Small Angle Detectors}
\label{SectionFCAL}
\subsection{Introduction}
%
% pair monitor
%
 At linear colliders it is expected that a large number of $e^+e^-$
pairs are created by the QED process
$\gamma \gamma$ $\rightarrow$ $e^+e^-$ where $\gamma$ may be
off-shell or near-on-shell~\cite{FCAL_eeee}. 
If the charge of the created and the on-coming bunch have the 
same sign, it is known~\cite{FCAL_pair} that  distributions of $p_T$ and 
azimuthal angle of the created particles can be used to
study $\sigma_x$ and $\sigma_y$ of the on-coming bunch. It should
be noted that this technique allows to measure 
independently the size of the two beams because the created
particles are deflected asymmetrically in the forward and backward
angular regions if the two beams have different parameters. The
goal of the pair monitor is to detect the $e^+e^-$ pairs in order
to extract the beam profile parameters.

%
% lum. monitor
%   
 Also, at linear colliders it is expected that a precise measurement
of the luminosity can be achieved by detecting electrons from
%%[AM]Bhabha scattering. calorimeters capable of detecting electrons at
Bhabha scattering. Calorimeters capable of detecting electrons at
the very forward regions are typically used for such luminosity
measurement. Another important role of such calorimeters is to veto 
forward leptons. It becomes important as it is closely 
related with the detection of sleptons \cite{FCAL_susy},
in the scenario where the lightest SUSY partile (LSP) is the lightest neutralino
$\tilde{\chi_1^0}$.%%[AM]
%%[AM] In the SUSY scenario 
%%[AM]with R-parity conservation, the lightest SUSY particle (LSP)
%%[AM]is the lightest neutralino $\chi$. 
This particle is considered as the best
candidate to satisfy the cosmological constraints on the dark matter
in the universe. The WMAP~\cite{FCAL_wmap}
results imply a very small difference between the
lightest slepton mass, the SUSY partner of the $\tau$ which will be called
$\tilde{\tau}_1$, and the LSP mass. Since this feature is quite general,
it is important to have an experimental access on the detection of sleptons
at the future linear collider. Experimentally, one needs to measure
$\tilde{\tau}_1$ decaying into a $\tau$ lepton with one or two neutrinos in the
final state which even further reduced the amount of visible energy.
Near threshold, the cross section is at the 10 $fb$ level with a potentially
very large background due to the four fermion final states, the so-called
$\gamma - \gamma$ background, which is at the $nb$ level. In usual cases
the standard backgrounds can be eliminated by vetoing forward electrons.
Therefore, another important
role of calorimeter detectors at small angles  
%%[AM]is to help in identifying leptons in the forward region.
is to veto background events by identifying electrons in the forward region.%%[AM]

 In this document, we call 
small angle detectors as a set of the following detectors: 
forward calorimeter (FCAL), the beamline calorimeter (BCAL), and
the pair monitor (PM).  
Silicon-Tungsten sampling calorimeter is the default technology
for both FCAL and BCAL.  As an option, diamond/W detector is
considered since the diamond is the radiation hard sensor.
FCAL can also serves as a beam mask.
PM serves as a beam profile monitor. 
A 3-dimensional silicon pixel technology is chosen for the PM.
In following sections, we discuss technology choice, configuration,
and performance of each detectors.

\subsection{Forward and Beamline Calorimeters}

\begin{figure}
\begin{center}
\includegraphics[width=14cm]{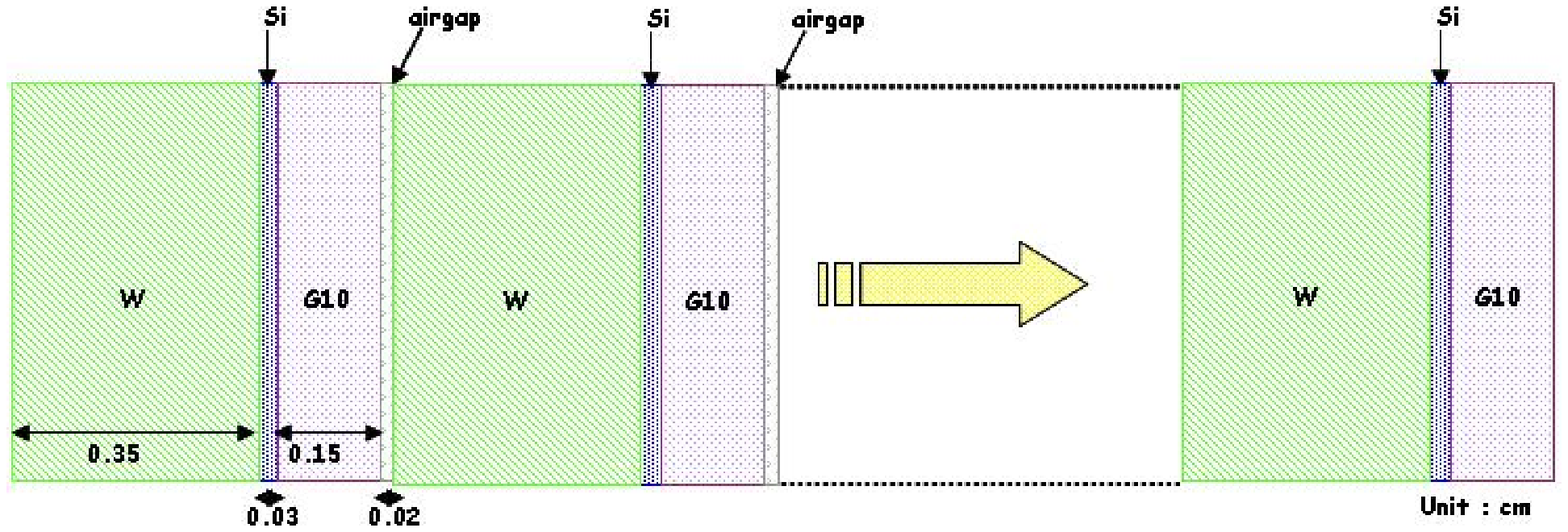}
\end{center}
\caption{\label{FCAL_fcal_geo}FCAL geometry}
\end{figure}

%%
%% detector
%%
 The FCAL is located behind the endcap of the TPC in $z$ direction. It
is a cylindrical structure with its axis coinciding with the $z$ 
%%[AM]axis. It has 41 layers of silicon-airgap-Tungsten sandwitch. The
axis ( see Figure~\ref{Con_GLDIR}). It has 45 layers of Silicon-air gap-Tungsten sandwich. 
The%%[AM]
baseline thickness of the detector is described in Table~\ref{FCALTable1}. 
\begin{table}
\begin{center}
\caption{Summary of parameters for the FCAL}
\begin{tabular}{rr}
\\
\hline
Material & Thickness (cm)\\
\hline
Silicon  & 0.03  \\
G10      & 0.15  \\
Air      & 0.02  \\
Tungsten & 0.35  \\
\hline
\end{tabular}
\label{FCALTable1}
\end{center}
\end{table}
The geometry of the FCAL is described in Figure~\ref{FCAL_fcal_geo}.
One sandwich is 5.5 mm thick and in total thickness of the FCAL becomes
250 mm. The FCAL covers from 260 cm to 285 cm in $z$ coordinates and 
from 12 cm to 36 cm in the radial coordinates. 
The segmentation in $\phi$ direction is eight. 
%[AM]Therefore,
%[AM]the total number of channels to read out becomes 8 $\times$ 20 $\times$ 2
%[AM]$\times$ 4 =
%[AM]1280 channels.

\begin{figure}
\begin{center}
\includegraphics[width=14cm]{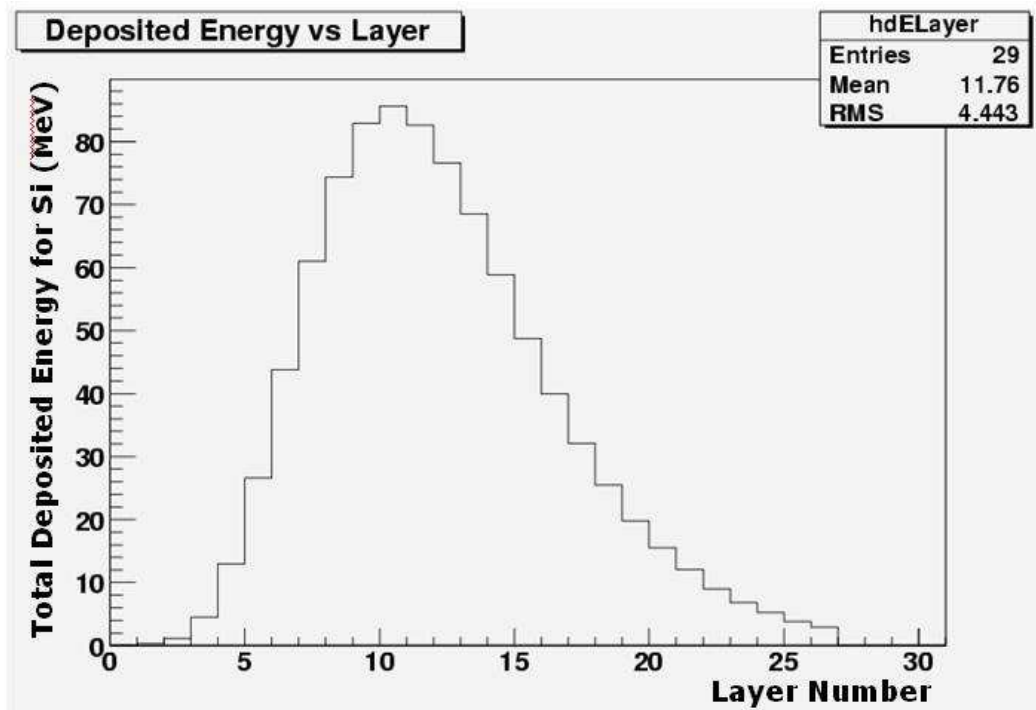}
\end{center}
\caption{\label{FCAL_fcal_energy}FCAL deposited energy as a function of
the layer}
\end{figure}

\begin{figure}
\begin{center}
\includegraphics[width=14cm]{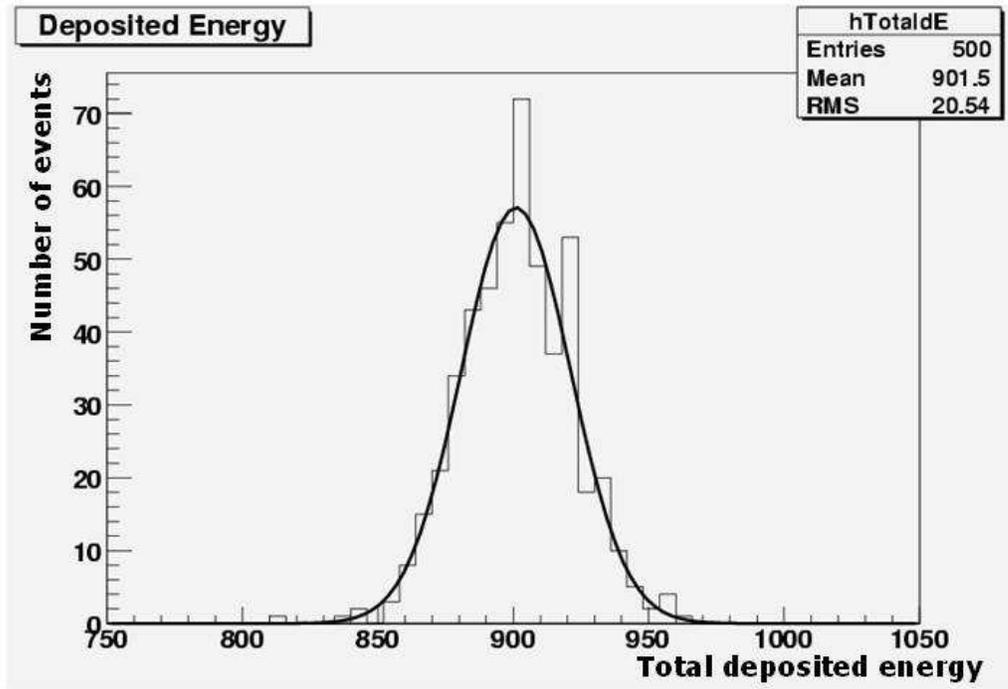}
\end{center}
\caption{\label{FCAL_fcal_deposit}FCAL total deposited energy}
\end{figure}

 In order to check the performance of FCAL, we used GEANT4 program
to simulate FCAL. Figure~\ref{FCAL_fcal_energy} shows the energy
deposited as a function of the layer for 100 GeV electrons hitting
the detector. 
The shower maximum occurs
at the layer 11 and small fraction of the energy is lost.
Figure~\ref{FCAL_fcal_deposit} shows the variations of the
%%[AM]total energy deposited in the FCAL. Approximately 20 GeV resolution
%%[AM]is seen at the 100 GeV electrons. (Is it right? The figure looks
%%[AM]strage. What is the unit of the $x$ axis ?) 
total energy deposited in the FCAL. Approximately the energy %%[AM]
resolution($\Delta E/E$) of $23\%/\sqrt{E}$ is achieved for the 100 GeV electrons.%%[AM] 

 The BCAL is located further downstream from the IP. It covers
from $z$ = 430 cm to $z$ = 450 cm. It has a sampling structure 
similar to FCAL, 
% Other parameters are same as FCAL
except the fact that the air gap is 0.7 mm.
%, bringing the BCAL is slightly
% thinner than the FCAL. 
% The BCAL also has 1280 channels to readout. 

%%
%% 20 mrad option 
%%
%%In the case the choice is made to have a large angle collision, the
%%design of endcap detectors should be changed. We discuss design
%%changes in such case below.

%%
%% readout electronics
%%
%%[AM]\begin{itemize}
%%[AM]\item electronics (missing, please give me data)
%%[AM]\item data size (missing, please give me data)
%%[AM]\end{itemize}

\subsection{Pair Monitor}

The current technology for PM adapts the 3-dimentional active pixel
sensor, forming a double layer of silicon disks. 
%%[AM]It is located in front of the FCAL. The $z$ position of the PM
It is located in front of the BCAL. The $z$ position of the PM%%[AM]
is 400 cm and 401 cm from the IP.
The inner radius of
the disk is 2 cm and the outer radius reaches to 8.5 cm. 
It consists of an array of trapezoidal wedges, each composed by a sensor
and a readout chip. There will be a large mount of $X$-ray hits from nearby
devices such as quadrupoles, luminosity monitors that need to be removed either
by requiring a coincidence between two adjacent wheels or by being able to
set a clean 70 keV threshold per hit~\cite{FCAL_threshold}, thus reducing
by 50 \% the overall number of needed channels. Most of the created pairs
%%[AM]will hit the central part of the monitor while $X$-rays are almost uniformly
will hit the central part of the monitor while $X$-rays are almost uniformly%%[AM]
distributed generating a background that needs to be removed in the
external region where the monitor should withstand an occupancy up to 30\%
of created pairs per train crossing.

  Beam instabilities are often caused by preceding bunches in a train:
%%[AM]it is thus desirable to measure the beam profile monitor for individual 
it is thus desirable to measure the beam profile for individual%%[AM] 
bunch or at least as  a function of time from the beginning of train. It is
also highly desirable that such information be available real time for machine
tuning. The high occupancy suggests that silicon strip detectors are
not suited for this application. A CCD sensor would be a good candidate 
in terms of occupancy; since the integrated charge deposit in each pixel is
read out at the end of each train, however, the information on the 
location in a train
is lost unless some external gating is applied. On the other hand, active
pixel sensors can handle high occupancies just as CCDs and timing of 
hit can be digitized on each pixel. The 3D pixel design is particularly suited
for this application for the following reasons:
\begin{itemize}
\item it is fast: the charge collection time is about 10 times faster than
typical pixel devices, for a 50 $\mu$m pitch 3D cell where the 
ionizing track is midway between the central $p$ and corner $n$ electrode, 
perpendicular to the surface of the silicon crystal.
\item it is radiation hard: the depletion voltage is of the order of 5 $V$
and has a large margin for increase by radiation.
\item it has a flexible geometry and it can be easily made in a trapezoidal 
shape and the sensor can be active very close to the edge.
\end{itemize}

 Each pixel sensor is bump-bonded to a readout chip which collects the
charge, discriminates and time-digitizes it. The proposed pixel size is
100$\times$100 $\mu$m, thus its readout electronics has to fit in such
%%[AM]a footprint and, given the bunch crossing rate, the readout time target for
%%[AM]the whole system has to be of the order of few ms.
a footprint.%%[AM]

%%[AM]\begin{itemize}
%%[AM]\item  I'm not sure what else to put in. I have no data than what I wrote so far.
%%[AM]\item  I have not received anything to fill in.
%%[AM]\end{itemize}

  % to be prepared by E.Won
\clearpage

%
% Magnet
%
\section{Detector Magnet and Structure}
\label{SectionMagnet}
\subsection{Detector Magnet}
\subsection*{Introduction}
 A superconducting solenoid magnet has been studied for the GLD detector. 
The required central magnetic field 
is 3~T in a cylindrical volume of 8~m in diameter and 9.5~m in total length. The stored 
energy is calculated to be 1.6~GJ. A 0.5~m length of correction coils is attached at both end 
of the solenoid coil to satisfy requirement of field uniformity. Because the magnet is planed to 
locate outside of the calorimeter, it is not necessary to design as a thin superconducting 
solenoid magnet. Therefore, the weight of the calorimeter estimated 2000 tons is planed 
to support at the inner wall of cryostat.

\subsection*{General Design}
\begin{figure}
\begin{center}
\includegraphics[width=8cm]{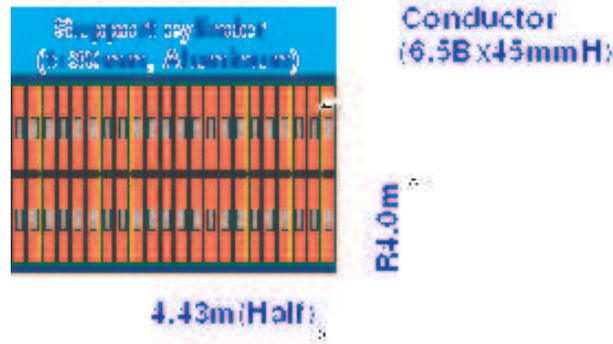}
\end{center}
\caption{Cross section of the GLD solenoid magnet. }
%\label{Resolutions}
\label{MagCoil}
\end{figure}

 The magnet provides a central magnetic field of 3~T at nominal current of 7956~A in a 
cylindrical volume of 8~m in diameter and 9.5~m in length. The current density is about 
60 percent of the critical current density. It is placed outside the calorimeter to achieve good 
hermeticity. The coil consists of double layers of aluminum stabilized superconductor 
wound around the inner surface of an aluminum support cylinder made of JIS-A5083(Figure~\ref{MagCoil}). 
According to the detector requirements, the integrated field uniformity 
has to be less than 2mm. Thus 0.5~m length of four-layer coil is adopted at both of the 
coil. Indirect cooling will be provided by liquid helium circulating through a single 
tube welded on the outer surface of the support cylinder. Main parameters of the GLD 
solenoid magnet are summarized in Table~\ref{Soltable}.
Radiation shields consist of 20~K shield and 80~K shield are placed between the coil and 
the vacuum vessel. These shields have to be decoupled electrically from both the coil 
and the vessel walls in order to avoid the effect of eddy current induced by the inner and 
the outer coaxial cylinders that are connected by flat annular bulkheads at each end.

\begin{table}[h]
\caption{Basic parameters of the solenoid magnet.}
\label{Soltable}
\begin{center}
\begin{tabular}{|l|c|}\hline 
Cryostat Inner Radius & 3.72~m \\
Cryostat Outer Radius & 4.4~m \\
Cryostat Half length & 4.75~m \\
Cryostat weight & 191~tons \\
\hline
Coil mean radius & 4~m \\
Coil Half length & 4.43~m \\
Cold mass & 78~tons \\
\hline
Central magnetic field & 3~T \\
Number of turns(2 layers) & 2880 \\
Current density & 741~A/mm2 \\
Nominal current & 7956~A \\
Stored energy & 1.6~GJ \\
\hline
\end{tabular}
\end{center}
\end{table}

\subsection*{Coil Design}
\begin{figure}
\begin{center}
\includegraphics[width=8cm]{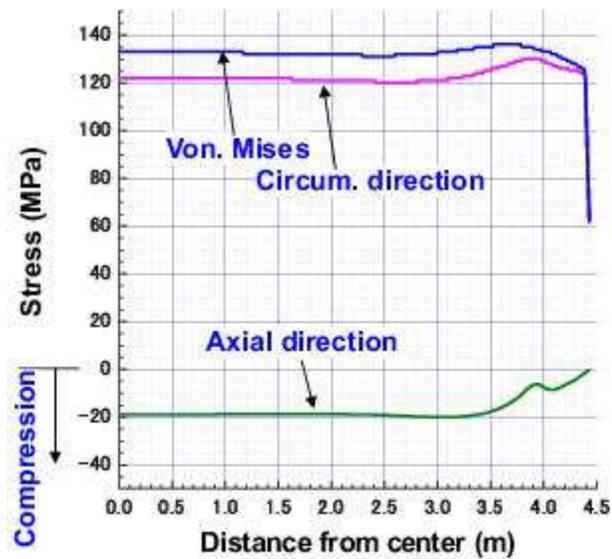}
\end{center}
\caption{Stress distribution in the coil. }
\label{Sol-Stress}
\end{figure}

\begin{figure}
\begin{center}
\includegraphics[width=8cm]{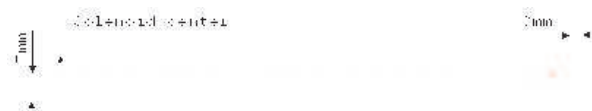}
\end{center}
\caption{Deformation of the coil due to magnetic force. }
\label{Sol-Deform}
\end{figure}

Superconducting wire for the GLD solenoid is based on the ATLAS superconducting 
wire. The size is scaled-up to 6.5mm width and 45mm height in order to absorb a magnetic 
force in the coil. The mechanical structure of the superconduction coil 
must sustain two main mechanical stresses, hoop stress and axial stress in the coil.
The calculation results are shown in Figure~\ref{Sol-Stress} and 
the deformation of the coil when the solenoid
excitation is shown in Figure~\ref{Sol-Deform}. 
Even though the yield strength of superconducting wire is 
achieved to around 250~MPa at recent high strength superconducting wire development, 
it is assumed to yield strength of 150~MPa in this design. The combined stress, Von Mises, in 
the coil when the central magnetic field of 3~T is calculated to be 130~MPa. 
According to the 
calculation result, required thickness to keep within yield strength in the coil is to be 
30~mm for support cylinder and 2 layers of 45~mm-thick for coil, respectively. From these 
required thickness, the cold mass is approximately 78 tons. The stored energy is 
calculated to be 1.6~GJ.
 The solenoid magnet will be supported at the innermost layers of barrel iron yoke by 
fixing the support structure on the outer vacuum vessel. The coil support system has to 
transmit both the weight of cold mass and the magnetic de-centering forces. If 
the geometrical center of the solenoid and the iron yoke are deviated, an unbalanced 
magnetic force is generated. So this force has to be taken into account on the coil 
support design. And another important requirement is that the coil supports have to 
follow the thermal shrinkage in the cooling down. The configuration of the coil supports 
is planed to adopt a rod type made of GFRP. In the de-centering calculation, a 380 tons in 
the axial direction and 130 tons in radial directions of de-centering forces act on the coil 
supports if the magnet and iron yoke is deviated 25~mm.
% The calculation result of stress distribution in the coil supports is shown in Figure~XX.

\subsection*{Cryostat Design}
\begin{figure}
\begin{center}
\includegraphics[width=16cm]{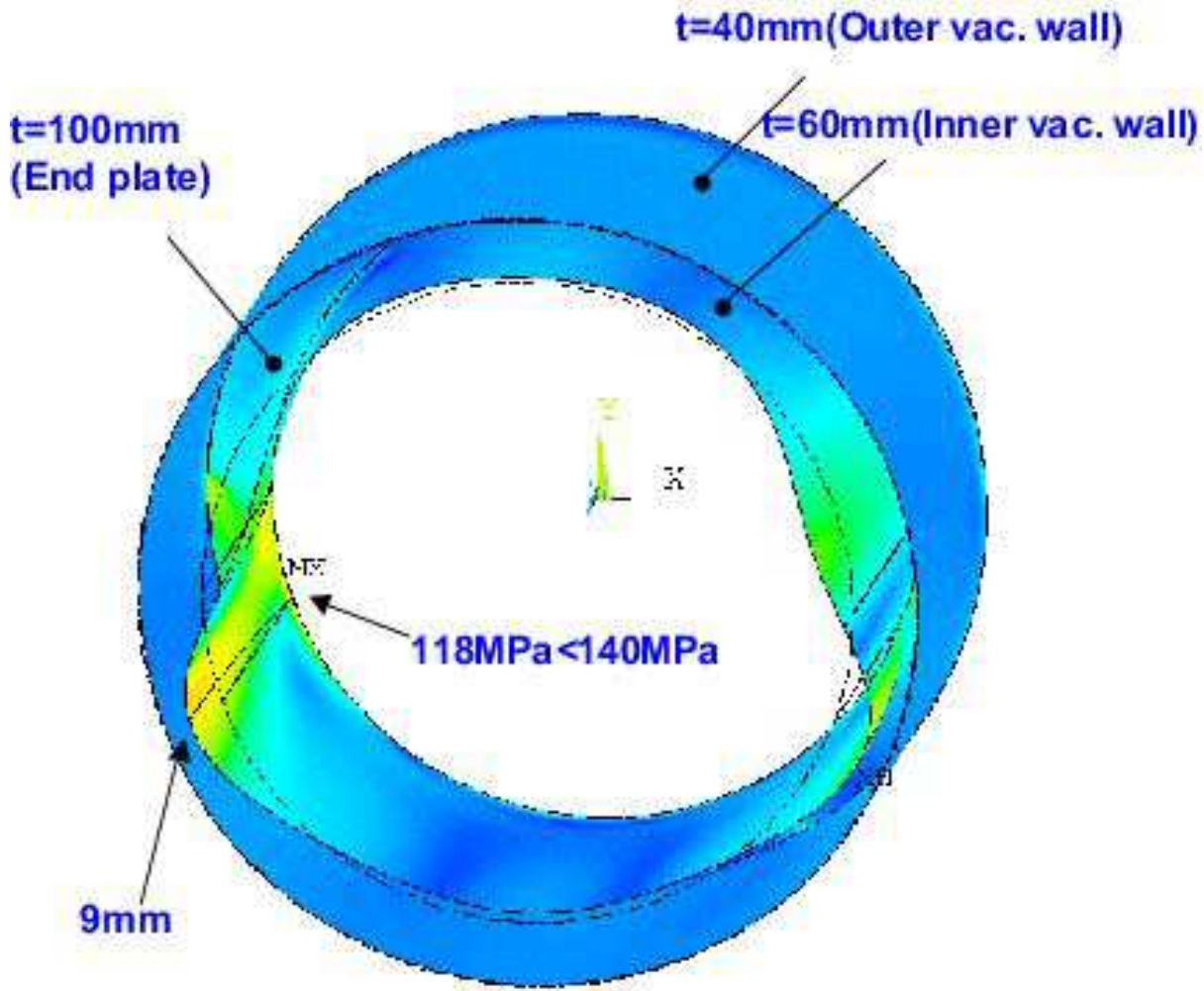}
\end{center}
\caption{Stress distribution on the cryostat when a 2000 tons of self-weight of the calorimeter
is supported at horizontal positions. }
\label{Cryo-stress}
\end{figure}

Since the magnet is located outside of the calorimeter, it is not necessary to design as a 
thin superconducting solenoid magnet. Therefore, the weight of the calorimeter which is 
estimated to be 2000 tons is planed to support at the inner wall of cryostat.
 Load conditions for cryostat design are 2000 tons of self-weight of the calorimeter, 
78 tons of cold mass, 380 tons of de-centering force on the bulk head and 0.1MPa of 
vacuum pressure. The cryostat material is assumed to be stainless steel. Thickness of 
the outer vacuum vessel can be calculated from the design guide line named 
NASA-SP8007. Then thickness of outer vacuum wall was calculated to be 60~mm with a 
safety margin of two.
 Figure~\ref{Cryo-stress} shows the calculation result when the weight of the calorimeter is suspended 
at horizontal positions of the inner wall. In this calculation, 60~mm of outer vacuum 
vessel, 100~mm of end plate and 60~mm of inner vessel were defined. The maximum 
stress is calculated to be about 118~MPa, this is below the allowable stress of stainless 
steel. The deformation is about 9~mm. In the seismic force calculation, load condition is 
0.3~G of the self-weight of calorimeter is input to the horizontal direction of the inner 
vacuum wall. This result is 125~MPa of maximum stress and 9mm of deformation. One 
important thing in these calculations is that the calorimeter has a stiff structure. Otherwise, 
the self-weight of the calorimeter is placed on the bottom of the inner wall of the 
cryostat, which will cause the high stress level in the cryostat.

\begin{figure}
\begin{center}
\includegraphics[width=16cm]{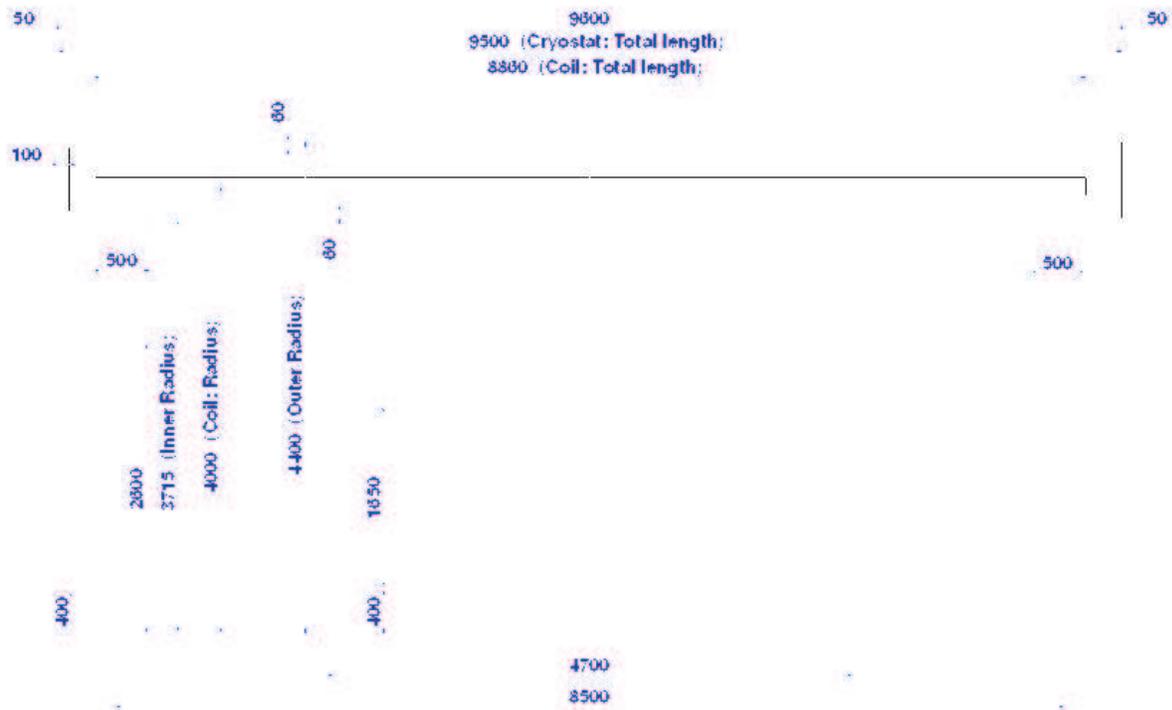}
\end{center}
\caption{Half view of the GLD solenoid magnet. }
\label{Sol-conf}
\end{figure}

\subsection*{Assembly Procedure}
Drawing of the solenoid magnet is shown in Figure~\ref{Sol-conf}. Assembly procedure of the solenoid magnet will be similar to the ATLAS solenoid 
magnet. In that procedure, the superconductor is wound to the temporal mandrel with a cover by insulator at first. Then the coil winding is carried out after the temporal mandrel is set 
to the winding machine located inside the support cylinder. After this process,  pure-al strips are mounted on the inner surface of the coil. The solenoid coil is completed after curing.
 Outer radiation shield is installed to the inside of the outer vacuum vessel. Then the 
coil with axial supports is installed. The inner radiation shield and inner vacuum vessel 
will be assembled. Then bulk heads are fixed. After the coil position is adjusted and the solenoid is laid down, the chimney is mounted. This work might be done after installation of the solenoid to the return yoke.

\subsection*{Conclusion}
The GLD superconducting solenoid magnet has been studied. The coil design and 
superconducting wire design has mainly reported in this section. The thermal design 
and cryogenic design should be more investigated.
In the mechanical design, support configuration of the solenoid magnet and the 
installation procedure should be carried out as the next step.

\subsection{Structure}
\subsection*{Introduction}
An iron structure for the GLD detector has been studied and designed.
 The structure consists of the barrel and two 
end-yoke sections with dodecagonal shape. The overall height is about 17~m from the 
floor level; the depth is about 16~m. The iron structure will be made from low-carbon 
steel (JIS-S10C). The total weight except for sub-detectors is 
approximately 16000~tons, and it will be placed on the transportation system for 
roll-in/out. Assembling of the iron yoke and installation of the sub-detector will be 
carried out at the roll-out position and then moved to the roll-in position for the 
experiment. The end-yoke separates at its center and then opens to access the 
sub-detectors.
 Some kind of design studies are required to optimize the iron yoke configuration. Those 
are the influence of the magnetic field, stability against the acted force such as 
self-weight of iron yoke and assembling/maintenance consideration. In the magnetic 
field study, the iron yoke is acted on the magnetic field of 3~T from the GLD solenoid 
magnet, amount of iron have to be determined for improving the field uniformity in the 
TPC volume, and leakage field outside of the iron yoke is kept minimize by absorbing 
the flux return. In the mechanical design, it is required to study on the deformation and 
stress level against heavy self-weight of iron and strong magnetic force. As the other 
important issue for the mechanical design, an earthquake resistant design is required.

\begin{figure}
\begin{center}
\includegraphics[width=8cm]{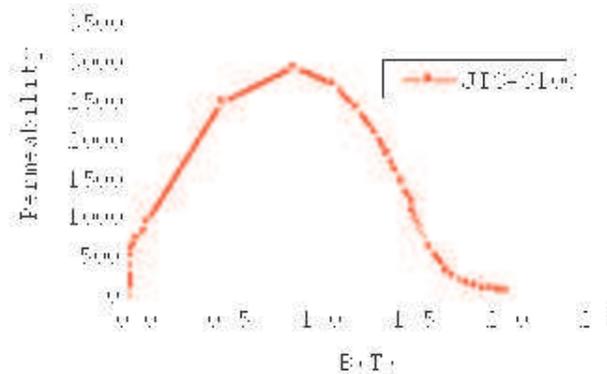}
\end{center}
\caption{Input permeability for the magnetic field calculation. }
\label{permeabiliy}
\end{figure}

\subsection*{Magnetic Field Design}
\begin{figure}
\begin{center}
\includegraphics[width=16cm]{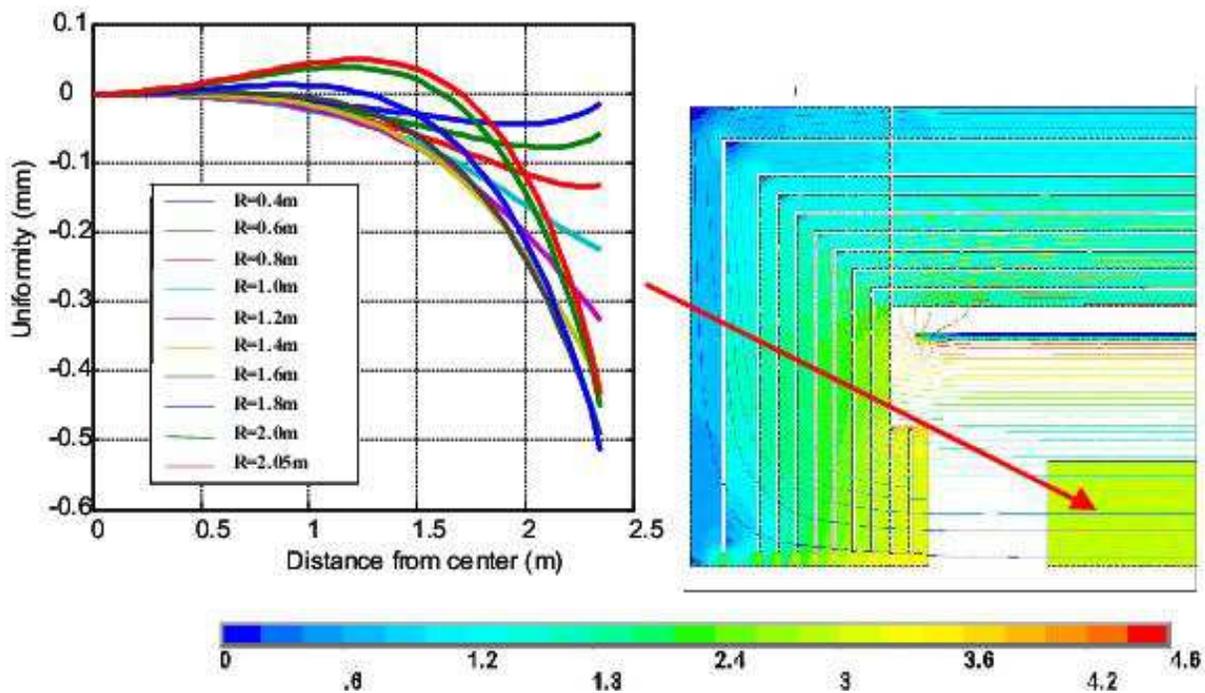}
\end{center}
\caption{Field uniformity in the TPC volume. }
\label{FieldUnif}
\end{figure}

\begin{figure}
\begin{center}
\includegraphics[width=8cm]{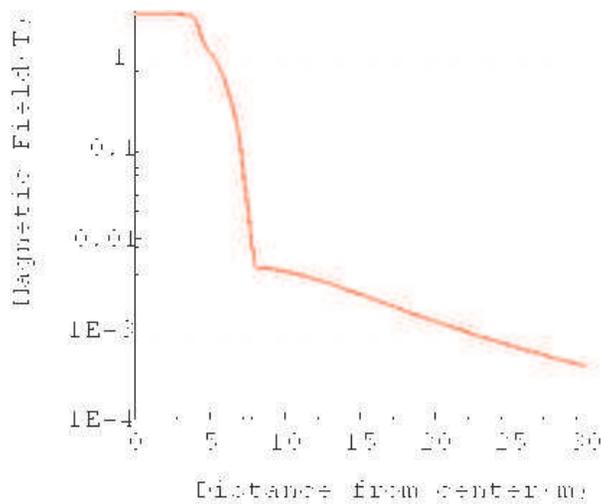}
\end{center}
\caption{Leakage field along the beam axis. }
\label{leakage}
\end{figure}

There are two criteria for the magnetic design. One is integrated field uniformity in the 
TPC volume is less than 2mm. And another is leakage field at 10~m away from the solenoid 
center should be less than 50 gauss from the requirement of accelerator design. Between the 
barrel-yoke and the end-yoke, a 50~mm of air-gap is provided for passing though the 
cables from the inner detectors. With keeping these requirements, thickness of iron 
plates were tried to minimize as much as possible. Permeability of iron plate used for the calculation is 
the measurement data of the BELLE iron yoke as shown in Figure~\ref{permeabiliy}. The ways to 
improve the field uniformity without increasing the amount of iron are are to change the 
parameters of correction coil, which is located at the both ends, such as current density and coil length. Other ways are to change the radius and thickness of iron inside the solenoid. 
The distribution of magnetic field density is show in right-hand side of Figure~\ref{FieldUnif}. The calculated 
field uniformity is from -0.5~mm to +0.05~mm as shown in left-hand side of
Figure~\ref{FieldUnif}. The magnetic field along the beam line is shown in Figure~\ref{leakage}.
The leakage field at 10~m away from the solenoid center is calculated to 
be 44 gauss. In this calculation, the outer two layers of return yoke
has to be 50~cm-thick of iron plate to reduce the leakage field.
 From this calculation, each iron plate thickness of barrel and end-yoke can be optimized.

\subsection*{Mechanical Design}
\begin{figure}
\begin{center}
\includegraphics[width=16cm]{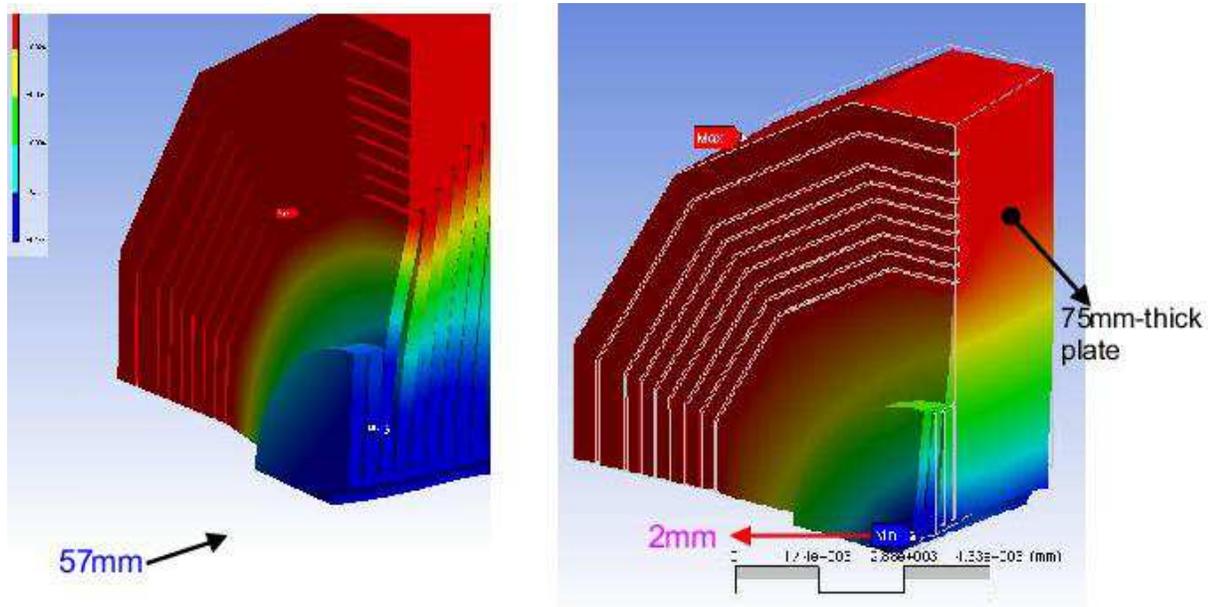}
\end{center}
\caption{Stress contour display against magneic force. }
\label{Magfor}
\end{figure}

The mechanical strength of JIS-S10C are 310~MPa of tensile strength and 205~MPa of
yield strength, respectively. The iron yoke has to 
withstand against three kinds of load conditions. Those are self-weight, magnetic force 
and seismic force. The self-weight of iron yoke is estimated to 16000 tons. The barrel 
yoke is about 7500 tons and end-yoke is 8500 tons, respectively. The magnetic force is 
calculated to be 18000~tons. An input acceleration for the seismic design is applied 0.3G.
The gravitational sag of barrel yoke is calculated to be 1.8~mm maximum. The stress level in the calculation was small enough. It is 
important to assemble with keeping the rigid structure. The gravitational sag of 
end-yoke is neglected. The deformation of the end-yoke due to the 
magnetic force of 18000 tons is shown in left-hand side of Figure~\ref{Magfor}. The maximum deformation was 
calculated to be 75~mm. Becausue this deformation can not be accepted, a support plate is 
required to reduce deformation and stress. The calculation result with 75~mm-thick of 
support plates are shown in the right-hand side of Figure~\ref{Magfor}. The support plate is mounted on every 90 degree. 
The deformation can be reduced to 1.8~mm. The result of seismic force is 2.8~mm in the horizontal direction.

\subsection*{Configurations}
\begin{figure}
\begin{center}
\includegraphics[width=16cm]{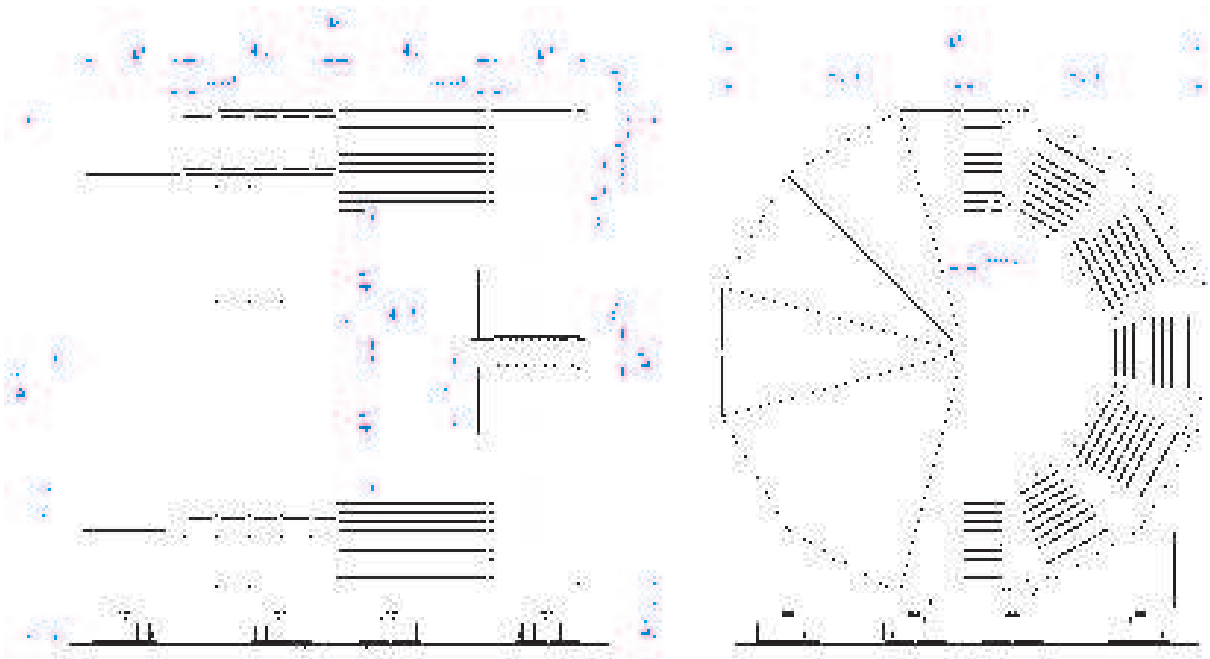}
\end{center}
\caption{Drawing of the GLD iron yoke. }
\label{YokeConf}
\end{figure}

Drawing of the iron yoke is shown in Figure~\ref{YokeConf}. The prefered shape of the yoke is 
a dodecagon from the viewpoint of the calorimeter design. There 
is a 50~mm air-gap between the barrel-yoke and the end-yoke as cable holes. So 
the barrel-yoke structure is constructed from twelve Muon modules. Each barrel Muon 
module consists of 9 layers with 50~mm thick instrumental gaps. Thicknesses of steel 
plates are 50~cm in outer two layers and 25cm in other layers, respectively. 
Few millimeters of flatness of steel plate should be taken into account on the Muon detector design.
 Each end-yoke is planned to separate into four quadrants containing eleven iron plates. 
Plate thickness of each layer is same as the corresponding plate of the barrel yoke. The end-yoke can slide out in order to provide an access to the inner detector. The iron yoke is sitting on an end-yoke transportation system.

\subsection*{Assembling}
\begin{figure}
\begin{center}
\includegraphics[width=16cm]{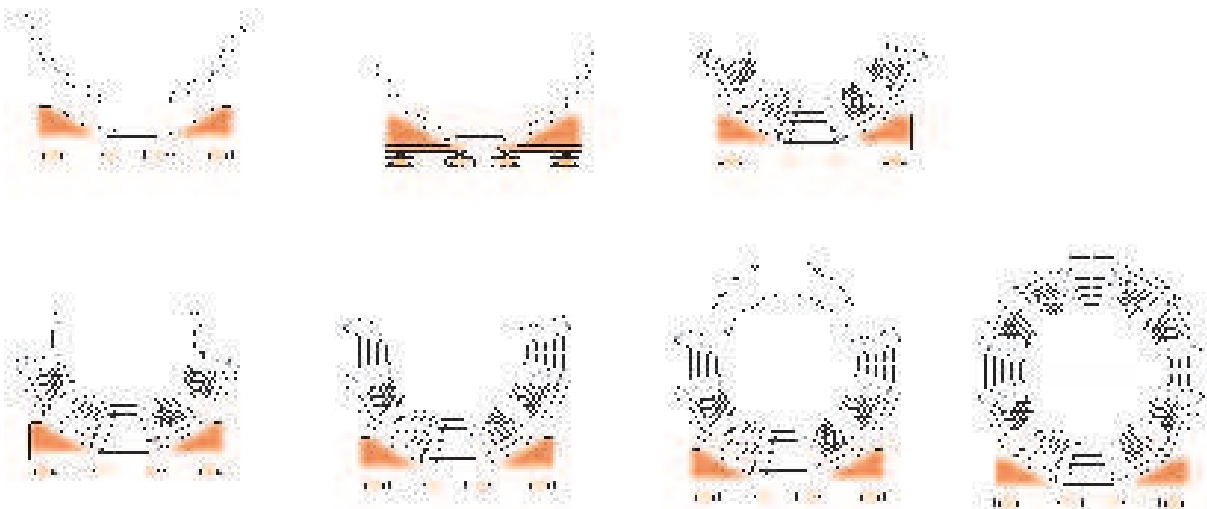}
\end{center}
\caption{Assembling procedure of the barrel-yoke. }
\label{AssemBY}
\end{figure}

\begin{figure}
\begin{center}
\includegraphics[width=16cm]{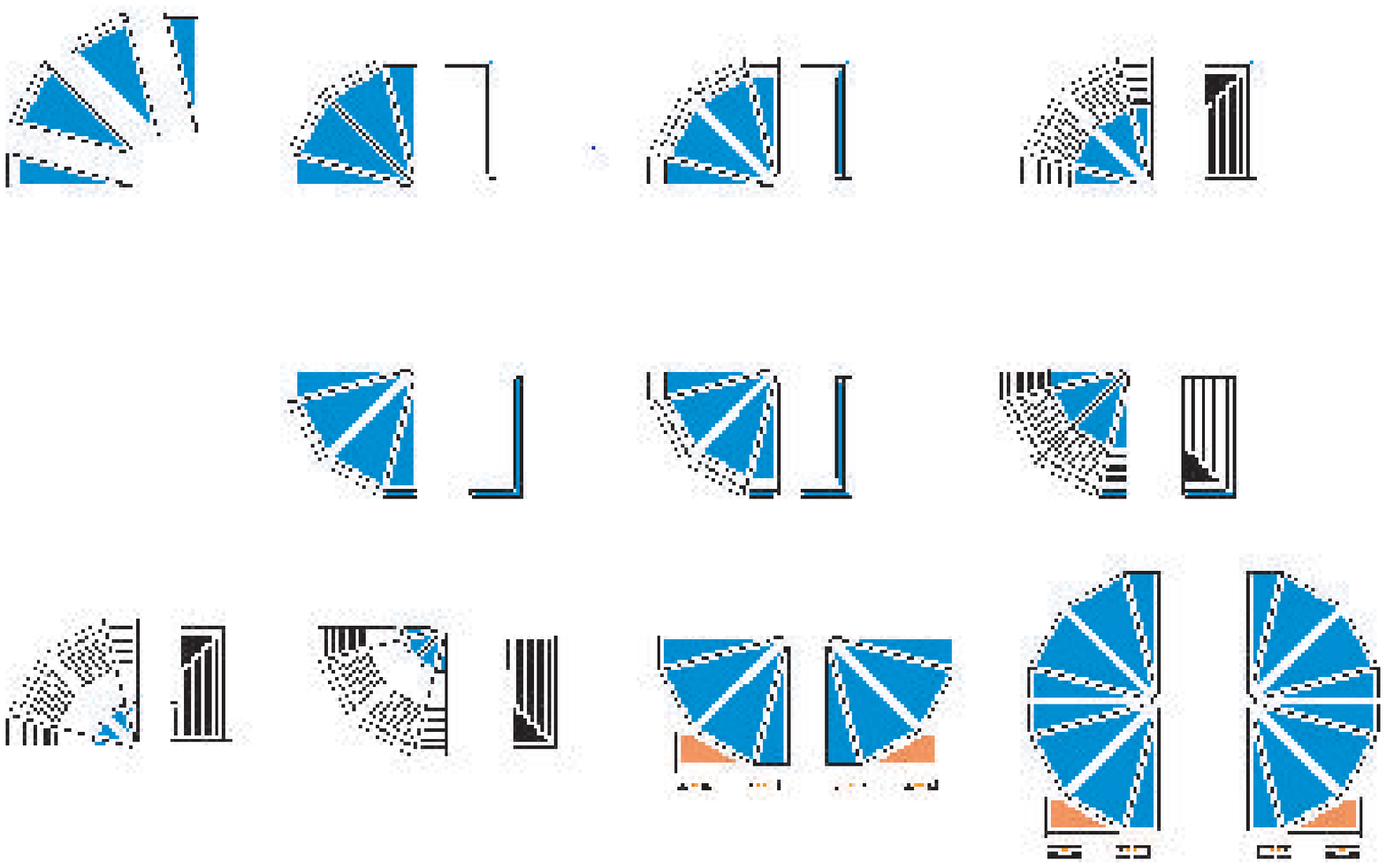}
\end{center}
\caption{Assembling procedure of the end-yoke. }
\label{AssemEY}
\end{figure}

Assembling procedure for the barrel-yoke is shown in Figure~\ref{AssemBY}. Each Iron plates are
bolted on the support frame. There are support jigs at each corner. Assembling will 
be done from the bottom side and the outer layer.
At end-yoke assembling, each pieces divided to eight are assembled by welding. Because 
the end-yoke must withstand the huge magnetic force, each piece should be fixed rigidly. 
After completing each segment, end-yoke is assembled by bolting joints as shown in Figure~\ref{AssemEY}.

\subsection*{Experimental Hall}
\begin{figure}
\begin{center}
\includegraphics[width=16cm]{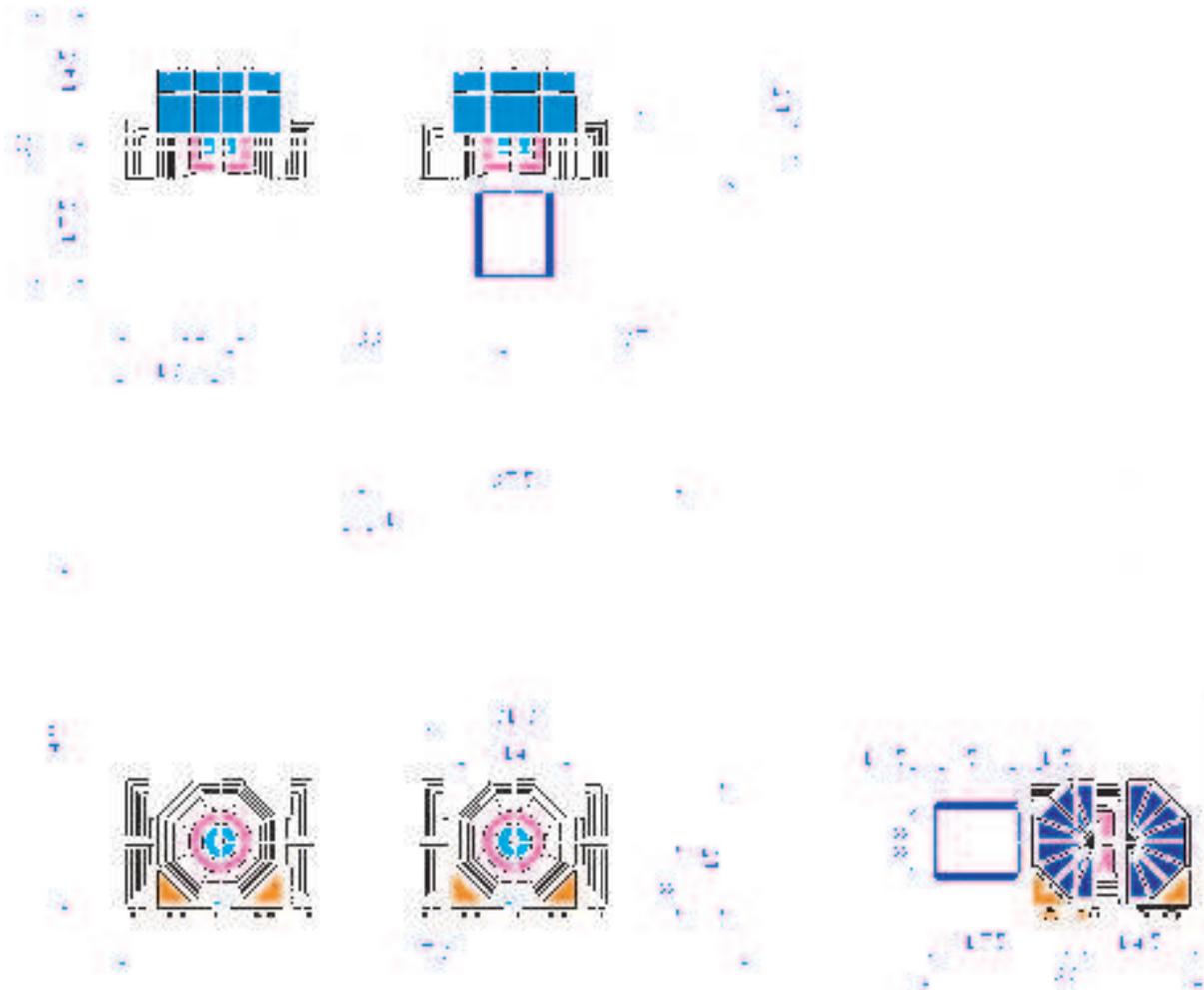}
\end{center}
\caption{Floor plan of the experimantal hall. }
% \label{Resolutions}
\label{ExpHall}
\end{figure}

One of the request from the facility study team is that the smaller experimental hall is preferred from the viewpoint of the cost. 
To meat this request, a mechanism to open the end 
yoke is important. The way to open the end yokes at the experimental hall is to rotate the 
end-yoke by 90 degree like a door at first, then slide it along the beam direction 
about 10~m. Then, an access to the inner detector is possible. There are several kinds of 
moving devices available in the  market, such as linear-guide system,  air-pad system and so on. It is preferred to use an spherical roller table. 
Because the movement of the end-yoke is not only in one direction, i.e., including rotation 
and sliding. The spherical roller table can be used to move multi directions and air 
supply system is unnecessary. 
 Figure~\ref{ExpHall} shows the experimental hall. The size of the hall is 72~m length, 32~m width and 
40~m high. Two area, experimental area and maintenance area, are planed. The distance 
from the beam line and the nearest wall is 9~m.

\subsection*{Conclusion}
 In this magnetic field design, field uniformity and leakage field can be satisfied.
Those are 0.2~mm and 40 gauss, respectively, and they are well below 
the criteria. To support the magnetic 
force on the end-yoke, 75~mm-thick of support plate is necessary even though effects of 
self-weight and seismic force are small.
Parameters of the GLD iron yoke is summarized in Table~\ref{yoketable}.
Based on the basic plan of the assembling procedure and the experimental hall presented here,
more detail design is necessary.

\begin{table}[h]
\caption{Parameters of the Iron Yoke.}
\label{yoketable}
\begin{center}
\begin{tabular}{|l|c|}\hline 
Shape & Dedecagon \\
\hline
Air gap for Muon & 5~cm \\
Uniformity(mm) & (-0.55) - (+0.05) \\
Leakage field & 43~gauss(at 10m)\\
Barrel yoke& 9~layers \\
End yoke & 11~layers \\
\hline
Width & 15.3~m \\
Length & 16~m \\
Height & 17~m \\
\hline
Barrel yoke & 8250~tons \\
End yoke & 4172x2~tons \\
Total & 16594~tons \\
\hline
\end{tabular}
\end{center}
\end{table}

    % to be prepared by H.Yamaoka
\clearpage

%
% Data Aquisition
%
\section{Data Acquisition}
\label{SectionDAQ}
\subsection{Introduction}
In a first look, the data acquisition (DAQ) system for a ILC detector
might be just a significant extrapolation, using more modern
technologies than those used in all prior large scale, high energy
physics collider experiments.

It should, a priori, be less demanding than the hadron colliders
(Tevatron and LHC). ILC detector doesn't have to cope with multiple
minimum bias events per bunch crossing (BX) and high rate triggering
for needles in a haystack, and radiation hardness is less critical.
Hence many more detector technologies are available and better
performance is possible. But, contrary to discovery machines, ILC
detector does have to record all the observable information;
measure jets and charged tracks with unparalleled precision, measure
beam energy and its spread, differential luminosity and polarization,
and tag all vertices, thus better performance and full hermiticity are
needed in order to do physics with all final states.

Compared to previous lepton colliders (LEP, SLC and B-Factories)
where jets and leptons are the fundamental quanta, they must be
identified and measured well enough to discriminate between $Z$'s,
$W$'s, $H$'s and possible new states. Challenging ILC physics goals
requires to improve the jet resolution by a factor of two, and
the excellent tracking capability for recoil mass measurements
in the Higgs boson search, which requires 10 times better momentum
resolution than LEP/SLD detectors and 1/3 better on the impact
parameter resolution than SLD. To catch multi-jet final states
(e.g. $t\bar{t}H$ can decay into 8 jets), the need of a real
4$\pi$ solid angle coverage with full detector capability is
fundamental. Furthermore, the paradigm of ``Particle Flow
Algorithm (PFA)'' for precision measurement of the jet energy
requires an unprecedented fine-granularity calorimeter for good
separation of charged and neutral particles. Such hermiticity
and granularity have never been done up to now. Finally, the PFA
also puts new challenge for real time analysis.

In summary, at the ILC the stringent requirements imposed by the high
precision physics goals need a detector concept structure with full
hermiticity, high granularity detectors with a DAQ able to select,
analyze and store standard as well as rare events in a huge background
environment. In addition, the high luminosity operation at the
interaction point of the linear accelerator with two long independent
arms needs to be optimized in real time using simultaneous information
of both machine and experiment. These features put new requirements
and challenges in this domain to be addressed.

From the conceptual point of view, the burst mode of the beam
structure of the ILC machine (3000 BXs for a duration of roughly
1 ms at 5 Hz frequency) yields a BX rate of 3 MHz for 1 ms followed
by a period of almost 200 ms without any interaction. That dictates
immediately an innovative DAQ architecture with no hardware trigger.

From the technical point of view, the rapid development of computing
and telecom technologies as well as the higher integration and lower
power consumption of electronic components fits nicely with the
requirements needed for all modern DAQ components. For such large
and long-term life system, the uniformity of the various interfaces
associated with standardized components is vital to achieve flexibility,
maintainability, scalability at an affordable effort, requiring
commodity hardware and industry standards to be used wherever possible.

Therefore, and partly because of the rather much dependence of the
detector design, the DAQ system presented at this stage is only a
conceptual model with some basic principles to be refined later,
for example, the implementation of special non machine synchronized
triggers such as Calibration, Cosmic. However, its aims is to define
clearly the common and uniform interfaces between the various logical
components and to show the feasibility of the system in general.

\subsection{Event Selection System (Trigger)}
The main goal of event selection system (Trigger) is to select
interesting events in the presence of several orders of magnitude
higher background without losing data of a possible, yet unknown,
physics process. In addition the rates of the known interesting
physics processes vary as well by several orders of magnitude.
The ILC operation conditions are quite different from those of
accelerators currently operating or being constructed. The main
ILC parameters of the 500 GeV present design, relevant for the
DAQ system, are:
\begin{itemize}
\item a long time interval between two bunch trains of 200 ms,
\item a separation of two bunches inside a train by 308 ns,
\item a train length of 868 $\mu$s.
\end{itemize}
These conditions drive the proposal of an event building without
any hardware trigger, followed by a fully software based event
selection, identification and analysis. This concept allows to
analyze each event one by one without taking into account a fixed
trigger latency as it is done in the standard experiments.

\begin{figure*}
\centering
\includegraphics[width=\textwidth]{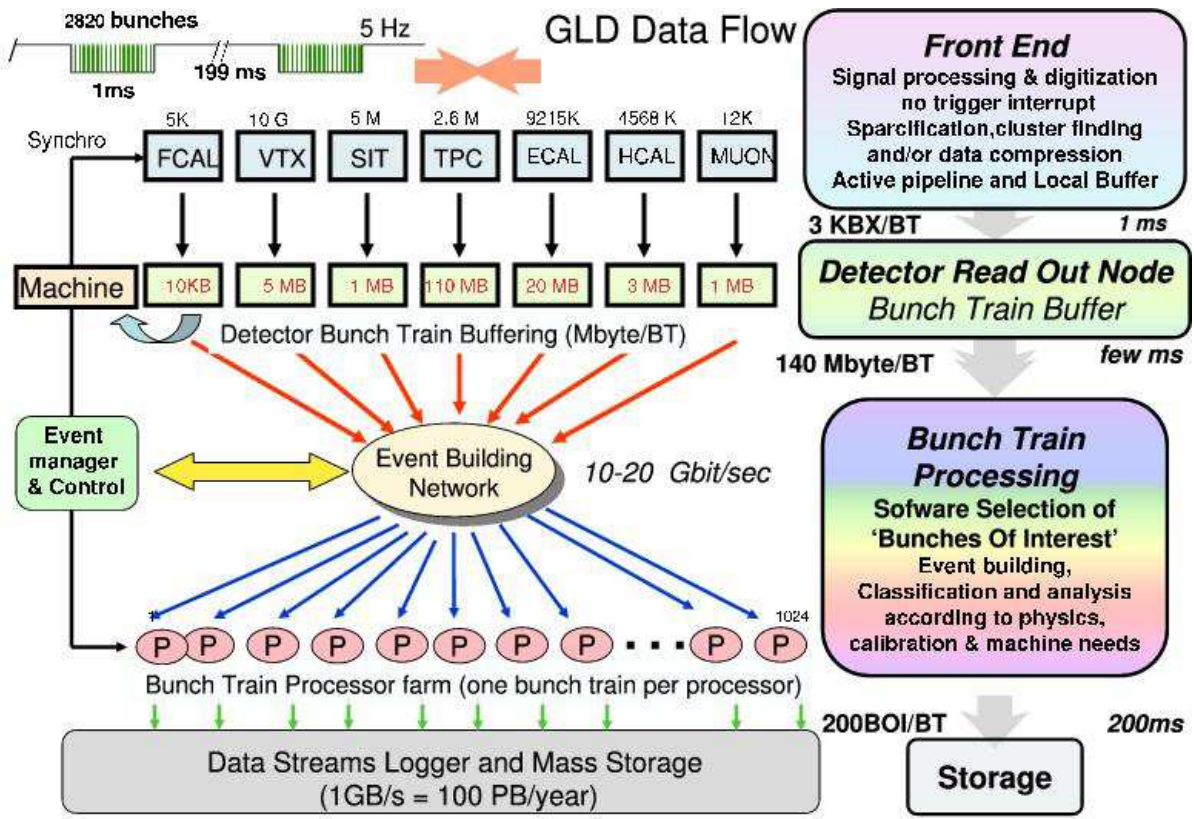}
\caption{Schematic view of the data paths and data flow in the DAQ system.}
\label{daq-dataflow}
\end{figure*}

To achieve a dead-time free data acquisition under the ILC operation
conditions, a trigger-free pipeline of 1 ms with readout of all data
in the pipeline within 200 ms is proposed. The full data paths and
data flow is presented in the Figure~\ref{daq-dataflow}.

In the first step, during the 1 ms bunch train of 3000 collisions,
each detector channel signal is digitally processed at its front end
without any trigger interrupt. During or immediately after this,
1 ms active pipeline, depending on the detector type, the raw data
throughput will be reduced directly at the subdetector front end level
by implementing the usual data compression tools and algorithms like
zero suppression, sparcification, hit detection or cluster finding
before storing the relevant data fragment in a local buffer.

In a second step, during the first part of 200 ms available with no beam,
a collection of all the data in the full bunch train is performed.
Each locally stored front end data fragment is digitally multiplexed
in a reasonable quantity of fast read out links and then pushed to
a series of Bunch Train read out Buffer (BTB) mapping each detector.
After the necessary time of data collection estimated few ms,
which is compatible with the power cycling of the font-end electronics,
the complete collection of the full bunch train data is performed
by one processing element, the Bunch Train Processor (BTP) of the online
farm, prior to a software selection of Bunches Of Interest (BOI) using a
specific set of raw data followed by an event classification according to
physics, calibration and machine needs. The only constraint in this model
is that the data collection task should be made within the $\sim$190 ms,
remaining several ms before the next bunch train, in order to free
all the read out buffers (BTB's), however the rest of the treatment
could be longer since the following bunch train will be analyzed in
another processor. This scheme of merging all the processing tasks
from the event builder and event finder in one processor unit, avoids
to move data several times and makes event management and control easier.

The evolution of embedded processing power and network will allow to
merge in this processor all the successive software tasks, accomplished
today in large experiments like the so-called Level 2 (L2), Level 3 (L3),
High Level Trigger and Filter (HLT), up to the first full reconstruction
and physics analysis using the PFA of each BOI event.
In addition the standard real time monitoring tasks should be performed
like calibration and alignment for the detectors as well as complete
analysis of sampled events. From the total hadronic cross section for
$e^{+}e^{-}$-annihilation, about 0.1 for high $Q^{2}$ events per train
are expected and 4 for Bhabha whereas about 200 events per train will
come from $\gamma\gamma$ reactions. All these events are potentially
interesting, e.g. $\gamma\gamma$ and low $\Delta M$ SUSY are very
difficult to distinguish, and a systematic storage of those candidates
might be necessary prior to a complex analysis carried out later on.
Assuming a rate of 200 BOI selected per bunch train and processing time
of 1 second per event in average to perform all these analysis tasks,
a farm of 1024 processors would be necessary.

\subsection{DAQ Boundaries and Read Out Architecture}
One of the most important tasks in the development of a DAQ system
for the GLD detector is the choice of a coherent architecture that
meets the requirements of all subdetectors. The system cost will be
minimized and the reliability and ease of debugging optimized if
an architecture is chosen that is uniformly applied. The requirements
for the GLD DAQ system fall into four broad categories. First are
the quantitative requirements such as those on the bandwidths of
data paths and the number of data sources and data volume to be
accommodated. Second are the important architectural requirements
that allow front end systems and event selection to be operated
together or independently. Third are the requirements for initializing,
controlling and monitoring the system. Last are the qualitative
requirements for items such as reliability, tolerable error rates
and maintainability.
All these requirements can be grouped in four functional blocks and
their interfaces.

\subsubsection{Detector Front End Electronics (DFE)}
The term of front end electronics is broadly used to denote the
electronics that is physically located within the GLD detector.
Its basic purpose is to convert sets of signals from each detector
element into digital event fragment. The signal processing typically
contains low-noise, high time resolution amplifiers and signal shapers.
Each amplifier receives a signal, almost always an analog signal,
from one sensor element such as pixel, Si strip, TPC, SiPM/MPPC.
The signal is shaped in order to extract the best quality time
and/or amplitude information from the input signal. Digitization or
quantization occurs immediately after this stage. The provision for
generating calibration signals is typically made. This feature can be
used for channel time and/or amplitude calibration by inserting
artificial sets of data to test all or part of the electronics chain.
The signal must be preserved within the front end system for the 1 ms
bunch train. The front end electronics system of the GLD detector
will be implemented in various ways, depending on the sensor type.

The impact of a 1 ms active pipeline without trigger interrupt and
therefore no possibility of a fast clear or buffer flushing for the
different subdetector pipelines has been studied and readout
technologies able to cope with this have been proposed for several
detector technologies. For the vertex detector built in CCD technology,
a column parallel readout at a speed of 50 MHz during the active
pipeline has been proposed to handle the expected data rates.
For the TPC several designs are being studied which allow the
collection of signals in an ungated mode for the full time of
the pipeline by limiting the ion feedback with Micro Pattern
Detector read out (GEM, Micromegas), and for the scintillator
devices (EM and hadron calorimeters and muon detector), built
out of silicon avalanche photodiode, a continuously running
pipeline with hit detection and multiplexing is foreseen.
The uniform interface between the DFE and DAQ system is
the Local Detector Buffer (LDB).

\subsubsection{Machine Detector Interface (MDI)}
The adjustment of the luminosity should be carried out at the beginning
of each bunch train. The current thinking is to use the very forward
calorimeter (BCAL) raw data of about 10 to 50 bunches to get an immediate
feedback for the machine. Accordingly, the forward calorimeters have to
be read out at each bunch crossing and the readout philosophy is
different from the rest of the detector elements. Other complex
information can be sent during the full bunch train analysis of
specific background events. Another feature connected to the machine
are the bunch synchronization and distribution.

\subsubsection{Data Acquisition (DAQ)}
This is the central part of the system. It is divided functionally
into three main architectural components, the read out node,
the network switch and the bunch train processor (BTP) farm.

%\paragraph{The Read Out Node}
The read out node associated with one detector collects all the data
fragments in one bunch train into a Bunch Train Buffer (BTB). This is
done by regrouping and multiplexing the various front end local buffer
scattered inside and around the GLD physical structure in a minimum
number of fast data links. Building the bunch train sub events as well
as checking the integrity of its content requires some intelligence
and fast connectivity at that level that controls all relevant tasks
associated with one detector. This is a networking hub where a uniform
interface across all the various detectors between the front end
electronics and the DAQ should be installed. Today, it will be done
using PCI mezzanines with powerful FPGA's embedded on VME or PC boards.

%\paragraph{The Network Switch}
The network switch with a typical capacity of 10-20 Gbit/sec will
be adequate to collect and distribute the full bunch train data,
estimated today at 140 MBytes, to the designated Bunch Train Processor.

%\paragraph{The Bunch Train Processor (BTP) Farm}
The event building will be performed at 1-2 GByte/sec and the
Bunch Of Interest selection can select up to 250 events including
the $\gamma\gamma$ physics processes as well as additional calibration
and background events for cross checks. The event size varies from
0.2 MBytes to 5 MBytes, depending on the amount of background included.
Assuming background data to be suppressed, the processed event size
will be on average about 1 MByte which leads to 250 MByte/sec output
to be transfer to the mass storage. This results in a total data volume
of roughly 1 GByte/sec or 10 PByte/year, e.g. 10 times the LHC production.
This seems a little bit too much and more study is needed to reduce this
data volume in an unbiased way. Considering that 90\% of the data
are produced by TPC, calorimeters and VTX and that most data volume
is background, some reduction would be possible to clean up event by
removing noise hits, using the time stamp on one bunch crossing and
recording only interesting regions or tracks. A rough qualitative
analysis of this issue would expect a reduction of a factor of 5 to 10.

Compared to systems built for the LHC experiments, the proposed DAQ
system for GLD is less demanding and key components like the fast
switching network and the computing units are available already today.

\subsubsection{Global Detector Network (GDN)}
The international worldwide structure of the ILC/GLD collaboration
combined with the fantastic development of fast communication tools
allows us today to imagine a new way in High Energy Physics to
gain control of the experiment from the detector calibration
and monitoring to the physics analysis.

With the exceptions of the detector integration and commissioning,
the rest of the detector operation, in particular data taking
and its associated services (ancillary controls, partitioning,
databases management, \ldots), can be distributed among several
virtual control rooms. It would be possible to select from these
remote stations the running modes between stand alone, cosmic test,
calibration and global runs without coming to the detector site
in person, as is done for many embedded satellite experiments.

From the physics analysis point of view, the boundaries between
so-called online and offline will disappear and the
development of GRID technologies shows that it will be possible
in the future to analyze data without knowing where the processing
is made.

\subsubsection{Technologies Forecast}
The fast development in the computing and network area makes it
difficult to predict the technologies which will finally be used
to build the DAQ system in several years from now. A short overview
of possible evolution of technologies is presented below, based on
past experiences, current running experiments and studies for
the experiments planned at the LHC.

Concerning the Standards: this is the end of traditional parallel
backplane bus paradigm (CAMAC, Fastbus, VME). It has been announced
every year since $\sim$1989, but still there at LHC with VME and PCI.
Recently, a new commercial standard has been emerging from the
telecom industry named Advanced Telecom Computing Architecture
(ATCA) designed for a very high system throughput (2 Tbit/sec)
and availability (99.999\%, failure rate of $\sim$5 min/year).
It has the technical features well suited for all our needs
particularly to host the detector bunch train read out hub
as well as the basic components of the machine control.
% The basic elements are the crate (sub rack or shelf) with a passive
% backplane, a shelf manager, cards (entry modules or blades),
% interconnections for servers, built-in air cooling and 48 Volts
% modular power supply. The ``Shelf Manager'' manages all module,
% crate and system utilities. The crate can contain up to 14 modules/board
% (blades), 14 to 16 units placed either vertically or horizontally
% (8U of 1,2 inch $\times$ 280 mm). The maximum power consumption available
% per board is 200 W. The passive backplane is a redundant dual star and full
% mesh (point to point) device that can use a large set of protocols like
% Ethernet, Fiber channel, PCI express, Infiniband, RapidIO. Backplane
% lines are also available for synchronization signals and up to 6 clock
% interfaces buses. Each board is associated with a Rear Transition Module
% for user up 20 W and can also be equipped with several customized carriers
% (Daughter Card, Plug-in Module, Advanced Mezzanine Card).
The basic software is already available under Linux.

The commercial networking products for Trigger/DAQ switches was already
available in the 90's: ATM, DS-Link, Fibre Channel, SCI, etc. Today,
the Gigabit Ethernet is the most popular with the rapid bandwidth
evolution from 1 Gbit/sec to 10 Gbit/sec and with the objective to
reach 30 Gbit/sec in 2010.

The ideal fast processing / memory / IO bandwidth devices existed
in the past: Transputers, DSP's, and RISC processors. Today, there are
powerful FPGA's that integrate everything: the LVDS receiver links,
DSP's, large memory, as well as more general purpose processors like
PowerPC.

For the point-to-point link technology, the old style used even recently
at LHC was parallel copper or serial optical. The recent evolution is
to go toward the serial copper and parallel optics. The link speed
is $>$ 3 Gbit/sec today and 10 Gbit/sec is in demonstration.

The continuous increasing of the computing power for the processor
power and clock (today 4GHz, 10 to 15 GHz foreseen in 2010) is not
anymore a critical issue. Idem for the memory size which is now
quasi unlimited (blocks of 256 MB today, few GB in 2010).

In conclusion, the evolution of technologies described above
shows that there will be adequate components when the final
decision has to be made.

\subsection{Strategy for the Future}
Previous experiences of DAQ systems demonstrate that the final
technologies and their associated commercial components should
be chosen as late as possible, e.g. 3--4 years before the first
beam collision. However, the architectural model with its
functional blocks and interfaces should be adopted as early as
possible in order that the various detectors could develop
the necessarily uniform interfaces at the front end electronic
level. Hence, the basic idea is to organize the DAQ development
around a common pilot project across the various concepts
and detector R\&D. This pilot project should address all the main
issues and also details attached to the conceptual model described
above, such as the integration and feedback of the machine and
its synchronization, slow controls, non machine synchronized
acquisition modes like calibration, test and cosmic trigger.
It is also the place where state of the art technologies like ATCA
could be practiced and evaluated. Finally, the worldwide aspect of
such pilot project will be a learning process for developing and
optimizing all the Global Detector Network functionalities.

% \subsection{Costing model}
% A rough estimation of the DAQ model based mainly on present components
% used in LHC (ATLAS) gives a superior limit for the hardware
% (read out hub, switch and data link, processor farm with 1000 units)
% of 20 M\$. This does not take into account the slow control,
% the software aspect, the Global Detector Network, the manpower and
% the general infrastructures.
    % to be prepared by P.Ledu
\clearpage

%
% MDI
% Modified. 6-March-2006 18:37  A.Miyamoto
%
% http://greentea.kek.jp/plone/groups/gld/internal/outline/glddod/
% http://ilcphys.kek.jp/internal/glddod/glddod.pdf
\section{Machine Detector Interface}
\label{SectionMDI}
%-- to be prepared by T.Tauchi

\subsection{Background Tolerances in GLD}
% (VTX, TPC, BCAL, LUMCAL, CAL etc.) 
%  T.Tauchi

Background tolerances of the vertex detector (VTX), the TPC,
and the calorimeter (CAL) are estimated in terms of tracking capabilities and radiation hardness.
The tolerances depends on readout time or sensitive time  of detectors, which are O(1) msec, 50~$\mu$sec and 100~nsec for FPCCD VTX, TPC and CAL, respectively.

To achiveve a good efficiency in track reconstruction, the occupancy 
in the vertex detector are required to be less  than 1\%.   Major background sources are $e^\pm$ pairs and neutrons backscattered at the extraction beam line.

The tolerable hit densities of pairs and neutrons of the vertex detector are $1\times 10^4$/cm$^2$hits/train and $1\times 10^{10}$n/cm$^2$/year, respectively, where a train have 2820 bunches with time separation of 307.7 nsec.   The hit rate corresponds to 1\% occupancy.

The TPC has $2.46 \times 10^{9}$ voxels in total, consisting of pads and time-buckets. Tolerance of background hits is estimated to be 0.5\% occupancy, that is $12.3 \times 10^{6}$ voxels/50$\mu$sec, where the 50 $\mu$sec is a full drift time in the TPC.  Assuming a hit has 5(pads)$ \times$5(buckets), the tolerance cab be expressed by $5.02 \times 10^{5}$ hits/50$\mu$sec.   From this evaluation, tolerance of muons can be estimated to be $1.23 \times 10^{3}$ muons/50 $\mu$sec assuming that  a muon creates 2000 hits in the bucket of a single pad.  Neutron tolerance can be estimated to be $4 \times 10^{4}$ n/50 $\mu$sec  assuming that  a neutron creates 10 hits assuming  100\% efficiency for neutrons to make hits which is so called neutron conversion efficiency.    Since the efficiency depends on the gas content, it has been estimated to 1\% with P10-gas in the TPC.

Muons can be serious background in the endcap calorimeters since they hit all the layers and mimic tracks. Signals can be read out  with 100 nsec gate in the calorimeters. 
Only one muon can be allowed in the endcap calorimeter of 30 m$^2$  within the gate.  Therefore, tolerance of muons is 0.03 $\mu$/m$^2$/100 nsec .  

The tolerances are listed in Table~\ref{tab:tolerance}.   They should be further investigated by the full simulation of Jupiter.  
%%%%%
\begin{table}[htdp]
\caption{Tolerances for background in VTX, TPC and CAL. }
\begin{center}
\begin{tabular}{|c|c|c|c|}\hline
Detector&Hits&Neutrons&Muons \\ \hline
VTX& $1\times 10^4$ hits/cm$^2$/train& $1\times 10^{10}$ n/cm$^2$/year & - \\
TPC& $5.02\times 10^5$ hits/50$\mu$sec& $4 \times 10^{4}$ n*/50 $\mu$sec & $1.23 \times 10^{3} \mu$/50 $\mu$sec \\
CAL& 1 hits/m$^2$/100 nsec &  -  & 0.03 $\mu$/m$^2$/100 nsec \\
\hline
%\hline
\end{tabular}
\end{center}
\label{tab:tolerance}
\end{table}
%%%%%

\subsection{Interaction Region (IR) Design}
%:  Y.Sugimoto, H.Yamamoto, E.Won	
% - beam pipe configuration
% - VTX, BCAL, LUMCAL, Pair monitor, Intra-train feedback system
% -  Performance of realtime luminosity monitor by BCAL and Pair monitor

% 2(2*) to be 2 (2*), *: there is an exit hole with the same radius.

\begin{figure}[ht] 
     \begin{center}
     \vspace*{.2cm}
      \includegraphics[width=10cm]{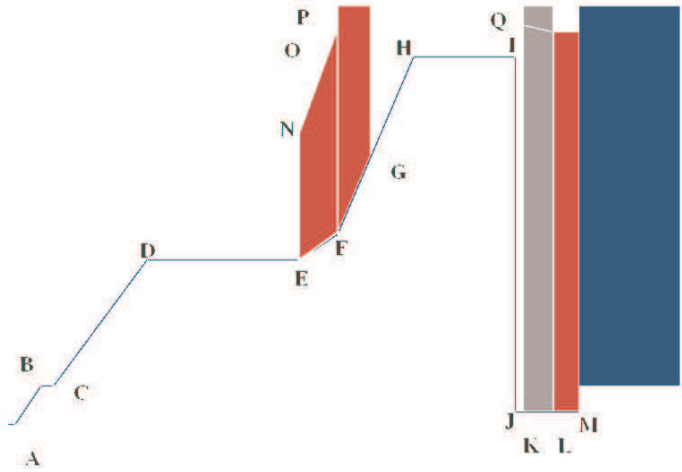}
     \end{center}
     \caption{IR Geometry, where red-region is calorimeter (FCAL and BCAL), the grey one is a CH$_2$ (low Z) mask and the blue is the final quadrupole magnet(QD0).  A pair monitor is set in front of the CH$_2$ mask, i.e. between J and K. } 
     \label{fig:beampipe}
     \end{figure}

IR geometries, especially the beam pipe and the innermost radius of VTX, depend on the machine parameters such as beam energy,  intensity and sizes at the interaction point (IP) as well as crossing angle, distance from IP to the final quadrupole magnet (L*) and the detector solenoid field (B).   While the accelerator can be seamlessly operated in the parameter space\cite{Con_raubenheimer,highlum},  the present designs were optimized with all the parameter sets at E$_{cm}=$500 GeV and 1 TeV, L* = 4.5 m and  B = 3 Tesla for physics and background studies.  
As explained in the VTX section, the beam pipe inner radius has been determined from the IP to the final focus quadrupole magnet (QD0) by the envelope of pairs \cite{sugimoto}.  
The configuration of beam pipe is shown in Figure~\ref{fig:beampipe}.
 The detailed geometrical data are listed together with the extreme cases of the high luminosity parameter sets and the improved one at E$_{cm}=$1 TeV \cite{highlum} in Table~\ref{tab:beampipe} .    The innermost VTX layer is located to be 0.4 or 0.5 cm from the beam pipe, e.g. its radius is 1.7 cm with the beam pipe of 1.3 cm radius at the nominal set.   The standard radii of beam pipe and the innermost VTX layer are 1.5 cm and 2.0 cm, respectively, for the Jupiter simulation.
%%%%
%%%%
%%%%%
\begin{table}[htdp]
\caption{IR geometrical data with 2 (20) mrad crossing angle; numbers in parentheses are those at 20 mrad crossing angle, while the others are common at the both angles. }
\begin{center}
\begin{tabular}{|c||c|c|c|c||c|c|}\hline
E$_{cm}$&\multicolumn{4}{|c||}{500 GeV}& \multicolumn{2}{|c|}{1 TeV}  \\
\hline %\hline
para.set&\multicolumn{2}{|c|}{Nominal}&\multicolumn{2}{|c||}{High Luminosity}&\multicolumn{2}{|c|}{High Luminosity-1} \\ \hline
position&R in cm&Z in cm&R in cm&Z in cm&R in cm&Z in cm\\ \hline %\hline
A& 1.3& 4.5& 1.9& 6.3&     1.5& 5 \\ \hline
B& 3(3.2)&25&4.2&25&    3.4(3.5)&25 \\ \hline
C& 3(3.2)&35&4.2&35&    3.4(3.5)&35 \\ \hline
D&8&110&9(10)&110&   8(9)&110  \\\hline
E&8&230&9(10)&230&   8(9)&230 \\ \hline
F&9.04&260&10.2(11.3)&260&   9.04&260 \\\hline
G&11.94&285&12.60(13.26)&285&  11.94(12.60)&285 \\\hline
H&16&320&16&320&  16&320 \\ \hline 
I&16&400&16&400&  16&400 \\ \hline
J&2(2*)&400&2(2*)&400& 2(2*)&400 \\\hline
K&2(2*)&405&2(2*)&405&2(2*)&405 \\ \hline
L&2(2*)&430&2(2*)&430&2(2*)&430 \\\hline
M&2(2*)&450&2(2*)&450&2(2*)&450 \\\hline
N&13&230&14(15)&230&  13(14)&230 \\\hline
O&17.70&260&18.83(19.96)&260&  17.70(18.83)&260 \\\hline
P&36&260&36&260&  36&260 \\\hline
Q&17.96&430&19.83(21.70)&430&  17.96(19.83)& 430 \\ \hline
%\hline
\end{tabular}
\end{center}
* : There are two holes with the same radius  for incoming and exit beams at the 20 mrad crossing angle.
\label{tab:beampipe}
\end{table}
%%%%%

\subsection{Collimation Aperture}
% with and without the tail holder (octupoles) : T.Abe
% 8 x 65 by A.Drozhdin, with initial halo distribution (5-13) x (36-93)
\subsubsection{Synchrotron Radiation}
       %-  innermost radius of the vertex detector
       %- masking system for the radiations
%Collimator depth is A.Drozhdin's ILCFF9, i.e. 5 x 65 , statistics is 500000 in LCBDS
%Ebeam:250 GeV, dp/p=0.3%, halo 50 x 200 in sigma's

Since the beams are accompanied by halos,  they should be sharply collimated at the collimation section at upstream in the beam delivery system.   The halo particles generate synchrotron radiations in magnets and muons at the collimators.   
Collimation apertures  have been optimized for the radiation profile to be small at IR region in order to prevent the direct hits at beam pipes and inner surfaces of the final focus magnets.   We adopt the collimation scheme by A. Drozhdin \cite{drozhdin}.  
The apertures are set to be  $8 \sigma_x \times  65 \sigma_y$ .  The properties of major collimators are listed in Table~\ref{tab:collimator}, where absorbers ( 30 X$_o$ at least )  are installed behind the spoilers of SP2, SP4 and SPEX .  Two masks of MSK1 and MSK2 are set to shield IR region against synchrotron radiations at downstream of the last bend magnets which are located at 96 m from IP.
%%%%%
\begin{table}[htdp]
\caption{Major collimators' location from IP, aperture, length and material (ILCFF9). }
\begin{center}
\begin{tabular}{|c|c|c|c|c|c|c|c|}\hline
name&Location&Thickness&Material&\multicolumn{4}{|c|}{Aperture}  \\
 & m & X$_o$ &  & x(mm)&y(mm)&x($\sigma_x $)&y($\sigma_y $)\\ \hline
 SP2 & 1483.27 & 0.6 & Copper & 0.9 & 0.5 & 8 & 65 \\
 SP4 & 1286.02 & 0.6 & Copper & 0.9 & 0.5 & 8 & 65 \\
 SPEX & 990.42 & 1 & Titanium & 0.5 & 0.8 & 10 & 62 \\
 MSK1 & 49.81 & 30 & Tungsten & 7.8 & 4.0 & 16 & 178 \\
 MSK2 & 13.02 & 30 & Tungsten & 7.4 & 4.5 & 12 & 151 \\ \hline
\end{tabular}
\end{center}
\label{tab:collimator}
\end{table}
%%%%%

\begin{figure}[ht] 
     \begin{center}
     \vspace*{.2cm}
      \includegraphics[width=14cm]{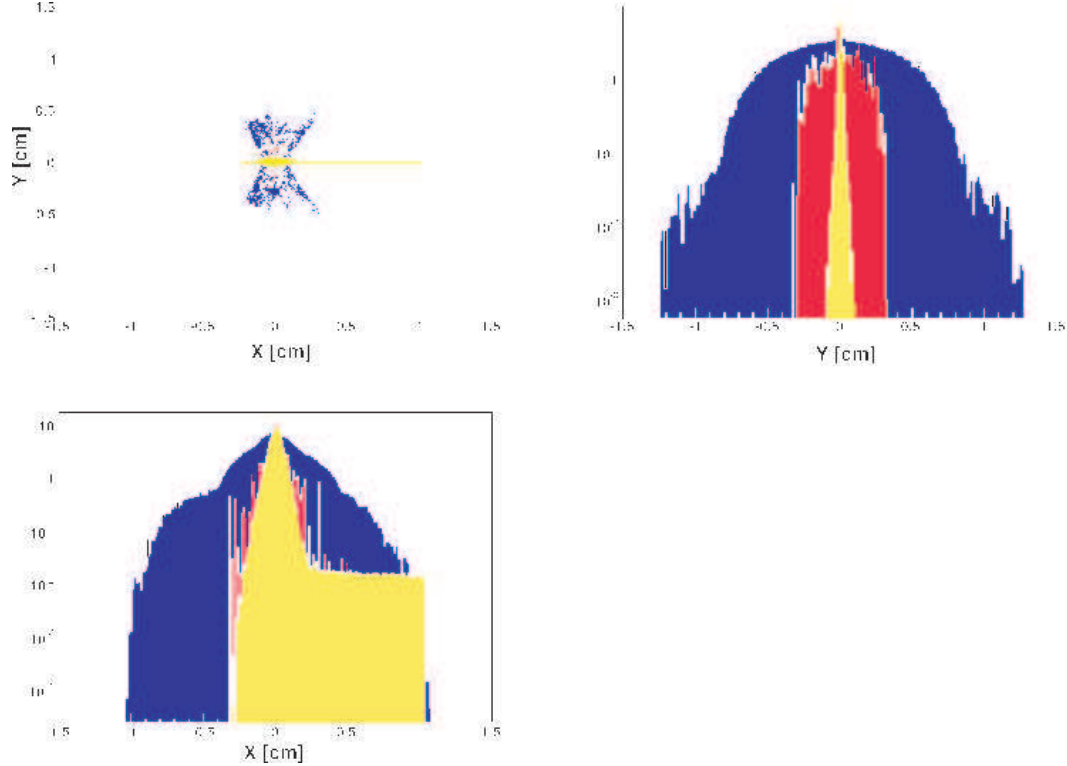}
     \end{center}
     \caption{Synchrotron radiation profiles at IP, where yellow and red parts show the profiles of  beam-core and -halo with the collimators, while blue part shows those without MSK1 and MSK2.  Vertical scale in projected figures (right and bottom) is an arbitrary unit. The nominal parameter set has been used for this study at E$_{cm}$ = 500 GeV, L* = 3.5 m and 20 mrad crossing angle.  } 
     \label{fig:synrad}
     \end{figure}

The synchrotron radiations have been simulated by a Monte Carlo program of LCBDS \cite{lcbds} based on GEANT4.  Transverse profiles of the radiations  are shown in Figure \ref{fig:synrad}, where beams have two components of core and halo.  The beam halo were created by Gaussian distributions with standard deviations of $50 \times \sigma_{x (x')} $ and $200 \times \sigma_{y (y')} $,  where $ \sigma_{x (x')} $ and $ \sigma_{y (y')} $ are those of the core distribution.   No halo particles hit the beam pipe at IP, since they are collimated within radius of 1 cm. 
%%%%
%%%%

%% to be optimized for the GLD case for larger beam pipe radius at IP.

\subsubsection{Muons}
% BDS is very similar to the Roadmap, KEK Report 2004-7, p.221-, muon attenuators are also similar; MUCALO results will be updated by LCBDS.
% two tunnel-filling (9 and 15 m long) spoilers at 660 and 350 m from IP -> 1/10000
 %      - spoilers, attenuators 
%We may refer the GLC Project for the muon background estimation by using muon attenuator which is an alternative to the muon spoiler of "tunnel filler".

Two muon spoilers are the baseline configuration to reduce muon flux by $1\times 10^{-4}$ in the ILC-BCD \cite{bcd}. They are so-called "tunnel fillers" of 9 and 15 m thick magnetized irons at 660 ad 350 m from IP, respectively. The rate has been estimated to be  0.8 muons/150 bunches in detectors with  the beam halo of $10^{-3}$ \cite{mokhov}.   

\begin{figure}[ht] 
     \begin{center}
     \vspace*{.2cm}
      \includegraphics[width=10cm]{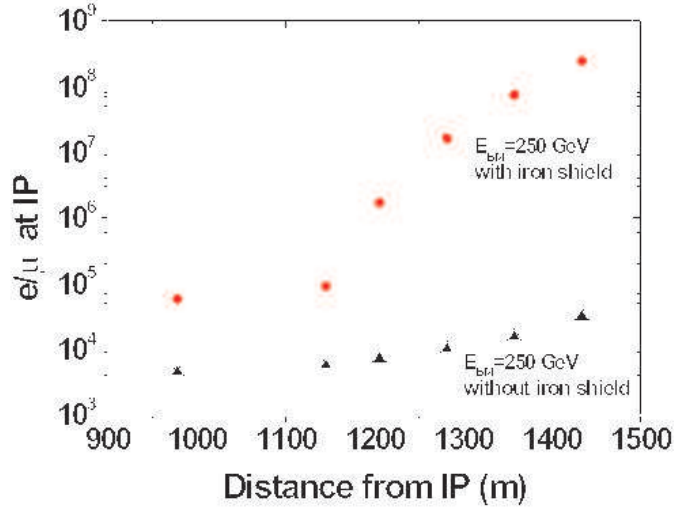}
     \end{center}
     \caption{Number of electrons to produce one muon reaching the IP as a function of muon creation point, this figure is the same one in the GLC Project, p.235 \cite{glc-project}.} 
     \label{fig:mu-bkg}
     \end{figure}

The alternative approach of muon attenuators has been investigated in the ACFA-LC working group \cite{Con_ACFA} .  Since the beam delivery system in the previous study \cite{glc-project}  is very similar to the ILC one, the major results can be borrowed  in this report.   All the empty beam line is covered by the attenuators  which is consisted of two coaxial iron pipes of 8 cm and 78 cm diameters around the beam pipe of 2 cm diameter.  The iron pipes of 244 m toal length are magnetized at 1 Tesla.   As shown in Figure~\ref{fig:mu-bkg}, the reduction factor of muon flux is close to that with the muon spoilers.
%%%%
%%%%

We are investigating the muon background by using LCBDS.

\subsection{Crossing Angle and Final Quadrupole Magnet}
% : 0, 2, 14 and 20mrad, and L* : T.Abe, A.Sugiyama, H.Fujishima,
% \subsubsection{Extraction Beam Line Configuration Including Beam Dumps}
% \subsubsection{ Background from  Pairs,  Disrupted beam,  Radiative Bhabha}
%  \subsubsection{Beam Loss (Density) in The Final Doublet}

%\subsection{Final Quadrupole Magnets}
% : T.Tauchi ( K.Tsuchiya, Y.Iwashita )
%\subsubsection{2mrad : Large Bore SC magnet}
%\subsubsection{2mrad : Permanent Magnet}
%\subsubsection{20 (14) mrad : Compact SC Magnet}
%\subsubsection{20 (14) mrad : Permanent Magnet}
% about 50W can be OK for the cryogetic power of the SC-QD0.

Two crossing angles have been set to be 2 and 20 mrad as the baseline configuration in the Baseline Configuration Document \cite{bcd}.  Also the baseline designs of final focus quadrupole magnet have been chosen to be a large bore super-conducting (SC) and a compact SC magnets at the 2 and 20 mrad angles, respectively, while  permanent magnets are the alternative ones for the both angles.   Their major geometries are listed in Table~\ref{tab:qd0} which  are used in the simulation studies.
%%%%%
\begin{table}[htdp]
\caption{Various types of QD0 and the first quadrupole magnets at the extraction line, which are relevant in IR region: properties used in the simulation, where L* is the distance from IP and QX1A and QDEX1A at the extraction line. }
\begin{center}
\begin{tabular}{|c|c|c|c|c|c|}\hline
type &Crossing angle&L* & Length & Aperture & Outer diameter  \\
 & mrad &  m & m & cm & cm \\ \hline
 Large SC & 2 & 4.51 & 2.5 & 7  & 42 \\
 Permanent & 2 & 4.51 & 2.5 & 2 & 18 \\ 
 Compact SC & 20 & 3.51 & 2.2 & 2 & 6.1 \\
 Permanent & 20 & 3.51 & 2.2 & 2 & 10 \\
 QX1A & 20 & 3.51 & 2.2 & 2.6 & 4.78 \\
 Compact SC & 14 & 4.51 & 2.2 & 2 & 7.2 \\
 QDEX1A & 14 & 6 & 1.64 & 3.6 & 9.2 \\
 \hline
\end{tabular}
\end{center}
\label{tab:qd0}
\end{table}
%%%%%

With small angle crossing, the most crucial issue is an energy deposit  at the SC-QD0 in the extraction beam line.  Sources of the energy deposit are pairs, disrupted beam and radiative Bhabha events.  The pairs  hit uniformly inside of the QD0 for their low energies, , while  the disrupted beam hit more at the downstream for higher energies.   
Table~\ref{tab:edepqd0} lists energy deposits due to the pairs at QD0 with 2 mrad crossing angle for the parameter sets, which were calculated by LCBDS. 
There should be no problem in the cryogenic power with these heat losses at any parameter sets.

Tolerance of the energy deposit in the SC magnets can be expressed in a unit weight, that is  0.5 mW/g to break the superconducting state. Since this tolerance has been estimated for the DC deposit, the ILC bunch-train structure may have more tolerable one.   The simulation shows that  pairs deposit energies of less than the tolerance for all the parameter sets even at E$_{cm}$ = 1 TeV while the disrupted beam exceeds it in cases of the High Lum a E$_{cm}$ = 500 GeV and the Low P, High Lum at E$_{cm}$=1 TeV.   The High Lum-1 has the marginal deposit energy at High Lum at E$_{cm}$ = 1 TeV.   This issue requires further study to optimize the configuration of SC-QD0 with 2mrad crossing angle. 
%%%%%
\begin{table}[htdp]
\caption{Energy deposits in unit of mW, at QD0, 2 mrad crossing angle, where values in parentheses are those for two beams to collide with horizontal offset of 200 $\sigma_x$. }
\begin{center}
\begin{tabular}{|c|c|c|c|c|c|c|}\hline
E$_{cm}$& Nominal & Low Q& Large Y & Low P & High Lum & High Lum-1  \\ \hline
500 GeV& 93(112) & 80(76) & 107(103) & 158(146) & 435(398) & - \\ 
1 TeV & 464(513) & 464(437) & 769(687) & 928(828) & 2180(2260) & 1170(1090) \\ \hline
\end{tabular}
\end{center}
\label{tab:edepqd0}
\end{table}
%%%%%

With  20 mrad crossing angle, energy deposits were calculated to be 43mW due to the pairs at first extraction quadrupole magnet ( QDEX1 ) which is located at 3.5 m from IP.  The QDEX1 has aperture of 2.6 cm, outer-diameter of 4.78 cm and length of 2.2 m as listed in 
Table~\ref{tab:qd0}.   Since disrupted beam can be  focused separately from incoming beam at the QDEX1,  it will not deposit energy more than the tolerance with all the parameter sets.

The permanent magnet must be very robust against radiations and the energy deposit.  
The typical demagnification is 0.3\% with $1 \times 10^{13}$n/cm$^2$ \cite{kawakubo}.
In addition, it does not generate artificial vibration for no cooling water and neither electricity.   Therefore, it must be a good alternative option for the QD0 \cite{mihara} .  
Since the permanent QD0 is compact, the disrupted beam can be extracted outside of it as similar to the compact SC one at 20 mrad crossing angle, where the extraction line has the same configuration as the baseline.  
In case of 2 mrad crossing angle, the energy deposit is expected to be greater than the large SC because of smaller aperture for pairs and disrupted beam.  However, the radiation damage must be  irrelevant. 
Such performances will be investigated by the full simulation together with the hardware R\&D.

%\subsection{Magnetic Field Configurations}
% : ( A.Miyamoto, H.Yamaoka and S.Andrei )
%\subsubsection{ 2mrad   L*=3.5m  v.s. 4.5m}
%\subsubsection{20mrad with DID}
%\subsubsection{14mrad with DID, or 14mrad with Anti-DID}
%\subsubsection{Anti-Solenoid for All The Final Q Types}

In the case of 20 mrad crossing angle, the Detector Integrated Dipole (DID) has been proposed in order to tune the beam orbits with level collisions, i.e. zero vertical angles,  at IP \cite{did} as well as to reduce the synchrotron radiation effects on the beam sizes. 
After SNOWMASS 2005 the intermediate crossing angle of 14mrad has been investigated as the alternative one  for less effects on the orbits and the radiations.  Taking account of these improvements, anti-DID, which has opposite sign of the DID magnetic field, was proposed at the 14 mrad crossing angle.  An optimization of interaction region was presented at the NANOBEAM 2005 \cite{anti-did}.

In next section, we evaluate the DID and anti-DID in terms of backgrounds by simulations.

\subsection{Background}

\subsubsection{ Hits in VTX and TPC}
% :  T.Abe, A.Sugiyama, H.Fujishima, Y.Sugimoto
%  - nominal, Low Q, Large Y, Low P and High Luminosity
%  neutrons created by energy deposit,  BCAL and beam dump at 350-400 m from IP with 20 and 14 mrad crossing,  650m with 2mrad crossing angle.
   
The incoherent pairs and disrupted beams have been generated with the machine parameter sets by CAIN. Background hits from them have been calculated by the full detector simulation of Jupiter as well as the LCBDS .  
Present results are very preliminary for limited statistics of the simulation and without optimization of detailed geometries,  where a single bunch collision has been simulated  with the parameter sets, 2 mrad and 20 mrad crossing angles and anti-DID at E$_{cm}$ = 500 GeV.

%Preliminary results of VTX and TPC hits  are shown in Figures~\ref{fig:vtx} and \ref{fig:tpc} at E$_{cm}$ = 500 GeV.
%%%%
%     \begin{figure}[ht] 
%     \begin{center}
%     \vspace*{.2cm}
%      \includegraphics[width=14cm]{mdi/VTX-hit.eps}
%     \end{center}
%     \caption{Hits density as a function of layer numbers in the VTX as well as the machine parameters, where the tolerance of $10^4$hits/cm$^2$/train is shown in red. } 
%     \label{fig:vtx}
%     \end{figure}
%%%%
%%%%
%     \begin{figure}[ht] 
%     \begin{center}
%     \vspace*{.2cm}
%      \includegraphics[width=6.5cm]{mdi/TPC-hit.eps}
%     \end{center}
%     \caption{Hits in the TPC as well as the machine parameters, where the hit number is normalized within the sensitive time of 50$\mu$sec . } 
%     \label{fig:tpc}
%     \end{figure}
%%%%
   
   The VTX hit density in the first layer may exceed the tolerance ( $10^4$/cm$^2$/ train ) at the high luminosity parameter sets, where the results have large statistical errors.    There are more hits at the outer layers at 20mrad than those of the 2 mrad.   The anti-DID actually reduced the hit densities at the outer layers close to the 2mrad cases .
   
    Number of TPC hits may exceed the tolerance ($2.46\times 10^5$hits/50 $\mu$sec) with the high luminosity and the low P sets and 
the tolerance is marginal with the 20 mrad crossing cases.   The anti-DID reduces the number of TPC hits below the tolerance, especially in case of the high luminosity.  The low Q parameter set has less hits by more than twice than the nominal one in VTX and TPC.
   
\subsubsection{Energy Deposit in BCAL, Beam Dumps and Neutrons}   
   
 Neutrons are created by electromagnetic energy deposit in materials with typical rate of 130 neutrons(n)/TeV, which corresponds to $10^9$n/J .   
 The pairs, disrupted beams and beamstrahlung photons hit the calorimeters (BCAL), magnets at the extraction line and the beam dumps, where  they deposit energies for  neutrons to be created.  The neutrons are backscattered into the VTX  around IP.  
 Assuming they scatter uniformly in all solid angle, their flux density at the VTX (IP)  is in inverse proportion to L$^2$, where L is distance from the creation point to IP.  The tolerance is estimated to be $10^{10}$/cm$^2$/year in the VTX.   
With the above consideration, tolerable energy deposit  can be expressed by E = 0.13 L$^2$  with E in Watt(W) and L in m, e.g. 0.13 W and 1 KW at L = 1 m and 88 m, respectively.  

Energy deposits in BCAL are listed with various conditions in Table~\ref{tab:bcal} .
 %%%%%
\begin{table}[htdp]
\caption{Energy deposits due to the pairs in BCAL with the 2 mrad crossing angle, where the unit is Watt (W) and values in parenthesis are those at the 20 mrad crossing angle.  The tolerance is also listed at BCAL, L = 4.3 m from IP, which is calculated by E = 0.13 L$^2$.}
\begin{center}
\begin{tabular}{|c|c|c|c|c|}\hline
E$_{cm}$ &Nominal&Low Q & High Lum & High Lum-1  \\ \hline
500 GeV&  0.05(0.10) & 0.03(0.07) & 0.27(0.42) & - \\
1000 GeV&  0.12(0.22) & 0.07(0.16) & 0.68(0.94) &  0.32\\ \hline
tolerance &\multicolumn{4}{|c||}{2.4}\\
 \hline
\end{tabular}
\end{center}
\label{tab:bcal}
\end{table}
%%%%%
   
In reality, a fraction of neutrons comes out in backward from the material.  Therefore, the above estimation is the conservative limit of energy deposit from the VTX damage.
For examples, it has been shown that neutrons were backscattered from the nearest quadrupole magnet in JLC detector with rates of 8.1 n/TeV and 36 n/TeV by simulations of GEANT3 and FLUKA, respectively \cite{Con_ACFA}.   In this case, the fractions are 0.06 and 0.28, respectively.    Also the fraction from the water beam dump was estimated to be $1.67 \times 10^{-4}$.

At ILC, total energy deposit is 23 MW at the beam dump.  Since the dump is located at 650 m and 300 m from IP at the 2 mrad and 20 mrad crossing angles, respectively, the tolerances are estimated to be 330 MW and 70 MW, respectively, with the fraction of the water dump. Therefore, the beam dumps should not be relevant in the VTX radiation damage.   Energy deposits will be investigated at the extraction line.

    % to be prepared by T.Tauchi
\clearpage

%%%%%%%%%%%%%%%%%%%%%%%%%%%%%%%%%
\chapter{Physics Performance}
\label{PART_performance}
%%
%
%
%\section{Introduction}
%In this chapter, we describe physics performance of the GLD detector 
%after describing 

\section{Tools for Studies of Physics Performances}
\begin{figure}
\begin{center}
\includegraphics[width=10cm]{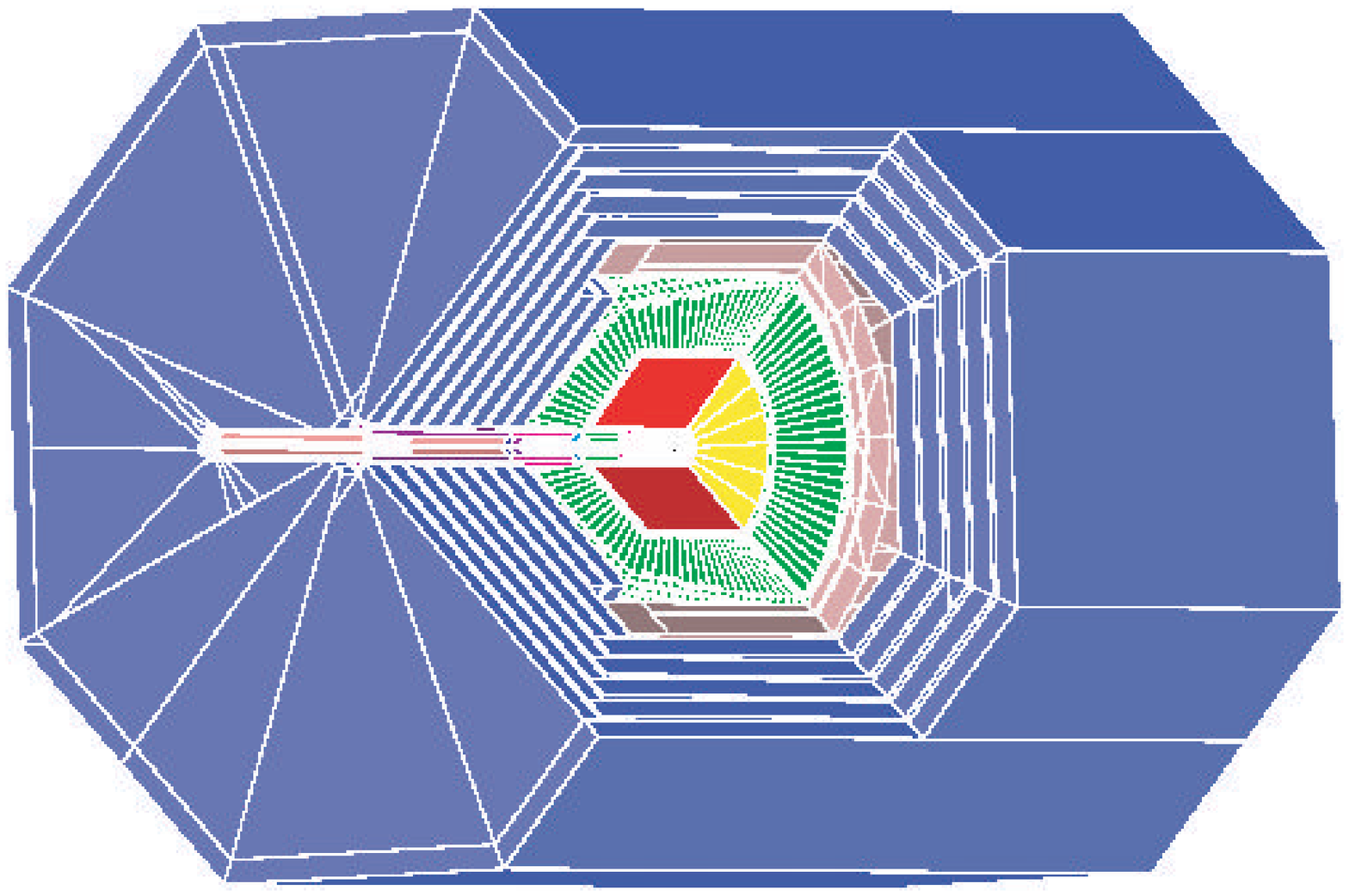}
\end{center}
\caption{\label{PP_CutView}GLD Detector Cut View}
\end{figure}

The detector configuration used for physics performance studies of the GLD detector presented in this chapter
is summarized in Table~\ref{PP_JUPITERGeom}. 
A cut view of the detector is shown in Figure~\ref{PP_CutView}.
The studies used the software modules QuickSim, JUPITER and SATELLITES implemented in the JSF \cite{SimTools}
framework unless specified otherwise. These tools are described 
briefly in this section before describing results in the 
subsequent sections.

JUPITER \cite{SimTools} is a full detector simulator based on Geant4 \cite{Geant4}.
For this study, we used version 7.0patch1.
JUPITER provides a modular structure for easy update and installation 
of sub-detector components.  To this end, powerful base classes 
have been developed including: a unified interface to facilitate easy 
installation of detector components,
classes to help implementation of detailed hierarchical
structures, automatic naming system and material management to minimize 
user-written source codes, and a mechanism to compose magnetic fields of 
many accelerator components.
Detector configurations are defined by a external text file to allow 
easy modification of detector configuration.

\begin{table}
\caption{\label{PP_JUPITERGeom}The standard geometry parameter used for JUPITER simulation.}
\begin{center}
\begin{tabular}{| l | l | }
\hline 
Component & Specification \\
\hline
Beam Pipe & Inner Radius 1.5 cm, Be, 250 $\mu$m$^t$  \\
  & Outside IP region($|Z| > 5.5$ cm, Al. 2 mm$^t$ \\
\hline
VTX & 6 layers of Si cylinders, 50 $\mu$m$^t$ each \\
  & Inner Radius  (cm) : 2.0, 2.2, 3.2, 3.4, 4.8, 5.0  \\
  & Half Z Length (cm) : 6.5, 6.5, 10.0, 10.0, 10.0, 10.0  \\
  & Resolution : $\sigma_{r\phi,z}$ = 4 $\mu$m\\
\hline
IT Barrel &  4 layers of Si cylinders, 560 $\mu$m$^t$ \\
  & Inner Radius  (cm) : 9.0, 18.5, 28.0, 37.5 \\
  & Half Z Length (cm) : 18.58, 38.20, 57.81, 77.43 \\
  & Resolution : $\sigma_{r\phi,z}$ = 10 $\mu$ m\\
\hline
IT Endcap & 7 layers of Si disks, 560 $\mu$m$^t$ \\
  & Inner Radius (cm) : 2.5, 4.5, 7.0, 8.0, 10.0, 12.0, 14.0 \\
  & Outer Radius (cm) : 7.0, 14.0, 21.0, 28.0, 36.0, 38.0, 38.0 \\
  & Z position (cm) : 14.5, 29.0, 43.5, 58.0, 72.5, 87.0, 101.5 \\
  & Resolution : $\sigma_{r\phi,z}$ = 10 $\mu$m\\
\hline
TPC & Sensitive Region : $ 43.715 cm < r < 197.765 cm $, $|z| < 255 cm$ \\
   & Gas: P10 \\
   & Inner TPC Support : 4.215 cm$^t$, 0.013$X_0$ \\
   & Outer TPC Support : 8.235 cm$^t$, 0.017$X_0$ \\
   & TPC End Plate : 5 cm$^t$, $0.1X_0$ \\
   & Number of radial sampling : 200 \\
   & Resolution : $\sigma_{r\phi}$ = 150 $\mu$m, $\sigma_z$ = 400 mm \\
\hline
CAL & Tower structure,  \\
  & Inner radius of Barrel : 210 cm \\
  & Front Z position of Endcap : 270 cm \\
  & Coverage : $| \cos\theta | < 0.98 $ \\
  & Tower size  EM 4 cm $\times $4 cm, HD 12 cm $\times$ 12 cm \\
  & EM Part : 27 $X_0$, 38 layers of 1 mm$^t$ Scinti. and 4 mm$^t$ of Lead \\
  & HD Part : 6.1 $\lambda$, 130 layers of 2 mm$^t$ Scinti. and 8 mm$^t$ of Lead \\
\hline
Solenoid & 3 Tesla \\
  & Field Radius : 157 cm \\
\hline
Muon Barrel & Octagonal shape, 460 cm $<$ r $<$ 780 cm \\
    & 4 Super layers, Active thickness 10 cm \\
Muon Endcap & 430 cm $<|$ Z $|<$ 535 cm : 2 super layers, cover 45 cm $<$ r $<$ 360 cm \\
 & 535 cm $<|$ Z $|<$ 845 cm : 5 super layers, cover 45 cm $<$ r $<$ 780 cm \\
 & Active thickness 10 cm \\
\hline
FCAL & 30 layers in Front and Tail part with conical shape\\
    & 1 layer consists of  Tungsten 0.3 cm$^t$, Si. 0.03 cm$^t$, Gap 0.67 cm$^t$ \\
   & Front part : 8.2 cm $<$ $R_{front}$ $<$ 13.0 cm, 230 cm $<|$ Z $|<$ 260 cm \\
   & Tail part : 9.2 cm $<$ $R_{front}$ $<$ 39.79 cm, 260 cm $<|$ Z $|<$ 275 cm \\      
\hline
\end{tabular}
\end{center}
\end{table}

JSF is an analysis framework based on ROOT \cite{ROOT}.
It provides a generic framework for event generation, simulation and 
analysis.  Event generator modules in JSF for Pythia \cite{PP:Pythia}
and other generator data in various format are used in our studies.
JUPITER was used within a JSF framework to create Monte Carlo simulated data in 
ROOT format. The analysis packages, SATELLITES and URANUS, for JUPITER data are prepared as 
JSF modules.
The URANUS package is a collection of analysis modules which would be applicable 
even for real data, while the SATELLITES package are modules derived from URANUS adding 
functions specific to analyses of simulation data.

Modules in SATELLITES includes those for Tracking, Calorimeter clustering,
particle flow analysis and jet clustering of particle flow objects.
In tracking, we make Kalman filter fits of smeared 
hit points simulated for the Vertex detector (VTX), Intermediate Tracker(IT) and Central Tracker(TPC) detectors. 
For the hit point selection, 
Monte Carlo truth information was used.  A description of 
a generic Kalman fitter program used for this analysis can be found 
in ref. \cite{Kalman}.  The particle flow analysis is described in 
the subsequent section.

For high statistic estimates of physics performance, we used a fast 
detector simulator, QuickSim \cite{SimTools}. Detector components included in QuickSim are
 for the tracking detectors, TPC, IT and VTX, and for the
calorimeter system.  In QuickSim, particles are swam through VTX, IT, and TPC to CAL, 
taking into accounts smearing of particle direction by multiple scattering in 
materials.  Hits in VTX and IT are created as a particle traverses them.
When a particle reaches the TPC, its momentum information at the entrance is 
used to generate 5 element helix parameters and error matrix taking into 
account smearing due to multiple scattering and measurement errors,
$\sigma_{r\phi}$ and $\sigma_z$, 
assuming equal-spacing hit point measurements along particle trajectory in TPC.
These smeared TPC tracks and smeared hit points in VTX and IT 
are used later to get a combined track parameter of charged particles.
When a particle reaches the calorimeter a lateral distribution of hit cells is 
generated as a sum of double Gaussian distribution, then the energy of each cells 
are smeared with given energy resolution parameter.  For simplicity, 
electrons and gammas create signals only in the electromagnetic calorimeter,
while muons and neutrinos do not generate signals in the calorimeter system. The remaining hadronic
particles create signals in the hadron calorimeter only.

\section{Performance for Single Particles}

\begin{figure}
%\begin{tabular}{cc}
%\includegraphics[width=6.7cm]{performance/Figs/Radiation_Length_total.eps}
%&
%\includegraphics[width=6.7cm]{performance/Figs/Radiation_Length_breakup.eps}\\
%(A)&(B)
%\end{tabular}
\begin{center}
\begin{tabular}{c}
\includegraphics[width=6.7cm]{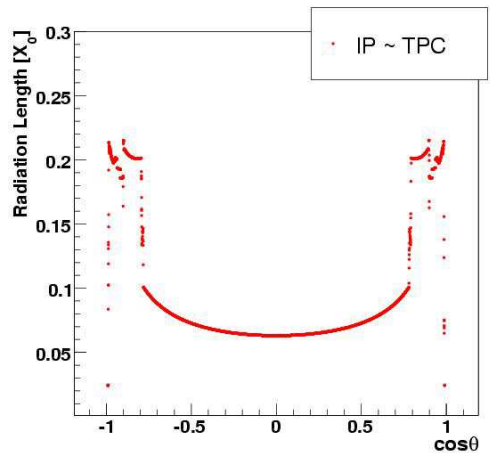}
\end{tabular}
\end{center}
\caption{\label{pp_material_radl}
The amount of material up to the EM calorimeter in terms of 
radiation lengths as a function of $\cos\theta$}
\end{figure}

\begin{figure}
\begin{center}
\begin{tabular}{cc}
\includegraphics[width=6.7cm]{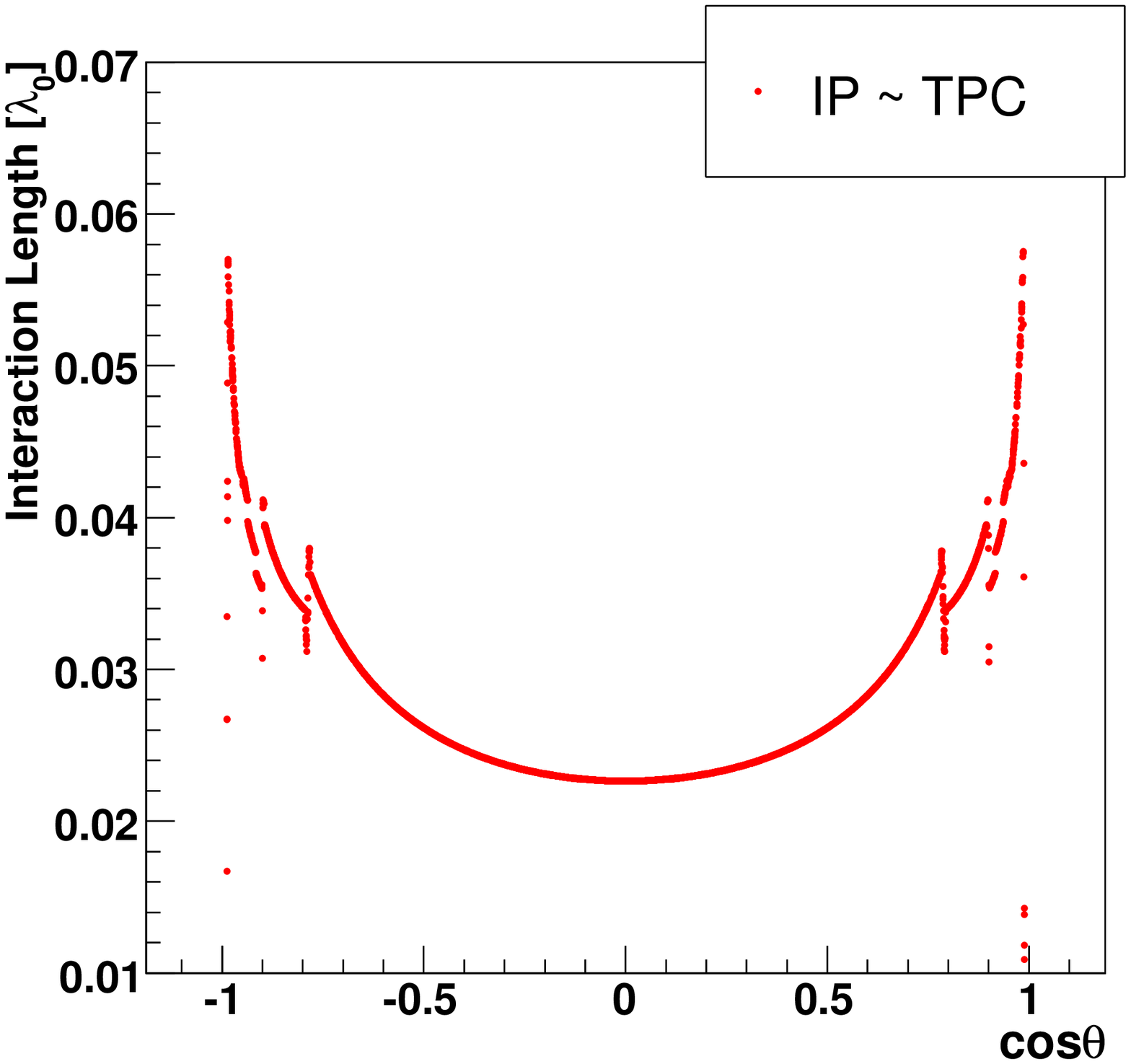}
&
\includegraphics[width=6.7cm]{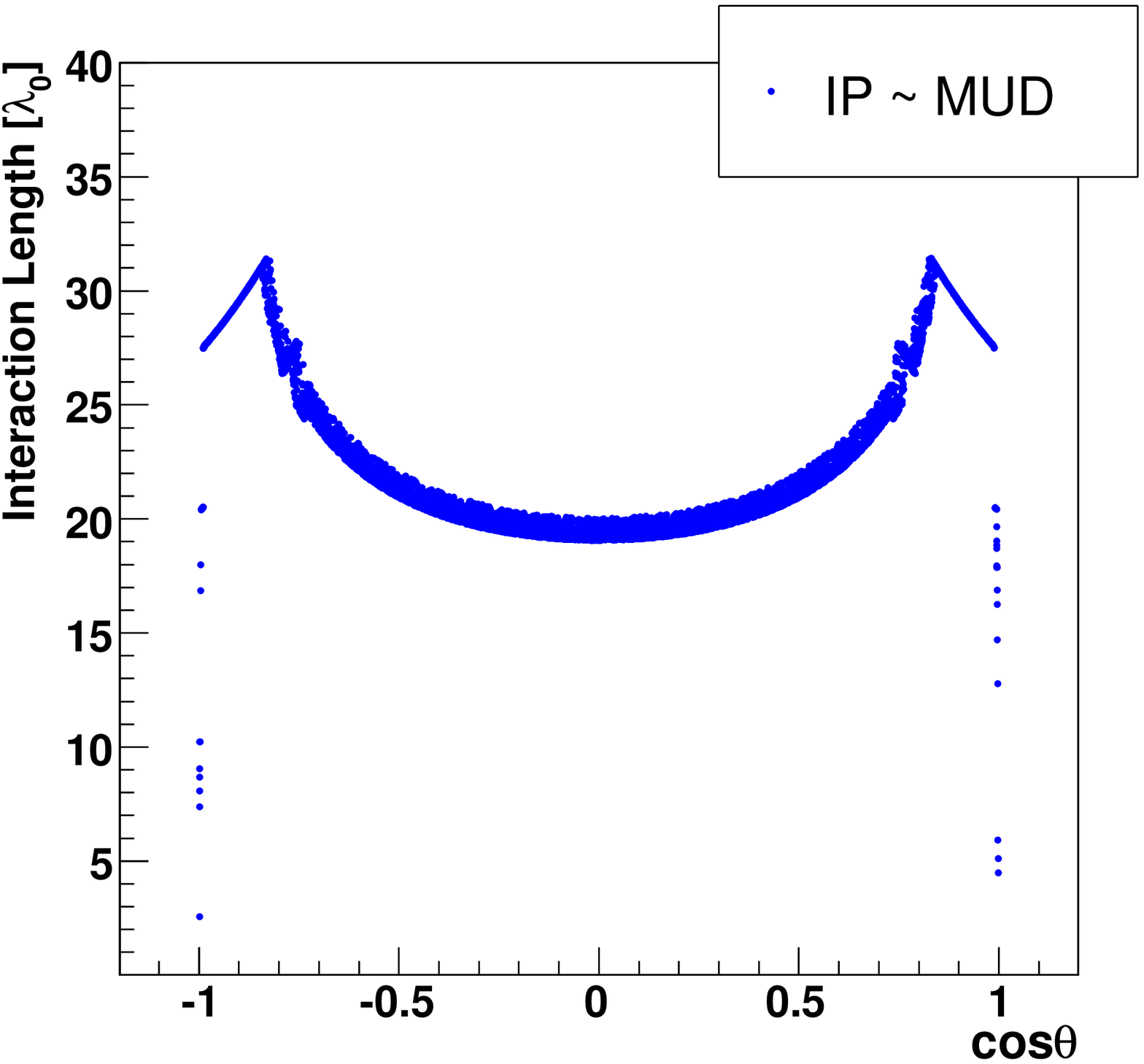}\\
(A)&(B)
\end{tabular}
\end{center}
\caption{\label{pp_material_int}
The amount of material in front of EM calorimeter in terms of 
nuclear interaction length (A) and to the end of Muon detector (B) as a function of $\cos\theta$}
\end{figure}

Materials for the GLD detector up to
the EM calorimeter is shown in Figure~\ref{pp_material_radl} in terms of radiation lengths.
The total material budget is about 0.1 to 0.2 X$_0$ depending on $\cos\theta$, except at 
$\cos\theta = 0$, where thin TPC central membrane ( $< 100$ $\mu m$ )
extends from radius of about 40 cm to about 200 cm resulting 5.4 X$_0$. In real experiment,
the probability of particles goes into this membrane is small and not shown in the 
figure.
Materials in terms of nuclear interaction lengths are shown 
in Figure~\ref{pp_material_int}.  The total is about 0.02 to 0.05 except at $\cos\theta = 0$
where TPC central membrane has about 2.8 $\lambda$ contribution.

\begin{figure}
\begin{tabular}{cc}
\includegraphics[width=6.7cm]{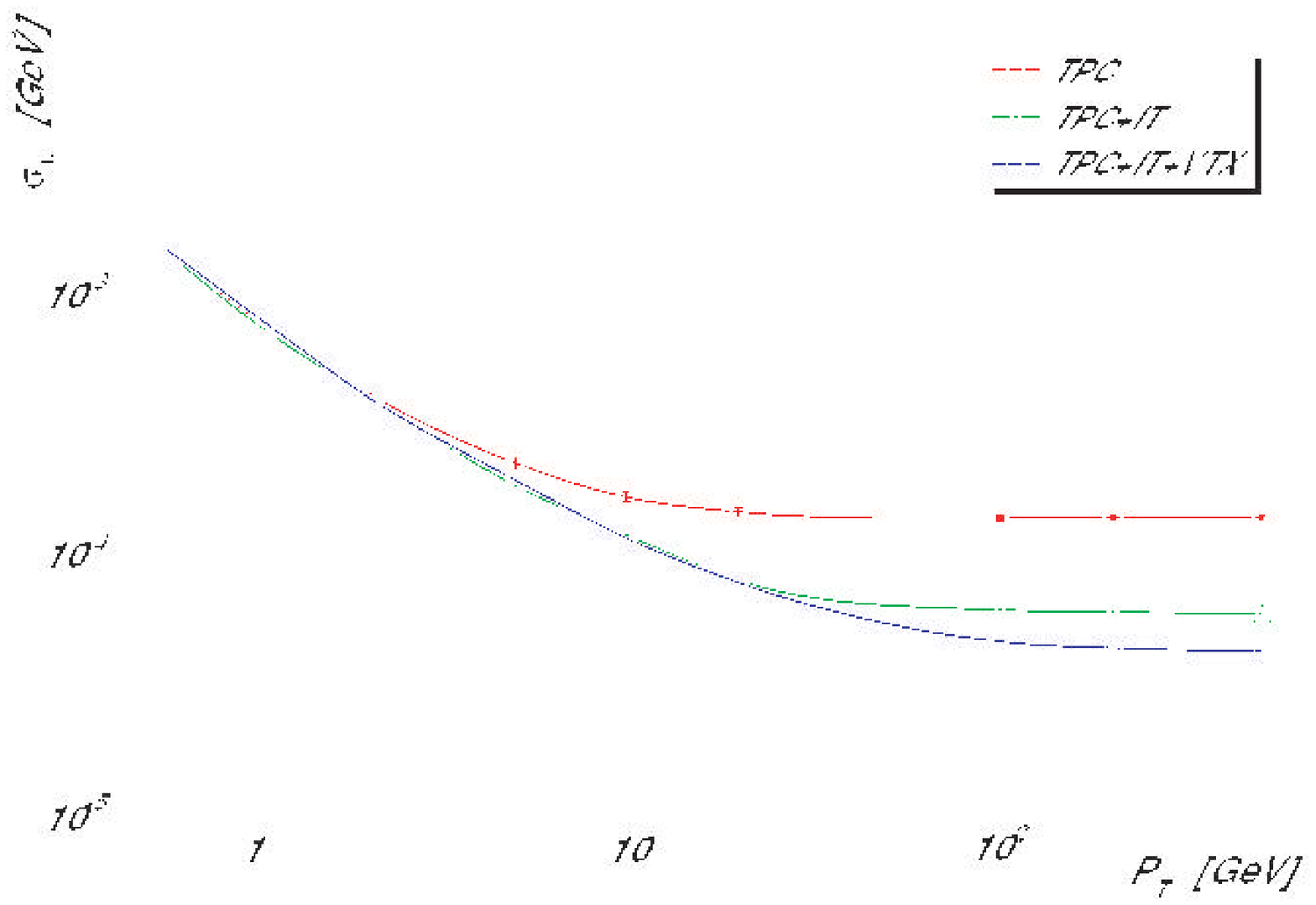}
&
\includegraphics[width=6.7cm]{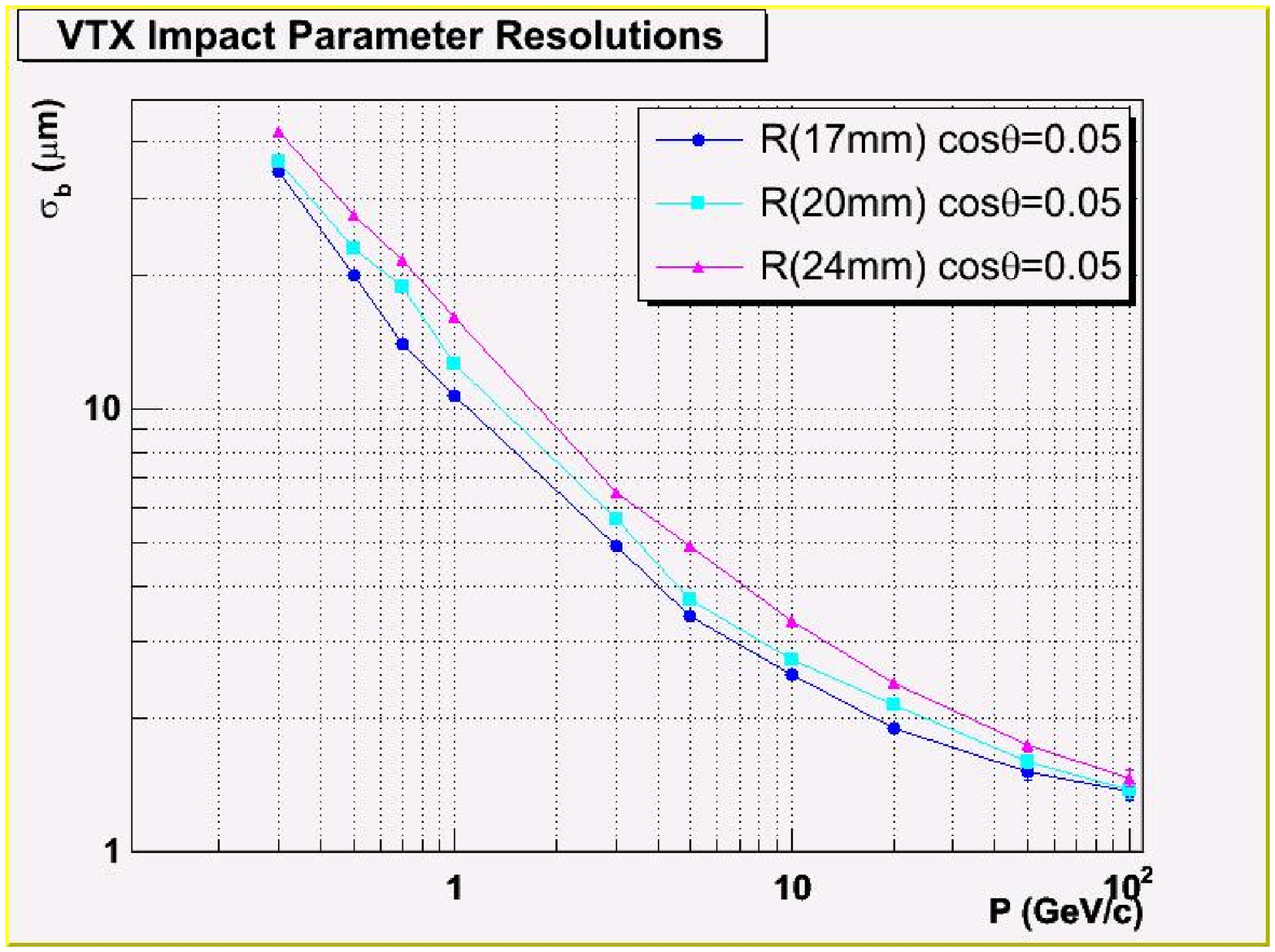}\\
(A)&(B)
\end{tabular}
\caption{\label{pp_ptresol}
Momentum resolution(A) and impact parameter resolution of GLD
as a function of Pt.  Studied using a track generated at $90^\circ$ 
from IP by JUPITER. }
\end{figure}

%%%%%%%%%%%%%%%%%%%%%%%%%%%%%%%%%%%%%%%%%%%%%%%%%%%%%%%%%%%%%%%%%%%%%%
The momentum resolution for muons produced at $90^\circ$ is 
shown in Figure~\ref{pp_ptresol}(A) for the cases of TPC only, TPC and IT, 
and TPC, IT and VTX combined \cite{PP:TrackerPerformance}. 
To avoid particles hitting the TPC central membrane, 
particles are generated from a point slightly away from IP in Z direction.
By combining TPC, IT and VTX, 
$\Delta p_t/p_t \sim 4 \times 10^{-5}$ is achieved for 100 GeV muons.

Figure~\ref{pp_ptresol}(B) shows the impact parameter resolution as 
a function of $p_t$ for different VTX detector configurations. 
The impact parameter is defined as a closest distance of tracks 
to the IP in the plane perpendicular to the Z axis. The cyan curve 
shows the performance in the case of the standard configuration. 
For the configuration shown by the blue curve, the inner radius of the innermost VTX layer is 
1.7 cm, those of 5th and 6th layers are same as the standard, 
and that of 3rd and 4th layers are adjusted so that to have equal distance 
between 2nd and 3rd, and 4th and 5th layers.  In the case of the purple 
curve, the inner radius of 1st layer is move to 2.4cm, while radius of remaining 
layers are adjusted similar to the blue curve case.

%

%\begin{figure}
%\begin{center}
%\includegraphics[width=7cm]{performance/Figs/MomentumResolution-vs-Pt.eps}
%\includegraphics[width=7cm]{performance/Figs/IP-Resolution.eps}
%\end{center}
%\caption{\label{pp_qsim_ptresol}Pt resolution of tracker by Quick Simulator}
%\end{figure}
%
%\begin{figure}
%\begin{center}
%\end{center}
%\caption{\label{pp_qsim_ipresol}Impact parameter resolution of GLD tracker system
%simulated by Quick Simulator}
%\end{figure}

%%%%%%%%%%%%%%%%%%%%%%%%%%%%%%%%%%%%%%%%%%%%%%%%%%%%%%%%%%%%%%%%%%%%%%%%%%%%

\begin{figure}
\begin{tabular}{cc}
\includegraphics[width=6.7cm]{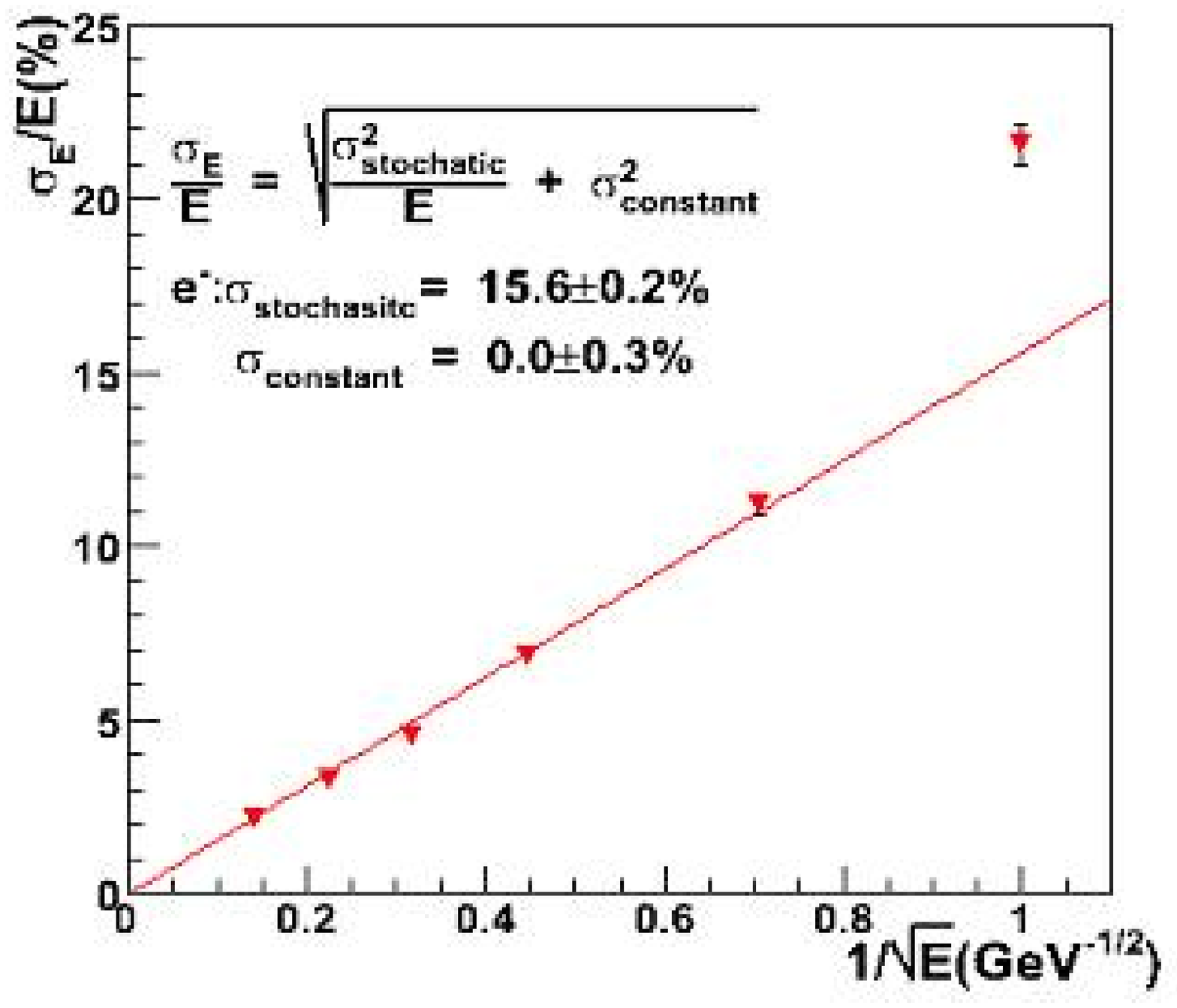}&
\includegraphics[width=6.7cm]{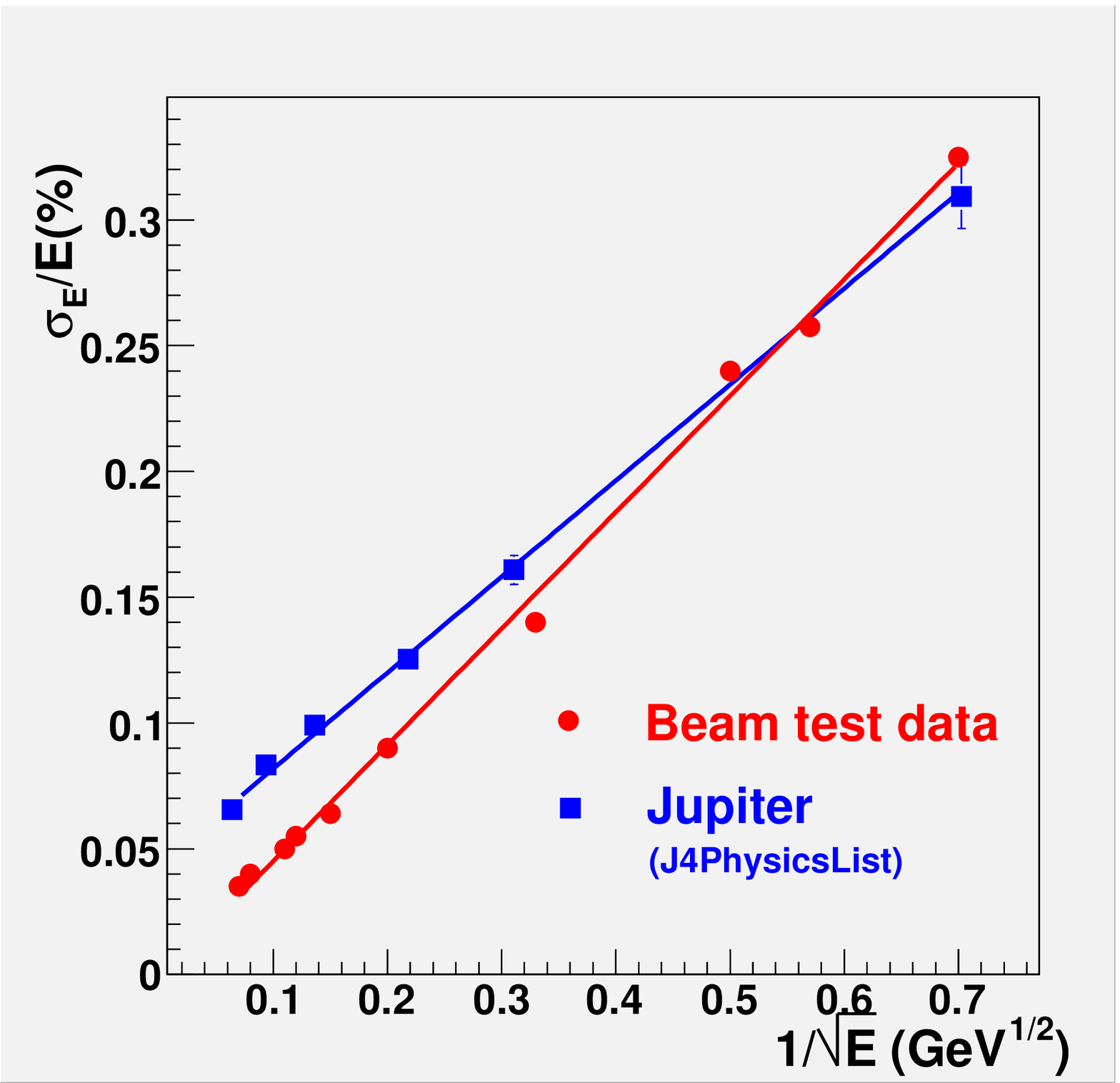}\\
(A)&(B)
\end{tabular}
\caption{\label{pp_CALResol}Energy resolution of EM (A) and HD(B) Calorimeter. Resolutions obtained by Jupiter simulation are compared with the beam test result in Figure (B)}
\end{figure}

%DESCRIBE REALSTIC PFA.
The energy resolutions of the EM and HD calorimeters are shown in Figure~\ref{pp_CALResol}
as a function of particle energies. 
In the study of the energy resolution of EM calorimeter, the range cut value for 
the Geant4 simulation was 10 $\mu m$, instead of default 1mm.  This was necessary to 
obtain the energy resolution consistent with the beam test results.
The energy resolution of the hadron calorimeter is show in Figure~\ref{pp_CALResol}B,
together with the beam test result with similar configuration.
It is well known that results of the hadron shower program in Geant4 
depends on the physics list.  We compared models such as QGSP, LHEP and Jupiter physics lists (J4PhysicsList), 
but none of them reproduced the beam data resolution at high energies.
In the Jupiter simulation, the Jupiter physics list derived from a Geant4 novice example is used as a default.

%%%%%%%%%%%%%%%%%%%%%%%%%%%%%%%%%%%%%%%%%%%%%%%%%%%%%%%%%%%%%%%%%%%%%%%%%%%%
\section{Jet Energy Resolution}
To achieve good resolution for jet energy measurements particle 
flow analysis (PFA) is crucial. In the PFA analysis, energies of charged particles 
are measured by trackers and neutral particle energies are measured by 
the calorimeter system.  Since the calorimeter is sensitive to charged particles, 
removal of charged particle energy depositions in the calorimeter is 
essential.  We used two methods for the PFA,
cheated method \cite{PP:CheatedPFA} and realistic method \cite{PP:CHDFinding1}, which will be discussed in the 
following subsections.

\subsection{Resolution by cheated PFA}
In the cheated method, 
cluster reconstruction and linking of charged track and calorimeter 
cluster for removal of charged track signals is pursued 
based on the Monte Carlo information. 
This gives optimistic estimates of the jet energy resolution,
but still includes effects such as shower fluctuation in calorimeter, signal 
overlaps in readout cells of calorimeter, particle decays in tracker, etc.

In this study, we generated $e^+e^- \rightarrow q\bar{q}$ events at center of mass energies 
of 91.17 GeV, 350 GeV and 500 GeV.
Only $u$, $d$, $s$ quarks were generated by Pythia \cite{PP:Pythia} without initial state radiation. 
Generated events were simulated by JUPITER 
and reconstructed by the SATELLITES analysis modules . In this analysis 
charged particles measured by the TPC were fitted 
and connected to IT and VTX hits by the Kalman track fitter to make a combined track parameter, called 
hybrid track parameter.  Hit points of candidate
tracks were selected based on the Monte Carlo truth information.
  
In the JUPITER simulation, we kept an ID of tracks at the exit of the TPC 
sensitive volume and the ID is attached to each Calorimeter cells.  This ID 
information was used to form a calorimeter cluster.  If a track 
with the same ID is found in the TPC, the cluster in the calorimeter is deleted.
If a track kinks in the TPC sensitive volume, we are left with a track before 
the kink (kink mother), a track after the kink (kink daughter) with 
a connected neutral cluster in the calorimeter, 
and the neutral cluster(neutral daughter) which might be produced at the kink.  
In this case, we used the track information of the kink mother and the neutral daughters
for the PFO(kink mother scheme), instead of using the kink daughter and 
neutral daughters(kink daughter scheme).  The kink mother scheme double counts 
the energies of neutral daughters, but the energy resolution was about 3\% 
better than the link daughter scheme according to our study.

\begin{figure}
\begin{tabular}{cc}
\multicolumn{2}{c}{
\includegraphics[width=8cm]{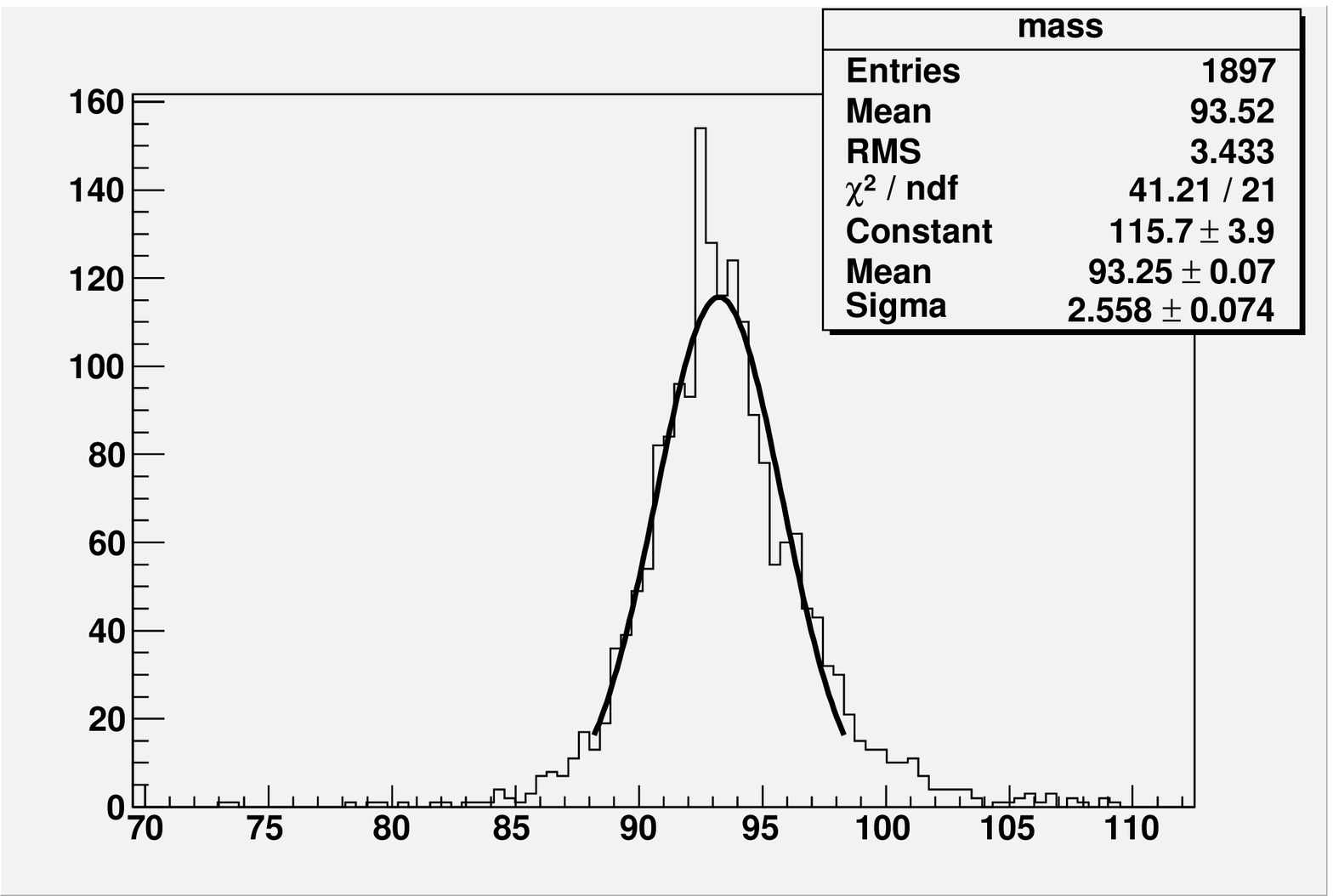}}\\
\multicolumn{2}{c}{(A)} \\
\includegraphics[width=8cm]{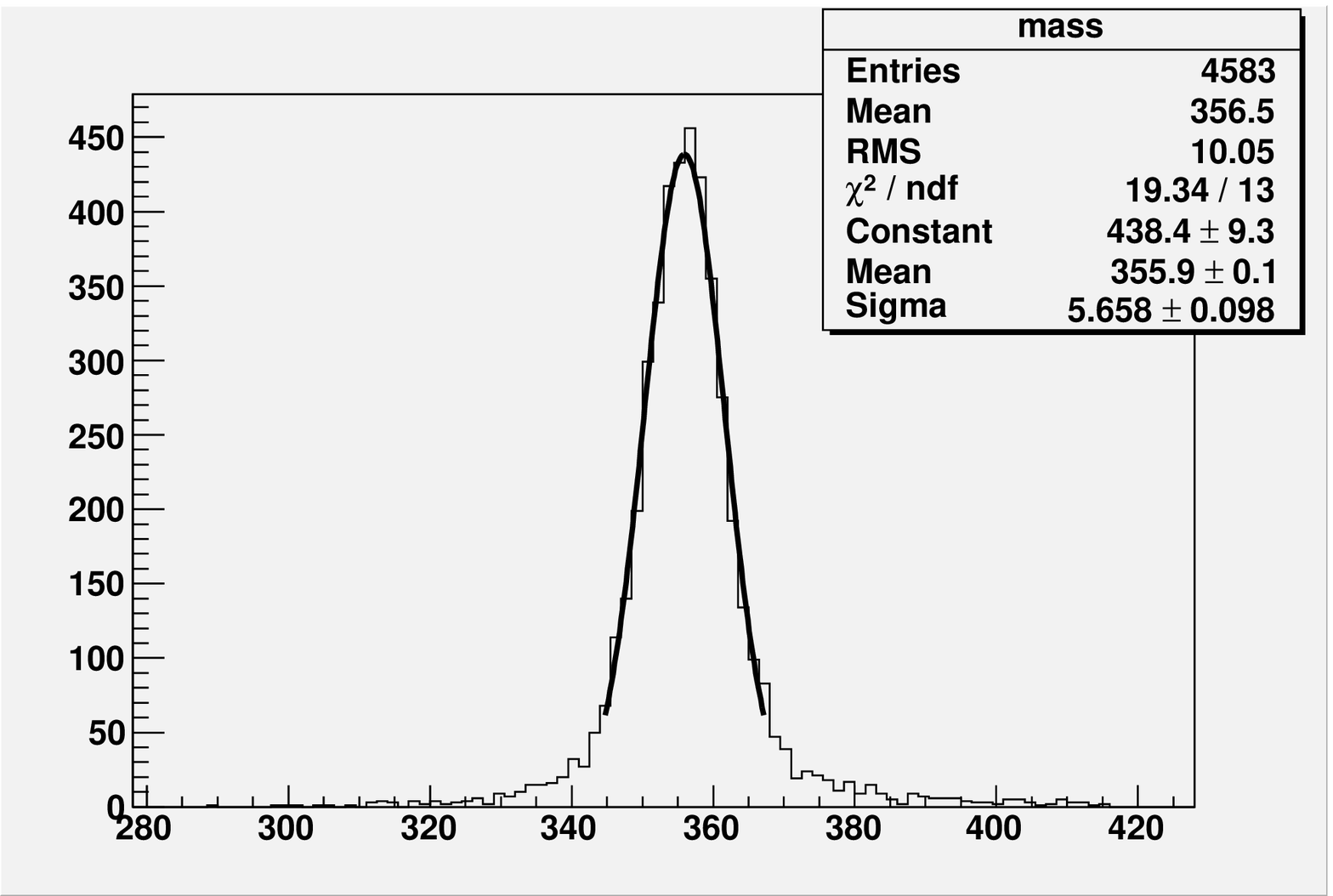}
&
\includegraphics[width=8cm]{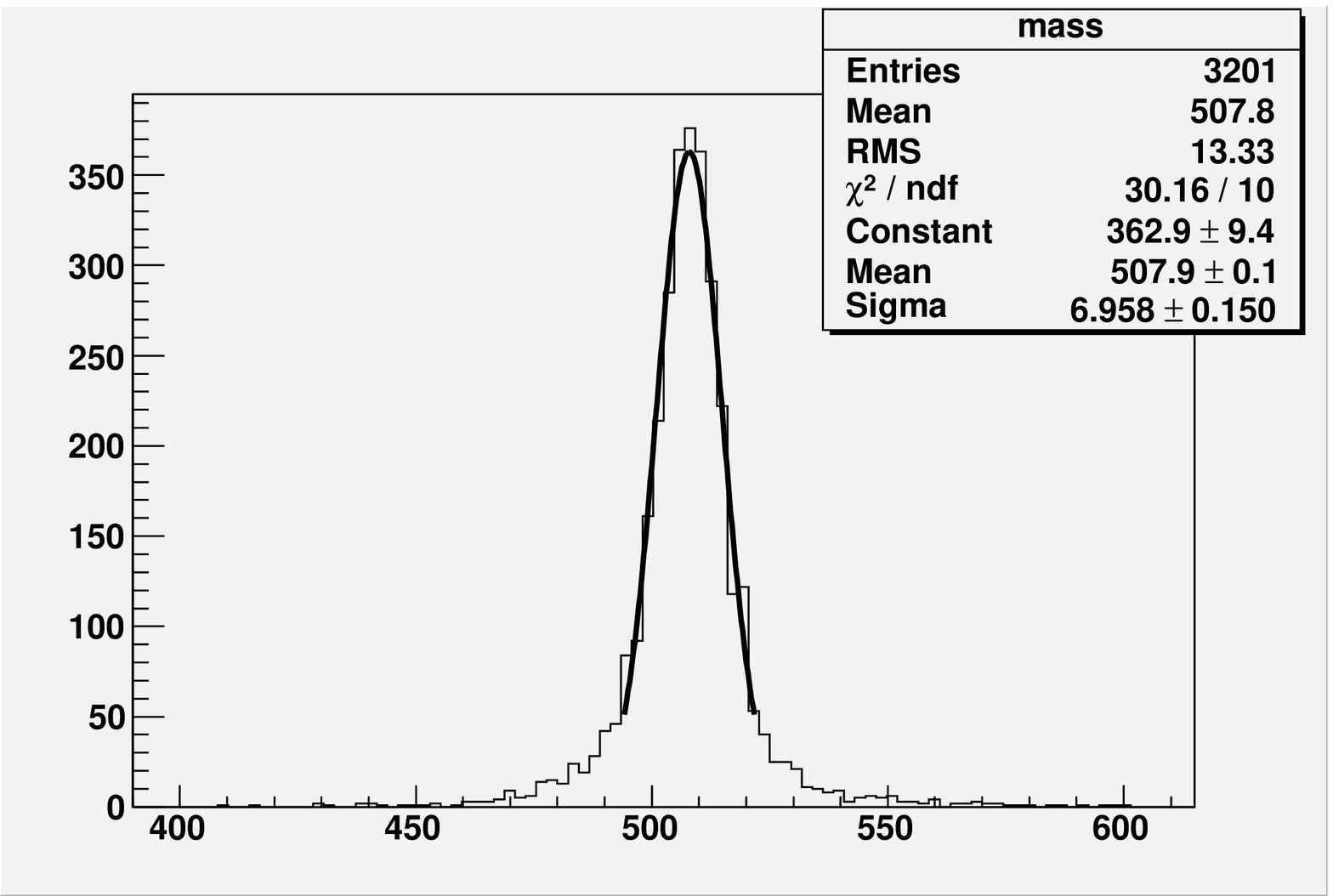}\\
(B) & (C)\\
\end{tabular}
\caption{\label{pp_jetmass_jupiter}The reconstructed jet energy distribution of the process, 
$e^{+}e^{-}\rightarrow u\bar{u}$, $d\bar{d}$, or $s\bar{s}$
at $\sqrt{s}=91.17$(A), $300$(B), and $500$ GeV(C), respectively.}
\end{figure}

The jet energy resolution obtained by the cheated PFA algorithm is shown in 
Figure~\ref{pp_jetmass_jupiter}.  In this study, we selected events 
where quarks were produced in the range $|\cos\theta_q|<0.8$.
The invariant mass of the sum of all particle flow objects is plotted in
the Figure. The curves shown in the figure are gaussian fits in the 
two $\sigma$ range from mean values.  Fits were repeated until 
the conversion of $\sigma$. 
The widths of the jet energy distribution obtained from the fits are summarized in the 
Table.\ref{PP_cheatedEresol}. As seen in the table, the energy resolution for 
jets at the Z pole is about $26\% / \sqrt{E}$ and worsens slightly to 
about $30\% / \sqrt{E}$ at 500 GeV.  
In the table, the resolution 
for the case of QuickSim are shown in the same table for comparison. 
For QuickSim, we prepared two parameter sets, the good one for 
performance similar to the cheated PFA performance and the bad one 
corresponding to the realistic PFA resolution which will be discussed in the 
next subsection.  The performance by the cheated PFA is the
ultimate jet energy resolution with the current detector configuration
which could be achieved by a clever PFA algorithm.  The performance of the realistic 
PFA is limited by the performance of the PFA algorithm which will be 
improved in future developments.  Final performance would be 
somewhere between two performances, thus, we will prepared two sets of 
QuickSim parameters to give optimistic and pessimistic estimates of 
physics performances.
\begin{table}
\caption{\label{PP_cheatedEresol}Summary of the jet energy 
resolution by the cheated PFA method of the process shown in Figure~\ref{pp_jetmass_jupiter}.
4th and 5th column shows the $\sigma$ in the case of QuickSim for good and bad parameter
sets, respectively. }
\begin{tabular}{| c | c | c | c | c | c| c |}  \hline
 & \multicolumn{2}{c|}{Cheated PFA} 
 & \multicolumn{2}{c|}{Realistic PFA} 
 & \multicolumn{2}{c|}{QuickSim} \\
\hline
$\sqrt{s}$(GeV)  & $\sigma_E$ (GeV) & ${\sigma_E \over \sqrt{E}}(\%)$ &
 $\sigma_E$ (GeV) & ${\sigma_E \over \sqrt{E}}(\%)$ &
 $\sigma_{good}$(GeV) & $\sigma_{bad}$(GeV) \\
\hline
91.2 & $ 2.56\pm0.07$ & $26.5\pm 0.7$ & $3.65\pm 0.07$ & $38.4\pm 0.7$&
$2.35 \pm 0.04$ & $3.30\pm 0.05$\\ \hline
350 & $5.66 \pm 0.10$ & $30.0 \pm 0.5$ & - & - & $5.75 \pm 0.09$ & $8.75 \pm 0.14$\\ \hline
500 & $6.96 \pm 0.15$ & $30.9 \pm 0.7$ & $19.8 \pm 0.65$ & $88.7 \pm 2.9$  & $7.21 \pm 0.11$ & $10.62 \pm 0.17$ \\ \hline
\end{tabular}
\end{table}

% New paragraph
Factors which affect the performance of the cheated PFA are 
summarized in Table.\ref{PP_CheatedResolution}.
In the table, the neutrino column shows the contribution of 
undetected-neutrino energies to the resolution of jet energies.
The angle cut column shows those due to particles produced 
within 200 mrad from the beam pipe, and 
the $p_t$ cut column shows those due to particles with 
$p_t$ less than 230 MeV.  These contributions were studied 
based on generator data of Z pole events.
In the GLD, particles with $p_t$
less than 230 MeV is not reconstructed by TPC.
If this cut off is reduced to 100 MeV by utilizing 
the track finding by IT, the contribution to the jet mass resolution from 
$p_t$ cut off is reduced to 0.4 GeV.
On the other hand, it increases to 1.2 GeV, 
if a 5 Tesla solenoid field is used instead of 3 Tesla and 
only TPC track finder is used.
Major contributions to the jet energy resolution are 
those by a fluctuation of measurement energies in 
EM Calorimeter and Hadron Calorimeter, which are 
indicated by EM CAL and HD CAL in the Table.\ref{PP_CheatedResolution}.
These contributions were estimated by using particle energies 
at the entrance of the calorimeter, instead of energies measured by 
the scintillators.
\begin{table}
\caption{\label{PP_CheatedResolution}
Breakdown of factors which affect the jet energy resolution.
The meanings of each column are described in the text}
\begin{center}
\begin{tabular}{| c || c | c | c | c | c | c | c |}
\hline
Items & Total & neutrino & $p_t$ cut & angle cut & EM CAL & HD CAL & TPC \\ \hline
$\sigma$(GeV) & 2.48 & 0.30  & 0.83 & 0.62 & 1.36 & 1.70 & 0.0 \\ 
\hline
\end{tabular}
\end{center}
\end{table}

%

%
%
%\section{Physics Performance}
%\label{SectionPerformance}

%\newpage

%\clearpage
\subsection{Realistic PFA}

% ==============================================================
% Realistic PFA part
% ==============================================================

\begin{figure}
  \begin{center}
    \includegraphics[width=8cm,clip]{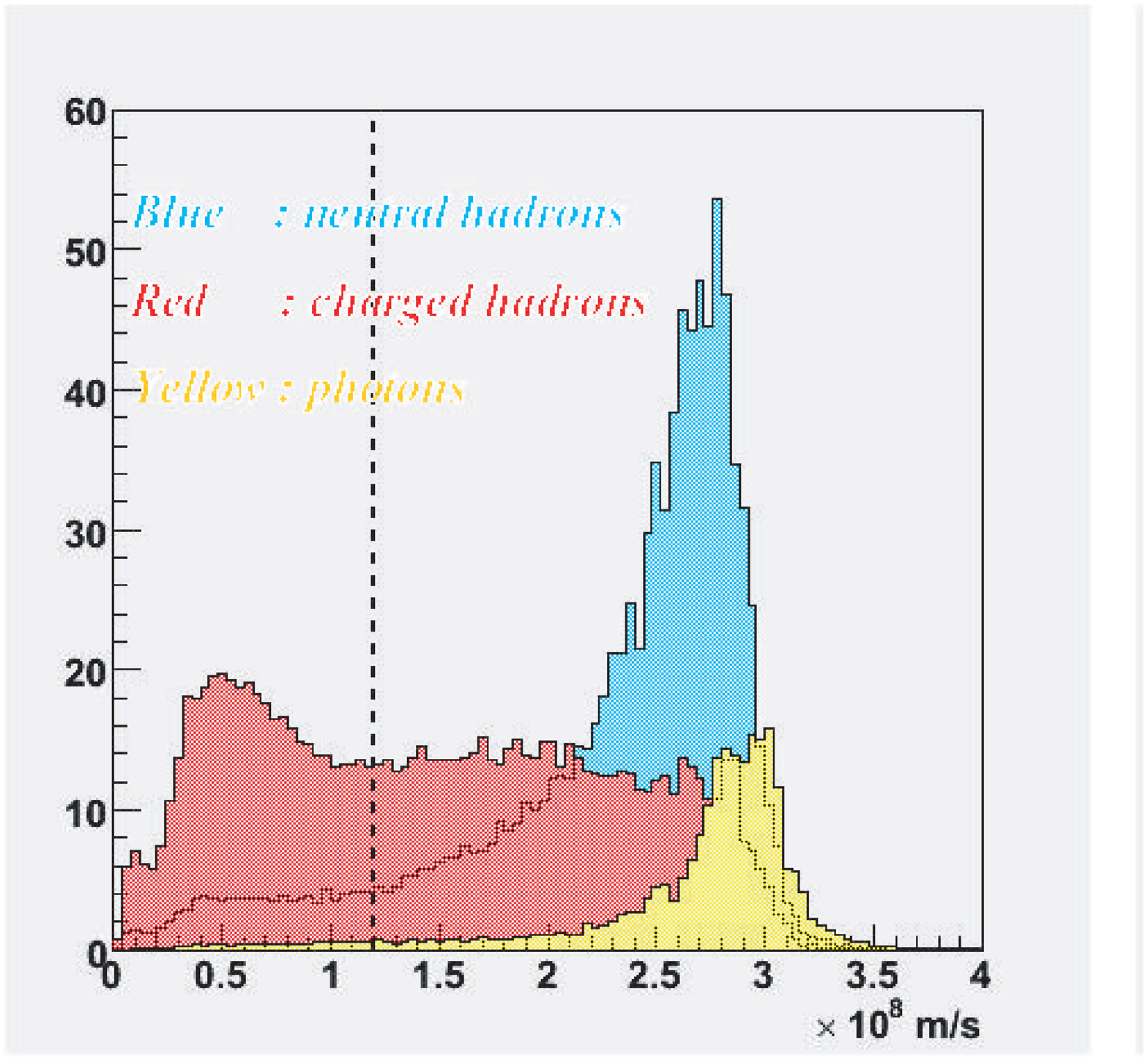}
    \includegraphics[width=8cm,clip]{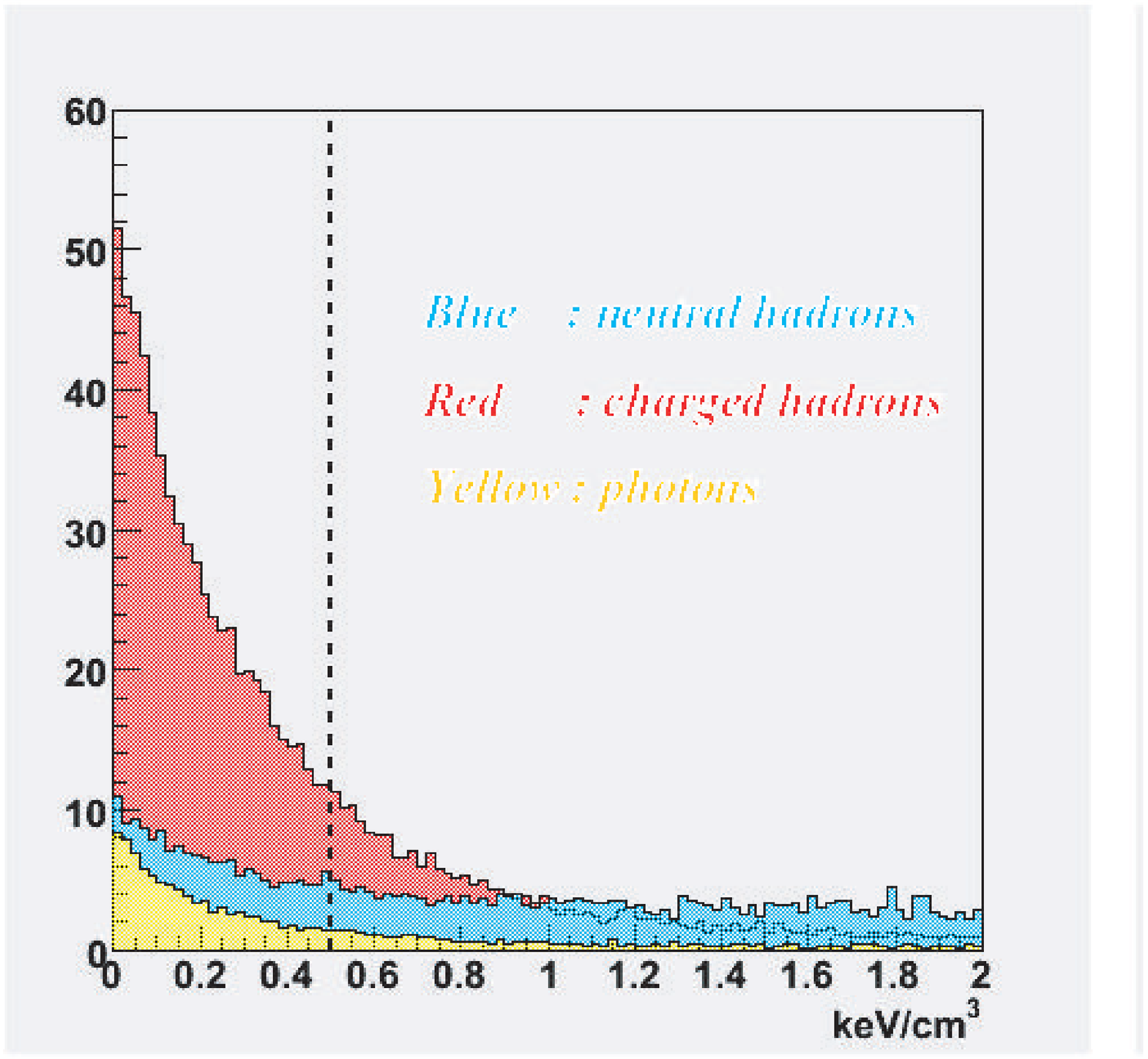}
  \end{center}
  \caption{\label{PP_RealPFA_NHD}Velocity (left) and energy density (right) 
distribution for the remaining clusters. Current cut positions are indicated
by the dotted line in the figures.}
\end{figure}

The GLD-PFA mainly consists of three parts: photon finding, 
charged hadron finding and neutral hadron finding.
The photon finding is based on a calorimeter hit cluster. 
A small cluster is formed by the so called nearest-neighbor 
clustering method. Further clustering is performed by 
collecting the small clusters within a certain tube region.
A cluster due to photon is then identified by using the
cluster information such as longitudinal energy profile.
Details of the photon finding can be found in \cite{PP:GammaFinding}.
After the photon finding, the charged hadron finding is performed.
A charged track is extrapolated to the calorimeter, and
distance between a calorimeter hit cell and the extrapolated
track is calculated. The distance is calculated for 
any track/calorimeter cell combination. 
The calorimeter hit cells within a certain tube radius are 
connected to form a charged hadron cluster.
Note that the tube radius for the ECAL and HCAL can be
changed separately. 
Details of the charged hadron finding can be found 
in \cite{PP:CHDFinding1, PP:CHDFinding2}.
After the charged hadron finding, the remaining calorimeter
hits are classified into a hit due to a neutral hadron or
a ``satellite'' hit due to a charged hadron.
This is done by using a velocity and energy density information.
The calculated velocity and energy density distribution
can be found in Fig. \ref{PP_RealPFA_NHD}.

\begin{figure}
  \begin{center}
    \includegraphics[width=10cm]{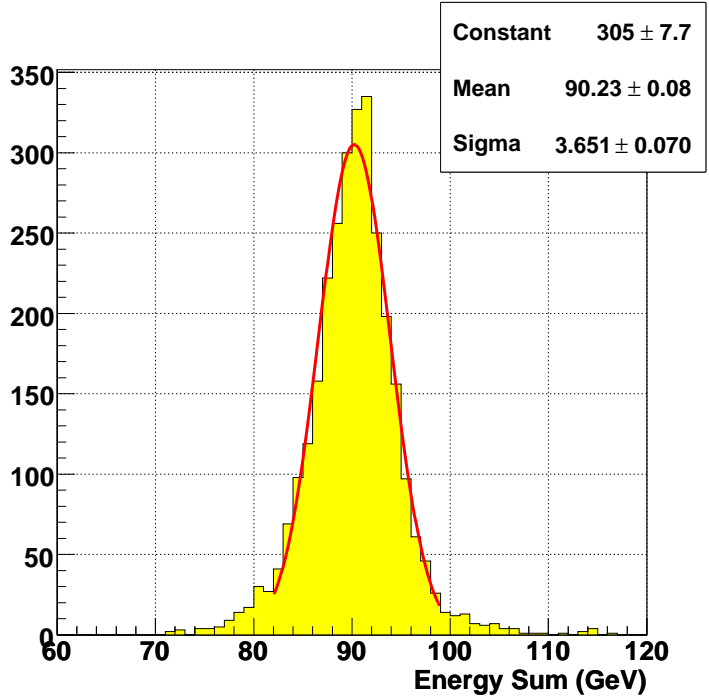}
  \end{center}
  \caption{\label{PP_RealPFA_Zpole}Energy sum for the Z pole events 
    when the GLD-PFA is applied. 38\%$/\sqrt{E}$ energy resolution 
    is achieved by using the PFA.}
\end{figure}

Performance of the GLD-PFA is studied by using $Z\to q\bar{q}$
events at the center of mass energy of 91.2GeV.
Note that the calorimeter configuration in the Jupiter
is tower geometry and cell size for ECAL and HCAL is
4 cm $\times$ 4 cm and 12 cm $\times$ 12 cm, respectively.
Performance of each tool described above 
is summarized in Table \ref{PP_RealPFA_Performance}.
The left most column shows a particle type identified by
the GLD-PFA. The energy weighted efficiency ($E_{XXX}$) for 
the photon finding, charged hadron finding and neutral 
hadron finding are 85.2\%, 84.4\% and 60.5\%, respectively.
The energy weighted purity ($P_{XXX}$) for the photon finding,
charged hadron finding and neutral hadron finding are
92.2\%, 91.9\% and 62.2\%, respectively. 
Note that if satellites hits are included in the charged
hadron, we gain 10\% efficiency at a cost of 3\% purity 
loss for the charged hadron finding.
Figure \ref{PP_RealPFA_Zpole} shows the energy sum
for the Z pole events when the GLD-PFA is applied.
38\%$/\sqrt{E}$ energy resolution is achieved by using 
the PFA while that of energy sum in the calorimeter 
is 60\%$/\sqrt{E}$.

\begin{table}[h]
    \caption{Energy weighted efficiency ($E_{XXX}$) and purity ($P_{XXX}$)
      for each particle type identified by the GLD-PFA.
      CHD and NHD represents charged hadron and neutral hadron, respectively.}
    \label{PP_RealPFA_Performance}
  \begin{center}
    \begin{tabular}{|c||c|c|c||c|c|c|}
      \hline
      & $E_{Photon}$(\%) & $E_{CHD}$(\%) & $E_{NHD}$(\%) 
      & $P_{Photon}$(\%) & $P_{CHD}$(\%) & $P_{NHD}$(\%) \\
      \hline \hline
      Photon & 85.2 & 0.626 & 8.19 & 92.2 & 2.03 & 5.11 \\
      \hline
      CHD    & 4.59 &  84.4 & 16.4 & 1.67 & 91.9 & 3.44 \\
      \hline
      NHD    & 6.27 & 4.51 & 60.5 & 11.2 & 24.1 & 62.2 \\
      \hline
      Satellite & 3.94 & 10.5 & 14.9 & 8.90 & 70.9 & 19.4 \\
      \hline
      CHD + Satellite & 8.53 & 94.9 & 31.3 & 2.67 & 89.0 & 5.64 \\
      \hline
    \end{tabular}
  \end{center}
\end{table}

% ==============================================================
% End of the Realistic PFA part
% ==============================================================

%%%%%%%%%%%%%%%%%%%%%%%%%%%%%%%%%%%%%%%%%%%%%%%%%%%%%%%%%%%%%%%%%%%%%%%
%\clearpage
\section{Performance by physics benchmark process}

\subsection{$e^+e^- \rightarrow Zh$}
The Higgsstrahlung ($e^+e^- \rightarrow Zh$) process which 
is one of the benchmark process is studied by using the QuickSim.
Data equivalent to $500 fb^{-1}$ have been generated for both 
signal and background processes ($e^+e^-\to ZZ,WW,$ and $e\nu W$).
The center of mass energy ($E_{CM}$) and the Higgs mass ($M_{h}$)
are set to be 350 GeV and 120 GeV, respectively.
The remaining part of this section is subdivided 
according to the different final state of the Higgsstrahlung process.
Events are generated by Pythia6.3\cite{PP:Pythia} including beamstrahlung spectrum 
generated by the BSGEN package\cite{SimTools}.

\subsubsection{$\bullet~Zh \to \nu\bar{\nu}b\bar{b}$ and $Zh \to q\bar{q}b\bar{b}$}

\begin{figure}[t]
  \begin{center}
     \includegraphics[width=16.5cm]{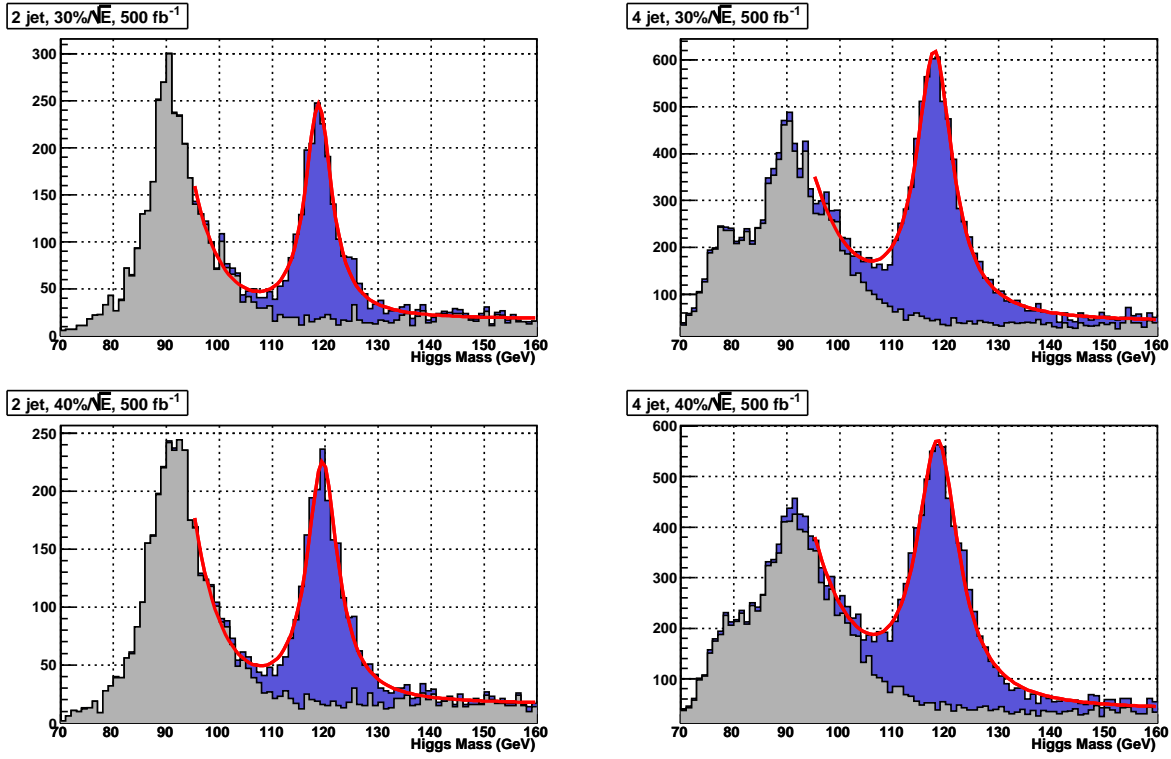}
  \end{center}
  \caption{\label{PP_ZH_Jet}Higgs mass distribution for 2-jet (left) and 
    4-jet (right) in the final state. Top (bottom) plot is for jet energy
    resolution of $30\%/\sqrt{E}$ ($40\%/\sqrt{E}$).}
\end{figure}

For $Zh \to \nu\bar{\nu}b\bar{b}$ mode, there are two jets
in the final state. In the event selection, it is required that 
the missing mass calculated from the two jets is consistent with 
the $Z$ boson mass. It is also required that the visible energy ($E_{vis}$)
is grater than 90 GeV and less than 200 GeV, and transverse 
momentum ($p_{t}$) is greater than 20 GeV/c.
In addition to the above selection criteria,
cut on number of off vertex track ($N_{\rm OffVertex}$)
is imposed for b tag. The $N_{\rm OffVertex}$ is required to be
greater than 6.
For $Zh \to q\bar{q}b\bar{b}$ mode, there are four jets in 
the final state and therefore six jet-pair combinations 
are exist in an event.
In order to select a correct jet pairing, both invariant 
mass of a jet-pair and missing mass calculated from the 
other jet-pair are required to be consistent with the $Z$ boson 
mass.
The event thrust is also calculated and required to be less than
0.9. In addition, $E_{vis}>240$GeV and $N_{\rm OffVertex}>6$
are imposed in the 4-jet event selection.
The distribution of invariant mass of two jets which are 
considered to come from the Higgs decay
is shown in Fig. \ref{PP_ZH_Jet} after imposing all the above
selection criteria.
As described in the previous sections,
current jet energy resolution is achieved to be $38\%/\sqrt{E}$
while the target resolution is $30\%/\sqrt{E}$.
Two different parameter sets that reproduce $30\%/\sqrt{E}$ 
and $40\%/\sqrt{E}$ jet energy resolution are therefore prepared.
Left (right) plot is for 2-jet (4-jet) event and top (bottom)
plot is for jet energy resolution of 
$30\%/\sqrt{E}$ ($40\%/\sqrt{E}$).
The blue area is signal events and gray area
is all background events.
In order to obtain accuracy for product of cross 
section ($\sigma$) and $h\to b\bar{b}$ branching ratio ($Br(h\to b\bar{b})$), 
the Higgs mass ($M_{h}$) and its accuracy ($\delta M_{h}$),
the histograms are fitted by a combined function of the
Lorenzian and exponential. The results are summarized
in Table \ref{PP_ZH_JetTable}.

\subsubsection{$\bullet~Zh \to l\bar{l}X$}

\begin{figure}
  \begin{center}
   \includegraphics[width=16.5cm]{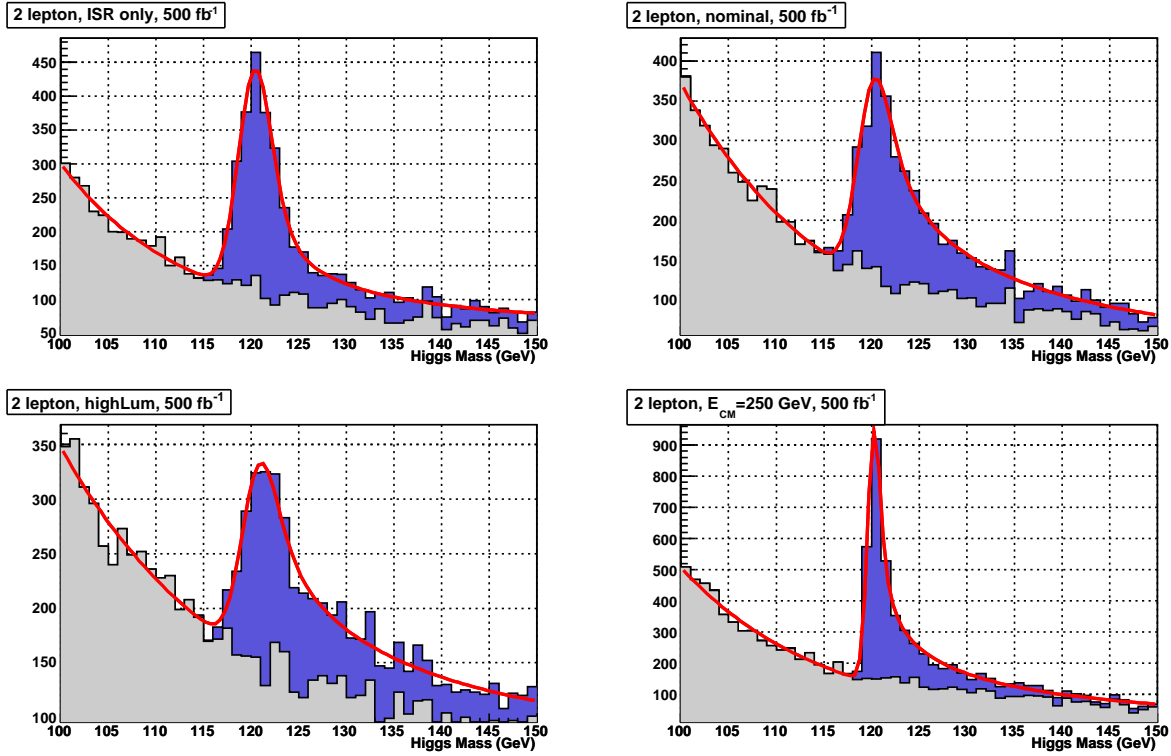}
  \end{center}
  \caption{\label{PP_ZH_BS}Comparison of higgs mass distribution between
  different machine parameters. The upper left, upper right, bottom left 
  and bottom right histogram show the case of ISR only, nominal, highLum 
  , respectively at $E_{CM}=350$ GeV. The right left histogram shows the case
  of  nominal spectrum at $E_{CM}$=250 GeV. Notice that cut value for visible
  energy ($E_{vis}$) is 150 GeV for $E_{CM}$=250 GeV case.}
\end{figure}

\begin{figure}
  \begin{center}
   \includegraphics[width=16.5cm]{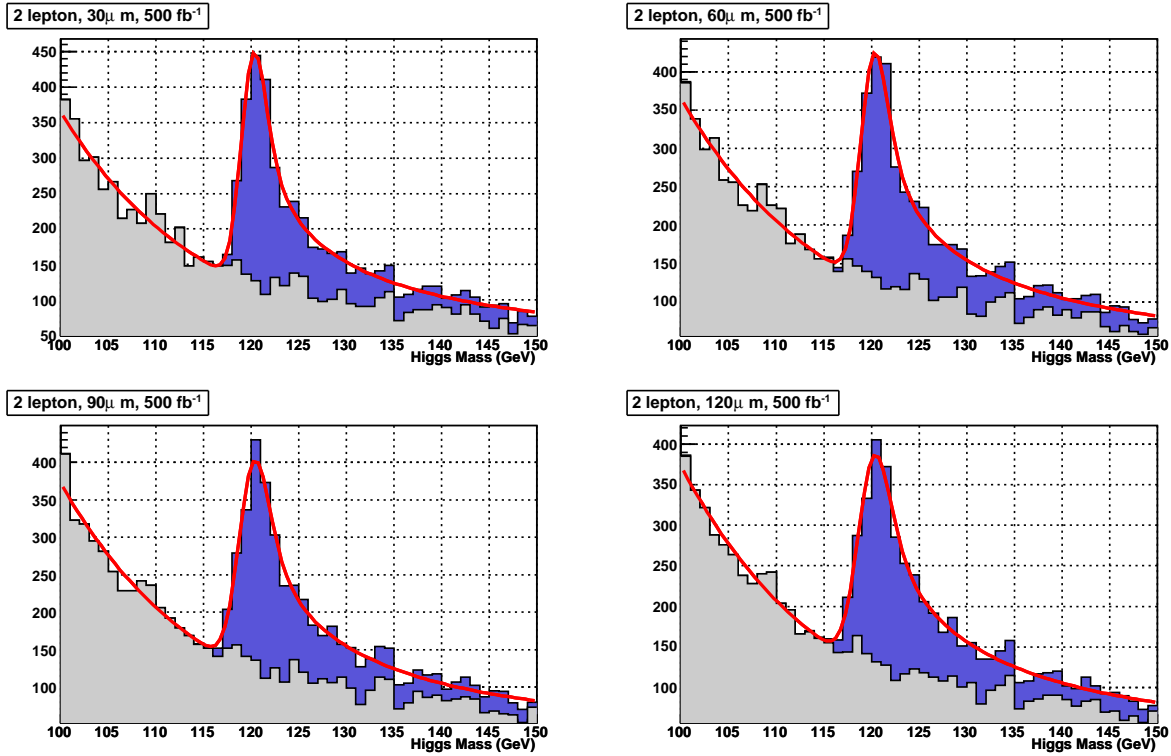}
  \end{center}
  \caption{\label{PP_ZH_TPC}Comparison of higgs mass distribution between
    different TPC point resolution. The upper left, upper right, bottom left 
    and bottom right histogram show the case of $30\mu m$, $60\mu m$, 
    $90\mu m$ and $120\mu m$, respectively. Notice that default resolution
    is $150\mu m$.}
\end{figure}

For $Zh \to l\bar{l}X$ event selection, $e^+e^-$ pair or $\mu^+\mu^-$ pair
is firstly identified. A particle whose energy deposit in the
ECAL is consistent with that of the TPC is identified with an electron
%%[AM]or positron. A particle whose energy deposit in the HCAL is consistent
%%[AM]with that of expected from the minimum ionizing particle or that
%%[AM]has a hit in the muon detector is identified with a muon.
or positron. A charged particle which does not deposit 
energy in calorimeters in QuickSim is identified as a muon
The invariant mass of the $e^+e^-$ pair and $\mu^+\mu-$ pair
($M_{Z}$) is then calculated and required to be consistent
with the Z boson mass.
For $Zh \to l\bar{l}X$ mode, performances between different beam condition
and between different TPC spatial resolution are compared.
The recoil mass distributions from the di-lepton pair after
imposing all selection criteria for different beam condition 
and different TPC spatial resolution
are shown in Figure \ref{PP_ZH_BS} and Figure \ref{PP_ZH_TPC}, respectively.
Each selection criteria together with the number of signal events after
imposing the cut and efficiency of the cut 
is summarized in Table \ref{PP_ZH_BStable} and
Table \ref{PP_ZH_TPCtable}. 
Note that $cos\theta_i$ is dip angle
of an $i-th$ lepton.
The histograms are fitted by the convoluted function of gaussian and 
the one inspired from the bremsstrahlung spectrum defined as follows;
\begin{eqnarray}
F(m) = N_H \int F_H(m,t)e^{-\frac{t^2}{2\sigma^2}}dt + F_Z(m) \nonumber 
\end{eqnarray}
where
$F_H(m,t) = \left(\frac{m+t-M_h}{\sqrt{s}-M_h}\right)^{\beta-1}$.
In the case of bremsstrahlung spectrum, 
$\beta = \frac{2\alpha}{2\pi\log{\sqrt{s}/m_e}-1}$.  
When fitting spectrums including beamstrahlung effects,
$\beta$ is determined by fits.
%%[AM]The $F_Z(m)$ is the background spectrum, $\alpha$ is fine structure 
$\alpha$ is fine structure 
constant and $m_{e}$ is the electron mass. $N_{H}$ is a
normalization factor for signal events.
The $F_Z(m)$ is a sum of an exponential function and a constant multiplied 
by a normalization factor,
whose parameters are determined by fitting the background spectrum 
in the fit range.
%%[AM]The expected performance is summarized in Table \ref{PP_ZH_BStable} and
%%[AM]Table \ref{PP_ZH_TPCtable}.
%%The result of the fit is summarized in Table \ref{PP_ZH_BStable} and
%%Table \ref{PP_ZH_TPCtable}. Note that it does not include a kinematical
%%constraint.

\begin{table}[h]
\caption{\label{PP_ZH_JetTable} Summary of expected performance
for 2-jet and 4-jet mode with different parameter set ($30\%/\sqrt{E}$
 and $40\%/\sqrt{E}$). Kinematical constraint is not included yet.}
\begin{center}
%\footnotesize
\begin{tabular}{|c||c|c||c|c|}
\hline
Mode & \multicolumn{2}{|c||}{$\nu\bar{\nu}b\bar{b}$} &
       \multicolumn{2}{|c|}{$q\bar{q}b\bar{b}$}  \\
\hline
Jet Energy Resolution & $30\%/\sqrt{E}$ & $40\%/\sqrt{E}$ 
                      & $30\%/\sqrt{E}$ & $40\%/\sqrt{E}$ \\
\hline\hline
$\delta(\sigma \times Br(h\to b\bar{b}))$ & 2.74\% & 2.85\% & 1.59\% & 1.67\%\\
\hline
%$M_{h}$ (GeV) & 118.735 & 119.480 & 117.907 & 118.501 \\
$M_{h}$ (GeV) & 118.7 & 119.5 & 117.9 & 118.5 \\
\hline 
%$\delta M_{h}$ (MeV) & 109.199 & 127.695 & 87.7258 & 103.584 \\
$\delta M_{h}$ (MeV) & 109.2 & 127.7 & 87.7 & 103.6 \\
\hline
\end{tabular}
\end{center}
\end{table}
\normalsize

\clearpage

\begin{table}[h]
\caption{\label{PP_ZH_BStable}Breakdown of efficiency for each selection criteria together with an expected accuracy for the Higgsstrahlung 
  cross-section ($\delta\sigma$) with and without b-tag, 
  the Higgs mass width ($M_{h}$) 
  and accuracy for the Higgs mass width ($\delta M_{h}$) 
  for different machine parameters.}
\begin{center}
\footnotesize
\begin{tabular}{|c||c|c|c|c|}
\hline
Machine Parameter & ISR only & Nominal & HighLum & $E_{CM}=250$ GeV  \\
\hline\hline
No Cut & 3299 (1.0000) & 3339 (1.0000) & 3170 (1.0000) & 4319 (1.0000) \\
\hline
$80.<M_Z<100.$ GeV & 2889 (0.8757) & 2907 (0.8706) & 2755 (0.8691) & 3745 (0.8671) \\
\hline
$|\cos\theta_{1,2}|<0.9$ & 2571 (0.8899) & 2528 (0.8696) & 2415 (0.8766) & 3073 (0.8206) \\
\hline
$E_{vis}>250.$ GeV & 2450 (0.9529) & 2411 (0.9537) & 2316 (0.9590) & 3032 (0.9867) \\
\hline
$N_{OffVertex}>4.$ & 1552 (0.6335) & 1524 (0.6321) & 1459 (0.6300) & 1876 (0.6187) \\
\hline
\hline
Total Efficiency & 0.4704 $\pm$ 0.0087 & 0.4564 $\pm$ 0.0086 & 0.4603 $\pm$ 0.0089 & 0.4344 $\pm$ 0.0075 \\
\hline
%$\delta\sigma$ & 2.857\% & 2.88\% & 2.939\% & 2.568\% \\
$\delta\sigma$ & 2.86\% & 2.88\% & 2.94\% & 2.57\% \\
\hline
$\delta(\sigma \times Br(h\to b\bar{b}))$
%& 3.026\% & 3.152\% & 3.347\% & 2.738\% \\
& 3.03\% & 3.15\% & 3.35\% & 2.74\% \\
\hline
%$M_{h}$ (GeV) & 119.956 & 119.511 & 119.882 & 119.955 \\ 
$M_{h}$ (GeV) & 120.0 & 119.5 & 119.9 & 120.0 \\ 
\hline
%$\delta M_{h}$ (MeV) & 79.5646 & 109.442 & 163.751 & 27.0773 \\
$\delta M_{h}$ (MeV) & 79.6 & 109.4 & 163.8 & 27.1 \\
\hline
\end{tabular}
\end{center}
\end{table}
\normalsize

\clearpage

\begin{table}[h]
\caption{\label{PP_ZH_TPCtable}Breakdown of efficiency for each selection criteria 
  together with an expected accuracy for the Higgsstrahlung 
  cross-section ($\delta\sigma$) with and without b-tag, 
  the Higgs mass width ($M_{h}$ ) 
  and accuracy for the Higgs mass width ($M_{h}$)
  for different TPC point resolution.}
\begin{center}
\footnotesize
\begin{tabular}{|c||c|c|c|c|}
\hline
TPC point resolution & 30 $\mu m$ & 60 $\mu m$ & 90 $\mu m$ & 120 $\mu m$ \\
\hline\hline
No Cut & 3382 (1.0000) & 3375 (1.0000) & 3361 (1.0000) & 3349 (1.0000) \\
\hline
$80.<M_Z<100.$ GeV & 2942 (0.8699) & 2932 (0.8687) & 2920 (0.8688) & 2915 (0.8704) \\
\hline
$|\cos\theta_{1,2}|<0.9$ & 2540 (0.8634) & 2538 (0.8656) & 2533 (0.8675) & 2530 (0.8679) \\
\hline
$E_{vis}>250.$ GeV & 2422 (0.9535) & 2421 (0.9539) & 2417 (0.9542) & 2414 (0.9542) \\
\hline
$N_{OffVertex}>4.$ & 1533 (0.6329) & 1531 (0.6324) & 1526 (0.6314) & 1526 (0.6321) \\
\hline
\hline
Total Efficiency & 0.4533 $\pm$ 0.0086 & 0.4536 $\pm$ 0.0086 & 0.4540 $\pm$ 0.0086 & 0.4557 $\pm$ 0.0086 \\
\hline
%$\delta\sigma$  & 2.874\% & 2.874\% & 2.877\% & 2.878\% \\
$\delta\sigma$  & 2.87\% & 2.87\% & 2.88\% & 2.88\% \\
\hline
$\delta(\sigma \times Br(h\to b\bar{b}))$
& 3.14\% & 3.14\% & 3.15\% & 3.15\% \\
\hline
$M_{h}$ (GeV) & 119.7 & 119.7 & 119.6 & 119.5 \\
\hline
$\delta M_{h}$ (MeV) & 77.7 & 86.3 & 96.4 & 105.0 \\
\hline
\end{tabular}
\end{center}
\end{table}
\normalsize

Beam parameters compared in Figures \ref{PP_ZH_BS} and \ref{PP_ZH_TPC}
and Tables \ref{PP_ZH_BStable} and \ref{PP_ZH_TPCtable}
are the case without the beamstrahlung effect
( ISR only), with the "Nominal" pararameter and "HighLum" 
parameter at $E_{CM}=350$ GeV, and  
the "Nominal" parameter case at $E_{CM}=250$ GeV. 
Beamstrahlung spectrum "Nominal"and "HighLum" 
at 350 GeV and 250 GeV are calculated by the program, CAIN, using 
the parameters defined in ref.\cite{Con_raubenheimer}, except 
the center of mass energy.  
%%The initial beam energy spread used for the spectrumis 0.5\%. 
The difference of the specific luminosity at different energy is not 
considered and all results based on the integrated luminosity 
of 500 fb$^{-1}$.  The width of Higgs distribution 
at 350 GeV in Table~\ref{PP_ZH_BStable} and Figure \ref{PP_ZH_BS}
show that the beamstralung effect increase the width significantly,
but it is small at 250 GeV thanks to the higher cross section 
and better momentum resolution due to lower lepton energies near threshold.

\clearpage
%\subsection{$e^+e^-\to \tau$}

%%%%%%%%%%%%%%%%%%%%%%%%%%%%%%%%%%%%%%%%%%%%%%%%%%%%%%%%%%%%%%%%%%%%%%%%%%%%%%

%\subsection{Variation of GLD}

\subsection{Vertex Charge Reconstruction}

Quark charge sign selection is an essential tool for a number of physics analyses
allowing e.g. the unfolding of cross-sections, measurement of spin correlations
or angular analysis \cite{PP:LCWS05Damerell}. 
It is among the benchmarks for vertex detector performance \cite{PP:Battaglia}. 
%%[AM]For $b$-flavour jets, the quark sign for the $40\,\%$ of jets that are charged can 
For $b$-flavour jets, the sign of quark charge for the 40\% of jets can 
be obtained by reconstructing the vertex charge, i.e.~the sum of the charges of all tracks 
assigned to the $B$-hadron decay chain, as described in detail elsewhere 
\cite{PP:LCWS05Hillert}. 
Depending on all tracks in the decay chain being correctly distinguished from $IP$ 
%%[AM]tracks, vertex charge is sensitive to the radius of the inner vertex detector 
tracks, an efficiency of vertex charge reconstruction is sensitive to the radius of the inner vertex detector 
layer and to multiple scattering effects \cite{PP:Snowmass05CJSDSH}.

Vertex charge reconstruction was studied using jets from $e^{+}e^{-}\rightarrow
b\bar{b}$ events generated using the fast simulation SGV \cite{PP:SGV}, interfaced
to PYTHIA version 6.1.52 \cite{PP:PYTHIA}. 
The vertex finder ZVTOP \cite{PP:ZVTOP} was run on all tracks assigned to a jet by
the JADE algorithm \cite{PP:JADE}.\\

\begin{figure}
\begin{center}
\begin{tabular}{cc}
\includegraphics[width=6.7cm]{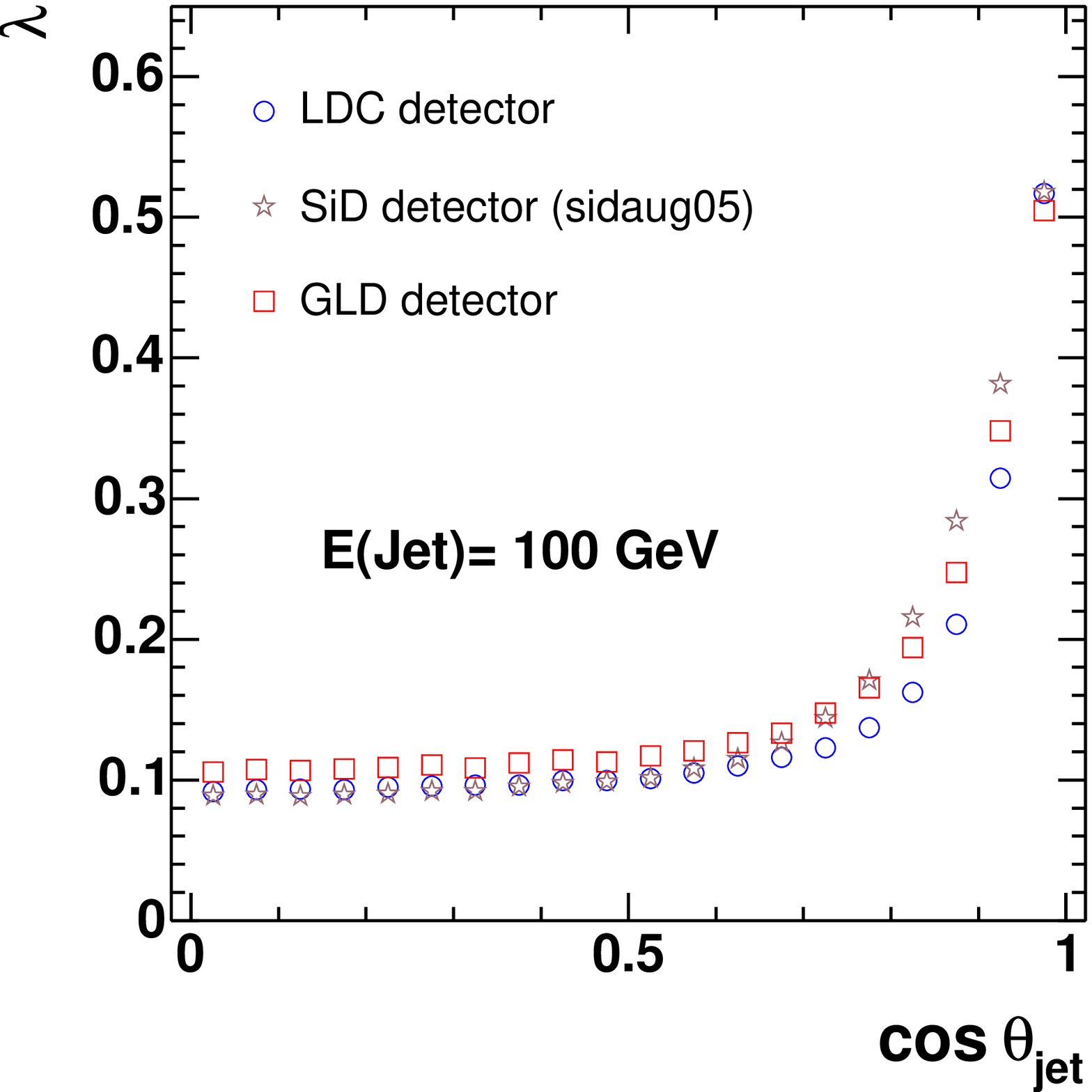} &
\includegraphics[width=6.7cm]{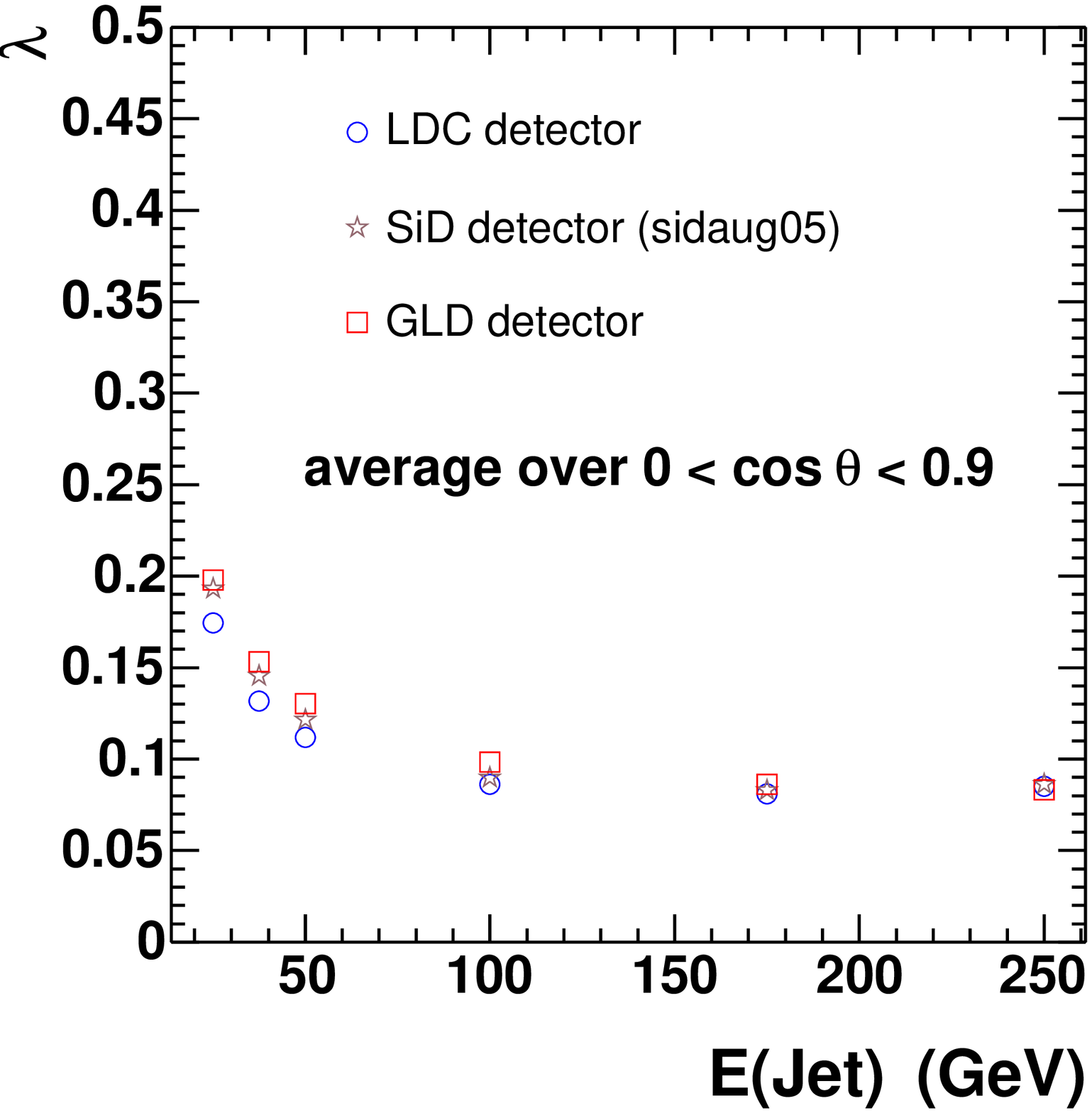} 
\end{tabular}
\end{center}
\caption{\label{PP_Qvtx}Comparison of the current vertex detector designs of 
the LDC, SiD and GLD concept studies in terms of the probability for reconstructing a 
neutral $B$ decay vertex as charged. The left figure shows the polar angle dependence 
for $50\,\mathrm{GeV}$ jets, and the right figure the energy dependence obtained when 
averaging over the polar angle range $0 < | \mathrm{cos}\, \theta | < 0.9$.}
\end{figure}

The performance of the GLD baseline vertex detector in terms of vertex charge 
reconstruction was studied as function of polar angle and jet energy.
The leakage rate $\lambda$, defined as probability of reconstructing a neutral
$B$ hadron as charged, was used to characterise detector performance.
Figure~\ref{PP_Qvtx} shows the angular and jet energy dependence for the GLD 
vertex detector compared to the vertex detectors of the SiD and LDC concept studies.
All three vertex detectors were embedded in the same 'global detector geometry',
corresponding to the TESLA detector design \cite{Con_TESLA}.
At a jet energy of $50\,\mathrm{GeV}$, typical in multi-jet processes at the ILC,
one finds a leakage rate of $\lambda \approx 10\,\%$ in the central region of the 
detector, degrading to $25\,\%$ for $0.85 < | \mathrm{cos}\, \theta | < 0.9$ due
to increased lever arm from the IP and multiple scattering effects.
When averaging over $0 < | \mathrm{cos}\, \theta | < 0.9$, values range from
$\lambda \approx 8\,\%$ at the highest jet energy considered, 
$E_{\mathrm{jet}} = 250\,\mathrm{GeV}$, to $\approx 20\,\%$ for $25\,\mathrm{GeV}$
jets.
Those low energy jets contain more low-momentum tracks, which are more easily
lost from the $B$ decay chain; also those jets are broader and hence more often 
contain tracks outside the angular acceptance of the detector.\\

Compared to the LDC vertex detector the current GLD detector design performs less
well. The main reason for this is the larger inner layer radius of the GLD
detector, which is needed because of its lower $B$ field of $3\,\mathrm{T}$.
Performance of the GLD and the SiD, averaged over polar angle, is similar,
with the GLD being worse than the SiD in the central region due to its larger
radius, but better as one moves to the detector edge, where SiD performance
is given by measurements in the vertex detector disks, which is impaired by the
amount of material at the end of the short SiD barrel staves.
While this study clearly shows some trends that ought to be taken into account
in the further design, it should be noted that it is currently based on a fast
simulation, and that differences between the global detectors were not considered.
Furthermore, the differences are important only for events with high jet
multiplicity. For two-jet processes such as $e^{+}e^{-}\rightarrow b\bar{b}$
at high energy, the measurement of vertex charge is insensitive to the
different vertex detector options.\\[5cm]

%%
  % to be prepared by A.Miyamoto
\clearpage

%%%%%%%%%%%%%%%%%%%%%%%%%%%%%%%%
%\chapter{Costing}
%\label{PART_costing}
%\input costing/costing.tex  % to be prepared by A.Maki
%\clearpage

%%%%%%%%%%%%%%%%%%%%%%%%%%%%%%%
%\chapter{Conclusion}
%\label{PART_conclusion}
%\input remain/conclusion.tex

%%%%%%%%%%%%%%%%%%%%%%%%%%%%%%%%%
%
%

\newpage
\pagestyle{empty}
\addcontentsline{toc}{chapter}{Acknowledgements}
%
% Acknowledgements
%

\begin{Large}
\begin{center}
{\bf 
Acknowledgements
}
\end{center}
\end{Large}

This work is supported in part by the following organizations and programs: 
Japan Society for the Promotion of Science (JSPS); JSPS-CAS-KOSEF Multilateral
Cooperative program under the Core University Program; 
JSPS Japanese-German Cooerative Program;
Japan-US Cooperative Program in High Energy Physics; 
Grant-in-Aid for Scientific Research No.17540282 from JSPS;
Grant-in-Aid for Scientific Research No.17043010 from Ministry of
Education and Science in Japan; U.S. Department of Energy;
Deutsches Elektronen Synchrotron, Hamburg; 
Max-Planck-Institut fuer Physik, Munich;
International Cooperative Research Program; Basic Atomic Energy Research
Institute Program and the CHEP SRC program of the Korean Science and
Engineering Foundation.

%\newpage
%\pagestyle{headings}
%\appendix
%\addcontentsline{toc}{chapter}{Appendix}

\end{document}